\pdfoutput=1

\documentclass[fleqn,usenatbib]{mnras}

\usepackage{mathtools}
\usepackage{lastpage}
\usepackage{totcount}
\usepackage{enumitem}
\usepackage{lineno}
\usepackage{color}
\usepackage{upgreek}
\usepackage{aas_macros}
\usepackage{natbib}
\usepackage[utf8]{inputenc}
\usepackage{csquotes}
\usepackage[version=4]{mhchem}
\usepackage{multirow}
\usepackage{amsmath}
\usepackage{amssymb}
\usepackage[acronym]{glossaries}
\usepackage{makeidx}
\usepackage{comment}

\title[Eta-Earth Revisited II]{Eta-Earth Revisited II: Deriving a Maximum Number of Earth-like Habitats in the Galactic Disk}

\author[Scherf et al.]{Manuel Scherf$^{1,2\ast}$, Helmut Lammer$^{1}$, Laurenz Spro{\ss}$^{1,2}$\\
{$^{1}$Space Research Institute, Austrian Academy of Sciences, Graz, Austria}\\
{$^{2}$Institute of Physics, University of Graz, Graz, Austria}\\
{$^\ast$Email Corresponding Author: manuel.scherf@oeaw.ac.at}
}

\pagerange{\pageref{firstpage}--\pageref{lastpage}} \pubyear{2024}

\begin{document}

\label{firstpage}

\maketitle
\begin{abstract}
In Paper I \citep{Paper1} of this study, we defined Earth-like Habitats (EH) as rocky exoplanets within the habitable zone of complex life (HZCL) on which Earth-like N$_2$-O$_2$-dominated atmospheres with minor amounts of CO$_2$ {can exist}, and derived a formulation for estimating the maximum number of EHs in the Galaxy by taking into account realistic probabilistic requirements that have to be met for an EH to evolve. Here, we apply this formulation to the galactic disk by considering only requirements that are already scientifically quantifiable. By implementing literature models for star formation rate, initial mass function, and the mass distribution of the Milky Way, we calculate the spatial distribution of disk stars as functions of stellar mass and birth age. For the stellar part of our formulation, we apply existing models for the galactic habitable zone and evaluate the thermal stability of nitrogen-dominated atmospheres {with different CO$_2$ mixing ratios inside} the HZCL by implementing the newest stellar evolution and upper atmosphere models. For the planetary part, we include the frequency of rocky exoplanets, the availability of surface water and subaerial land, and the potential requirement of hosting a large moon by evaluating their importance and implementing these criteria from minima to maxima values as found in the scientific literature. We also discuss further factors that are not yet scientifically quantifiable but may be requirements for EHs to evolve. Based on such an approach, we find that EHs are relatively rare by obtaining {plausible} maximum numbers of {{$2.5^{+71.6}_{-2.4}\times10^{5}$ and $0.6^{+27.1}_{-0.59}\times10^{5}$}} planets that can potentially host N$_2$-O$_2$-dominated atmospheres with {maximum} CO$_2$ mixing ratios of 10\% and 1\%, respectively,{ implying that, {on average,} a minimum of $\sim 10^3 - 10^6$ rocky exoplanets in the HZCL are needed for 1\,EH to evolve}. The actual number of EHs, however, may be substantially lower than our maximum ranges since several requirements {with unknown occurrence rates} are not included in our model {(e.g., the origin of life, working carbon-silicate and nitrogen cycles)}; this also implies extraterrestrial intelligence (ETI) to be significantly rarer still. Our results illustrate that neither every star can host EHs, nor that each rocky exoplanet within the HZCL will evolve such that it might be able to host complex {animal-like} life or even ETIs. The {Copernican Principle of Mediocrity} therefore cannot be applied {to infer that such life will be common in the Galaxy.}
\end{abstract}

\begin{keywords}
Eta-Earth, Earth-like Habitats, habitability, astrobiology, complex life, atmospheric evolution
\end{keywords}

\tableofcontents

\section{Introduction}\label{sec:intro}

Whether we are alone in the universe, or life might be common within the Milky Way and beyond is a fundamental question that has occupied mankind for centuries. As Giordano Bruno puts it in his famous work `De l'infinito universo et mondi' \citep{Bruno1584}:

\begin{displayquote}
In space there are countless constellations, suns and planets; we see only the suns because they give light; the planets remain invisible, for they are small and dark. There are also numberless earths circling around their suns, no worse and no less than this globe of ours. For no reasonable mind can assume that heavenly bodies that may be far more magnificent than ours would not bear upon them creatures similar or even superior to those upon our human earth.
\end{displayquote}

More than 500 years later, astronomers were by now able to discover well over 5\,000 of the once-invisible planets. Whether at least some of these heavenly bodies are indeed like Earth, and bear any kind of living creatures upon them, however, remains unknown until this day.\footnote{See, e.g., \url{https://exoplanetarchive.ipac.caltech.edu/} for an up-to-date list of all discovered exoplanets.}

Even more so, as already pointed out within part one of our study \citep[][thereafter called Paper I]{Paper1}, not even a clear and unambiguous definition of the expression `Earth-like' can be found within the scientific literature. The related term `Eta-Earth' ($\eta_{\oplus}$) loosely defines the mean number of any rocky exoplanet that orbits within (some definition of) the habitable zone (HZ) of its host star \citep[e.g.,][]{Haghighipour2014}, and covers planetary radii between 0.5\,R$_{\oplus}$ \citep[e.g.,][]{Kopparapu2018} to 1.0\,R$_{\oplus}$ \citep[e.g.,][]{Petigura2013,ForemanMackey2014} as lower, and 1.2\,R$_{\oplus}$ \citep[e.g.,][]{Burke2015,Bryson2020} to 2.0\,R$_{\oplus}$ \citep[e.g.,][]{Petigura2013,ForemanMackey2014} as upper boundary\footnote{{For a historical account of the HZ and on estimating $\eta_{\oplus}$ based on the Principle of Mediocrity one may be referred to \citet{Lingam2021HZ}.}}. It therefore comes as no surprise that estimated values for $\eta_{\oplus}$ range over almost two orders of magnitude, i.e., from $\eta_{\oplus}\approx$0.02 \citep{Savel2020} to $\eta_{\oplus}\approx$1.25 per star \citep{Garrett2018}. Any search for Earth-like exoplanets would benefit from a more precise and quantifiable definition of what Earth-like actually means.

Within Paper\,I, we define the term Earth-like Habitat (EH) as a rocky exoplanet within the HZCL, i.e, the habitable zone for complex life\footnote{{As we further discuss in Section~\ref{sec:caveats}, `complex life' in this context refers to organisms similar to advanced metazoans, i.e., millimeter to meter-sized animals with a blood circulatory system.}} \citep[][see also Sections~\ref{sec:XUVEffect} and \ref{sec:betaHZCL}]{Schwieterman2019,Ramirez2020} that evolved an N$_2$-O$_2$-dominated atmosphere with minor amounts of CO$_2$ as a result of geologic activity and the emergence and evolution of (microbial) life\footnote{Such atmosphere can also contain other trace species (e.g., CH$_4$, O$_3$ and N$_2$O). Here we focus on N$_2$, O$_2$ and CO$_2$, however.}. Such an Earth-like atmosphere would most likely constitute a geo- and bio-signature \citep[see Paper I,][]{Stueeken2016,Lammer2019,Spross2021} since this particular combination of atmospheric gases would not be stable over geologic timescales without working carbon-silicate and nitrogen cycles, and without the prevalence of microbial life that can recycle fixed nitrogen back into the atmosphere. {Such an atmosphere also presents a biogenic disequilibrium \citep{KrissansenTotton2018}, and hence again, a remotely detectable biosignature}.

Even if it turns out that abiotic pathways can lead to N$_2$-O$_2$-dominated atmospheres, it will nevertheless constitute an Earth-like Habitat as these conditions are crucial for most of present-day life on our planet. Finding such a world would therefore, in any case, signify an immense milestone towards better understanding the prevalence of life in the universe.

Since the term $\eta_{\oplus}$ can only provide some estimate on the frequency of rocky exoplanets in the HZ(CL), but can state nothing about whether these worlds indeed provide habitable conditions for \textit{life as we know it}, we propose in Paper\,I to substitute $\eta_{\oplus}$ through $\eta_{\rm EH}$, the mean number of EHs orbiting in the HZCL. Introducing $\eta_{\rm EH}$ will allow us to include specific quantifiable parameters that are important for the evolution and long-term stability of Earth-like Habitats, and it will further enable us to derive a maximum number of EHs presently existing in the Milky Way.

While Paper I presented our hypothesis and formula to calculate such a maximum number, the present work will apply this equation by including our current knowledge about stellar evolution, galactic habitable environments and the evolution and stability of Earth-like Habitats. Since the present status of research can only quantify some of the relevant parameters, while others remain poorly constrained, and therefore neglected within our estimate, our calculation can only provide a maximum number of EHs. This result, however, will be further refinable in the future when observatories like JWST, ELT, and potentially LIFE and HWO, provide us with spectroscopic data of rocky exoplanet atmospheres within the HZs of their host stars. While earlier approaches to estimate habitable worlds or even intelligent life in the Galaxy such as those based on the Drake equation \citep[e.g.,][]{Drake1965,Vakoch2015} and/or on $\eta_{\oplus}$ were relatively speculative \citep[e.g.,][]{Westby2020,Cai2021}, our estimate will for the first time give a quantifiable upper limit on EHs existing in the Milky Way.

In the next section, we briefly summarize our formula for deriving the maximum number of Earth-like Habitats, $N_{\rm EH}$, and the formalism for calculating $\eta_{\rm EH}$. {This is followed by an outline of our model approach and a discussion on caveats and methodological issues in Section~\ref{sec:model}.} In Section~\ref{sec:stars} we calculate the number of stars, $N_{\star}$, in the galactic disk, followed by a step-by-step calculation of relevant and presently quantifiable parameters that feed into estimating the maximum number of \textit{suitable} disk stars (Section~\ref{sec:NEStar}), and into $\eta_{\rm EH}$ itself (Section~\ref{sec:max}). {Section~\ref{sec:discussion} provides a general discussion on our results, while a comprehensive summary of our model inputs and concluding remarks can be found in Sections~\ref{app:inputTable} and \ref{sec:conclusion}, respectively. In the following appendices, we review and discuss stellar evolution and its importance for the stability of Earth-like atmospheres (Appendix~\ref{sec:appStellar}), the planet occurrence rate of rocky exoplanets in the HZ (Appendix~\ref{sec:etaHZ}), the right amount of H$_2$O and its respective occurrence rate (Appendix~\ref{sec:h2oMain}), and the potential importance and occurrence rate of large moons (Appendix~\ref{sec:appMoon}).}  A discussion on other relevant factors that affect the number of EHs within the Milky Way but are yet too poorly constrained or understood to be included into our estimate follows in {Appendix~\ref{sec:other}.}

\section{A Formula for Estimating the Maximum Number of Earth-like Habitats}\label{sec:eq}

A famous way to estimate the number of communicating extraterrestrial civilizations in the Milky Way was proposed by Frank Drake in the 1960s \citep[e.g.,][]{Drake1965}, i.e.,
\begin{equation}\label{eq:drake}
  N_{\rm civ} = R \times f_{\rm pl} \times n_{\rm e} \times f_{\rm life} \times f_{\rm in} \times f_{\rm civ} \times L,
\end{equation}
where $N_{\rm civ}$ is the number of civilizations in the Galaxy that communicate through electromagnetic waves, $R$ is the star formation rate, $f_{\rm pl}$ is the fraction of stars that host planets, $n_{\rm e}$ the fraction of those planets that are suitable for life, and $f_{\rm life}$, $f_{\rm in}$, and $f_{\rm civ}$ are those planet fractions at which life, intelligent life and electromagnetically communicating civilizations evolve. The factor $L$ finally prescribes the average lifetime of such a civilization. Although the Drake equation certainly stimulated fruitful discussions and is an important conceptualization of life in the Galaxy \citep[see, e.g.,][for an extensive discussion]{Vakoch2015}, it is still relatively speculative and its outcome basically arbitrary due to our limited scientific knowledge. Each term from $f_{\rm life}$ onwards is currently based on pure guesswork.

Our approach to estimating the maximum number of Earth-like Habitats does neither take into account $f_{\rm in}$, nor $f_{\rm civ}$ and $L$. However, since life is likely needed to develop a stable N$_2$-O$_2$-dominated atmosphere, the emergence of \textit{life as we know it}, still remains a crucial factor. As we do not know how likely the origin of life on any given planet might be, this term will later be set equal to one; logically, such an approach can only result in a maximum number of EHs.

As outlined in Paper~I, our basic equation to derive $N_{\rm EH}$, the maximum number of EHs, can be written as follows:
\begin{equation}\label{eq:NEH}
  N_{\rm EH} \leq N_{\star} \times \eta_{\star} \times \eta_{\rm EH},
\end{equation}
where $N_{\star}$ is the number of stars in the Galaxy or galactic disk, $\eta_{\star}$ the fraction of stars that can presently sustain an Earth-like Habitat, and $\eta_{\rm EH}$ is the fraction of EHs around such stars.

The parameters $\eta_{\star}$ and $\eta_{\rm EH}$ depend on several necessary requirements that affect the general habitability of the stellar system and the planet itself. All of these must be met for a planet to evolve into an Earth-like Habitat. Some of these factors are already quantifiable at present, some are not; others are debatable or might not even be known yet. All of them, however, can be subsumed by the following formalism. For $\eta_{\star}$ we can write
\begin{equation}\label{eq:etaStar}
  \eta_{\star} \leq A_{\rm GHZ}(\prod_{i=1}^n \alpha_{\rm GHZ}^{i}) \times A_{\rm at}(\prod_{i=1}^n\alpha_{\rm at}^{i}),
\end{equation}
%
where $A_{\rm GHZ}$\footnote{The Greek characters $A$ (capital letter) and $\alpha$ for stellar requirements were chosen based on the Greek translation of star, i.e., $\upalpha \upsigma \uptau \acute{\upepsilon} \uprho \upiota$.} is the fraction of stars that are within galactic habitable environments \citep[i.e., within the so-called galactic habitable zone; see, e.g.,][]{Lineweaver2004,Kaib2018}. This is in itself dependent on several different requirements such as $\alpha_{\rm GHZ}^{\rm SN}$, which denotes sterilizing effects by nearby supernovae \citep[e.g.,][]{Gowanlock2011,Spitoni2017}, and $\alpha_{\rm GHZ}^{\rm met}$, which refers to the metallicity threshold below which {(and potentially above which)} rocky exoplanets cannot form \citep[e.g.,][]{Johnson2012,Jiang2021}, and the respective distribution of metals in the Galaxy. Binary systems might be another factor that feeds into $A_{\rm GHZ}$ since several studies indicate that some but not all of the binary systems might provide habitable conditions \citep[e.g.,][]{Eggl2013,Eggl2020,Simonetti2020}. Other criteria that might be subsumed within $A_{\rm GHZ}$ are sterilizing effects by gamma-ray bursts and active galactic nuclei, as well as orbital instabilities due to close encounters with other stars (see discussion in Section~\ref{sec:other}).

The second term, $A_{\rm at}(\prod_{i=1}^n\alpha_{\rm at}^{i})$ denotes the fraction of stars that presently permit a rocky exoplanet within its HZCL to sustain the long-term stability of an N$_2$-O$_2$-dominated atmosphere against atmospheric escape and moist or runaway greenhouse within the pressure limits discussed in Paper~I. This includes a lower limit, $\alpha_{\rm at}^{\rm ll}$, which is defined by the radiation and plasma environment of a star, and an upper limit, $\alpha_{\rm at}^{\rm ul}$, that is set by the evolution of a star's bolometric luminosity. At least the lower limit could be further divided into several different necessary requirements such as $\alpha_{\rm at}^{\rm XUV}$, $\alpha_{\rm at}^{\rm flare}$, i.e., the need of an Earth-like atmosphere to survive the quiet short-wavelength radiation (i.e., X-ray and extreme ultraviolet, together XUV) of its host star and its potentially occurring highly-energetic flares, respectively. Both denote the protection against atmospheric erosion through strong thermal escape. Survival against non-thermal losses, induced through strong {coronal mass ejections (CMEs)} and the intense stellar wind of a star is another necessary requirement within $\alpha_{\rm at}^{\rm ll}$ and can be denoted as $\alpha_{\rm at}^{\rm sw}$.  The parameters $N_{\star}$, $\eta_{\star}$ and its requirements will be discussed and calculated within Sections~\ref{sec:stars} and \ref{sec:etaStar}, respectively.

Finally, the term $\eta_{\rm EH}$ can be written as
\begin{equation}\begin{split}\label{eq:etaEH}
  \eta_{\rm EH} & \leq  \beta_{\rm HZCL} \times B_{\rm pc}(\prod_{i=1}^n \beta_{\rm pc}^i )\\ & \times B_{\rm env}(\prod_{i=1}^n \beta_{\rm env}^i ) \times B_{\rm life}(\prod_{i=1}^n \beta_{\rm life}^i ).
\end{split}\end{equation}
Here, $\beta_{\rm HZCL}$\footnote{The Greek characters $B$ (capital letter) and $\beta$ for planetary requirements were chosen based on the Greek translation of habitat, i.e., $\upbeta \upiota \acute{\mathrm{o}} \uptau \mathrm{o} \uppi \mathrm{o}$.} is almost equivalent with $\eta_{\oplus}$, except that it scales the frequency of rocky exoplanets from the HZ to the HZCL; it may therefore also be written as $\beta_{\rm HZCL} = \delta_{\rm HZCL}\eta_{\oplus}$, where $\delta_{\rm HZCL}$ is a scaling factor for applying $\eta_{\oplus}$ from the HZ to the HZCL. The term $B_{\rm pc}$ signifies the necessary requirement for an appropriate planetary compositional and mineralogical set-up that allows for aerobic life and an N$_2$-O$_2$-dominated atmosphere to evolve. It comprises several different factors, $\beta_{\rm pc}^i$, such as having the right amount of the so-called CHNOPS elements carbon, hydrogen, nitrogen, oxygen, phosphorus, and sulfur, i.e., the most common elements needed by \textit{life as we know it}, and the need to have a balanced amount of volatiles and radioactive isotopes. Among others, $B_{\rm pc}$ also includes
\begin{description}
  \item[\textbf{Right amount of water, $\beta_{\rm pc}^{\rm H_2O}$:}] Life as we know it needs water \citep[e.g.,][]{Westall2018}. Planets with too little or too much H$_2$O will remain uninhabitable (see Appendix~\ref{sec:h2o}).
  \item[\textbf{Accretion parameter, $\beta_{\rm pc}^{\rm accr}$:}] Planets that grow too massive within the protoplanetary disk accrete a thick H$_2$-He-dominated atmosphere; if not completely eroded by atmospheric erosion \citep[see Paper~I and, e.g.,][]{Lammer2020a,Owen2020,Erkaev2022} such primordial atmosphere will prohibit the evolution of an EH.
  \item[\textbf{C/O ratio, $\beta_{\rm pc}^{\rm C/O}$:}]  Planets with C/O ratios above $\approx$0.65 can evolve a graphite outer shell, potentially rendering its surface uninhabitable \citep[][]{Hakim2018,Hakim2019a,AllenSutter2020}.
\end{description}
Several other factors that may be subsumed within $B_{\rm pc}$ could be important for a planet to evolve into an EH, although further research is needed for better understanding the role and importance of some of them. The bio-availability of phosphorus is one of these \citep[e.g.,][]{Wordsworth2013,Lingam2019WaterFrac,Hinkel2020,Syverson2021}; the retention of nitrogen through the active phase of a star and its degassing afterward is another crucial factor, particularly for M stars \citep[e.g.,][]{Scalo2007,Lammer2007,Lammer2011,Johnstone2019}.



The second term in Equation~\ref{eq:etaEH}, $B_{\rm env}(\prod_{i=1}^n \beta_{\rm env}^i )$, signifies the necessary requirement of long-term environmental stability of a planet's habitability. Some potential factors are:

\begin{description}
  \item[\textbf{Working C-Si and N cycles, $\beta_{\rm env}^{\rm cycle}$:}] Without carbon-silicate and nitrogen-cycles, neither an N$_2$-O$_2$-dominated atmosphere could be stable or even build-up \citep[e.g.,][]{Lammer2019}, nor could the climate of the planet be appropriately regulated \citep[e.g.,][]{Foley2016,Rushby2018,Kasting2019,Osterloo2021}. 
  \item[\textbf{Existence of a large moon, $\beta_{\rm env}^{\rm moon}$}:] Over recent years, various arguments in favor of a large moon were discussed in the literature but its importance is still highly debated \citep[see, e.g.,][]{Waltham2019}. We discuss all arguments in depth in Appendix~\ref{sec:moon1}. 
  \item[\textbf{Existence of an intrinsic magnetic field, $\beta_{\rm env}^{\rm mag}$:}] The presence of an intrinsic magnetic field was long believed to be an essential requirement for habitability \citep[e.g.,][]{Dehant2007,Lammer2009,Tarduno2014} but critical debates on this paradigm emerged in recent years \citep[e.g.,][]{Blackman2018,Gunell2018,Egan2019,Ramstad2021,Way2022}. We further discuss its potential importance in Section~\ref{sec:mag}.
\end{description}


Another factor feeding into $B_{\rm env}$ might be $\beta_{\rm env}^{\rm tl}$, the potential requirement of not being tidally locked to the host star, a fate that will befall most of the exoplanets within the HZs of M and low-mass K stars \citep[e.g.,][]{Barnes2017}. Whether tidally-locked planets can provide a stable climate and habitable conditions, however, is yet disputed \citep[e.g.,][also see Section~\ref{sec:locked}]{Pierrehumbert2019,Kane2022}. 
Other factors that might contribute to $B_{\rm env}$ could be the orbital eccentricity of a planet \citep[e.g.,][]{Way2017} since this can affect a planet's mean equilibrium temperature and tidal heating, or the chance of being rendered uninhabitable by catastrophic events such as the sterilizing collision with a moon or asteroids \citep[e.g.,][]{Siraj2020,Kokaia2022}.

The final term within Equation~\ref{eq:etaEH} constitutes $B_{\rm life}(\prod_{i=1}^n \beta_{\rm life}^i )$, i.e., the origin and co-evolution of life capable of denitrification, anammox and similar processes that release N$_2$ back into the atmosphere, and aerobic life capable of oxygenic photosynthesis. An N$_2$-O$_2$-dominated atmosphere likely constitutes a biosignature \citep[][see also Paper~I]{Stueeken2016,Lammer2019,Spross2021} and is a necessary requirement for an EH to evolve. Specific microbes can recycle N$_2$ back into the atmosphere due to biochemical processes called denitrification and anammox \citep[e.g.,][]{Zerkle2017b,Lammer2018,Lammer2019}. Without these important pathways N$_2$ and O$_2$ would likely become devoid due to fixation via high-energy processes such as lightning \citep[e.g.,][]{NavarroGonzalez1998}, cosmic rays \citep[e.g.,][]{TabatabaVakili2016}, and meteoroid impacts \citep[e.g.,][]{Kasting1990,Heays2022}, and, if only anaerobic life exists at the planet, due to biofixation \citep[e.g.,][]{Zerkle2017b}.

Volcanic degassing might not be able to resupply enough N$_2$ if no tectonic processes exist that provide the right parameter range for oxygen fugacity, pressure and temperature within the mantle to convert NH$_3$ and NH$_4^+$ into highly volatile N$_2$ \citep{Mikhail2014}. Other potential recycling processes such as Fe$^{2+}$-induced nitrogen outgassing \citep{Ranjan2019} have yet to be studied in more detail to be fully understood \citep[e.g.,][]{Spross2021}. Besides, life is also needed to {produce and }maintain O$_2$ within a stable N$_2$-O$_2$-dominated atmosphere {via oxygenic photosynthesis \citep[e.g.,][]{Och2012,Lyons2014,Planavsky2014,Fischer2016}}. If O$_2$ builds up abiotically through XUV flux-induced dissociation of H$_2$O, an N$_2$-dominated atmosphere will not be stable due to the high energy input into the thermosphere \citep{Lammer2019}. Both gases combined, therefore, require $B_{\rm life}$.

So, the origin of life, $\beta_{\rm life}^{\rm origin}$ could certainly be seen as one of the necessary requirements that feed into $B_{\rm life}$; the origin of aerobic life constitutes another crucial factor. Besides, a working phosphorus cycle \citep[e.g.,][]{Wordsworth2013,Dohm2014,Glaser2020,Lingam2019WaterFrac} may also be subsumed within $B_{\rm life}$ in addition to further biological needs such as the oxygenation time for releasing enough O$_2$ into the air for complex life to evolve \citep{Catling2005}. {This also comprises the life-essential} biogeochemical sulfur {and trace metal cycles, the latter of which includes Fe, Mo, Cu, Zn, and Cd \citep[e.g.,][]{Anbar2002Metals,Morel2003}}. Although sulfur is no limiting nutrient, it is a bio-essential element and its cycle is intimately connected to several other life-essential cycles \citep[e.g.,][]{Wasmund2017} and the evolution of the Earth's redox state \citep{Fike2015}. {Trace metals, for instance, affect the rate of photosynthesis \citep[e.g.,][]{Morel2003}, and Fe and Mo are further important for the biological fixation of N$_2$, thereby playing a crucial role in limiting its biological availability \citep[e.g.,][]{Anbar2002Metals}.}

We will discuss most of the requirements above in more detail within Sections~\ref{sec:NEStar}, \ref{sec:max} and Appendix~\ref{sec:other}. {Next, we will outline our general model approach and discuss related methodological issues, e.g., the potential positive or negative correlation between different requirements.}

\section{Model Outline and Caveats}\label{sec:model}

\subsection{Outline and Model Approach}

The main aim of our study is to apply our formula, as derived in Paper~I, for estimating the maximum number of EHs in the Galaxy. By now, several studies calculated the number of habitable zone rocky exoplanets within the Milky Way \citep[e.g.,][]{Bryson2021} or estimated the amount of communicating extraterrestrial intelligences (CETI) that may presently exist in the Galaxy \citep[e.g.,][]{Westby2020}. However, no study so far has applied the current knowledge about scientifically quantifiable stellar or planetary requirements that are needed for EHs to evolve onto the distribution of presently existing stars and planets within the Milky Way to derive a maximum number of EHs, on which life as we know it might indeed be able to originate and evolve.

To do so, we will first calculate the distribution of stars, $N_{\star}$, within the galactic disk as functions of galactic radial distance and height, stellar mass and stellar birth age. For this, we implement and combine different stellar initial mass functions (IMFs), star formation histories (SFHs), stellar main-sequence lifetimes and galactic mass distribution models from the scientific literature in Section~\ref{sec:stars}.  Next, we will implement requirements into our model that feed into $\eta_{\star}$ and are already scientifically quantifiable. These are the concept of the galactic habitable zone (by considering the galactic metallicity distribution/evolution, and sterilization events by supernovae), and the effects of stellar short-wavelength radiation and bolometric luminosity evolution on planetary atmospheres, as we will describe in detail in Section~\ref{sec:NEStar} and the related appended Section~\ref{sec:appStellar}. In the subsequent Section~\ref{sec:max} and its related appended Sections~\ref{sec:etaHZ}, \ref{sec:h2oMain} and \ref{sec:appMoon}, we will further discuss and implement the only requirements feeding into $\eta_{\rm EH}$, which are already scientifically quantifiable to a certain extent. These include, in decreasing order of quantifiability, (i) the planet occurrence rate, (ii) the right amount of water, and (iii) the importance of a large moon. We will do so for stars within stellar masses of $M_{\star}=0.10-1.25$\,M$_{\odot}$ by implementing the stellar evolution model \textit{Mors}\footnote{The latest version of the model can be found at \url{https://github.com/ColinPhilipJohnstone/Mors}.} (see Section~\ref{sec:XUVEffect}) by \citet{Johnstone2021Stars} which provides stellar evolutionary tracks for this specific stellar mass range.

All the galactic, stellar, and planetary requirements that need to be met for EHs to evolve will be applied to the distribution of stars within the galactic disk. This will finally give us the distribution of EHs as functions of galactic spatial location, stellar mass, and stellar birth age. Crucially, the received numbers will be {plausible} maximum numbers since several necessary and potentially necessary requirements are not implemented in our model. These criteria are evaluated further in the appended Section~\ref{sec:other}. Our results will, finally, be summarized and discussed within Section~\ref{sec:discussion}.

For deriving our Earth-like Habitat distribution and maximum numbers, we perform 6 different model runs for which a summary of all input parameters, including their scientific sources, can be found in Section~\ref{app:inputTable}. First, we will calculate our distribution for two types of atmospheres, i.e.,
\begin{enumerate}
  \item for N$_2$-O$_2$-dominated atmospheres that have a maximum mixing ratio of 10\% CO$_2$ (i.e., $x_{\rm CO_2,max} = 10\%$) as this is often considered to be the toxicity limit for complex animal life \citep[see, e.g.,][ and Paper~I]{Schwieterman2019,Ramirez2020};
  \item for N$_2$-O$_2$-dominated atmospheres that have a maximum mixing ratio of just 1\% CO$_2$ (i.e., $x_{\rm CO_2,max} = 1\%$), an atmospheric composition that is closer to the Earth's (with $\sim$400\,ppm CO$_2$).
\end{enumerate}


In addition, we will perform 3 different model runs for the two atmospheric cases, i.e.,
\begin{enumerate}
  \item a nominal case, which always implements mean values for the input parameters from the scientific literature -- this often coincides with the values assumed to be most realistic and/or reliable by the published studies we investigated;
  \item a maximum case, for which we always implement reasonable maximum values;
  \item and a minimum case that conversely takes into account minimum values.
\end{enumerate}

Combining our two atmosphere scenarios with nominal, maximum and minimum cases results in the aforementioned total of 6 different model cases. Atmospheric composition does not feed into our calculations before investigating the effects of stellar short-wavelength radiation; our model cases will therefore increase from 3 to 6 not before Section~\ref{sec:fat}. We will, however, from time to time discuss other potential scenarios in case there are alternative input parameters.


Within each section {and their related appendices}, we will review the importance of the different requirements, discuss, and estimate their occurrence rates, prescribe their implementation into our model cases, and present their effects on the sample of stars/planets in the galactic disk. For this, we will start with the entire distribution of stars (Section~\ref{sec:stars}) and then apply one necessary requirement at a time to the (remaining) distribution. With each implemented criterion, the sample of remaining stars/planets will therefore decrease until we receive our final maximum distribution of EHs in the galactic disk. At the end of Sections~\ref{sec:stars}, \ref{sec:NEStar}, and \ref{sec:max} we present a summary section where we discuss our derived results and distributions for $N_{\star}$ (Section~\ref{sec:NStar}), $\eta_{\star}$ (Section~\ref{sec:etaStar}), and $\eta_{\rm EH}$ together with $N_{\rm EH}$ (Section~\ref{sec:NEH}). All of input parameters and main results are further summarized within {six} tables in Section~\ref{app:inputTable}.


\subsection{Caveats, methodological \& philosophical issues}\label{sec:caveats}

This model approach comes with several caveats, as well as implicit and explicit assumptions. These can be divided into five broad categories as described in the following.

\subsubsection{Uncertainties in the magnitude of the involved variables}\label{sec:uncertainties}

Many factors that feed into the emergence and evolution of EHs cannot be scientifically quantified at present (see specifically Appendix~\ref{sec:other}) while others are only quantifiable up to a certain, very low extent. Although our approach allows to set most of them equal to 1, several implemented parameters remain that are still highly uncertain. A prominent example is `Eta-Earth' itself, the occurrence rate of rocky exoplanets in the HZ of solar-like stars (see Sections~\ref{sec:betaHZCL} and \ref{sec:etaHZ}); an important variable through which we estimate the occurrence rate of rocky exoplanets in the HZCL. Literature values for $\eta_{\oplus}$ since 2019 range over more than an order of magnitude (see Table~\ref{tab:etaEarth}), i.e., from as low as $\eta_{\oplus}\sim 0.01$ \citep{Neil2020} up to as high as $\eta_{\oplus}\sim 0.3$ \citep{Bryson2021}, thereby still neglecting the specific error bars of these studies. This discrepancy partially stems from a relatively poor definition with no fixed lower and upper boundaries for orbital periods and planetary radii, indicating that scaling to the same boundary conditions slightly reduces this margin \citep[see, e.g.,][and again Table~\ref{tab:etaEarth}]{Kunimoto2020}.

However, further difficulties come into play here. The most crucial one relates to the fact that no rocky exoplanets were yet discovered in the HZ of solar-like stars, implying that observations from either lower-mass stars or tighter orbital periods must be extrapolated. This is non-trivial since additional effects such as dependencies with stellar mass or temperature must be accounted for. Another, often neglected effect that is increasingly taken into account in recent years relates to atmospheric erosion {of primordial atmospheres} on close-in orbits \citep[e.g.,][]{Pascucci2019,Neil2020}, an important process that can lead to substantial overestimates of $\eta_{\oplus}$ due to accidentally factoring in so-called evaporated cores. These are planets that were observed to be rocky close to the star but will be dominated by H$_2$-He-atmospheres in the HZ \citep[see, e.g.,][]{Lopez2018,Pascucci2019}. 

Scaling from a specific range of orbital periods and radii to different parameter ranges, or from a specific stellar mass to another, as we need to perform for deriving the occurrence rate of rocky exoplanets in the HZCL, brings in further uncertainties, although these will be less pronounced than the inherent uncertainties in Eta-Earth itself. However, it will take quite some time until new instrumentation will discover rocky exoplanets in the HZs of solar-like stars and it will even take longer until robust occurrence rate statistics will exist on $\eta_{\oplus}$. Until then, we have to accept and implement relatively large uncertainties but need to highlight them properly.

{Certainly, `Eta-Earth' is not the only parameter with high uncertainties. While the stellar parameters in our model are comparably well defined, any planetary parameters must be taken with caution. This is also true for the occurrence rate of planets with an appropriate water mass fraction and the frequency of large moons as both are mostly based on theoretical models but not on direct observations\footnote{Their actual importance is discussed in the next category.}. Also, atmosphere and GHZ models are mostly based on theoretical considerations and on inherent assumptions that can lead to high uncertainties. As an example, are all relevant cooling agents implemented in an atmospheric model or are important molecules omitted, forgotten or even unknown that may lead to a change in thermal stability (Section~\ref{sec:appStellar})? What actually quantifies as a supernova (SN) `sterilization' event is another relevant example.}

{These uncertainties are one of the reasons why we include literature reviews on various related topics (as mostly found in the appendices) and implement each of the parameters from comprehensive literature-based minimum to maximum values. They also highlight the possibility that our estimates can significantly change in the future when new models and observations are available.}

\subsubsection{Uncertainties in the choice of requirements themselves}\label{sec:choices}

Our derived formulation relies on the implementation of clusters of necessary requirements. However, whether a `necessary requirement' is indeed a necessary requirement is a category of caveat in itself \citep[e.g.,][]{Cirkovic2012}. While some of them are very well accepted, others are not. The existence of a host star around which the rocky exoplanet orbits in the habitable zone is certainly well accepted, at least for the type of habitat we are considering, i.e., for an `Earth-like Habitat'. We define this as \textit{a rocky exoplanet within the habitable zone of complex life (HZCL) on which Earth-like N$_2$-O$_2$-dominated atmospheres with minor amounts of CO$_2$ can exist}. Although this sounds almost tautological, this is not necessarily the case for other potential habitats such as the recently suggested Hycean Worlds \citep{Madhusudhan2021} for which no host star would be needed\footnote{However, they will have their own specific problems, see \citet{Innes2023}.}, for subsurface ocean worlds \citep[e.g.,][]{Nimmo2016Ocean} {and planets with deep biospheres \citep[e.g.,][]{McMahon2013,Lingam2020Deep}, both of which} do not have to orbit in the HZ(CL) at all, or for brown dwarf habitats which neither need a host star nor a conventional HZ, a rocky surface or an N$_2$-dominated atmosphere \citep[e.g.,][]{Lingam2019BrownDwarf}. It is, however, less simple for other requirements.

{Another example relates to the `right amount of water', $\beta_{\rm pc}^{\rm H_2O}$. It is very well accepted that \textit{life as we know it} (see discussion below) needs H$_2$O to evolve and thrive \citep[e.g.,][]{Westall2018} even though this can already be quite subtle for certain extremophiles \citep[e.g.,][]{Merino2019Extrem}. However, it is certainly less accepted whether it is additionally required to have both subaerial land and oceans at the same time for `complex life' to evolve (another issue discussed below). In the appended Section~\ref{sec:h2o}, we give several plausible arguments for this assumption, but it is certainly not a well-established fact. Even if all the listed arguments in this section turn out to be true, there will still be the potential to resolve all of them by other means. For instance, if abiogenesis necessarily needs subaerial land (see appended Section~\ref{sec:h2o}), panspermia could in principle still transport life to such a planet \citep[see, e.g.,][and Section~\ref{sec:originLife}]{Lingam2022Panspermia}. Our knowledge to provide definitive conclusions is yet certainly too poor. This in turn also relates to the first category described above: in case that complex life can commonly evolve on water worlds, our parameter range as based only on planets that have subaerial land and oceans may be too narrow.}

{There is another type of category, i.e., the ones that are strongly debated and may finally not turn out to be a requirement at all. Most of them we did not implement into our model (some of these are discussed in Appendix~\ref{sec:other}) but one specific example we did, i.e., the potential requirement of possessing a large moon. The initial argument in favor of the `Rare Moon' relates to the need for a large satellite to stabilize obliquity variations \citep[e.g.,][]{Laskar1993}. As it turned out, such an argument was not confirmed by simulations afterward \citep[see, e.g.,][]{Waltham2019}, thereby illustrating that specific arguments must not necessarily be true. However, even though several further arguments in favor of a large moon exist (see appended Section~\ref{sec:moon1}), one could apply a similar reasoning as discussed above related to the water issue. Although we implemented a relatively `optimistic' parameter range for the occurrence rate of large moons (see Sections~\ref{sec:moonEff} and \ref{sec:moonFreq}) it may still overestimate its relevance if the requirement turns out to be negligible. On the other hand, if its importance will be confirmed in the future and if large moons are indeed rare, then our implemented parameter range could be too optimistic. Future observations can give important hints on this issue and on the uncertainties in the various parameters in general.}

\subsubsection{Methodological issues}\label{sec:limitations}

{Our formulation closely relates to the famous Drake Equation \citep[e.g.,][]{Drake1965} and therefore inherits some of its difficulties to a certain extent. One crucial problem with this type of equation is its lack of temporal structure \citep{Cirkovic2004,Lingam2021Life}. In its usual version, the Drake Equation assumes a uniformity in time. As an example, the star formation rate (SFR) of the Galaxy feeds into the Drake Equation via $R_{\star}$, the mean rate of SFR in the Galaxy. However, the galactic SFR is not temporally uniform, as it was significantly higher in the past \citep[see, e.g.,][and Section~\ref{sec:SFR}]{Snaith2015,Xiang2018}. Assuming a mean rate could therefore lead to subsequent misconceptions. This is even more evident, if one takes the chemical `evolution' of the Galaxy, as pointed out by \citet{Cirkovic2004}. The metallicity in the galactic disk was much lower in the past making the formation of planets much more effective at present-day (see Section~\ref{sec:metallicity}). Averaging over the entire galactic history would therefore lead to wrong results. As a work around, one could divide the history of the Galaxy into a stepwise function, one with too little metallicity in the past and another one with sufficient metallicity at present.

A similar assumption could be introduced for the GHZ by assuming that the Milky Way may have been uninhabitable in the past due to frequent SNs, again suggesting a stepwise function of habitability. In this case, $f_{\rm life}$, the fraction of planets on which life originates and evolves, could equal $f_{\rm life} = 0$ (or any other very small number) for $0<t\leq t_{\rm p}$, where $t_{\rm p}$ signifies the phase transition from uninhabitable towards inhabitable with $f_{\rm life}$ suddenly becoming a non-negligible number for  the time interval $t_{\rm p}<t\leq t_{\rm 0}$, where $t_0$ relates to the present day \citep[see, e.g.,][]{Annis1999,Cirkovic2004,Lingam2021Life}. These issues can lead to an under- or over-estimate of habitable planets and life in today's Galaxy. It is therefore important to be accounted for properly.}

{However, our approach is not entirely equivalent to a uniform, step-less Drake Equation since we effectively try to tackle this problem for specific parameters. As described in the following sections, we will indeed implement star formation rate, initial mass function, and stellar main-sequence lifetime into our model to arrive at a stellar age distribution for the Milky Way. This is of fundamental importance due to the aforementioned metallicity evolution and the thermal stability of N$_2$-dominated atmospheres. Because of the latter we must account for stellar evolution, otherwise we could not calculate which stars allow for such atmospheres to be thermally stable (which is roughly favored for older stars; see appended Section~\ref{sec:appStellar}). In the same manner, we also implement the evolution of galactic metallicity for considering the temporal distribution of rocky exoplanets. If we neglected these evolutionary parameters, we would overestimate the number of stars that can in principle host Earth-like atmospheres.}

{We do, however, not consider the evolution of SN rates which indeed introduces a methodological uncertainty due to the spatial dynamics in the galactic disk. Drake Equation-like formalisms cannot properly account for such dynamics which is also true for our present framework. As an example, even though we only need to know the probability distribution of being sterilized by supernovae for the present-day, this does not assure us to choose the correct stellar systems for being sterilized because of neglecting stellar migration within the disk. A hypothetical star could, for example, migrate from a denser, more metal-rich region in the inner disk outwards towards a metal-poorer region with far less detrimental supernovae. This star could therefore have a sufficient metallicity for forming planets without being sterilized by supernovae exactly because of its migration towards a quieter region. Since we ignore such migration, however, the same hypothetical star could have ended up being sterilized as it would have been if it had remained in the inner disk.}

{There are further evolutionary aspects we do not currently consider in our framework. One important example relates to the evolution of planetary atmospheres. Their composition (and pressure) can significantly change over geological timescales, thereby strongly affecting their thermal stability against atmospheric escape into space (see Sections~\ref{sec:XUVEffect} and \ref{sec:XUV}). Conceptually this is a crucial point in our formulation, although we do not consider it explicitly (but assume it implicitly). After the active phase of a star, when an N$_2$-dominated atmosphere theoretically becomes thermally stable at a planet, it must not necessarily be the case that such atmosphere will indeed evolve. We presently assume that any planet around a sufficiently weakly active star will host such an atmosphere -- certainly a clear overestimate (see appended Section~\ref{sec:C-N-Availability}). We therefore set this requirement equal to unity, as we do with all requirements that are not implemented into our model.}

Another crucial limitation of a Drake Equation-like approach is the simple multiplication of parameters that may or may not be independent of each other \citep[e.g.,][]{Cirkovic2012}. In reality, some of the parameters feeding into Relations~\ref{eq:NEH}, \ref{eq:etaStar} and \ref{eq:etaEH} can be {positively or negatively correlated, implying that the outcome could be either an overestimate or underestimate, if known correlations are not addressed properly. We give some examples in Paper~I, but here we emphasize} an obvious {positive correlation} between two well accepted necessary requirements that we did indeed implement. These are the stellar metallicity and the occurrence rate, $\beta_{\rm HZCL}$, of rocky exoplanets in the HZCL. Since we have knowledge about this {positive correlation} we tackle it in our framework by weighting $\beta_{\rm HZCL}$ for regions in the disk that show different metallicities than the Solar Neighborhood from which the occurrence rate is initially derived (see Section~\ref{sec:metallicity}). Applying this methodology, as well as implementing evolutionary aspects like the SFH and the evolution of metallicity, illustrates that our formalism allows for more complex calculations than simple multiplication of uniform values.


{Potential {correlations} could in principle be tested and explored by performing statistical tests. This also includes the implementation of probability distribution functions instead of mostly taking point values or ranges (although these variables vary over space, time, and stellar mass in our model). {This was already pointed out by several different authors who highlighted point variables instead of probability distribution functions to be another major limitation of the classical Drake Equation \citep[e.g.,][]{Maccone2010Drake,Maccone2011Drake,Glade2012Drake,Lingam2021Life}. However, implementing probability distribution functions} is beyond the scope of the present study.

This brings us to a final methodological issue. As written above, we run 3 different cases for two types of atmospheres, a nominal case for which we always take mean values from the scientific literature, and minimum and maximum cases for which we always take minimum and maximum values, respectively. Therefore, both minimum and maximum cases are statistically highly unlikely since it is improbable that any chosen parameter will always reach either its theoretical minimum or maximum in reality. Based only on the implemented requirements, this would necessarily imply that the minimum cases are too low, and the maximum cases are too high with the actual values being closer toward the nominal cases. We further emphasize that this methodology still only results in a variation of the maximum number because several necessary requirements are not implemented into our framework but simply set equal to unity. Given this, it makes sense to not only consider maximum values for calculating the actual maximum number of EHs within the Galaxy but to use the entire value range of each parameter from minimum to maximum for illustrating the uncertainties within the present scientific knowledge. If we were to only calculate the maximum number of EHs with strict maximum values from the scientific literature, the results would (i) pretend a strict theoretical maximum by neglecting any uncertainty range, and (ii) give unrealistically high maximum numbers as just described above.

\subsubsection{Neglected factors}

{As already mention above, our approach ignores any potential requirements whose occurrence rates cannot be properly assessed with our current scientific knowledge. The occurrence rate of all these requirements is set equal to 1, which implies that our derived number of Earth-like Habitats must represent a maximum number by definition. The actual occurrence rate of EHs should therefore always be lower than the rate found by our formulation. An extensive list of potential galactic, stellar, planetary and biological requirements ignored by our present study can be found in Appendix~\ref{sec:other}.}

Our present formulation is further restricted in terms of potential habitats that it can presently cover, as it is designed for investigating the occurrence rate of Earth-like Habitats. It therefore does not cover any other hypothetical habitats such as subsurface ocean worlds, or more exotic ones such as aforementioned Hycean Worlds. However, this restriction is again by design. We are specifically interested in EHs because we can evaluate their existence at least partially based on some hard physical parameters. The most crucial of these parameters is the thermal stability of Earth-like atmospheres, i.e., N$_2$-O$_2$-dominated atmospheres with a minor amount of CO$_2$. Here, we emphasize the specific importance of the minor species CO$_2$ which not only relates to atmospheric stability but also to toxicity limits of complex life (see below). Based on such relatively well understood atmospheric parameter, we can show that by far not all stars are able to presently host EHs. This is a crucial point to emphasize since, based on specific known requirements, our formulation allows to derive some information on the prevalence of EHs. Even though this may seem like an anthropocentric ansatz, it is less clear how we can retrieve similar information on other types of habitats since our knowledge about them is much more restricted.

Another advantage of our approach relates to observability. Even though we are presently not at a stage where we can directly observe N$_2$-dominated atmospheres, we will be able to do so in the future. In the meantime, we can already start doing atmospheric statistics that will give important insights into the prevalence of EHs: If most planets in the HZCL have H$_2$- and/or CO$_2$-dominated atmospheres or even none at all, we can induce that N$_2$-dominated atmospheres will likely be rare. One can also infer that finding an N$_2$-O$_2$-dominated atmosphere will likely be an indication for life since these atmospheres function as a biosignature (see Paper~I). This is an exciting prospect that makes deriving the maximum number of EHs in unison with their stellar/galactic distribution a worthwhile study to endeavour.

{We should further explicitly highlight that we do not exclude the actual existence of other forms of habitats. If EHs are rare compared to the number of stars, it may well be that other forms of habitats could be more abundant. Although we are at present simply agnostic about them, we will discuss a pathway for extending our formalism towards other forms of habitats in Section~\ref{sec:exohab}. We also point out that our formulation does not cover communicating extraterrestrial intelligence like the Drake Equation does. Also for this, we will suggest a potential extension in Section~\ref{sec:eti}.}

{Finally, we only investigate the galactic disk and neither the galactic bulge nor the galactic halo (as detailed in Section~\ref{sec:MWcomp}). We further neglect any space beyond our Galaxy as it is not entirely trivial to expand our formalism towards a broader region of the Universe \citep[e.g.,][]{Gonzalez2005}. In addition, we will neither calculate the evolution of habitability nor the maximum number of EHs throughout galactic history. Our calculations only relate to the present day. Finally, we want to emphasize that we do not consider parameters that may increase the maximum number of EHs. Obvious examples could be panspermia or the existence of hypothetically habitable exomoons. We do, however, discuss these issues in some detail in Section~\ref{sec:discussion} and Appendix~\ref{sec:other}.}

\subsubsection{Some philosophical issues}\label{sec:phil}

As a last category, we briefly discuss certain philosophical issues related  to our methodology of which the definition of complex aerobic life is the most critical. We already discuss this issue in detail in Paper~I and therefore keep it relatively short but it is of importance to reiterate it here. Our definition closely follows three studies that discuss certain atmospheric toxicity limits for complex life. These are \citet{Catling2005} related to O$_2$, \citet{Schwieterman2019} related to CO$_2$, and \citet{Ramirez2020} related to CO$_2$ and N$_2$, i.e., the three prime atmospheric species for our framework.

{Based on these studies, we use the terms `complex (aerobic) life', 'life as we know it', `advanced metazoans', and `animal life' synonymously to mean millimetre to meter-sized carbon-based heterotrophs that are mobile and contain a blood-circulatory-like system comparable to advanced metazoans\footnote{We note, however, that the term complex life is conventionally used in a broader sense \citep[e.g,][]{Knoll2011}.}. Whenever we mean other forms of life in our study, we spell it out explicitly. {We also note that focusing on habitats suitable to animal-like complex life, albeit it may seem arbitrary at first glance, may actually be well justified. On Earth, animals are the only complex multicellular eukaryotes capable of phagocytosis and animals are effective `ecosystem engineers'. For these reasons, \citet{Lingam2021Life} suggest that the evolution of animal multicellularity is one of the five, potentially universal, key critical steps for the emergence of technological intelligence on Earth and on exoplanets in general\footnote{According to these authors, the others are abiogenesis, oxygenic photosynthesis, eukaryogenesis, and technological intelligence).}.}

\citet{Catling2005} precisely outline the universal importance of atmospheric O$_2$ for such complex life by showing that a certain O$_2$ partial pressure of $p$O$_2\gtrsim0.1$ PAL (with PAL as Earth's present atmospheric level of O$_2$) is needed to allow for the emergence of complex life above a size of about 1\,cm in diameter based on a crossover of the maximum physical size for O$_2$ diffusion and the minimum physical size for the emergence of a blood circulatory system. Based on such physical and on energetic arguments, one can therefore expect that complex life needs similar O$_2$ partial pressures also on other worlds \citep{Catling2005}. Note, however, that the lower O$_2$ partial pressure threshold for the emergence of technological species may be even more restrictive than the one for complex aerobic life, since this likely needs {$p$O$_2\gtrsim0.86$\,PAL (i.e., a mixing ratio of $x_{\rm O_2}\geq18\%$ for a 1 bar atmosphere)} based on certain combustion limits {\citep{Balbi2023}}.} The CO$_2$ toxicity levels of advanced metazoans on Earth are further outlined in \citet{Schwieterman2019} and \citet{Ramirez2020} who both find the maximum level of $p$CO$_2$ to be about 0.1\,bar (i.e., 10\% CO$_2$ in a 1\,bar atmosphere). \citet{Ramirez2020} additionally investigates N$_2$ toxicity limits for complex life and finds maximum pressures of about $p$N$_2\sim 2$\,bar to be lethal. 

Here, we need to highlight, however, that extraterrestrial life must not necessarily evolve towards similar toxicity limits but may find additional pathways to cope with toxic atmospheres, especially if the evolutionary pressure of survival is high. We nevertheless find these pressure limits to be a useful starting point for evaluating the prevalence of EHs in the Galaxy even if such limits may vary substantially on other worlds. Until no other biospheres are found and investigated, it is reasonable to evaluate atmospheres within aforementioned limits as long as it is made clear that these can be substantially different for extraterrestrial complex life.

{Whenever we talk about Earth-like atmospheres we implicitly mean N$_2$-dominated atmospheres with O$_2$ as a second, less abundant main species, and with a minor amount of CO$_2$. Such atmospheres can certainly include other trace species and the exact ratios of the three main species are not fixed as long as the CO$_2$ mixing ratio is either below $x_{\rm CO_2}$\,=\,10\% or $x_{\rm CO_2}\,=$\,1\%. We have chosen these specific values because (i) $x_{\rm CO_2}$\,=\,10\% CO$_2$ coincides with $p$CO$_2 = 0.1$\,bar for 1\,bar total pressure, i.e., the aforementioned toxicity limit for complex life, (ii) HZCL boundaries exist for exactly these mixing ratios related to a 1\,bar atmosphere, and (iii) atmospheric models of N$_2$-dominated atmospheres exist for the same mixing ratios as well. In addition, the $x_{\rm CO_2}$\,=\,1\% CO$_2$ case more closely resembles the actual atmosphere of the Earth which is interesting in itself. We further emphasize that 10\% and 1\% CO$_2$ mixing ratios can always be regarded as maximum values and we will henceforth use 10\% and 1\% CO$_2$ synonymously to 0.1\,bar and 0.01\,bar CO$_2$.}

This specific definition of an atmosphere also relates to our definition of what an EH actually constitutes. As we have already stated, we define an Earth-like Habitat as a \textit{rocky exoplanet within the HZCL on which Earth-like N$_2$-O$_2$-dominated atmospheres with minor amounts of CO$_2$ can exist}. This definition in principle allows for the possible existence of complex aerobic life that obeys to similar limits as advanced metazoans here on Earth, but it is important to note that such putative organisms do not have to actually live on such a planet. {However, the proposition that the Earth-like atmospheres themselves, {i.e., the simultaneous existence of N$_2$ and O$_2$ with minor amounts of CO$_2$,} act as a biosignature makes EHs extremely relevant for astrobiology in itself, regardless of whether complex life actually exists there or not.}

Although EHs are close to the definition of Class~I habitats in \citet{Lammer2009}\footnote{\citet{Lammer2009} define Class~I habitats as: ``Class~I habitats represent bodies on which stellar and geophysical conditions allow Earth-analog planets to evolve so that complex multi-cellular life forms may originate.''}, both are not entirely the same since the latter does not directly relate to atmospheric composition. It does, however, relate to certain \textit{stellar and geophysical conditions} needed for such habitats to evolve, which can be regarded to be semantically equivalent with our meaning of \textit{necessary requirement}. Class~I habitats further exclude ocean worlds explicitly on which the ocean is not in direct contact with the mantle silicates due to the formation of a high-pressure (HP) ice layer between both \citep[such planets are defined as Class~IV habitats in][]{Lammer2009}. We also exclude them from our definition, but this does not relate to ocean planets without subaerial land on which an HP ice layer may not exist. We discuss arguments on whether these planets could constitute an EH in detail in appended Section~\ref{sec:h2o}. However, we also note that \citet{Lammer2013} refined the definition of Class~I habitats, so that (i) it does exclude ocean worlds without subaerial land entirely (with ocean worlds constituting Class~V habitats) and (ii) N$_2$ as main species should be present in their atmospheres\footnote{In \citet{Lammer2013} it is defined as: ``Class I habitats represent planetary bodies on which stellar and geophysical conditions allow Earth-analogue planets to evolve so that complex multicellular life forms may originate and inhabit the planets hydrosphere, surface and subsurface environments.''}.}

{If one takes the definition of EHs (or Class I habitats), it seems logical that it becomes narrower the more is scientifically known about said certain stellar and geophysical conditions. If we, for instance, could exclude planets without subaerial land based on future scientific knowledge, this will narrow down its definition and hence the number of planets that can evolve into an Earth-like Habitat. One must, however, take care to not over-specify its definition, otherwise the definition of Earth-like Habitat will be indistinguishable from Earth, and Earth will by definition be the only Earth-like Habitat in existence. It is therefore important to keep a balance between a too narrow, tautological definition and a broader useful one that allows searching for planets on which complex aerobic life may evolve based on identifying and testing relevant and scientifically quantifiable requirements.}

{Another issue is related to the GHZ. It is important to highlight that the models we utilize for the sterilization rates of SNs were performed for complex life as defined above. As we discuss in Section~\ref{sec:GHZ}, the effects of supernovae would be less pronounced if we would only consider extremophiles or microbial life. We also want to mention that the definition of rocky exoplanet, and hence of `Earth-like' itself, seems to be slightly contentious, as there is no fixed mass or radius range that applies to the term rocky exoplanet. We discuss this in detail in Sections~\ref{sec:etaHZ} and \ref{sec:bHZCLMain}. Finally, we also emphasize that our results have certain philosophical implications such as for the Copernican Principle, the Fermi Paradox, and the Search for and Messaging to Extraterrestrial Life, i.e., for SETI and METI. We discuss these potential implications in Section~\ref{sec:discussion}.} 

\section{The amount of stars within the Milky Way, $N_{\star}$}\label{sec:stars}

\subsection{The components of the Milky Way and their relevance for estimating $N_{\star}$}\label{sec:MWcomp}

\subsubsection{The various components of the Milky Way}\label{sec:MWComp2}

The Milky Way \citep[see, e.g., reviews by][]{BlandHawthorn2016,Helmi2020} is a typical disk galaxy and can be divided into different regions, i.e.,
\begin{description}
 \item[\textbf{Galactic bulge:}] The galactic bulge is the high-density inner region of the Milky Way which extends to about 2\,kpc from the center \citep[see, e.g.,][for a recent review of the bulge and bar region]{Barbuy2018} with an average stellar density of about 14 stars per {cubic} pc \citep[e.g.,][]{Robin2003}. Its structure is strongly barred with a long lower-density bar extending outwards from the bulge into the inner disk with a half-length of $5.0\pm 0.2$\,kpc \citep{Wegg2015}. The bulge itself holds a stellar mass of $1.34\pm0.04 \times 10^{10}\rm M_{\odot}$ within $\sim$2.2\,kpc \citep{Portail2017} that is mainly populated by a generation of old stars distributed over a relatively wide range of metallicities ranging from about $\rm[Fe/H]\sim -1.0$ to $+0.4$ \citep[e.g.,][]{Zoccali2003}, and with many of them being $>$8 billion years (Gyr) old \citep[e.g.,][]{Hasselquist2020,Meredith2022}.
  \item[\textbf{Galactic halo:}] The halo \citep[e.g.,][]{Belokurov2018,Helmi2018,Helmi2020} is the extended spherical part of the Milky Way and is populated by lone stars and globular clusters with metallicities that are mostly clearly below $\rm [Fe/H]\sim-1.0$ \citep[e.g.,][]{An2015,Zuo2017,Liu2018}, and out of which many have ages of $>$12\,Gyr \citep{Jofre2011,Carollo2016}. The stellar mass of the halo was recently estimated to be $\approx1.3\times10^9 \rm M_{\odot}$ \citep{Deason2019,Mackereth2019}.
  \item[\textbf{Galactic disk:}] The galactic stellar disk \citep[e.g.,][]{Rix2013,BlandHawthorn2016,Bovy2016,Helmi2020} is commonly decomposed into two different components, e.g., the thin and the thick disk \citep[e.g.,][]{Gilmore1983}. Here, the thin disk is the main component of the Milky Way and the place of ongoing star formation with a current star formation rate (SFR) of $\approx1.6\,\rm M_{\odot} year^{-1}$ \citep{Licquia2015}. It has a compact scale height, a wide spread of different stellar ages and metallicities, and its origin \citep[e.g.,][]{Kilic2017} can be traced back to about 8 billion years ago (Ga). The thick disk, on the other hand, has a larger scale height, but is much more diffuse than the thin disk \citep[e.g.,][]{McMillan2017}, and its age is of the order of 10\,Gyr \citep{Kilic2017}. Its metallicity distribution function peaks lower than for the thin disk at $\rm [Fe/H]\sim-0.5$, and its chemical abundance is often attributed to being different from the thin disk \citep[see, e.g.,][for a discussion]{Helmi2020}. As we will discuss below in more detail, we will focus on the stellar mass within the galactic disk which is generally calculated to be between $2.7 \times 10^{10}\,\rm M_{\odot}$ \citep{Naab2006} and $6.5 \times 10^{10}\,\rm M_{\odot}$ \citep{Sofue2009} with a recent calculated value being $4.1 \times 10^{10}\,\rm M_{\odot}$ \citep{Xiang2018}. About one-quarter to one-third of this mass might be part of the thick disk \citep{McMillan2011,Kubryk2015,McMillan2017}.
\end{description}
In total, the Milky Way holds a stellar mass of about $\approx 4.5 - 7.2 \times 10^{10}\,\rm M_{\odot}$ \citep{Flynn2006,McMillan2011,Licquia2015,McMillan2017,Cautun2020}, which coincides with the total amount of stars and stellar remnants within the Galaxy. Even though this stellar mass is the relevant parameter for our study, we note that it is just a fraction of the entire virial mass of the Galaxy, that is, the whole mass enclosed within the gravitationally bound system of the Milky Way (including gas, dust and dark matter) as this amounts to $\sim10^{12}\,\rm M_{\odot}$ \citep[see][for further references and a review on our Galaxy's mass]{Wang2020}.

For calculating $N_{\star}$, we restrict ourselves to the galactic disk, {thereby neglecting the galactic bulge and halo. We will first briefly discuss our reasoning for excluding the halo followed by our arguments for excluding the bulge.}

\subsubsection{Excluding the galactic halo}\label{sec:halo}

{As written above, the galactic halo mainly consists of very metal-poor and old stars. \citet{Zuo2017}, for example, can reproduce the metallicity distribution function (MDF) in the galactic halo well by separating it into three distinct groups with peak metallicities of [Fe/H]\,$\sim$\,-0.63, -1.45, and -2.0, where the two metal-poorer components correspond to the inner and outer halo, respectively\footnote{\citet{Zuo2017} surveyed the southern galactic halo based on the Sloan Digital Sky Survey (SDSS) and the South Galactic Cap u-band Sky Survey' (SCUSS; see \url{http://batc.bao.ac.cn/Uband}) and define the inner and outer halo through galactocentric heights of $-8 < z < -4$\,kpc and $-12 < z < -8$\,kpc, respectively.}. The additional, relatively metal-rich component, which corresponds to substructures within the galactic halo such as clouds and streams like the Sagittarius stream \citep[e.g.,][]{Belokourov2007,Grillmair2016,Koposov2012}, covers about 10\% of halo stars \citep{Zuo2017}. Similar results were found by other surveys, e.g., [Fe/H]\,$\sim$\,-1.2\,$\pm 0.3$, and -2.0\,$\pm 0.2$ by \citet{Liu2018} for the inner and outer galactic halos, respectively.}

It seems unlikely that many of them will provide habitable conditions at present, however. As already pointed out by \citet{Lineweaver2004}, many of the halo stars will have metallicities that are too low to host rocky exoplanets, a threshold that might be somewhere around $\rm [Fe/H]\sim-0.5$ \citep[e.g.,][see Section~\ref{sec:metallicity}]{Johnson2012}, {a value that is roughly similar to the aforementioned metal-rich component of the galactic halo. If we assume a stellar mass of $\approx1.3\times10^9 \rm M_{\odot}$ \citep{Deason2019,Mackereth2019} for the entire halo, the 10\% component of metal-rich stars will equate to $\approx1.3\times10^8 \rm M_{\odot}$. That number is only about $\sim$0.3\% of the entire stellar disk mass, if we take the recently calculated value of $4.1 \times 10^{10}\,\rm M_{\odot}$ by \citet{Xiang2018}, which is around the average stellar mass of all disk models \citep[see,][]{Wang2020}. Roughly 3 out of 1000 stars\footnote{This estimate assumes that the average mass of a single star is similar in the disk and in the halo, an assumption that might not be entirely correct.} that could in principle host rocky exoplanets in the combined region of disk and halo will therefore belong to the halo. However, this is a relatively optimistic estimate, as the peak metallicity of [Fe/H]\,$\sim$\,-0.63 for the metal-rich halo component still suggests a high amount of stars lie below [Fe/H]\,=\,-0.5.}

In addition, the high majority of stars within the galactic halo, including the inner halo, are of an age of $\sim$10-12\,Gyr \citep[e.g.,][]{Jofre2011} while the metal-rich stellar component of the Sagittarius stream holds ages of $\sim$9.5\,-\,11\,Gyr \citep{Carollo2016}. Similar ages can be found for other metal-rich halo regions such as the Styx and Orphan streams, both of which were found to cover ages of $\sim$10\,-\,11\,Gyr \citep{Carollo2016}. These stellar ages are significant, and one can expect that many of the rare planets that formed around these stars do not show geological activity at the present day, specifically if one considers that planets with an Earth-like radiogenic heat budget will be geologically active for about 6\,Gyr \citep[e.g.,][]{Mojzsis2021}. These worlds would therefore provide relatively limited conditions for the evolution and maintenance of complex life. In fact, these stellar ages are even too old to generally allow inhabited planets around G-type stars based on their average main-sequence lifetime and bolometric luminosity evolution. EHs in the galactic halo must therefore logically be restricted either to very old low-mass stars, which pose further difficulties (as discussed in later sections), or to one of the very few younger ones that were either ejected out of the galactic disk \citep[e.g.,][]{Faltova2023} or formed at younger ages under as of yet poorly constrained conditions \citep[e.g.,][]{Bellazzini2019}.

Besides, no study has by now explicitly modeled the habitability of the galactic halo, so no potential distribution can be incorporated within our model (as we do for the disk in the following subsections). However, even if there are stars within the halo that currently allow for the existence of an Earth-like Habitat, their numbers will be minuscule, most likely significantly below 1 permille of the entire population, at least based on the considerations above. Due to their average distance, the potential to observe and characterize some of them in the near future may further be relatively elusive. Based on this reasoning, we will therefore not consider the galactic halo within this study.

\subsubsection{Excluding the galactic bulge}\label{sec:bulge}

The story looks a bit different for the galactic bulge. Depending on the specific study \citep[e.g.,][]{McMillan2011,Licquia2015,McMillan2017}, its stellar mass amounts to about $\sim$20\% of the entire galactic stellar mass. So, if we assume the bulge to be habitable, it cannot be neglected as long as we want to estimate the maximum number of EHs within the entire Milky Way. Whether the bulge is indeed habitable, however, is debated \citep[e.g.,][]{Balbi2020}, as outlined in the following.

The habitability of different galactic regions, which is often subsumed under the concept of the so-called galactic habitable zone (GHZ), depends on various factors \citep[e.g.,][]{Kaib2018} that are usually broadly divided into three {(non-exhaustive)} areas, i.e.,
\begin{enumerate}
  \item high-energetic events such as active galactic nuclei, supernovae, and gamma-ray bursts (GRB) that both sterilize a planet and erode its atmosphere;
  \item close stellar encounters that perturb the orbit of a planet and/or trigger comet and asteroid bombardments;
  \item and the metallicity of the local interstellar medium (ISM) which may hinder the accretion of rocky exoplanets around newly forming stars.
\end{enumerate}
Several specific problems may arise for the bulge, particularly within (i) and (ii).

Since the stellar density in the bulge is significantly higher than in the disk and halo regions, it can be expected that a planet suffers from detrimental supernovae more often than a planet in the other disk regions. As estimated by \citet{Gehrels2003}, a supernova within $r \leq 8$\,kpc can significantly affect a planet's habitability by depleting about 30\% of its ozone layer and, thereby, doubling the incident UV irradiation; a value that can trigger mass extinctions of land-based life \citep[e.g.,][]{Melott2011}. While the occurrence rate of such an event at Earth is calculated to be $\sim 1.5$\,Gyr$^{-1}$ based on the average galactic rate of core-collapse supernovae \citep{Gehrels2003}, \citet{Balbi2020} estimated that such extinction events may have a frequency of $\sim40 - 110$\,Gyr$^{-1}$ at an average planet within the bulge. However, based on the resilience of Ecdysozoa, a group of protostome animals that includes Milnesium tardigradum (i.e., the highly resilient tardigrades, also known by the name of water bears), \citet{Sloan2017} calculated that it needs a supernova within $\leq$0.04\,pc to completely sterilize a habitable planet. By considering such boundary condition, \citet{Balbi2020} estimated that the expected number of such events closer than 2\,kpc from the center are significantly smaller, i.e., $\sim 0.6 \times 10^{-6}$\,Gyr$^{-1}$ to $\sim 1.4 \times 10^{-5}$\,Gyr$^{-1}$. However, it should be further noted that tardigrades are exceptionally resilient survivors and are clearly non-representative of the average animal. While the radiation threshold LD$_{50}$, at which 50\% of the entire investigated population dies, is in the order of 5000 Gray for Milnesium tardigradum within 48 hours \citep[e.g.,][]{Horikawa2006}, this value is around a few Gray for typical animals such as rats, frogs and deer \citep[e.g.,][]{Radiation2008}, i.e., $\sim$1000 times lower.

Besides supernovae, GRBs \citep[e.g.,][]{Scalo2002,Melott2004,Thomas2005a,Thomas2005b,Piran2014,Gowanlock2016,Thomas2021} could be another {threat} within the bulge. \citet{Piran2014}, for example, found that these could render the galactic center inhospitable to life. However, a study by \citet{Gowanlock2016} that considers the star formation history (SFH) and metallicity evolution of the Galaxy suggests a more complex picture. If GRBs correlate with low-metallicity environments \citep[see, e.g.,][]{Jimenez2013}, predominantly the metal-poor outskirts of the Milky Way will presently be affected by GRBs. If GRBs, on the other hand, exclusively follow the galactic SFH, most of them will occur in the galactic center.

The supermassive black hole (SMBH) Sagittarius A$^{\star}$ (Sgr~A$^{\star}$) presents another potential {threat} to habitability \citep[e.g.,][]{Balbi2017,Forbes2018,Lingam2019,Wislocka2019,Balbi2020,Ambrifi2022,Heinz2022}. Even though its early phase as an active galactic nucleus (AGN) had a likely initial duration of $\lesssim10^8$\,Myr \citep{Marconi2004}, other black hole-related processes may still present a threat to the habitability of the bulge. The frequency of so-called tidal disruption events (TDE), i.e., the disruption of a star that passes too close to the SMBH, is estimated to be $\lesssim100$\,Myr$^{-1}$ \citep[e.g.,][]{Komossa2015,Stone2020}. \citet{Pacetti2020} found that the induced radiation from such an event could significantly harm habitability within distances of $\sim0.1-1.0$\,kpc from the galactic center. In addition, stochastic processes such as gravitational instabilities can frequently activate SMBHs in the mass range of Sgr~A$^{\star}$ every $10^7 - 10^8$ years with an outburst duration of $\sim10^5$ years \citep[e.g.,][]{Hopkins2006}. Such outbursts could again render planets uninhabitable within $\lesssim1$\,kpc \citep{AmaroSeoane2019,Ambrifi2022}. There is indeed evidence that an outburst of X-ray radiation took place around 2 to 8 Myr ago \citep[e.g.,][]{Nayakshin2005,BlandHawthorne2013}, illustrating that Sgr~A$^{\star}$ cannot be neglected when studying the habitability of the bulge, not even at present-day.

As a side note, it should be mentioned that the effect of the XUV flux from an SMBH on the atmospheric mass loss of a rocky exoplanet may be underestimated. \citet{Balbi2017} and \citet{Balbi2020} calculate this process with the well-known energy-limited escape esquation  (see Equation~\ref{eq:elc} in Section~\ref{sec:h2oEffect}), {in which the square of the photospheric radius, $R_{\rm ph}$, and the EUV flux absorption radius, $R_{\rm XUV}$, enter the numerator\footnote{For rocky exoplanets with a non-opaque secondary atmosphere, $R_{\rm ph}$ is equal to the planetary radius, $R_{\rm pl}$, i.e., to the radius of the rocky surface of the planet. For planets with opaque hydrogen-dominated atmospheres, $R_{\rm ph}$ equals the radius where the visible part of the stellar flux is absorbed in the atmosphere. The radius $R_{\rm XUV}$, on the other hand, equals the radius where the XUV flux is absorbed in the thermosphere. For highly irradiated planets, $R_{\rm XUV}$ can be larger than $R_{\rm ph}$ by up to several times.} \citep[e.g.,][]{Watson1981}.} However, {by simply assuming $R_{\rm ph}^3$ instead of $R_{\rm ph}^2R_{\rm XUV}$, as many studies do}, they do not consider the XUV flux-induced extreme expansion of a hydrodynamically expanding atmosphere. This leads to a substantial underestimate of this loss process, {implying that the deleterious effects from an SMBH could be substantially higher than calculated by these studies.} (see also Appendix~\ref{sec:appStellar}).

Because of the high stellar density in the galactic center, stellar systems are susceptible to orbital perturbations by nearby stars \citep[e.g.,][]{Ovelar2012,JimenezTorres2013,Balbi2020,McTier2020,Arbab2021}. \citet{McTier2020}, for example found that 8 out of 10 bulge stars experience stellar encounters within 1000\,AU at a rate of $\sim1$~Gyr$^{-1}$, while about half of all bulge stars even experience $>$35~Gyr$^{-1}$. Such close encounters can either significantly perturb protoplanetary disks around such stars \citep{Bhandare2019,Li2019,Cai2019}, or destabilize and even eject planets from their hosts \citep{Winter2018a,Winter2018b,Vincke2018}. See appended Section~\ref{sec:perturb} for a more detailed discussion.

{Finally, we point out a potentially positive aspect of habitability related to the bulge. Apart from its generally high density of stars, which can be regarded as a positive aspect in itself at least related to the number of systems that can be examined per volume of space, the advantage for panspermia, i.e., for transferring life from one stellar system to another, should particularly be highlighted \citep[e.g.,][]{Adams2005Panspermia,Belbruno2012,Chen2018Panspermia,Balbi2020,Gobat2021,Chen2021Panspermia}\footnote{We discuss panspermia, specifically `lithopanspermia', in more detail in Section~\ref{sec:originLife}.}. In the bulge, the travel time of an ejecta for reaching the nearest star can be smaller by at least an order of magnitude compared to the galactic disk; it may hence be within the estimated survival time of certain extremophiles \citep{Ginsburg2018,Balbi2020}.}

The arguments of the last paragraphs do not paint an entirely coherent picture or generally prove the complete non-inhabitability of the galactic bulge, at least not for microbial life and/or extremophiles. However, they indicate that other galactic regions, particularly within the disk, might provide far more favorable habitable environments than the galactic center. Because of this and the lack of studies on the SFH and the distribution of SNs and metallicity within the bulge, we presently restrict ourselves to estimating the maximum number of EHs in the galactic disk.

To derive such a maximum number, however, we need to calculate the spatial distribution, age, and total number of presently existing stars in the galactic disk. To achieve such a distribution, we first need to implement the initial mass function (IMF) and star formation history (SFH) into our model.

\subsection{Implementing IMF, main-sequence lifetime, and SFH}\label{sec:NStarImp}

\subsubsection{The Initial Mass Function (IMF)}\label{sec:IMF}

The IMF describes the initial distribution of stellar masses, $M_{\star}$, for a population of stars, $N_{\star}$. It was first prominently introduced by \citet{Salpeter1955} and can in its simplest form be written as a power law such that
\begin{equation}\label{eq:imfS55}
  \frac{dN_{\star}}{dM_{\star}} = \xi_0 M_{\star}^{\alpha},
\end{equation}
where $\xi_0$ is a normalization constant {and $\alpha$ is the power law index}, which is $\alpha = -2.35$ in the case of the Salpeter-IMF \citep{Salpeter1955}. By integrating Equation~\ref{eq:imfS55}, i.e.,
\begin{equation}\label{eq:imfS55Int}
  \int dN_{\star} = \xi_0 \int_{M_{_{\star}\rm low}}^{M_{_{\star}\rm up}} M_{\star}^{\alpha} dM_{\star} = \frac{\xi_0}{\alpha + 1} [M_{\star}^{\alpha + 1}]^{M_{_{\star}\rm up}}_{M_{_{\star}\rm low}},
\end{equation}
%
one will get the fraction of stars between the lower and upper stellar masses, $M_{_{\star}\rm low}$ and $M_{_{\star}\rm up}$, respectively. For obtaining the distribution of {the number of} stars, $\delta N_{\star}$, within certain mass bins, $\delta M_{\star}$, we have to set $N_{\star} = 1$ and calculate $\xi_0$ by inserting the minimum and maximum stellar masses for $M_{_{\star}\rm low}$ and $M_{_{\star}\rm up}$.

The lower stellar mass limit, $M_{_{\star}\rm low}$, is usually constrained by the hydrogen burning limit to be around 0.07 and 0.08\,M$_{\rm \odot}$ \citep[e.g.,][]{Kroupa2001,Kroupa2013} or set to be 0.1\,M$_{\rm \odot}$ \citep[e.g.,][]{Salpeter1955,Chabrier2003,Chabrier2005}\footnote{{Note that Trappist-1, for instance, has a stellar mass of $M_{\star}=0.0898\pm0.0023$\,M$_{\odot}$ \citep{Agol2021}, which is below $M_{\star} = 0.1\,$M$_{\odot}$.}}. The upper stellar mass limit, $M_{\star\rm up}$, usually resides between 100\,M$_{\rm \odot}$ and 150\,M$_{\rm \odot}$ \citep[e.g.,][]{Kroupa2013} but its exact chosen value has no notable influence on the distribution since higher mass stars are becoming increasingly rare. The lower limit, on the other hand, shifts the distribution towards smaller stellar masses, because low-mass stars are dominating the distribution (see also Table~\ref{tab:spectralClass}). Such a shift leads to a lower average stellar mass, and, therefore, to more stars within the disk. However, since we will only consider stellar masses from $0.1$\,M$_{\odot}$ to $1.25$\,M$_{\odot}$ (see Section~\ref{sec:fat}), varying $M_{\star \rm low}$ between 0.07\,M$_{\odot}$ and 0.1\,M$_{\odot}$ does not substantially alter the final number of stars that we consider for deriving a maximum number of EHs (see also Table~\ref{tab:AppNStar})\footnote{With this choice, we entirely exclude any EHs that may potentially exist around brown dwarfs \citep{Barnes2013,Lingam2020BrownDwarfs}; whether they can even exist there, we will discuss in Section~\ref{sec:increase}. This choice also excludes potential habitats within the atmospheres of brown dwarfs \citep{Yates2017,Lingam2019BrownDwarf}. However, since these are not covered by our definition of Earth-like Habitats, the maximum number of EHs cannot be increased by including them. Besides, we are completely agnostic about their potential existence.}.


\begin{table*}\footnotesize
  \begin{center}
    \caption{The initial distribution of stars within the stellar spectral classes for different IMFs and $M_{\rm low} = 0.08^{+0.02}_{-0.01}$.}
    \label{tab:spectralClass}
    \resizebox{\textwidth}{!}{%
\begin{tabular}{l|c|c|c|c|c|c|c}
  \hline
   & M stars & K stars & G stars & F stars & A stars & B stars & O stars \\
   \hline
   mass range [$M_{\odot}$]$^a$ & 0.08-0.44 & 0.45-0.79 & 0.80-1.03 & 1.04-1.39 & 1.40-2.09 & 2.10-16 & $>$16 \\
   S55$^b$ & 0.903$^{+0.016}_{-0.034}$ & 0.053$^{+0.018}_{-0.009}$ & 0.013$^{+0.005}_{-0.002}$ & 0.010$^{+0.004}_{-0.001}$ & 0.009$^{+0.003}_{-0.002}$ & 0.011$^{+0.004}_{-0.001}$ & $\sim 0.001$ \\
   MS79$^c$ & 0.682$^{+0.022}_{-0.042}$ & 0.141$^{+0.018}_{-0.010}$ & 0.054$^{+0.007}_{-0.004}$ & 0.045$^{+0.006}_{-0.003}$ & 0.037$^{+0.004}_{-0.003}$ & 0.041$^{+0.006}_{-0.003}$ & $\sim 0.001$ \\
   %
   %
   K01/13$^d$ & 0.727$^{+0.023}_{-0.029}$ & 0.143$^{+0.019}_{-0.010}$ & 0.038$^{+0.005}_{-0.003}$ & 0.030$^{+0.004}_{-0.002}$ & 0.026$^{+0.003}_{-0.002}$ & 0.034$^{+0.005}_{-0.002}$ & $\sim 0.002$ \\
   C03$^e$ & 0.728$^{+0.017}_{-0.034}$ & 0.130$^{+0.016}_{-0.008}$ & 0.041$^{+0.005}_{-0.003}$ & 0.033$^{+0.004}_{-0.002}$ & 0.029$^{+0.003}_{-0.002}$ & 0.037$^{+0.005}_{-0.002}$ & $\sim 0.002$ \\
   \hline
\end{tabular}}
\end{center}
$^a$mass range after \citep{Habets1981}; $^b$\citet{Salpeter1955}; $^c$\citet{Miller1979}; $^d$\citet{Kroupa2001} and \citet{Kroupa2013};
$^e$\citet{Chabrier2003}
\end{table*}

\begin{figure*}
\centering
\includegraphics[width = 0.7\linewidth, page=1]{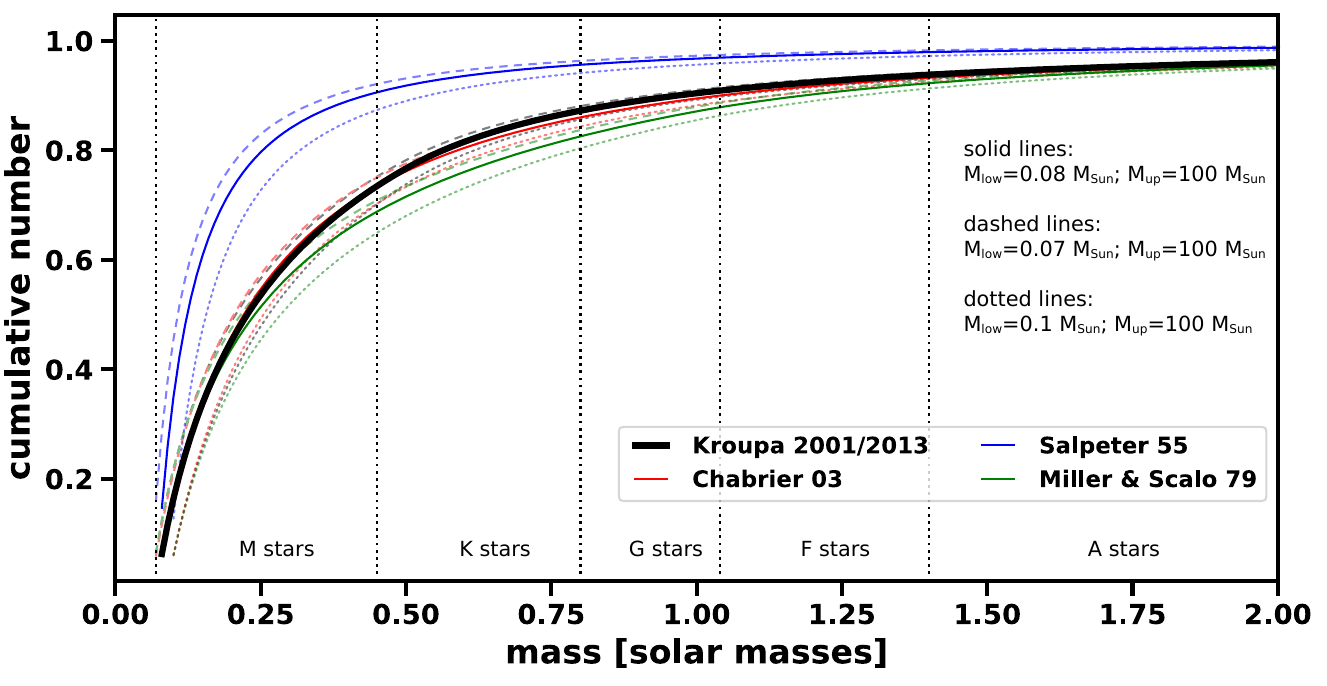}
\caption{Different cumulative stellar mass distributions for different Initial Mass Functions (IMFs), i.e., of an initial population of stars, as calculated through the empirical power laws by \citet{Salpeter1955}, \citet{Miller1979}, \citet{Kroupa2001}/\citet{Kroupa2013}, and \citet{Chabrier2003}, displayed for stellar masses up to 2.0\,$M_{\odot}$. The solid lines were calculated between stellar mass limits of $0.07 - 100\,M_{\odot}$, while the dashed and dotted lines show calculations for $0.08 - 100\,M_{\odot}$, and $0.1 - 100\,M_{\odot}$, respectively. The dotted vertical lines give the borders of the different stellar spectral classes \citep[after][]{Habets1981}.}
\label{fig:imfs}
\end{figure*}

The IMF was subsequently further investigated by several different authors \citep[e.g.,][]{Miller1979,Kroupa1993,Kroupa2013,Kroupa2001,Chabrier2003,Chabrier2005} to better fit the refined observational data. \citet{Kroupa2013} provide a thorough review of the IMF and propose a two-part power-law `canonical IMF' for masses between $M_{\star \rm low} = 0.07$\,M$_{\odot}$ and $M_{\star \rm up} = 150$\,M$_{\odot}$ based on the IMF by \citet{Kroupa2001}. We will implement this canonical IMF in our nominal case since it is commonly used within various relevant models of the galactic disk mass or SFH. It can be written as \citep{Kroupa2013}
\begin{eqnarray}
\begin{split}\label{eq:imfC05}
  \frac{dN_{\star}}{dM_{\star}} = \\ \xi_0
  \begin{cases}
    \left(\frac{M_{\star}}{0.07}\right)^{-1.3 \pm 0.3}, & 0.07 \leq M_{\star} \leq 0.5\,M_{\odot} \\
    \left[\left(\frac{0.5}{0.07}\right)^{-1.3 \pm 0.3}\right]\left(\frac{M_{\star}}{0.5}\right)^{-2.3 \pm 0.36}, & 0.5 < M_{\star} \leq 150\,M_{\odot}.
  \end{cases}
\end{split}
\end{eqnarray}
By integrating this relation, we obtain the number of stars, $\delta N_{\star}$, within each stellar mass bin, $\delta M_{\star}$. This can be displayed as a cumulative mass fraction of stars, as plotted in Figure~\ref{fig:imfs} for stars with stellar masses up to $M_{\star} = 2.0$\,M$_{\odot}$. There, the solid lines depict mass distributions based on various IMFs with $M_{\star\rm low} = 0.07$\,M$_{\odot}$ and $M_{\star\rm up} = 100$\,M$_{\odot}$ as boundary conditions, while the dashed and dotted lines show the same IMFs but for $M_{\star\rm low} = 0.08$\,M$_{\star\odot}$ and $M_{\star\rm low} = 0.1$\,M$_{\odot}$, respectively. As can be seen, the change in  $M_{\star\rm low}$ mostly affects the distribution within the spectral class of M dwarfs (see also Table~\ref{tab:spectralClass}). For our nominal case we use $M_{\star\rm low} = 0.08$\,M$_{\odot}$ and $M_{\star\rm up} = 100$\,M$_{\odot}$ as suggested by \citet{Kroupa2001}, while we implement $M_{\star\rm low} = 0.07$\,M$_{\odot}$ and $M_{\star\rm low} = 0.1$\,M$_{\odot}$ for our minimum and maximum cases, respectively (with $M_{\star\rm up}$ always set to be 100\,M$_{\odot}$). We further restrict ourselves to the IMFs by \citet{Kroupa2001} for the nominal and minimum cases, and \citet{Chabrier2003} for the maximum case. We will not consider the early, seminal IMF by \citet{Salpeter1955} since it was already shown by \citet{Paresce2000} that a single power-law mass function cannot reproduce the whole stellar distribution but overestimates stars below $M_{\star}\sim$0.5\,M$_{\odot}$.

We should finally mention that the IMF may not be constant but varies over time. \citet{Li2023}, for instance, found that the relative fraction of low-mass M dwarfs that form during an episode of star formation increases with stellar metallicity and hence galactic age. Based on correction factors by \citet{Liu2019} and \citet{Moe2019}, \citet{Li2023} also provide an IMF correction for binary systems that may further increase the relative fraction of young low-mass stars. This indicates that a higher number of M dwarfs form at present than at earlier ages of the Galaxy. Since we implement the canonical IMF and do not consider such variability, this may slightly affect our results. If we implemented a variable IMF instead, our stellar distribution would contain a higher number of young M dwarfs while the fraction of stars in the other spectral classes would be slightly lower. Such alteration would potentially lead to a slightly lower maximum number of EHs since a larger number of stars would fail to meet $A_{\rm at}^i$, i.e., the necessary requirement of permitting the long-term stability of an N$_2$-O$_2$-dominated atmosphere (see specifically Section~\ref{sec:fat}).

\subsubsection{The main-sequence lifetime of stars}\label{sec:MSL}

Since the IMF describes the initial stellar distribution, we need to implement the stellar main-sequence lifetime as a function of stellar mass to calculate the potential survival of a star until the present day. For the main-sequence lifetime, we follow the same approach as \citet{Westby2020} who combine the estimated main-sequence lifetime of the Sun of about 10\,Gyr \citep{Schroder2008} with the luminosity-mass correlation by \citet{Salaris2005}. This gives the following relation between main-sequence lifetime, $\tau_{\rm MS}$, and stellar mass, $M_{\star}$, i.e. \citep{Westby2020},
\begin{eqnarray}\label{eq:ageLimits}
  \tau_{\rm MS} &\approx&
  \begin{cases}
    7.1 M_{\star}^{-2.5}, & 2.0 \leq M_{\star} \leq 20 \,M_{\odot} \\
    10 M_{\star}^{-3}, & 0.43 \leq M_{\star} < 2.0 \,M_{\odot} \\
    43 M_{\star}^{-1.3}, & M_{\star} < 0.43 \,M_{\odot}.
  \end{cases}
\end{eqnarray}

\begin{figure}
\centering
\includegraphics[width = 1.0\linewidth, page=1]{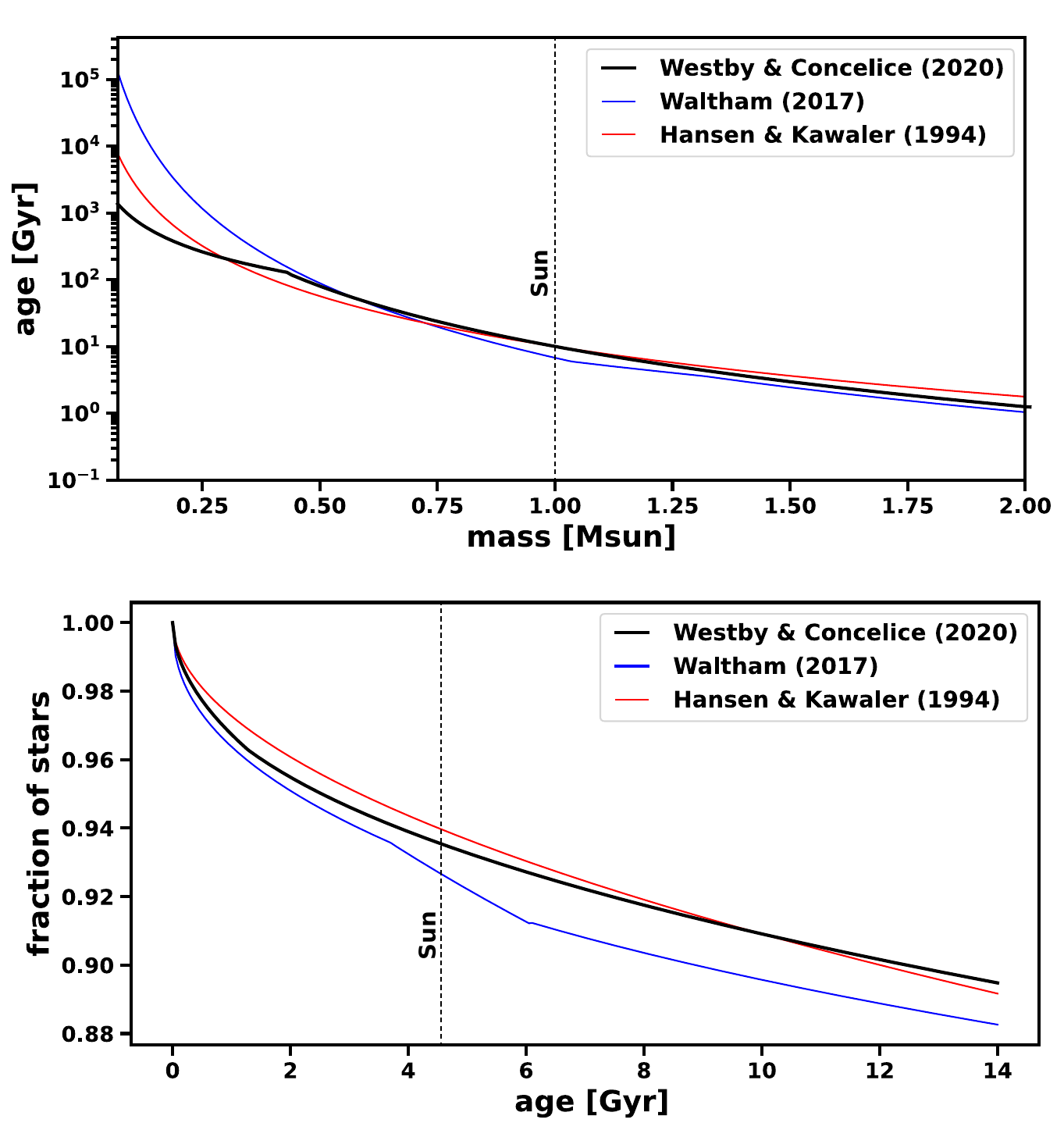}
\caption{Upper panel: Main-sequence lifetime of stars with masses between 0.1\,M$_{\odot}$ and 2.0\,M$_{\odot}$. The solid black line shows the lifetime as calculated through the approach by \citet{Westby2020} and as described in Equation~\ref{eq:ageLimits}. {For comparison, the red line is calculated with a similar age-mass relation as suggested by \citet{Hansen1994} while the blue line shows the \textit{mean habitable lifetime for planets that may possess intelligent observers} as proposed by \citet{Waltham2017}; note, however, that this estimate gives larger ages for low-mass stars than their respective main-sequence lifetime}. Lower panel: The fraction of stars that are still on the main sequence as a function of age for the same three relations.}
\label{fig:ageLimits}
\end{figure}

By applying this relation, we obtain the black solid line in the upper panel of Figure~\ref{fig:ageLimits} which illustrates the main-sequence lifetime of stars as a function of stellar mass. For comparison, the red line corresponds to a calculation performed with a similar relation by \citet{Hansen1994}. The blue line additionally illustrates the `mean habitable lifetime for planets that may possess intelligent observers' by \citet{Waltham2017} and was calculated with the power-law fit from the same study. \citet{Waltham2017} base their mean habitable lifetime on a probabilistic combination of the Copernican and Anthropic principles and therefore assume planets to be inhabitable for intelligent life if these provide conditions similar to the Earth's. Note, however, that (i) this lifetime is larger than the main-sequence lifetime for stars with masses below $\sim$0.6\,M$_{\rm \odot}$, {and (ii) that it is intended for intelligent observers and not for our definition of complex life (see Section~\ref{sec:caveats})}; this lifetime should hence be taken with caution. The lower panel of Figure~\ref{fig:ageLimits} additionally shows the fraction of stars that are still on the main sequence as a function of stellar age for the same relations.

While we implement the relation by \citet{Westby2020} as an upper limit of stellar age into our nominal case (the white area in Figure~\ref{fig:distStars} derives from this limit), we will see in Section~\ref{sec:fat} that our results are insensitive to the chosen stellar main-sequence lifetime relation. This is because the upper age limit of stars, which allow for the existence of EHs, will always be set by the maximum bolometric luminosity that still permits the survival of complex life in the HZCL, a value that will always be lower than the actual main-sequence lifetime \citep[but may intersect with the mean habitable lifetime by][]{Waltham2017}.

\subsubsection{The galactic star formation history}\label{sec:SFR}

As a final step, we need to implement the actual star formation history (SFH), i.e., the evolution of the star formation rate (SFR) of the galactic disk to obtain the distribution of presently existing main-sequence stars as a function of stellar mass and age. For this, we implement the SFH as reconstructed by \citet{Snaith2015} into our nominal case. These authors developed a chemical evolution model to reconstruct the SFH of the galactic disk over time from present-day chemical abundances of the Milky Way's inner ($<7-8$\,kpc) and outer disk. We chose this specific SFH by \citet{Snaith2015} over others, since their model also provides the metallicity evolution of the disk within one consistent framework. In addition, most of the other studies reconstruct a so-called cosmic SFH that is based on observations of star formation rates in various galaxies throughout the Universe \citep[e.g.,][]{VanDokkum2013,Madau2014,Stanway2018,Westby2020}. Recent studies by \citet{RuizLara2020} and \citet{Alzate2021SFH}, however, that reconstructed the star formation history of the Solar Neighborhood based on the Gaia Data Release 2 \citep{Brown2018}, find an SFH that also shows bursts of star formation. {The formation of the Solar System took place during one of these major bursts that peaked at $\sim 5.5$\,Ga \citep{Madau2023} and was likely triggered by an external perturbation \citep{Mor2019} such as the first passage of the Sagittarius dwarf galaxy \citep{RuizLara2020} {or a nearby supernova \citep[e.g.,][]{Cameron1977SN,Boss2008SN,Banerjee2016SN}.}}

\begin{figure}
\centering
\includegraphics[width = 1.0\linewidth, page=1]{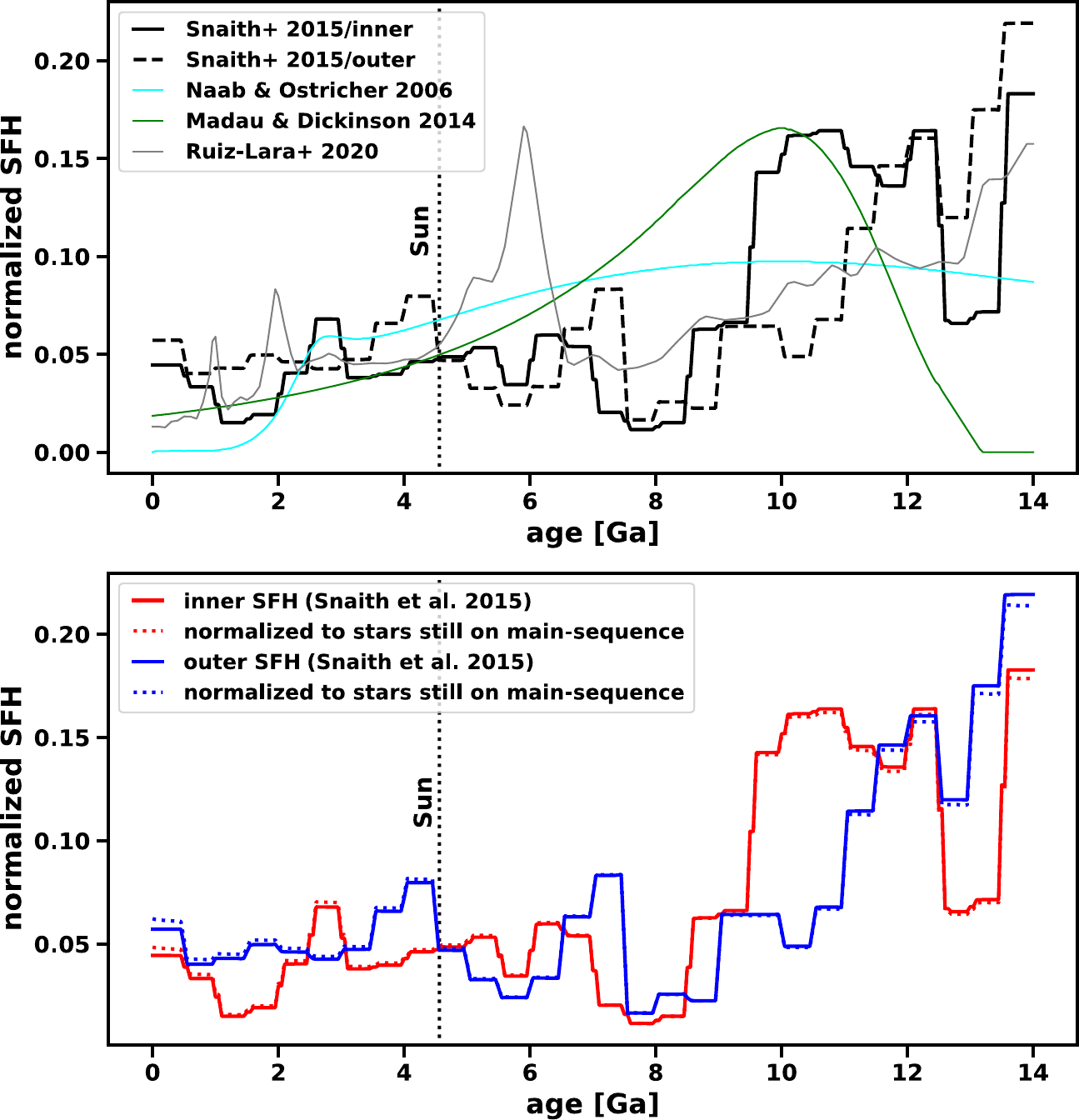}
\caption{Upper panel: The normalized best-fit SFH for inner (solid black line) and outer disk (dashed black line) by \citet{Snaith2015}, the reconstructed SFH of the Milky Way by \citet{Naab2006}, the cosmic SFH (green line) by \citet{Madau2014}, and the SFH of a 2\,kpc wide bubble around the Sun (grey line) by \citet{RuizLara2020}. Lower panel: Inner and outer SFH by \citet{Snaith2015} renormalized (dotted lines) to only include stars that still reside on the main sequence.}
\label{fig:sfr}
\end{figure}

The upper panel of Figure~\ref{fig:sfr} shows the best-fit SFH for the inner (solid black line) and outer disks (dashed black line) by \citet{Snaith2015} that we use within our nominal case. For comparison, we also display the reconstructed SFH for the Milky Way by \citet{Naab2006}, the cosmic SFH (green line) by \citet{Madau2014}, and the SFH of a 2\,kpc wide bubble around the Sun (grey line) by \citet{RuizLara2020}. We will evaluate how a change in SFH might affect the maximum number of EHs in Section~\ref{sec:max}.

The SFH in the upper panel of Figure~\ref{fig:sfr}, however, shows any star that ever existed in the galactic disk. Since we are only interested in those that are still on the main sequence, we need to renormalize the distribution. This can easily be done by excluding any star that already diverged from the main sequence via Equation~\ref{eq:ageLimits}. The lower panel of Figure~\ref{fig:sfr} consequently shows the renormalized SFH for the inner and outer disk as implemented in our nominal case. The distribution, as expected, slightly shifts towards the present day.

As a side note, be aware that the SFH by \citet{Snaith2015} starts at an age of 14\,Ga, even though the actual age of the Universe is currently estimated to be around 13.8\,Gyr \citep{Planck2020} and the epoch of reionization with the emergence of the first stars might date some few 100 Myr later still \citep[e.g.,][]{Robertson2021} with the oldest galaxies having emerged earlier than about 400\,Myr after the Big Bang \citep{Robertson2022}\footnote{{The most distant, earliest galaxy reported to be observed in May 2024 was JADES-GS-z14-0, which is suggested to have been in place at least 300\,Myr after the Big Bang \citep{Carniani2024}.}}. By using this specific SFH, we are stuck with this age determination, but this will not affect our results significantly. A later starting age would potentially decrease the maximum number of EHs further since the earliest stars would have had less time for their XUV flux to decrease (see Section~\ref{sec:fat}).

\subsubsection{First derived stellar properties}\label{sec:starProps}

By implementing IMF, SFH, and main-sequence lifetime into our model, we can already derive some interesting numbers even though we did not yet calculate $N_{\star}$ itself (see also Table~\ref{tab:AppNStar} for all input parameters).

From all the stars that ever formed within the disk, 91.15(+0.60/-2.04)\%\footnote{{From} here onwards, the main value in such notation always denotes the nominal case while the upper and lower limits are derived from the minima and maxima cases or vice versa.} are yet residing on the main sequence. Within a stellar mass range of $M_{\star}=0.1-1.25$\,M$_{\odot}$, even 97.39(+0.10/-0.0)\% survived until the present day. For our nominal case with $M_{\star\rm low} = 0.08$\,M$_{\odot}$, the IMF from \citet{Kroupa2001} and the SFH from \citet{Snaith2015}, only 0.45\% of all presently existing stars have a mass that is $>$1.25\,M$_{\odot}$, while 12.78\% are below 0.1\,M$_{\odot}$. Consequently, 86.77\% of all stars in the disk reside between $M_{\star}=0.1-1.25$\,M$_{\odot}$. For our maximum case with the IMF from \citet{Chabrier2003} and the SFH from \citet{Naab2006}, none of the stars have a mass below 0.1\,M${_\odot}$ since $M_{\star\rm low}$ was chosen to be exactly 0.1\,M${_\odot}$. Here, 0.27\% of all presently existing stars have a stellar mass that is $>$1.25\,M$_{\odot}$ while 99.73\% reside within $M_{\star}=0.1-1.25$\,M$_{\odot}$. For our minimum case with $M_{\rm low} = 0.07$\,M$_{\odot}$, the IMF from \citet{Kroupa2001} and the SFR from \citet{Snaith2015}, 83.39\% of all presently existing stars reside within $M_{\star}=0.1-1.25$\,M$_{\odot}$ while only 0.22\% are above that mass range.

The average initial stellar mass calculated through the IMF is $\bar{M}_{\star}$\,=\,0.57(+0.08/-0.04)\,M$_{\odot}$, while the average stellar mass for the present day is significantly smaller due to massive stars living much shorter, i.e., $\bar{M}_{\star}$\,=\,0.29(+0.03/-0.01)\,M$_{\odot}$. For comparison, \citet{Chabrier2003} gives a present-day stellar number density in the disk of $n_{\star}\sim0.134$\,pc$^{-3}$ and a mass density of $\rho_{\star}\sim4.1 \times 10^{-2}\, \rm M_{\odot} pc^{-3}$ which results in an average stellar mass of $\bar{M}_{\star} \sim 0.31$\,M$_{\odot}$. As initial values, these authors further give $n_{\star}\sim0.145$\,pc$^{-3}$ and $\rho_{\star} \sim 8.2 \times 10^{-2}\, \rm M_{\odot} pc^{-3}$ equating to $\bar{M}_{\rm \odot} \sim 0.57$\,M$_{\star}$\footnote{These present and initial values for $n_{\star}$ and $\rho_{\star}$, as taken from \citet{Chabrier2003}, are combined values, i.e., the sums over the respective densities for stars with $M_{\star}\leq$1.0\,M$_{\odot}$ and $M_{\star}\geq$1.0\,M$_{\odot}$ as listed in their Tables~3 and 4.}.


Finally, only 46.86(+0.48/-3.33)\% of the entire stellar mass that has ever been produced within the disk is yet residing on the main sequence which is in stark contrast to the aforementioned stellar survival rate of 91.15(+0.60/-2.04)\%. This illustrates a strong shift towards lower-mass stars over time as can also be seen through a change in the relative stellar distribution within the different spectral classes. By considering the canonical IMF with $M_{\rm low} = 0.08$\,M$_{\odot}$, the respective fractions of M and K stars increase from their initial values of 72.74\% and 14.28\% to 79.80\% and 15.67\%, respectively. All the other spectral classes decrease in relative size, i.e., G and F stars from 3.76\% and 2.96\% to 3.38\% and 0.84\%, respectively. Higher spectral classes show an even stronger decrease due to their very short lifetimes with present-day values of 0.23\%, 0.07\%, and $<$0.001\% for A, B, and O stars, respectively.

\begin{figure}
\centering
\includegraphics[width = 1.0\linewidth, page=1]{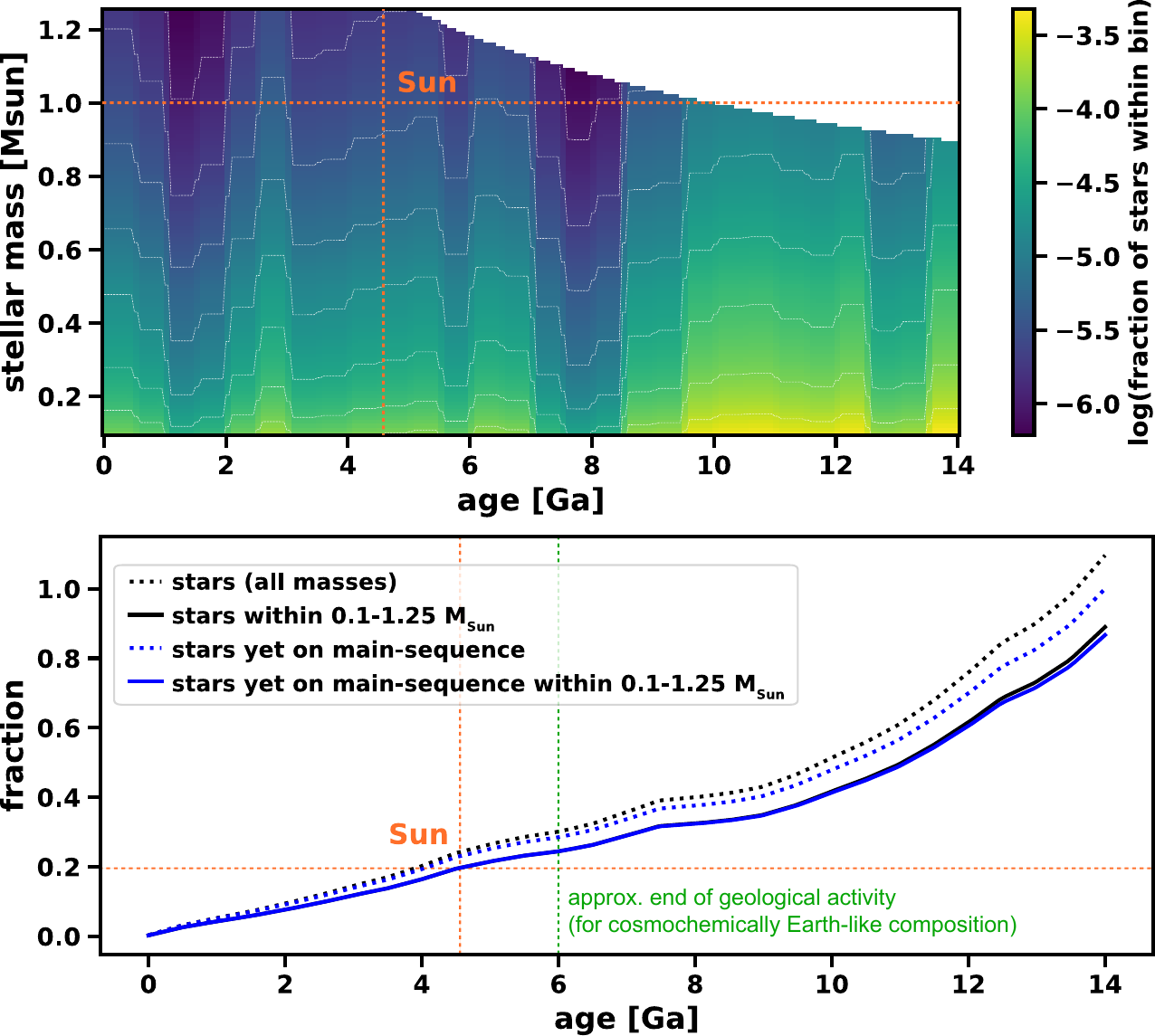}
\caption{Upper panel: {The relative fraction} of today's main-sequence stars between 0.1-1.25\,$M_{\odot}$ as a function of birth age and stellar mass within bins of 50\,Myr and 0.01\,$M_{\odot}$ for our nominal case. The varying densities at different ages are an effect of the implemented SFR by \citet{Snaith2015} and depict higher and lower star formation rates. The dashed orange lines indicate the birth age and mass of the Sun. Lower panel: The fraction of stars that are younger than the respective age, i.e., for all stars ever existing over the entire stellar mass range (dotted black line), for stars between 0.1\,M$_{\odot}$ and 1.25\,M$_{\odot}$ (solid black line), and the same for today's main-sequence stars (dotted and solid blue lines). Only 23\% of stars between 0.1\,M$_{\odot}$ and 1.25\,M$_{\odot}$ are younger than the Sun. {The green dashed line indicates an age of 6\,Gyr, the approx. age geological activity will cease for cosmochemically Earth-like planets. A total of 71.5\% of all stars are older than that age.}}
\label{fig:distStars}
\end{figure}

The relative distribution of stars at the present day as a function of stellar birth age and mass can be seen in Figure~\ref{fig:distStars} for our nominal case. The upper panel shows the relative fraction of stars within age and mass bins of $\delta t\,$=\,50\,Myr times $\delta M\,$=\,0.01\,M$_{\odot}$ and within $M_{\star} = 0.1 - 1.25$\,M$_{\odot}$. One can see that low-mass stars are dominating the distribution, and the ages of stronger star formation rates from the implemented SFH by \citet{Snaith2015} can easily be identified. There are no stars in the white area at the upper right side of this figure since their main-sequence lifetime is too short to still exist at the present day. The lower panel depicts the fraction of stars born later than the respective age for all stars ever existing over the entire stellar mass range (dotted black line) and within $M_{\star} = 0.1 - 1.25$\,M$_{\odot}$ (solid black line), respectively, and the same for today's main sequence stars (dotted and solid blue lines). There, it can be seen that only 23.0\% of all main sequence stars and 19.7\% of those within $M_{\star} = 0.1 - 1.25$\,M$_{\odot}$ are younger than the Sun.

According to our nominal case, about 71.5\% of all main sequence stars are older than 6\,Gyr. This age is of particular interest since data from coupled galactic chemical evolution and geophysical thermal models indicate that the window for geologically active, cosmochemically Earth-like planets is around 6\,Gyr \citep{Frank2014,Mojzsis2021}. Older planets might therefore show strong restrictions in maintaining carbon-silicate and nitrogen cycles.

\subsection{Final values for $N_{\star}$}\label{sec:NStar}

Since the evolution of an EH also depends on its location within the galactic disk, we will not only need to calculate the number of stars but also its distribution within the disk. For this purpose, we will make use of the widely accepted axisymmetric mass distribution and gravitational potential model of the Milky Way by \citet{McMillan2011,McMillan2017}. Their model divides the galaxy into 6 axisymmetric components, i.e., the bulge, the dark-matter halo, thin and thick stellar disks, \textsc{Hi} and molecular gas disks. Out of these, we make use of the thin and thick stellar disk models which we will subsequently combine and treat as just one galactic disk.

The density distribution, $\rho_{\rm d}$ of both disks can be modeled by an exponential distribution and can be written as \citep{McMillan2017}
\begin{equation}\label{eq:diskdist}
  \rho_{\rm d}(r,z) = \frac{\Sigma_0}{2 z_{\rm d}} \exp \left( - \frac{|z|}{z_{\rm d}} - \frac{r}{R_{\rm d}}\right),
\end{equation}
where $\Sigma_0$ is the central surface density, $z$ the height from the galactic plane, $r$ the distance from the galactic center, and  $z_{\rm d}$ and $R_{\rm d}$ are the scale heights and lengths, respectively. The total disk mass $M_{\rm d}$ can further be written as \citep{McMillan2017}
\begin{equation}\label{eq:diskmass}
  M_{\rm d} = 2 \pi \Sigma_0 R_{\rm d}^2.
\end{equation}
In both equations, $\Sigma_0$, $z_{\rm d}$ and $R_{\rm d}$ are specific for the thick and thin disk, respectively, and its values based on the best-fitting model by \citet{McMillan2017} can be found in Table~\ref{tab:disk}.

By using Equation~\ref{eq:diskmass}, we can calculate the whole stellar mass, $M_{\rm d}$, of the thin and thick disk to be $M_{\rm d, thin} = 3.52 \times 10^{10}$\,M$_{\odot}$ and $M_{\rm d, thick} = 1.05 \times 10^{10}$\,M$_{\odot}$, respectively, which gives a total stellar disk mass of $M_{\rm d} = 4.57 \times 10^{10}$\,M$_{\odot}$ for the best-fitting model. If one also includes the bulge, the total stellar mass of the galaxy would be $5.43 \pm 0.57 \times 10^{10}$\,M$_{\odot}$ \citep{McMillan2017} which fits very well to other stellar mass estimates as described in Section~\ref{sec:MWcomp}.


\begin{table*}
  \begin{center}
    \caption{Parameters of the best-fitting model for the galactic stellar disks by \citet{McMillan2017}, stellar disk mass and stars (nominal case).}
    \label{tab:disk}
    \resizebox{0.8\textwidth}{!}{%
\begin{tabular}{l|c|c|c}
  \hline
  & thin disk & thick disk & total \\
  \hline
  $\Sigma_0$ [M$_{\odot}$\,pc$^{-2}$] & 896 & 183 & \\
  $R_{\rm d}$ [pc] & 2500 & 3020 & \\
  $z_{\rm d}$ [pc] & 300 & 900 & \\
  \hline
  $M_{\rm d}$ (incl. 20\% stellar remnants) [M$_{\odot}$] & $3.52 \times 10^{10}$ &  $1.05 \times 10^{10}$ & $4.57 \times 10^{10}$ \\
  $M_{\rm d}$ (excl. stellar remnants) [M$_{\odot}$] & $2.82 \times 10^{10}$ & $0.84 \times 10^{10}$ & $3.65 \times 10^{10}$ \\
  \hline
  stars (whole mass range; entire disk) & $9.60 \times 10^{10}$ & $2.86 \times 10^{10}$ &  $12.46 \times 10^{10}$ \\
  stars (whole mass range; $r = 2.0-21.0$\,kpc; $z_1 = 2.5$\,kpc) & $7.83 \times 10^{10}$ & $2.32 \times 10^{10}$ & $10.15 \times 10^{10}$ \\
  stars (within $M = 0.1-1.25$\,M$_{\odot}$;$r = 2.0-21.0$\,kpc; $z_1 = 2.5$\,kpc) & $6.80 \times 10^{10}$ & $2.01 \times 10^{10}$ & $8.81 \times 10^{10}$ \\
  \hline
\end{tabular}}
\end{center}
\end{table*}


Here, it has to be noted that the model by \citet{McMillan2017} assumes no ``central hole'' for the thin and thick disk. It presumes that the disks extend into the bulge until the center of the galaxy at $r = 0$\,kpc. We, therefore, have to exclude the central part and start our calculation at $r_0 = 2$\,kpc to take account of the bulge.

We further need to clarify, which class of objects are included within the stellar mass of the disk. Since \citet{McMillan2011,McMillan2017} give no further description of the stellar disk mass, $M_{\rm d}$, we implement the definition commonly used in other Milky Way disk models \citep[e.g.,][]{Licquia2015,Xiang2018} and for the measurement of galaxies in the MPA-JHU catalogs \citep{Brinchmann2004}\footnote{{The data catalogs from SDSS (Sloan Digital Sky Survey) studies at the Max-Planck-Institute for Astrophysics (MPA) and the Johns Hopkins University (JHU); see \url{https://wwwmpa.mpa-garching.mpg.de/SDSS}.}}. This excludes substellar objects such as brown dwarfs (BD) but includes stellar remnants. Even though we obtain the entire stellar mass that was ever produced within the Milky Way together with the stellar fraction that still exists at the present day (i.e., 46.86\% for our nominal case), it is not correct to assume that 53.14\% of the entire stellar mass within the disk is, consequently, concentrated within stellar remnants, since massive stars lose a substantial amount of mass over their lifetime, particularly during their collapse into a white dwarf, neutron star or black hole \citep[e.g.,][]{Xiang2018}. \citet{Flynn2006} found that $\sim20.3\%$ of the stellar mass within the disk resides within stellar remnants which is in good agreement with subsequent studies by \citet{McKee2015} and \citet{Xiang2018} who obtained $\sim19.2\%$ and 20.9\%, respectively. We will, therefore, assume that 20\% of the stellar disk mass relates to stellar remnants. From a total stellar mass within the disk of $M_{\rm d} = 4.57 \times 10^{10}$\,M$_{\odot}$, 80\% therefore belong to stars that are still on the main sequence, i.e., $M_{\star} = 3.66 \times 10^{10}$\,M$_{\odot}$.

If we divide these numbers by our calculated average stellar mass of $\bar{M}_{\odot} = 0.29$\,M$_{\odot}$ for our nominal case, we obtain $12.46 \times 10^{10}$ main-sequence stars within the entire disk and $14.81 \times 10^{11}$ main-sequence stars within the whole Milky Way including bulge and halo. From all such disk stars, $10.80 \times 10^{10}$ stars are within a mass range of $M_{\star} = 0.1-1.25$\,$M_{\odot}$ with most of the remaining stars being below 0.1\,$M_{\odot}$. As a side-note, if one simply assumes that stars with minimum masses of initially $M_{\star} \geq 15-20\,$M$_{\odot}$ will end up as black holes (BH), we obtain $\sim2-3 \times 10^{8}$ BHs that are presently existing within the entire disk. This simple estimate is in relatively good agreement with a recent thorough calculation of $\sim1.2 \times 10^8$ BHs by \citet{Olejak2020}.

To actually calculate the stellar distribution in the galactic disk, we have to integrate Equation~\ref{eq:diskdist} {over $dV_{\rm d} = 2\pi r dr dz$ and multiply the integral by a factor of 2 since we assume axial symmetry and only integrate for positive $z$ values}, i.e.,
\begin{equation}\label{eq:diskdist2}
\begin{aligned}
  \int\int dM_{\rm d} = {} & \frac{2 \Sigma_0 \pi}{z_{\rm d}} \int_0^{z_1}\int_{r_0}^{r_1} r \exp \left( - \frac{|z|}{z_{\rm d}} - \frac{r}{R_{\rm d}}\right) dz dr \\
  = {} & 2{\pi}\Sigma_0 \int_{r_0}^{r_1} r \exp\left(\frac{z_1}{z_{\rm d}}-1\right)\\
  & \times\exp\left(-\frac{z_1}{z_{\rm d}}-\frac{r}{R_{\rm d}}\right) dr \\
  = {} & -2\pi R_{\rm d} \Sigma_0 \left[\exp\left(\frac{z_1}{z_{\rm d}}-1\right)\right]\\
  & \times(r + R_{\rm d})\exp\left[-\frac{r}{R_{\rm d}}-\frac{z_1}{z_{\rm d}}\right]_{r_0}^{r_1}.
  \end{aligned}
\end{equation} %
We will set $r_1 = 21.0$\,kpc and $z_1 = 2.5$\,kpc. Since the distribution of this Milky Way Model is exponential, this will leave a small part of stars outside of our box (as will any chosen boundary) but the remaining fraction of stars is $<$1.7\% and located in a very metal-poor environment \citep[e.g.,][]{Hayden2015}.

\begin{figure}
\centering
\includegraphics[width = 1.0\linewidth, page=1]{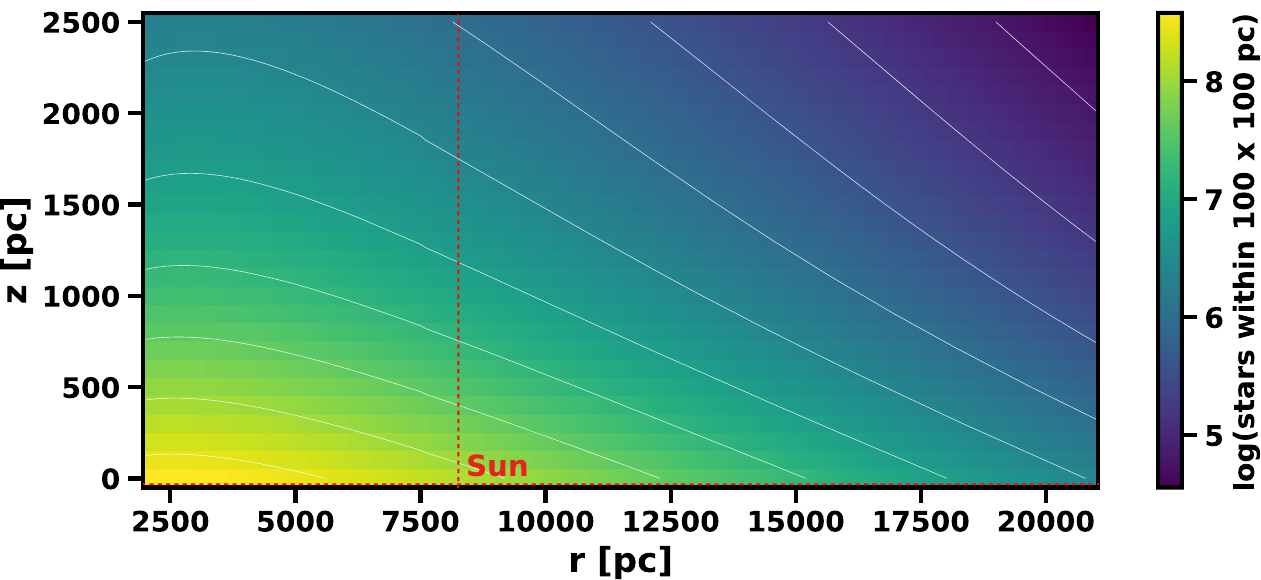}
\caption{The radial and vertical distribution of disk stars from $r_0 = 2000$\,pc to $r_1 = 21000$\,pc, and $z_0 = 0$\,pc to $z_1 = 2500$\,pc. The Sun is located at a distance from the center of $r_{\odot} = 8178 \pm 35$\,pc \citep{Abuter2019} and $z_{\odot} = 17 \pm 5$\,pc above the galactic mid-plane \citep{Karim2017}.}
\label{fig:starsDiskRZ}
\end{figure}

Within such boundary conditions (i.e., $r = 2.0-21.0$\,kpc; $z = 0.0-2.5$\,kpc), we calculate a total of $10.15 \times 10^{10}$ stars within the galactic disk out of which $7.83 \times 10^{10}$ and $2.32 \times 10^{10}$ stars are part of the thin and thick disk, respectively. Their distribution within the galactic disk, as based on Equation~\ref{eq:diskdist2}, can be seen in Figure~\ref{fig:starsDiskRZ}.\footnote{We will subsequently not distinguish between thin and thick disk stars but treat them as just one set.}. Since 86.77\% of these stars are within our chosen stellar mass range ($M_{\star} = 0.1-1.25$\,M$_{\odot}$), we obtain a total of $8.81 \times 10^{10}$ stars for our subsequent nominal case calculation (see Table~\ref{tab:AppNStar} for maximum and minimum cases). From maximum to minimum case we obtain an entire range of $8.81(+0.54/-0.17) \times 10^{10}$ stars within $r = 2.0-21.0$\,kpc, $z = 0.0-2.5$\,kpc, and $M_{\star} = 0.1-1.25$\,M$_{\odot}$.

\section{The Fraction of Stars, $\eta_{\star}$, able to host Earth-like Habitats}\label{sec:NEStar}

\subsection{Galactic Habitable Environment, $A_{\rm GHZ}^{i}$}\label{sec:GHZ}

The concept of the galactic habitable zone (GHZ) was first discussed in detail by \citet{Gonzalez2001} and \citet{Lineweaver2004}, and refers to regions within a galaxy in which planets can provide a ``long-term habitat for animal-like aerobic life'' \citep{Gonzalez2001}. As already discussed in Section~\ref{sec:MWcomp} related to the bulge, simulations for the GHZ mostly consider three different criteria, i.e., (i) high-energetic events that can sterilize a planet and erode its atmosphere, (ii) orbital perturbations of a stellar system, and (iii) the metallicity of the ISM.

\begin{figure}
\centering
\includegraphics[width = 1.0\linewidth, page=1]{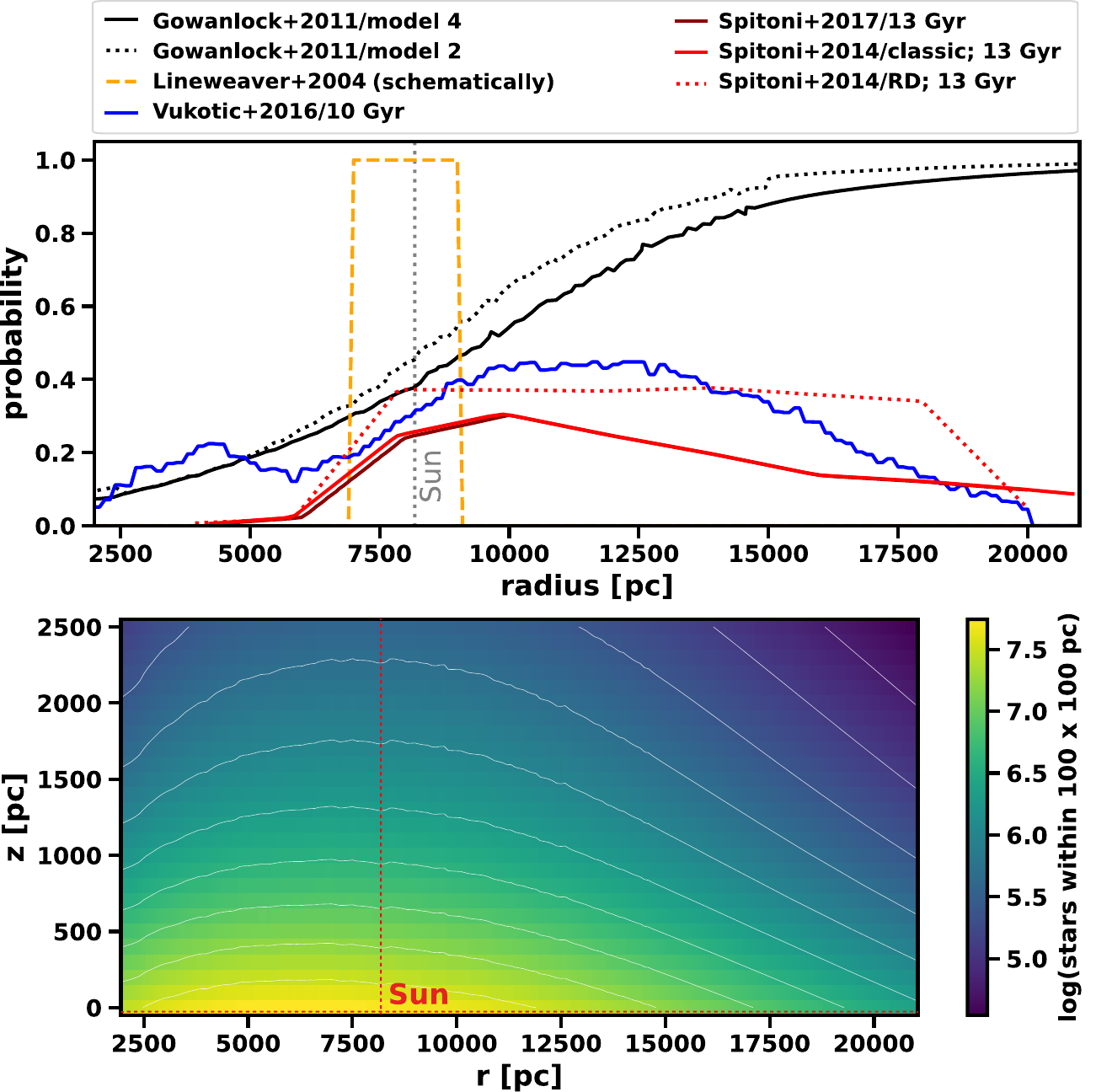}
\caption{Upper panel: Galactic habitable zones (GHZ) from various studies, which show the probability of being habitable as a function of galactocentric distance. The black lines show the fraction of stellar systems that are {not} sterilized by SNs in two of the GHZ models (both of them extrapolated from 15000\,pc up to 21000\,pc) by \citet{Gowanlock2011}. The dashed orange line {schematically and in a simplified manner illustrates} the annular region of 7000-9000\,pc from the landmark study by \citet{Lineweaver2004} within which habitable planets are the most likely; note, however, that the GHZ in \citet{Lineweaver2004} {is much more complex (see their Fig.~3) and that habitable planets are not restricted to this annulus.} The blue line indicates results by \citet{Vukotic2016} for the 10\,Gyr old Galaxy based on SN explosions, metallicity, and orbital stability. The red and dark-red lines show different models by \citet{Spitoni2014} and \citet{Spitoni2017} that include SN explosions and the chemical evolution of the Milky Way. {Lower panel: Distribution of stars in the r-z plane of the galactic disk that were not sterilized by supernovae during the last 4\,Gyr based on `model~4' by \citet{Gowanlock2011}, i.e., the model implemented in our nominal case.} Data extracted from: upper panel of Fig.~6 by \citet{Gowanlock2011} for the solid black lines (model~2 and model~4);  bottom right panel of Fig.~5 in \citet{Vukotic2016} for the blue line;  Fig.~4 and right panel of Fig.~11 (dark red lines, both) by \citet{Spitoni2014} for the red solid and dotted lines, respectively;  right panel of Fig~5 (yellow line) by \citet{Spitoni2017} for the dark red line; GHZ by \citet{Lineweaver2004} as stated within their article.}
\label{fig:ghz}
\end{figure}

The upper panel of Figure~\ref{fig:ghz} shows the probability of a planet being habitable as a function of galactocentric distance for several different GHZ models from the scientific literature. \citet{Lineweaver2004} modeled the evolution of habitability in the Milky Way and considered stellar distribution, metallicity, sufficient time for biological evolution, and the effect of supernovae. They found that most habitable environments are within a narrow annulus of 7-9\,kpc from the galactic center as schematically and in a simplified manner illustrated by the dashed orange line in Figure~\ref{fig:ghz}\footnote{The actual GHZ region in \citet{Lineweaver2004} is much more complex; see their Fig.~3}. Between this region and the galactic center, the high-energetic radiation of supernovae becomes too strong and frequent to allow for complex life. Farther outside the metallicity becomes so low that the probability of forming rocky exoplanets becomes low as well.

The black lines in Figure~\ref{fig:ghz} show the probability of not being sterilized by supernovae of types Ia (SNIa) and II (SNII) {during the last 4\,Gyr} in two of the four GHZ models provided by \citet{Gowanlock2011}, as derived from the SFH and metallicity gradient modeled by \citet{Naab2006}. The dotted black line shows `model~2' by \citet{Gowanlock2011}, which is additionally based on the Kroupa IMF and the stellar number density from \citet{Carroll2006}, while the solid black line illustrates their `model~4', which is also based on the Kroupa IMF but takes the stellar number density from \citet{Juric2008}\footnote{We do not display the other two models by \citet{Gowanlock2011} which take the same densities but include the Salpeter IMF.}. By additionally excluding stars with a metallicity too low to host rocky exoplanets (not displayed in Figure~\ref{fig:ghz}), \citet{Gowanlock2011} further conclude that most habitable planets will be towards the inner Galaxy since this galactic region is much more densely populated by stars that have a sufficiently high metallicity to form rocky exoplanets.

{To derive the probability distribution of not being sterilized by SNs, \citet{Gowanlock2011} assume that a stellar system must not be sterilized within the last 4\,Gyr, so that metazoan life can potentially arise on a planet, a timescale that is simply based on an analogy with the emergence timescale of metazoan life on the Earth. The spatial and temporal occurrence rates of SNIa and SNII type supernovae in their galactic models were calculated by assuming that stars with a stellar mass of $\gtrsim$8\,M$_{\odot}$ become SNII \citep[based on][]{Kennicutt1984} and that 1\% of white dwarfs with progenitor masses of 0.08\,M$_{\odot} < M_{\star} < 8\,$M$_{\odot}$ become SNIa \citep[based on][]{Pritchet2008}, both at the end of their main-sequence lifetimes. Instead of a given sterilization distance for all SNs, \citet{Gowanlock2011} further assume sterilization distributions for both types of supernovae, depending on the absolute magnitude distributions for SNII by \citet{Richardson2002} and SNIa by \citet{Wang2006SNIa}, respectively. With an assumed sterilization distance of 8\,pc for an average SNII \citep[based on][]{Gehrels2003}, \citet{Gowanlock2011} calculate that sterilization distances (see their Fig.~3) range from $\sim$1-27\,pc and $\sim$13-27\,pc for SNII and SNIa, respectively. An average individual SNIa is hence more lethal than an average individual SNII, but all SNII supernovae taken together are about twice as lethal since their occurrence rate is about an order of magnitude higher than the occurrence rate of SNIa supernovae.}

{Here we note that such probability distribution may provide relatively conservative assumptions on the lethal effects of supernovae. If the putative organisms on a planet evolved and adapted to already harsh radiation environments and/or the emergence of complex life needs less time, it may overestimate the lethal role of high-energetic galactic events. In addition, it must be noted that complete sterilization of a biosphere is relatively unlikely since some extremophiles can withstand very high amounts of radiation \citep[e.g,][]{Balbi2020}. On the other hand, new simulations, based on the evaluation of SN-emitted cosmic rays \citep{Thomas2023} and X-rays \citep{Brunton2023}{\footnote{We note that these authors only calculate their lethal effects based on a certain type of supernovae, i.e., X-ray luminous SNe.}}, find that supernovae might be lethal over even larger galactic distances than initially thought, which could also indicate an underestimate of lethal distances within existing GHZ models. However, the important aspect of high-energetic events for a biosphere relates to the question of whether the planetary conditions can be restored towards habitability in such a way that an extended biosphere can develop afterward. That means that not the entire planet has to be sterilized (including extremophiles) but its biosphere must not recover (see also discussion in Section~\ref{sec:MWcomp}).}

The upper panel of Figure~\ref{fig:ghz} also shows simulations by \citet{Spitoni2014,Spitoni2017} who considered supernovae and the chemical evolution of the Milky Way with \citep[solid red line; ][]{Spitoni2014} and without radial gas flow \citep[dotted red line; ][]{Spitoni2014}, as well as the effect of dust \citep[solid dark-red line; ][]{Spitoni2017} on the GHZ. After 13\,Gyr of evolution, the model without consideration of radial gas flow (solid red line) finds the most habitable region to be within 8 and 12\,kpc, while the model that includes radial gas flow enhances the habitability of the outer galactic disk. Similarly, \citet{Spitoni2017} find the galactic disk to be most habitable around 8\,kpc within their updated model.

All of the above-mentioned models, as well as further research by \citet{Morrison2015}, \citet{Forgan2017}, and \citet{Spinelli2021}, broadly agree that the Milky Way has the highest probability of providing habitable conditions either around the location of the Sun or, particularly in terms of numbers of habitable planets, towards the inner disk. However, the solid blue line in Figure~\ref{fig:ghz} shows another study by \citet{Vukotic2016} who found that habitability might increase further towards the outer disk with the peak value to be around 10-15\,kpc. As pointed out by these authors, the reason for this discrepancy might be due to the chosen stellar threshold density, star formation rate, and particularly due to dynamical effects that cause outward migration of metal-rich stars \citep[see also,][]{Gowanlock2018}.

To implement galactic habitable environments into our model, we will first consider the effect of supernovae on the habitability of the galactic disk based on the GHZ results by \citet{Gowanlock2011} and others. The evolution of metallicity will then be implemented separately afterward, as outlined in Section~\ref{sec:metallicity}. We, further, assume orbital perturbations from nearby stars to be negligible within the disk and do not consider the effect of GRBs for estimating the maximum number of EHs (see appended Section~\ref{sec:otherAGHZ}).

{Finally, we need to note that we do not consider radial mixing of gas and stellar migration within the galactic disk although such processes were likely important for galactic chemical evolution \citep[e.g.,][]{Sellwood2002,Roskar2008,Schoenrich2009,Minchev2013,Spitoni2015}. This could affect our results since stars can migrate from more hostile towards more habitable regions and vice versa, thereby affecting the conditions for life within these systems. Based on age-metallicity relations, it was even suggested \citep{Lu2022Sun,Baba2023} that the Sun itself migrated outwards from a potential galactocentric birth location around 5\,kpc towards its present position at around 8.1\,kpc. If so, the Sun would have been born in a region with higher star formation and supernova rates, affecting the evolution of habitability in the Solar System. Enhanced supernovae and Wolf-Rayet winds, for instance, could explain the increased implantation of short-lived radionuclides into the protoplanetary disk of the Sun \citep{Baba2023}, a potential beneficial feature related to the evolution of EHs (see appended Section~\ref{sec:26Al}). If the Solar System remained in the inner disk, however, it could have affected its habitability negatively due to the increased SN rates\footnote{However, it also was suggested that the radial migration could have induced Snowball-Earth episodes via the Sun's repeated passage through the galactic spiral arms \citep{Tsujimoto2020}.}.}

\subsubsection{The supernova requirement, $\alpha_{\rm GHZ}^{\rm SN}$}\label{sec:SN}

\paragraph{Implementing $\alpha_{\rm GHZ}^{\rm SN}$:}\label{sec:SNImp}

For considering the effect of supernovae as part of $A_{\rm GHZ}$, we simply implement the fraction of stellar systems not sterilized by supernovae as a function of galactocentric distance, $r$. Because modeling the evolution of the galactic supernova rate, and implementing it self-consistently into our model is beyond the scope of the present work, we rely on the different models and results provided by other studies. Since most of these models also consider the evolution of metallicity, we focus on the results provided by \citet{Gowanlock2011}. Even though these authors also implement the effect of metallicity into their model, they additionally provide a simple probability distribution of stellar systems that are not affected by nearby supernovae over the last 4\,Gyr (see discussion in the last section)\footnote{We could have also utilized the supernova rates provided by \citet{Snaith2015} to be consistent with our chosen SFH and metallicity evolution (see Section~\ref{sec:metallicity}). However, as these authors only provide rates for the inner and outer disk and not for smaller spatial scales, this would significantly affect our spatial resolution. In addition, we would also need to implement our own sterilization model, a potential task for a future study.}.

For our nominal case, we implement the probability distribution of `model 4' by \citet{Gowanlock2011}, which makes use of the Kroupa IMF \citep{Kroupa2001} with $M_{\rm low} = 0.08$\,M$_{\odot}$, and the stellar number density provided by \citet{Juric2008}. The derived stellar distribution of our nominal case within the disk's r-z plane can be seen in the lower panel of Figure~\ref{fig:ghz}. Even though the relative fraction of stars that are not affected by supernovae is significantly higher towards the outskirts of the galaxy, the inner part still hosts a significantly larger absolute number of non-affected stars, which is due to the much higher stellar number density towards the galactic center.

\paragraph{The effect of $\alpha_{\rm GHZ}^{\rm SN}$:}\label{sec:SNEff}

The highest density of stars that are not affected by supernovae within our nominal case can be found at a galactocentric distance of 7.0\,kpc, and about 49\% of such stars are located farther away from the galactic center than the Sun. In total, $2.38 \times 10^{10}$ stars are yet fulfilling our requirements, that is, 23.45\% of all stars within the galactic disk between $r = 2.0-21.0$\,kpc. If we keep all input parameters the same but only change from `model~4' to `model~3' by \citet{Gowanlock2011}, the number of stars not affected by SN explosions rises towards $2.65 \times 10^{10}$ stars; by also varying IMF, SFH, and $M_{\star\rm low}$, we find a total range of $2.38(+0.45/-0.04)\times 10^{10}$ stars that reside within galactic habitable environments.

If we change the SFH within our nominal case to the one by \citet{Naab2006} to be in line with the SFH used by \citet{Gowanlock2011} but keep everything else the same, we obtain $2.42 \times 10^{10}$ stars which is an increase of $\sim1.5\%$. If we would take the IMF by \citet{Salpeter1955} together with the SFH by \citet{Naab2006} and combine this with `model~1' from \citet{Gowanlock2011}, which is indeed based on the Salpeter IMF and has the highest probabilities for not being sterilized by supernovae, the number of remaining stars rises significantly towards $3.91 \times 10^{10}$ stars. This, however, must be an overestimate since the Salpeter IMF feeds into this number twice, i.e., through the IMF used in both models, ours and `model~1' by \citet{Gowanlock2011}. There would consequently be too few SNs on the one hand but too many low-mass stars on the other.

{As noted above, recent studies by \citet{Thomas2023} and \citet{Brunton2023} found that supernovae may be lethal over larger distances than initially thought. \citet{Thomas2023} calculated that the cosmic ray flux from an average SN can be lethal up to 20\,pc instead of 8-10\,pc as usually assumed. Depending on the total energy of the supernova, its conversion efficiency to cosmic rays and variations in interstellar transport, the lethal distance estimated by \citet{Thomas2023} can vary between 4 and 160\,pc. \citet{Gowanlock2011} take 8\,pc from \citet{Gehrels2003} as the average lethal distance for an SN with lethal distances varying between about 2 and 27\,pc depending on supernova type and luminosity. One can therefore expect that implementing the new calculations by \citet{Thomas2023} would decrease the number of stellar systems that are not affected by SNs in our model. In addition, \citet{Brunton2023} estimate that the emitted X-ray flux of supernovae can be lethal up to 50\,pc, an effect that may likely lead to an additional decrease of unaffected systems. However, implementing these very recent results into our model is beyond the scope of the present study.}

If we finally take into account studies that also include metallicity within their GHZ, the number of stars that can still host EHs decreases significantly and ranges from $0.70 \times 10^{10}$ stars for the \citet{Spitoni2017} model to $2.00 \times 10^{10}$ stars for the \citet{Vukotic2016} model. We will compare these numbers with our implemented metallicity distribution in the next section.

\subsubsection{The metallicity requirement, $\alpha_{\rm GHZ}^{\rm met}$}\label{sec:metallicity}

\paragraph{Calculating and implementing $\alpha_{\rm GHZ}^{\rm met}$:}\label{sec:MetImp}

Planet occurrence rates seem to be correlated with the metallicity of the host stars \citep[e.g.,][]{Adibekyan2019,Zhu2019,Hansen2021}. {Based on observations, \citet{Fischer2005} were the first to present a probability relation between a star's metallicity, [Fe/H], and the occurrence rate of gas giant planets with orbital periods shorter than 4 years around FGK main-sequence stars, i.e.,}
\begin{equation}\label{eq:metGas}
  \mathcal{P}(\rm GGP) = 0.03 \times 10^{2.0[\rm Fe/H]},
\end{equation}
{where $\mathcal{P}$(GGP) denotes a star's probability of hosting such gas giant planet over the metallicity range $-0.5 < \rm [Fe/H] < 0.5$. A correlation between giant plants and metallicity was later also confirmed by planet population synthesis models \citep[e.g.,][]{Mordasini2012,Emsenhuber2021}}. In addition, \citet{Johnson2012} proposed a critical metallicity value needed for a planet to form that also depends on the orbital distance, $r_{\rm pl}$, from the star, i.e.,
\begin{equation}\label{eq:planetForm}
  \mathrm{ [Fe/H]_{\rm crit}} \backsimeq -1.5 + \log(r_{\rm pl}/\rm 1\,AU).
\end{equation}
Based on this relation \citet{Johnson2012} further suggested that the first Earth-like planets should have formed around stars with a metallicity of $Z_{\star} \gtrsim 0.1$\,Z$_{\odot}$\footnote{Here, $Z_{\odot} = 0.0196$ \citep[e.g.,][]{Vagnozzi2019} is the metallicity of the Sun. Stellar metallicity, $Z_{\star}$, can be converted to [Fe/H] via the simple relation $Z_{\star}/Z_{\odot} =  10^{\rm [Fe/H]}$.}. {Planet population synthesis models \citep{Emsenhuber2021} later found that the occurrence rate of Earth-like planets may peak at a metallicity around [Fe/H]\,=\,-0.2 since their formation is either limited by too few building blocks for lower metallicities or negatively affected for higher metallicities by more massive and potentially dynamic planets that either accrete or eject terrestrial planets. {Earlier observations found an average metallicity for rocky exoplanet hosting stars of [Fe/H]\,=\,-0.02, i.e., about solar \citep{Buchhave2014}.} For low-mass stars, the same synthesis models obtained a strong positive dependence between metallicity and the occurrence rate of terrestrial habitable zone planets \citep{Burn2021}.}

\citet{Zhu2019} and \citet{Hansen2021} further investigated the relation between planetary radius, $R_{\rm pl}$, and metallicity of the host star and found that the {positive} correlation between planet occurrence rate and metallicity still holds for small-sized planets below $R_{\rm pl} < 4$\,R$_{\oplus}$, where R$_{\oplus}$ denotes the radius of the Earth. Based on the Kepler sample, all confirmed KOI planets with such radii are found around stars with $Z = Z_{\star}/Z_{\odot} \gtrsim 0.25$, i.e., with $\rm [Fe/H] \geq -0.6$ \citep{Adibekyan2019,Hansen2021}. \citet{Jiang2021} further found a relation between planet mass, $M_{\rm pl}$, and stellar metallicity; almost no planets with $M_{\rm pl}\lesssim 2$\,M$_{\oplus}$ can be found for metallicities below $\rm [Fe/H] \sim -0.3$, i.e., for $Z \sim 0.5$. However, predefining a definitive metallicity cutoff, $Z_{\rm min}$, for rocky exoplanets might at present be elusive. We will, therefore, implement $Z_{\rm min} = 0.3$ for our nominal case, $Z_{\rm min} = 0.1$ for our maximum case and $Z_{\rm min} = 0.5 - 0.75$ for our minimum case, respectively.

\begin{figure}
\centering
\includegraphics[width = 1.0\linewidth, page=1]{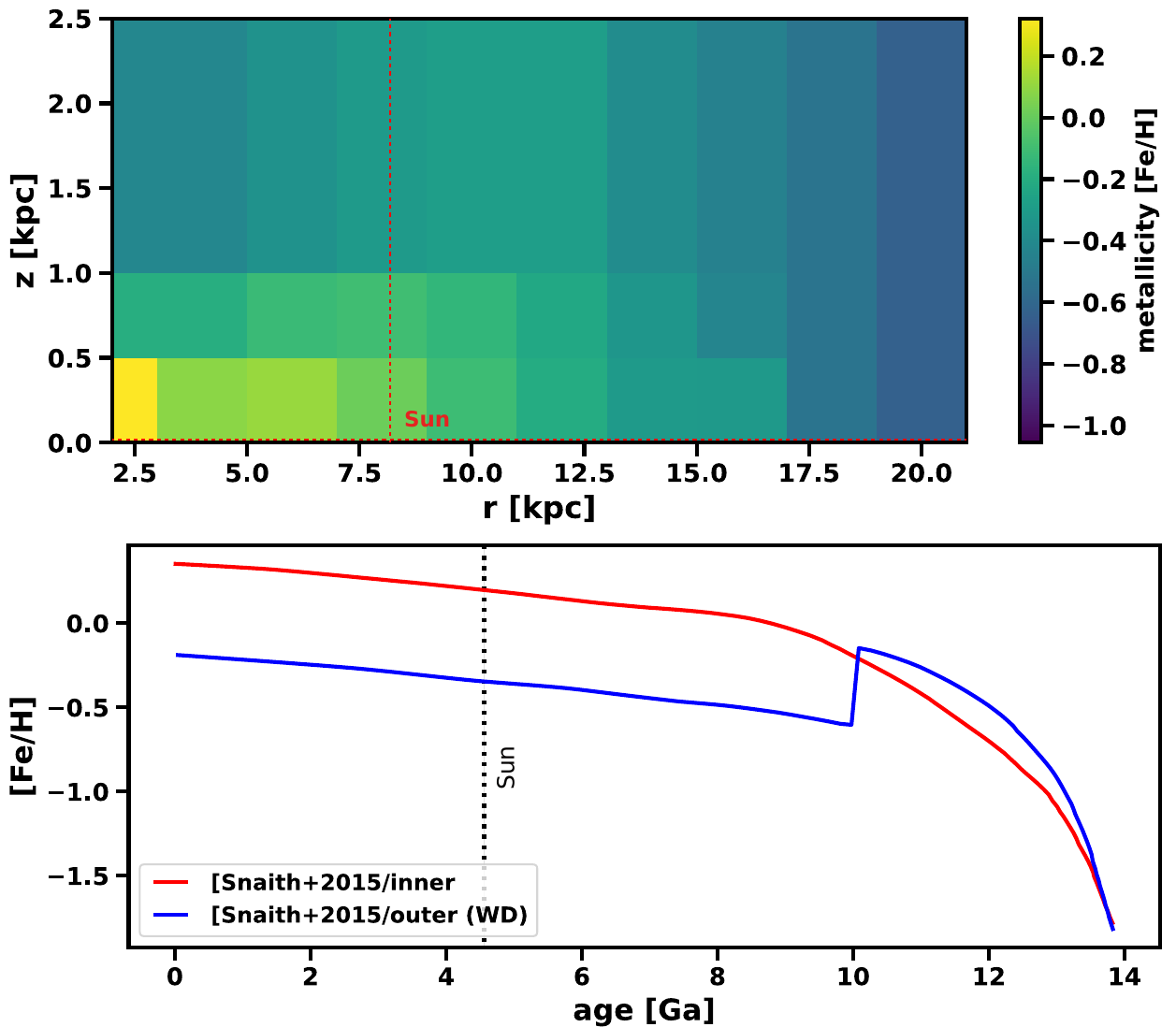}
\caption{Upper panel: The metallicity [Fe/H] distribution of the present-day galactic disk according to \citet{Hayden2015}. Closer than $r = 3$\,kpc and beyond $r = 15$\,kpc the value for [Fe/H] was extrapolated with the fitting function provided by \citet{Hayden2015}; for $z>2$\,kpc it was further assumed to be the same as for $z<2$\,kpc. Lower panel: Evolution of [Fe/H] for the inner (red; model with dilution; see text) and outer disk (blue) according to the best-fit model by \citet{Snaith2015}.}
\label{fig:met1}
\end{figure}

To implement the importance of metallicity (i.e., $\alpha_{\rm GHZ}^{\rm met}$) into our model, we have to consider
\begin{enumerate}
  \item the spatial metallicity distribution within the present galactic disk (see Figure~\ref{fig:met1}, upper panel), and
  \item the temporal metallicity evolution of the galactic disk over time (see Figure~\ref{fig:met1}, lower panel).
\end{enumerate}
If we did not account for the evolution of galactic metallicity, we would overestimate G and F-type stars with lower metallicities but at the same time underestimate M and K dwarfs with lower metallicities, since old, low-metallicity stars are predominantly low-mass stars.

To take account of the present-day metallicity, we will closely follow \citet{Westby2020} and implement the galactic metallicity distribution functions (MDFs) suggested by \citet{Hayden2015}. These authors provide skewed Gaussian metallicity distributions, including mean values, standard deviation, and the respective skewness, for different galactocentric distances within the disk. Similar to \citet{Westby2020}, we will use these values to calculate the galactic probability distribution of stars above a certain metallicity threshold. Since \citet{Hayden2015} only provide data between 3\,kpc and 15\,kpc, we will make use of their fitting relation for the inner and outer skirts of the disk, i.e., $\mathrm{[Fe/H]} = -0.053 \times r + 0.428$. For $z = 2.0-2.5$\,kpc, we assume the same mean metallicity value as given by \citet{Hayden2015} for $z = 1.0-2.0$\,kpc, which may slightly overestimate the number of stars above our metallicity thresholds. Similarly, skewness and standard deviations are assumed to stay constant outwards into the disk. The upper panel of Figure~\ref{fig:met1} shows the mean values for the metallicity distribution according to \citet{Hayden2015} with the extensions discussed in this paragraph. Within $r=3-15$\,kpc and $z=0-2$\,kpc, the range for which these authors provide observational data, the highest mean value for the metallicity of $\rm [Fe/H] = + 0.11$ was measured at a galactocentric distance and height of $5 < r < 7$\,kpc and $0 < z < 0.5 $\,kpc, respectively, while the lowest measured value of $\rm [Fe/H] = -0.42$ was found surprisingly close to the center at $3 < r < 5$\,kpc but for $1 < z < 2 $\,kpc.

Since we will further account for the metallicity evolution of the Milky Way, we will implement the best-fit [Fe/H]-age tracks provided by \citet{Snaith2015} for the inner and outer disk, which correlate with the best-fit SFH from their chemical evolution model. These tracks can be seen in the lower panel of Figure~\ref{fig:met1} where `WD' (with dilution) denotes their best-fit model for the outer disk. For this, \citet{Snaith2015} assumed a galactic accretion event at 10\,Ga that diluted the in-situ gas of the outer disk with primordial gas to match the age-metallicity and age-$\alpha$ observations by \citet{Haywood2013}. {It is worth noting that a similar dilution effect was later also considered by \citet{Spitoni2019Dilution,Spitoni2020Dilution,Spitoni2021Dilution} and \citet{Lian2020Dilution1,Lian2020Dilution2} within their chemical evolution models for reproducing the abundance ratio of [$\alpha$/Fe]\footnote{{The ratio of $\alpha$-process elements vs iron, see, e.g., \citet{Burbidge1957Synthesis} and \citet{Woosley1992Alpha} for nucleosynthesis via the $\alpha$-process.}} vs [Fe/H] in APOGEE\footnote{The Apache Point Observatory Galactic Evolution Experiment, see, e.g., \citet{Wilson2010}.} stars in the outer disk.}


To combine the evolutionary tracks from \citet{Snaith2015} with the present-day distribution from \citet{Hayden2015}, we re-normalize the [Fe/H]-age tracks for inner and outer disk with the present-day values of the different Milky Way regions so that the evolving metallicity of the best-fit model by \citet{Snaith2015} will reach the certain regional [Fe/H]-values as observed by \citet{Hayden2015}. The average metallicity of all stars within a certain region will hence be re-normalized to meet the present-day metallicity value of this respective region. {If we generally average over the [Fe/H]-values provided in \citet{Hayden2015} for the outer disk region between 7 and 15\,kpc (for a disk height $0<z<0.5$\,kpc), we receive [Fe/H]\,=\,-0.19, which is very close to [Fe/H]\,$\sim$\,-0.20, the value \citet{Snaith2015} receive for their outer disk for the present day (see their Fig.~9d). For the inner disk, \citet{Snaith2015} find [Fe/H]\,$\sim$\,+0.35 which is slightly higher than [Fe/H]\,=\,+0.23, the average value in \citet{Hayden2015} for $3<r<7$\,kpc and $0<z<0.5$\,kpc. Note, however, that the inner disk in \citet{Snaith2015} covers the entire disk with radii $r<7-8$\,kpc, which may explain such a higher value.} To calculate the Gaussian distribution, we will assume that mean, standard, and skewness remained constant over the entire age of the Milky Way.

\paragraph{The effect of $\alpha_{\rm GHZ}^{\rm met}$:}\label{sec:MetEff}

\begin{figure}
\centering
\includegraphics[width = 1.0\linewidth, page=1]{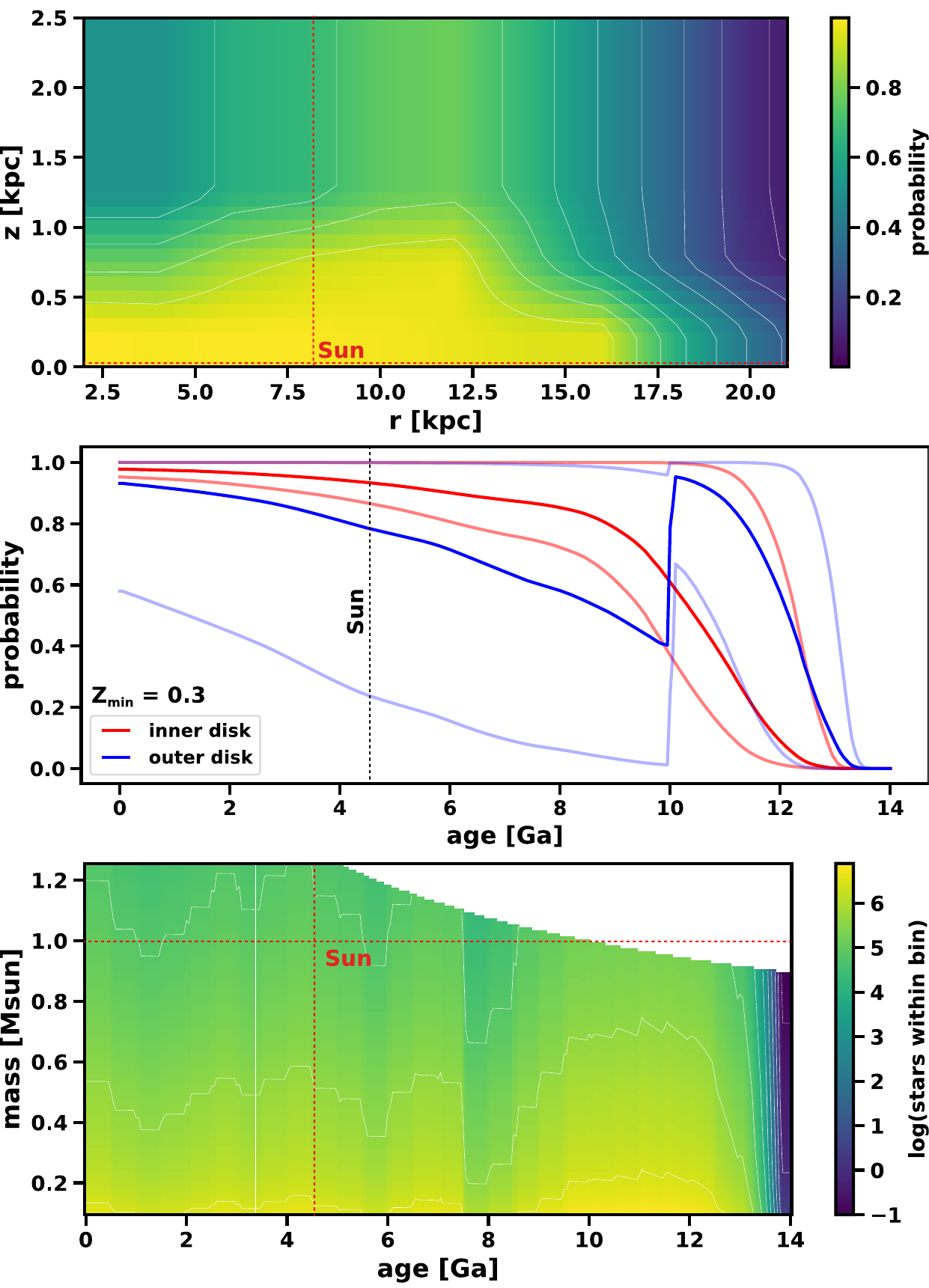}
\caption{Upper panel: The present-day probability for a certain star to be above a metallicity threshold of $Z_{\rm min} = 0.3$ (nominal case) as a function of galactocentric distance $r$ and height $z$. This probability is based on the metallicity distribution functions by \citet{Hayden2015}. Middle panel: The probability of being above the same threshold ($Z_{\rm min} = 0.3$) for each time step for the inner (non-transparent red line) and outer disk (non-transparent blue line). The probability is averaged over each region (i.e., spatial bin within our model) that holds a different present-day value of [Fe/H] within the inner and outer disk, respectively. {The probabilities to be above $Z_{\rm min} = 0.3$ of those regions with the lowest and highest [Fe/H]-values in the inner and outer disk are additionally visualized by the transparent red and blue lines, respectively, to illustrate the entire probability spread in the disk.} Here, the probability is based on the evolutionary [Fe/H]-tracks from the best-fit model by \citet{Snaith2015} but re-normalized to the present-day values of each region. Lower panel: The mass-age distribution of stars within the GHZ that have a metallicity of $Z > 0.3$ (within a bin size of 0.1\,M$_{\odot}$ times 50\,Myr). As can be seen, most of the oldest stars do not fulfill the metallicity requirement, which the middle panel can easily explain.}
\label{fig:met2}
\end{figure}

The upper panel of Figure~\ref{fig:met2} shows the probability of any presently existing star to be above the threshold value of our nominal case, i.e., $Z_{\rm min} = 0.3$, as a function of galactocentric distance $r$ and height $z$ and based on today's galactic [Fe/H]-values observed by \citet{Hayden2015}.  As described above, each region\footnote{Here, a galactic region is defined as a spatial bin of 0.1\,kpc times 0.1\,kpc in all our simulations. The present-day [Fe/H] values were interpolated onto these smaller galactic regions via Python's \texttt{scipy.interpolate} package, see \url{https://docs.scipy.org/doc/scipy/reference/interpolate.html}.} with a different present-day [Fe/H]-value will have a slightly different [Fe/H] evolution over time, since we require that the average metallicity of all stars presently existing within such a region has to reach the present-day value of [Fe/H] finally. The middle panel of Figure~\ref{fig:met2}, therefore, illustrates the probability of being above such threshold for each time step averaged over all galactic regions for the inner (red line) and outer disk (blue line). One can see that the galactic accretion event at 10\,Ga significantly reduced the probability of a star being above the metallicity threshold. The light red and blue lines further show those galactic regions in the inner and outer disks that presently have the minimum and maximum probability values, respectively.

The lower panel of Figure~\ref{fig:met2} illustrates the age distribution of remaining stars that fulfill our nominal case metallicity threshold of $Z_{\rm min} = 0.3$. There is a strong increase in stars above the threshold immediately after the galaxy's birth as many of the oldest stars have a too low metallicity to allow the formation of rocky exoplanets. However, since the [Fe/H]-tracks are steeply rising (as seen in this figure's middle panel), this effect is mostly restricted to very early ages. At the Sun's birth age, for instance, the mean metallicity of stars born at the same age and region measures $\rm [Fe/H] \sim 0.12$, a value that is above the solar value.

In total, 69.3\% of all remaining stars meet the metallicity threshold of $Z_{\rm min} = 0.3$, that is, $1.65 \times 10^{10}$ stars. For $Z_{\rm min} = 0.1$, this value rises toward $2.42 \times 10^{10}$ stars while it decreases toward $1.30 \times 10^{10}$ stars for $Z_{\rm min} = 0.5$. If we increase the metallicity threshold to $Z_{\rm min} = 0.75$, only $0.94 \times 10^{10}$ stellar candidates remain. By comparing the latter result with GHZ studies that include the effect of metallicity, this value closely resembles the range of stars we obtain by implementing the probability distributions of \citet{Spitoni2014} and \citet{Spitoni2017} for which we find $0.70 \times 10^{10}$ to $1.16 \times 10^{10}$ stars. By implementing the one by \citet{Vukotic2016}, we obtain $\sim 2.0 \times 10^{10}$ stars, a value that is relatively close to $2.42 \times 10^{10}$ stars for our case with $Z_{\rm min} = 0.1$. It is, therefore, reasonable to take $Z_{\rm min} = 0.1$ as our maximum and $Z_{\rm min} = 0.75$ as our minimum case\footnote{As comparison, \citet{Westby2020} take $Z_{\rm min} = 1.0$ for their `strong scenario'. If we take the same, only $\sim0.65 \times 10^{10}$ stars remain in our model.}. If we also vary SFH, IMF, etc., we obtain a total range of $1.65(+0.77/-0.72)\times 10^{10}$ stars.

\begin{figure}
\centering
\includegraphics[width = 1.0\linewidth, page=1]{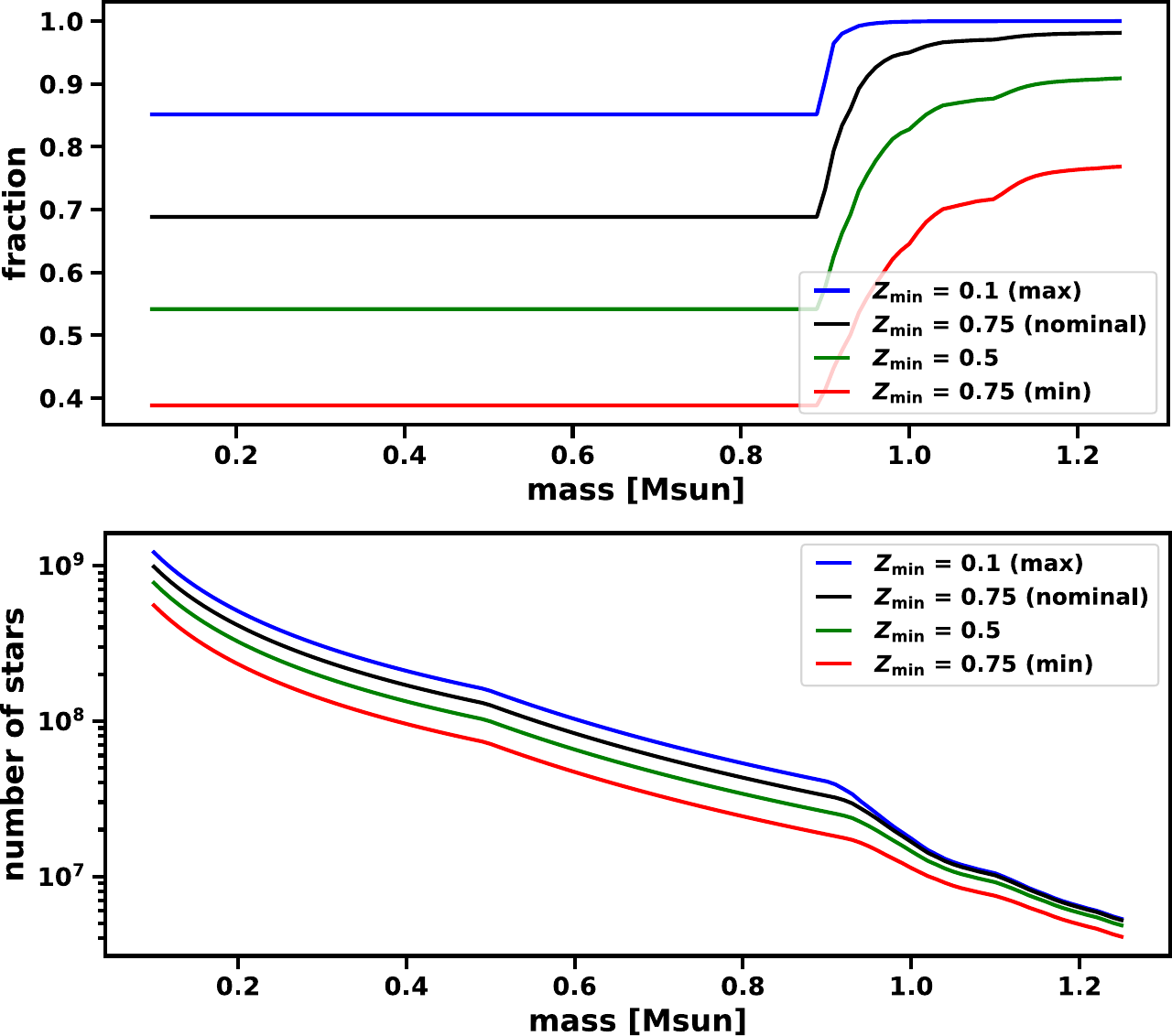}
\caption{Upper panel: The fraction of stars above the different metallicity thresholds $Z_{\rm min}$ as a function of stellar mass. Lower panel: The number of stars above $Z_{\rm min}$, again as a function of stellar mass.}
\label{fig:metFrac}
\end{figure}

Finally, 76.69(+1.00/-0.45)\% of the remaining stars are low-mass M dwarfs which is a slightly lower value than for the entirety of GHZ stars, for which it is 77.27(+0.77/-0.0)\%. This small shift relates to the fact that the metallicity of the ISM has been increasing since the galaxy's birth, thereby implying that M dwarfs are, on average, below the threshold more often than heavier stars. This can be seen in Figure~\ref{fig:metFrac}, where the upper panel illustrates the stellar fraction above the respective metallicity threshold $Z_{\rm min}$. This fraction remains constant for stars with a main-sequence lifetime larger than the present age of the Galaxy. As soon as the main-sequence lifetime becomes shorter, however, the stellar fraction starts to increase since stars born close to the birth age of the Milky Way already left the main sequence. This is also exemplified in this figure's lower panel, which shows the number of stars above the respective $Z_{\rm min}$, again as a function of stellar mass. For larger $Z_{\rm min}$, more M and K dwarfs are below the threshold and the relative fraction of G and F stars is consequently rising. A total of 17.83(+0.10/-1.54)\% of the remaining stellar sample are hence K dwarfs, while 4.43(+0.34/-0.17)\% and 1.15(+0.34/-0.03)\% are G and F stars, respectively.

As a final note related to metallicity, we point out that the occurrence rate of rocky exoplanets will not only depend on a lower metallicity cutoff. Their formation probability may also decrease for high metallicities. This seems to be likely due to (i) a rise in the occurrence rates of Hot Jupiters \citep[e.g.,][]{Buchhave2018,Osborn2019} and warm, dynamically active giant planets \citep{Schlecker2021} with increasing metallicity, and (ii) an increase in rocky building blocks that may lead to the growth of more massive planets with hydrogen-dominated atmospheres. Through assuming (i) and the related destruction of rocky exoplanets by migrating Hot Jupiters, \citet{Prantzos2008} and subsequently \citet{Carigi2013}, used Equation~\ref{eq:metGas} by \citet{Fischer2005} to calculate the probability of a star hosting Earth-like planets but not Hot Jupiters, thereby taking into account a potential lower probability of rocky exoplanets around high-metallicity stars. However, \citet{Spitoni2014} already pointed out that this assumption might be questionable since the relation by \citet{Fischer2005} covers any gas giant planets (GGPs) with orbital periods shorter than 4 years\footnote{\citet{Spitoni2014} used the same relation but correctly implemented it as the probability of forming an Earth-like planet and not a gas giant (instead of Hot Jupiter).}. Since GGPs on wider orbits (as is the case in the Solar System) will not necessarily destroy or disturb Earth-like planets in circumstellar habitable zones, one may expect that applying such relation could slightly underestimate the occurrence rate of rocky exoplanets around high-metallicity stars \citep[see also Fig.~8 in][]{Carigi2013} while the opposite may hold for neglecting any such relation (as in our model). This reasoning, however, is only based on argument (i) while the relevance of argument (ii) cannot be properly assessed with our present scientific knowledge. 


\subsection{The thermal stability of N$_2$-O$_2$-dominated atmosphere, $A_{\rm at}^i$}\label{sec:fat}

One of the most important requirements for an Earth-like Habitat, even though it is often overseen or ignored, is the long-term stability of its N$_2$-O$_2$-dominated atmosphere against atmospheric escape to space on the one hand (that is, the lower limit $\alpha_{\rm at}^{\rm ll}$) and its heat death on the other hand (the upper limit $\alpha_{\rm at}^{\rm ul}$).

To meet the lower limit, any atmospheric sinks such as thermal and non-thermal escape to space must be either negligible or at least smaller than any atmospheric sources (e.g., volcanic degassing or biological processes) that would otherwise replenish nitrogen and oxygen into the atmosphere. Here, thermal escape is mostly driven by the incident XUV flux\footnote{As XUV flux we understand the X-ray and extreme ultraviolet (EUV) wavelengths that range from 0.517-12.4\,nm and 10.0-92.0\,nm, respectively; see \citet{Johnstone2021Stars}.} of the host star, {which heats and expands the upper atmosphere (see appended Section~\ref{sec:XUV}),} and thereby removes neutral gas via a hydrostatic or hydrodynamic flow {into space} \citep[e.g.,][]{Tian2008,Erkaev2013,Kubyshkina2021,Johnstone2019,Johnstone2021}. Non-thermal escape, on the other hand, mostly removes ionized gas from the atmosphere and is mainly driven by CMEs and the stellar wind of the star and/or the polar outflow from a planet's own magnetosphere \citep[][]{Khodachenko2007,Lammer2007,Airapetian2017,GarciaSage2017,Dong2017Proxima,Dong2018,Dong2019,Scherf2021}.

The upper limit, $\alpha_{\rm at}^{ul}$, of the necessary requirement, $A_{\rm at}^i$, is set by the stellar bolometric luminosity, $L_{\rm bol}$, of the star and the connected stellar effective surface flux, $S_{\rm eff,\star}$, at a planet's orbit, which is slowly increasing over a star's main-sequence lifetime. At some specific point, its luminosity will become so strong that the greenhouse effect will enter the moist and/or runaway greenhouse stage, and life will cease to exist. For the Sun, this was first estimated to be at a solar age of about 5.5\,Gyr \citep{Kasting1993} and it coincides with the time when the Earth crosses the inner HZ boundary. Based on the HZ model by \citet{Kopparapu2014} and the solar evolution model by \citet{Girardi2000}, \citet{Waltham2019} calculated this value to be between 5.45-5.74\,Gyr. From here on, we will denote the upper limit as $\alpha_{\rm at}^{S_{\rm eff,\star}}$.

{In the appended Sections~\ref{sec:StellarEvo} and \ref{sec:XUV}}, we discuss the scientific background of stellar evolution, {with a particular focus on the relationship between stellar mass, rotational and XUV flux evolution} {(Section~\ref{sec:StellarEvo})}, and the fundamental role of the stellar radiation (XUV flux and flares) and plasma environment (stellar winds and CMEs) for the stability of Earth-like atmospheres and the destruction of ozone {(Section~\ref{sec:XUV}). The results presented in these appended sections indicate the importance of considering the evolution of the entire radiation and plasma environment of the various stellar spectral classes. Here, however, we will only consider the role of the stellar XUV flux, but will neither include (super)flares nor the role of non-thermal escape into our model. For the latter, the role of intrinsic magnetic fields for atmospheric protection is by now insufficiently investigated and there are no quantifiable model approaches that we can include in our framework, particularly for the evolution of stellar winds and CMEs on stars other than solar-like ones. However, this is an important topic, and we therefore strongly encourage the reader to consult these appended sections.

In the next section, we derive a distribution of stars that can presently host HZCL planets with thermally and climatically stable N$_2$-O$_2$-dominated atmosphere by including the aforementioned lower and upper limits. The stellar XUV flux evolution as a function of stellar mass will serve as a lower threshold by providing lower age limits, for which the XUV flux has declined below a specific threshold level. The bolometric luminosity, $L_{\rm bol}$, on the other hand, will serve as an upper age limit by removing those stars from our sample, for which $S_{\rm eff}$ becomes too strong to support complex life.

\subsubsection{The lower stellar limit, $\alpha_{\rm at}^{\rm XUV}$}\label{sec:XUVEffect}

\paragraph{Implementing $\alpha_{\rm at}^{\rm XUV}$:}\label{sec:XUVImp}

{A star's XUV irradiation is absorbed in the upper atmosphere of a planet, which leads to the heating and expansion of its thermosphere (see Figure~\ref{fig:LxLimit} in the Appendix). As we detail in the appended Section~\ref{sec:XUV}, this effect can be specifically important for Earth-like atmospheres, for which XUV surface fluxes only 5 to 6 times as high as received by today's Earth will be enough for the atmosphere to expand adiabatically and to erode into space within a few Myr \citep[e.g.,][]{Tian2008,Johnstone2021}. Since CO$_2$ serves as an infrared coolant \citep[e.g.,][]{Kulikov2007,Lichtenegger2010,Johnstone2021}, it can prolong the thermal stability of Earth-like atmospheres and we derive certain stability thresholds for nitrogen-dominated atmospheres with maximum mixing ratios of $x_{\rm CO_2,max}$\,=\,1\% and $x_{\rm CO_2,max}$\,=\,10\% CO$_2$ in the appended Section~\ref{sec:XUV}. This important phenomenon implies that stars with certain XUV surface fluxes in the HZCL cannot host planets with stable N$_2$-O$_2$-dominated atmospheres\footnote{{One has to highlight that planetary atmospheres evolve over time as does a star's XUV flux. If a higher CO$_2$ concentration is therefore prevalent on a young planet it could have offset the higher XUV flux of the younger star, at least to some certain extent, because of CO$_2$ being an infrared coolant. However, for XUV surface fluxes above $\sim15-20\,\rm erg\,s^{-1}\,cm^{-2}$, even pure CO$_2$ atmospheres may not survive \citep[e.g.,][]{Johnstone2021}.}}. The lower stellar limit, $\alpha_{\rm at}^{\rm ll}$ of $A_{\rm at}(\prod_{i=1}^n\alpha_{\rm at}^{i})$, is therefore of crucial importance for the distribution of EHs in the Galaxy.}

For implementing $\alpha_{\rm at}^{\rm ll}$, which we simply implement as XUV limit, $\alpha_{\rm at}^{\rm XUV}$, we will calculate the X-ray and/or XUV surface flux, $F_{\rm X}$ and $F_{\rm XUV}$, in the middle of the HZCL for a range of stellar masses. For this, we will make use of the Python package Mors\footnote{The latest version of the model can be found at \url{https://github.com/ColinPhilipJohnstone/Mors}.}, which solves the stellar evolution model developed by \citet{Johnstone2021}. The package implements a grid of stellar isochrones by \citet{Spada2013} as input for the basic physical parameters and it can be used to simulate derived stellar parameters such as X-ray and EUV luminosities as functions of stellar mass and age for any percentile of the stellar rotational distribution\footnote{See appended Section~\ref{sec:StellarEvo} for a discussion on the relation between stellar rotational and XUV flux evolution.}. Noteworthy, \citet{Kubyshkina2022} were able to reproduce the mass-radius distribution of presently known exoplanets as a function of stellar activity and evolution by implementing Mors into their model. Since Mors can simulate any star between $M_{\star} = 0.1-1.25$\,M$_{\odot}$, this also predefines the stellar mass range we consider within our model. {Although this approach excludes any low-mass M dwarfs below $M_{\star}=0.1\,$M$_{\odot}$ such as Trappist-1, this lower limit does not significantly affect our derived results since, among other reasons (see Section~\ref{sec:mdwarfs}), Earth-like atmospheres are unlikely to be thermally stable around these low-mass dwarfs, as we derive in the following sections.}

Through this approach, we will exclude any star for which $F_{\rm X}$ and/or $F_{\rm XUV}$ will be too high to permit the long-term stability of an N$_2$-O$_2$-dominated atmosphere. This statement, however, deserves some further explanation.

We will focus on the HZCL since this concept provides important boundary conditions for our model. \citet{Schwieterman2019}, and more recently \citet{Ramirez2020}, define the HZCL \citep[or CLHZ as abbreviated by][]{Ramirez2020} as the zone around a star where (i) liquid water can exist at a planet's surface and (ii) the CO$_2$ partial pressure is below the toxicity limit for complex life (see Section~\ref{sec:phil}). Both studies find a partial pressure of $p$CO$_2 \sim 0.1$\,bar to be the upper toxicity limit for {advanced metazoans, i.e., all animals except sponges\footnote{{Interestingly, sponges can also survive under very low $p$O$_2$ values as low as 0.5-4\% of Earth's present $p$O$_2$, although they will be limited to small size under such extreme conditions \citep{Mills2014Sponges,Knoll2014}.}}}. This pressure limit refines the outer HZ boundary since the farther away a planet might be from its host star, the more greenhouse gases will be needed to counterbalance the decrease in the stellar surface flux. A maximum amount of $p$CO$_2$ therefore restricts such boundary and defines the HZCL.

\citet{Schwieterman2019} provide HZCL boundaries for three different values of the CO$_2$ partial surface pressure, i.e., $p$CO$_2 = \{0.01,0.1,1.0\}$\,bar, and show that for a solar-like star, the HZCL is only 21\%, 32\%, and 50\% as wide as the conventional HZ, respectively. \citet{Ramirez2020} refines the HZCL for $p$CO$_2$\,=\,0.1\,bar by using an advanced energy balance model that also includes the effect of clouds and finds that it would be $\sim$35\% wider than initially found by \citet{Schwieterman2019}. Trace species such as N$_2$O and CO$_2$, pressure broadening by N$_2$, planetary and atmospheric mass, continental distribution, surface gravity, the surface water content or planetary rotation rate, however, can further alter HZ or HZCL boundaries \citep[e.g.,][]{Vladilo2013,Kopparapu2014,Schwieterman2019,Kodama2019}.

Since we are interested in the maximum number of EHs, i.e., HZCL planets that can host N$_2$-O$_2$-dominated atmospheres with minor amounts of CO$_2$, we can combine the results obtained by \citet{Johnstone2021} on the stability of N$_2$-dominated atmospheres with the concept of the HZCL. For this, we assume any planet to be in the middle of the HZCL even though the actual positions of HZCL planets will certainly vary over its entire orbital range. However, planets inside the mean HZCL distance, $d_{\rm \langle HZCL \rangle}$, will receive a correspondingly higher amount of $F_{\rm XUV}$. In contrast, planets farther out will receive a correspondingly lower one. By assuming that all the HZCL planets are evenly distributed over the entire HZCL, we can simplify the problem to the mean value, $d_{\rm \langle HZCL \rangle}$. A smaller HZCL will further result in $d_{\rm \langle HZCL \rangle}$ being closer to the star.

Similarly, we will assume that all planets have a mass of $M_{\rm pl}\,=\,$1.0\,M$_{\oplus}$ since this is the mass for which reliable simulations are available. Planets with higher masses than the Earth will tend to grow faster to reach such a high mass, and the more massive a planet, the likelier it will become that any accreted primordial atmosphere cannot be lost afterward. The so-called `Fulton Gap' \citep[e.g.,][see also Section~\ref{sec:XUV}]{Fulton2017,Fulton2018} further supports the hypothesis that planets of a certain mass/radius will tend to host a primordial atmosphere while planets with a radius of $R_{\rm pl} \lesssim 1.5$ \,R$_{\oplus}$ \citep{Rogers2015}, or even as low as $R_{\rm pl} \lesssim 1.23$ \,R$_{\oplus}$ \citep{Chen2017}, likely end up being rocky exoplanets\footnote{See also Section~\ref{sec:betaHZCL} for a more detailed discussion.}, at least for orbital periods with $P_{\rm pl} < 100$\,days \citep[e.g.,][]{Fulton2017}. Such a radius would give an upper limit for the mass of a rocky exoplanet of $M_{\rm pl}\lesssim4\,$M$_{\oplus}$. However, for planets orbiting farther out from their host star this value might become significantly lower since photoevaporation will be substantially weaker in the HZCL of solar-like stars. For a planet that orbits a G-type star at 1\,AU, an accreted H$_2$-dominated atmosphere might not be eroded subsequently, if the planet grows to a mass of only $M_{\rm pl}\sim1$\,M$_{\oplus}$ within the protoplanetary gas disk \citep{Erkaev2022}. This mass, however, will certainly be higher for lower-mass stars, and there are indications that planets around more massive stars indeed tend to have larger primordial atmospheres \citep{Lozovsky2021}.

Another upper mass limit can be derived from the degassing of volatiles from a planet's interior and, potentially, its tectonic regime. While \citet{Valencia2007} and \citet{VanHeck2011} find that plate tectonics might be equally or more frequent on super-Earths, several other studies conclude that the opposite might be true  \citep[e.g.,][]{ONeill2007,Noack2014,Miyagoshi2015}. In addition, \citet{Noack2017} and \citet{Dorn2018} for stagnant-lid, and \citet{Kruijver2021} for plate tectonic planets, found that the degassing of volatiles will start to be significantly reduced for planets with $M_{\rm pl} \gtrsim 3-4\,$M$_{\oplus}$. This provides another indication that the upper planetary mass limit may be found below $M_{\rm pl}\sim4$\,M$_{\oplus}$. A further constraint on the upper mass limit for habitable planets can be derived from the lifetime of a primordial continent. \citet{Dohm2018Size} argue that a primordial anorthositic continent on super-Earths will quickly be transported into the interior due to intense mantle convection, thereby strongly limiting the supply of nutrients needed for the emergence of life. If so, the origin of life would have to happen earlier on heavier planets than on lighter, Earth-like, ones.

On the other side of the mass scale, small-mass planets will not be able to hold on to an N$_2$-dominated atmosphere. A Mars-mass planet around a G-type star might already be too small to assure the long-term stability of such an envelope and even more massive planets might struggle to keep such an atmosphere, particularly for K and M dwarfs. {In fact, non-thermal atmospheric escape tends to reach maximum loss rates for a planetary radius of $R_{\rm pl} \sim 0.7$\,R$_{\oplus}$ \citep{Chin2024}\footnote{This gives a planetary mass of $M_{\rm pl}\sim 0.3$\,M$_{\oplus}$, calculated with the upcoming Equation~\ref{eq:MRrel}}, indicating that the lower radius of a habitable planet may approach $R_{\rm pl} \gtrsim 0.7$\,R$_{\oplus}$}. Small-mass planets will also need longer to provide stable conditions for an N$_2$-O$_2$-dominated atmosphere against escape into space but, at the same time, will become geologically inactive within a shorter duration. At some age, these limits will start to overlap, and these will overlap faster for small-mass planets than for higher-mass ones.

These arguments exemplify that for both sides of the mass distribution planets will become inhospitable to complex life for certain mass limits. However, these upper and lower limits are not (yet) strictly definable, and their mean value might diverge from $M_{\rm pl}$\,=\,1.0\,M$_{\oplus}$. However, since the distribution of planetary masses will likely have a peak value between $\log M_{\rm pl}/\rm M_{\oplus} = 0.6-1.0$ \citep{Malhotra2015}, assuming a mean value of $M_{\rm pl}\,=\,$1\,M$_{\oplus}$ might potentially be an overestimate, as long as the minimum mass is not very close to or at 1\,M$_{\oplus}$. 

As a first step for implementing the XUV limit, $\alpha_{\rm at}^{\rm XUV}$, however, we need to calculate the HZ boundaries. For this, \citet{Kopparapu2013} give the following relationship between the stellar surface flux, $S_{\rm eff}$, and the stellar effective temperature, $T_{\rm eff}$, i.e.,
\begin{equation}\label{eq:HZ}
S_{\rm eff} = S_{\rm eff,\odot} + a T_{\star} + b T_{\star}^2 + c T_{\star}^3 + d T_{\star}^4,
\end{equation}
where $T_{\star} = T_{\rm eff} - 5780$\,K. The coefficients are dependent on the chosen inner and outer boundary, which typically resemble limits such as \textit{Water Loss}, \textit{Recent Venus} or \textit{Runaway Greenhouse} for the inner, and \textit{Moist Greenhouse}, \textit{Maximum Greenhouse}, and \textit{Recent Mars} for the outer boundary \citep[see, e.g.,][for a discussion]{Kasting1993,Kopparapu2013,Kopparapu2014}. The parameter $S_{\rm eff,\odot}$ further gives the solar effective surface flux at the respective HZ boundaries in units of the solar surface flux, S$_{\rm eff,\oplus}$, that reaches the top of today's Earth's atmosphere.

We will vary between \textit{Moist Greenhouse} and \textit{Runaway Greenhouse} limits as the inner boundary, which coincide with $S_{\rm eff,\odot} \sim 1.014\,$S$_{\rm eff,\oplus}$ and $S_{\rm eff,\odot} \sim 1.0512\,$S$_{\rm eff,\oplus}$, respectively, in \citet{Kopparapu2013}, and with $S_{\rm eff,\odot} \sim 1.107\,$S$_{\rm eff,\oplus}$ for the \textit{Runaway Greenhouse} in \citet{Kopparapu2014}. At the \textit{Moist Greenhouse} limit the planetary atmospheric temperature increases to such an extent ($T_{\rm pl} \gtrsim 330$\,K) that the stratosphere becomes dominated by H$_2$O and an Earth-ocean can be lost on Gyr timescales \citep[e.g.,][]{Kasting1993,Kopparapu2013,Kasting2015,Wolf2017}. At the \textit{Moist Greenhouse} limit, on the other hand, the Sun/star induced heating reaches a certain threshold at which the thermal equilibrium can only be maintained by the evaporation of the entire surface water \citep[e.g.,][]{Nakajima1992,Goldblatt2013,Leconte2013}. These boundaries were calculated by \citet{Kopparapu2013} to be at 0.99\,AU and 0.97\,AU, respectively, for the present-day Sun.

\begin{figure*}
\centering
\includegraphics[width = 0.7\linewidth, page=1]{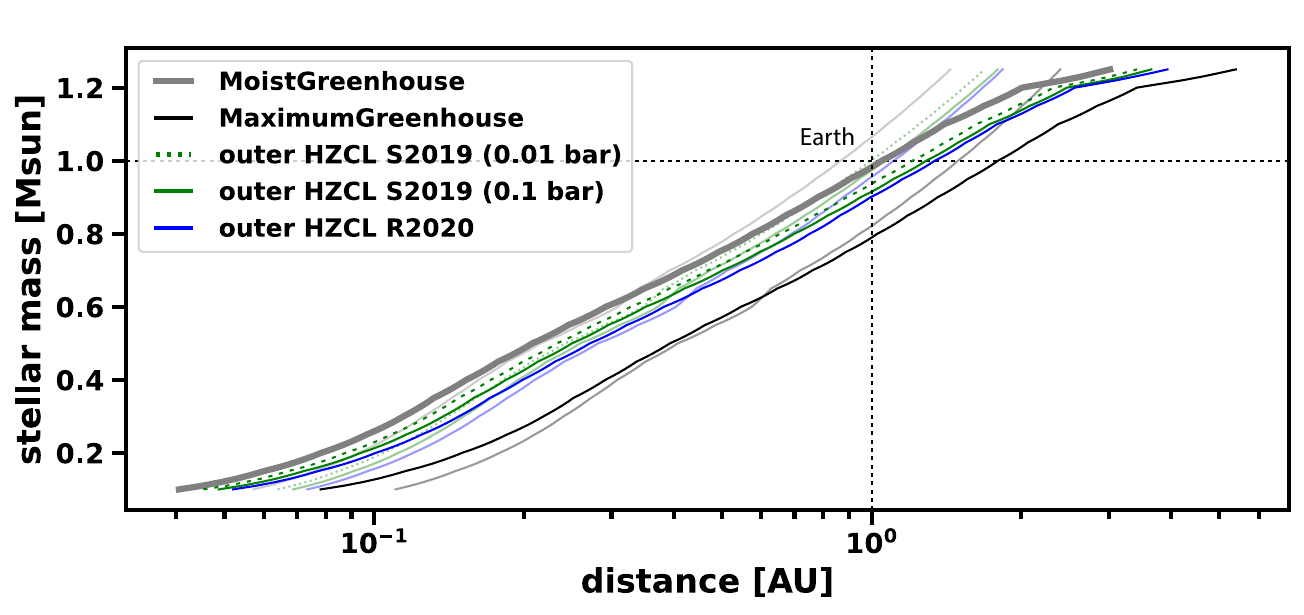}
\caption{The inner HZ boundary (thick grey line) of the HZ according to the \textit{Runaway Greenhouse} threshold and several different outer boundaries, i.e., the HZCL for $p$CO$_2 = 0.1$\,bar (green line) and $p$CO$_2 = 0.01$\,bar (green dashed line), both according to \citet{Schwieterman2019}, for $p$CO$_2 = 0.1$\,bar (blue line) according to \citet{Ramirez2020}, and for the \textit{Maximum Greenhouse} threshold (black line). All these boundaries are calculated through Mors and are for an age of 5\,Gyr. The respective light colors in the background illustrate the same HZ boundaries but for 100\,Myr. The dashed black lines illustrate the position of the Earth.}
\label{fig:HZ}
\end{figure*}

We will substitute the outer boundary of the HZ with the HZCL definitions by \citet{Schwieterman2019} and \citet{Ramirez2020}. While \citet{Schwieterman2019} gives the corresponding values for \textit{a}, \textit{b}, \textit{c}, and \textit{d}, we will directly take $S_{\rm eff}$ for a 1\,bar atmosphere from Fig.~2 in \citet{Ramirez2020}. The parameter $T_{\rm eff}$ we will further calculate with Mors for stellar masses between $M_{\star} = 0.1-1.25$\,M$_{\odot}$ and for the entire lifetime of the respective stars (with the present age of the Universe, i.e., the Hubble time of $t_0\sim13.8$\,Gyr, as maximum age). The corresponding habitable zone distances can then be calculated via \citep{Kopparapu2013,Kopparapu2014}
\begin{equation}\label{eq:HZdist}
  d_{\rm HZ(CL)} = \left(\frac{L_{\rm bol}/\mathrm{L}_{\rm bol,\odot}}{S_{\rm eff}}\right)^{0.5},
\end{equation}
where $L_{\rm bol,\odot}$ is the solar and $L_{\rm bol}$ the stellar bolometric luminosity, of which the latter will also be simulated through Mors, the Python package provided by \citet{Johnstone2021Stars}. The mean HZCL distance, $d_{\rm \langle HZCL \rangle}$, for which we will calculate our threshold, is then simply defined as the average between the inner and outer HZCL boundaries.

For our nominal case, we take the outer HZCL boundary by \citet{Ramirez2020}. By additionally taking the \textit{Runaway Greenhouse} limit by \citet{Kopparapu2014} as the inner boundary, this results in a mean HZCL distance of $d_{\rm \langle HZCL \rangle} = 1.17$\,AU for a solar-like star at 5\,Gyr. Figure~\ref{fig:HZ} further shows several different HZ boundaries for an age of 5\,Gyr calculated with Mors and Equations~\ref{eq:HZ} and \ref{eq:HZdist}. The half-transparent lines in the figure's background illustrate the same boundaries but for an age of 100\,Myr. Note that for 5\,Gyr the Earth position (dashed black lines) is already slightly outside the HZCL. This is because the bolometric luminosity of the stellar evolutionary track that closest resembles the Sun in the model of \citet{Spada2013} is slightly higher at 5\,Gyr than for the real Sun (i.e., $L_{\rm bol}/L_{\rm bol, \odot} = 1.077$).

To be consistent with the outer HZCL boundary in \citet{Ramirez2020}, we assume a maximum of $p$CO$_2 = 0.1$\,bar for the atmosphere. As can be seen in Figure~\ref{fig:LxLimit} of appended Section~\ref{sec:XUV}, this threshold coincides with a maximum value for the X-ray surface flux at $d_{\rm \langle HZCL \rangle}$ of $F_{\rm X} \sim 5$\,\rm erg\,s$^{-1}$\,cm$^{-2}$ or $F_{\rm X} \sim 7.7\,\mathrm{F}_{\rm X,\oplus}$\footnote{Thereafter, we will denote the Sun's present mean values for the X-ray, EUV, and XUV surface fluxes at Earth's orbit as $F_{\rm X,\oplus}$, $F_{\rm EUV,\oplus}$, $F_{\rm XUV,\oplus}$, respectively.}, respectively (vertical solid black line). For such a flux, the exobase of an $N_2$-O$_2$-dominated atmosphere with a partial pressure of $p$CO$_2 = 0.1$\,bar (solid violet line) is already expanded to such an extent that a complete erosion of the atmosphere will take place within less than $\sim$10\,Myr \citet{Johnstone2021}. Here, it has to be noted that the simulations by \citet{Johnstone2021} were performed with reconstructed synthetic solar spectra. As outlined in more detail in appended Section~\ref{sec:StellarEvo}, however, the Sun seems to be particularly weak in the X-ray wavelength part of the spectrum but slightly less weak in its EUV part when compared to other stars \citep{Johnstone2021Stars}. Only taking into account $F_{\rm X}$  as a threshold might, therefore, underestimate the number of stars that can presently hold an N$_2$-O$_2$-dominated atmosphere since the ratio of $F_{\rm X}$/$F_{\rm EUV}$ is on average lower for the entire rotational distribution of stellar masses between 0.1 and 1.25\,M$_{\odot}$ than for the Sun. To counter this effect, we will take into account the whole XUV wavelength range (that is, $F_{\rm XUV} = F_{\rm X} + F_{\rm EUV}$) for our nominal case which results in a threshold value for the XUV surface flux of $F_{\rm XUV,max} = 35$\rm\,erg\,s$^{-1}$\,cm$^{-2}$ for a wavelength range of 0.5-92\,nm and based on a ratio of $F_{\rm EUV} = 6 F_{\rm X}$ \citep[e.g.,][]{Tu2015,Johnstone2021Stars}. In the lower panel of Figure~\ref{fig:Lx}, one can see that such a threshold value coincides very well with the XUV flux of a slow rotator for the same age of $\sim$600\,Myr. One can, therefore, expect that this threshold value might even be slightly `optimistic' since $F_{\rm X}/F_{\rm EUV}$ is smaller for the Sun than for the average stellar distribution.

For our minimum case, we, consequently, take into account only the X-ray part of the spectrum with $F_{\rm X,max} = 5$\,\rm erg\,s$^{-1}$\,cm$^{-2}$. For the maximum case, on the other hand, we assume $F_{\rm X,max} = 8$\,\rm erg\,s$^{-1}$\,cm$^{-2}$ (vertical dashed line in Figure~\ref{fig:LxLimit} and the entire XUV spectrum with $F_{\rm XUV,max} = 56$\,\rm erg\,s$^{-1}$\,cm$^{-2}$. As can be seen in Figure~\ref{fig:LxLimit}, not even an N$_2$-O$_2$ atmosphere that contains 25\% CO$_2$ would be stable for these conditions according to the results of \citet{Johnstone2021Stars}. However, such an `optimistic' case might cover any potential parameters that might generally enhance the stability of an N$_2$-dominated atmosphere such as unknown cooling agents \citep[e.g., additional atomic line cooling by oxygen as in][]{Nakajima2022}, an average mass for habitable rocky exoplanets that is significantly higher than 1.0\,M$_{\oplus}$, and/or the unlikely case that the actual short-wavelength activity of the stellar distribution is significantly weaker than found by \citet{Johnstone2021Stars}.

\begin{figure*}
\centering
\includegraphics[width = 0.7\linewidth, page=1]{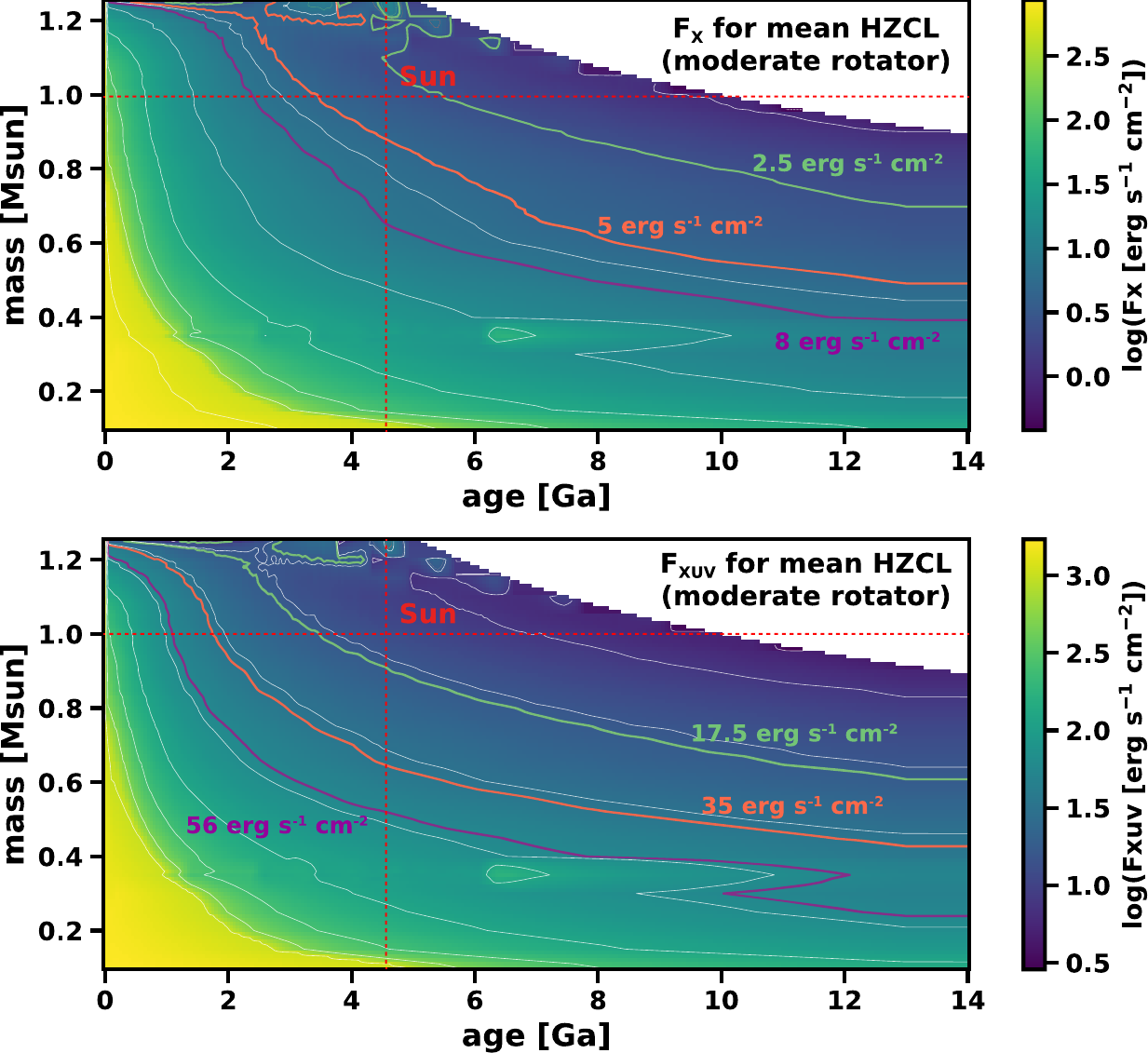}
\caption{Upper panel: The stellar X-ray surface flux, $F_{\rm X}$, in the middle of the HZCL for moderately rotating stars (i.e., the 50th percentile of the stellar rotational distribution) with stellar masses between $M_{\star} = 0.1-1.25$\,M$_{\odot}$ as a function of stellar mass and age, calculated with Mors \citep{Johnstone2021Stars} and based on a grid of stellar isochrones by \citet{Spada2013}. The plot shows three colored contour lines with different threshold values for the X-ray surface flux, i.e., $F_{\rm X, max} = \{2.5,5.0,8.0\}\,\rm erg\,s^{-1}\,cm^{-2}$. Lower panel: The same but for the XUV surface flux with colored contour lines at $F_{\rm XUV,max} = \{17.5,35.0,56.0\}\,\rm erg\,s^{-1}\,cm^{-2}$.}
\label{fig:LLplot}
\end{figure*}

\paragraph{The effect of $\alpha_{\rm at}^{\rm XUV}$:}\label{sec:XUVEff2}

Figure~\ref{fig:LLplot} shows the X-ray (upper panel) and XUV surface flux (lower panel) at $d_{\rm \langle HZCL \rangle}$ for stars at the 50th percentile of the stellar rotational distribution as a function of stellar mass and birth age, calculated through the stellar evolution model Mors from \citet{Johnstone2021Stars}. The orange contour lines illustrate the threshold values of our minimum (upper panel) and nominal cases (lower panel) for atmospheres with $x_{\rm CO_2,max}$\,=\,10\%, i.e., $F_{\rm X,max} = 5$\,\rm erg\,s$^{-1}$\,cm$^{-2}$, and $F_{\rm XUV,max} = 35$\,\rm erg\,s$^{-1}$\,cm$^{-2}$, respectively. All stars that are placed \textit{above} the respective contour lines have a surface flux that is below the associated threshold value, implying that under such assumption, an N$_2$-O$_2$-dominated atmosphere with 10\% CO$_2$ is stable. Here, it can be seen that a higher number of stars is fulfilling the $F_{\rm XUV,max}$-levels than the related $F_{\rm X,max}$-levels, which stems from the finding by \citet{Johnstone2021Stars} that the Sun's $F_{\rm X}$/$F_{\rm EUV}$ ratio is lower than the average stellar ratio. The violet contour line in the lower panel further illustrates our maximum case for $x_{\rm CO_2,max}=10\%$ with $F_{\rm XUV,max} = 56$\,\rm erg\,s$^{-1}$\,cm$^{-2}$ and its respective value for the X-ray surface flux of $F_{\rm X} = 8$\,\rm erg\,s$^{-1}$\,cm$^{-2}$ can be seen in the upper panel. The grey lines show the threshold values of our minimum case for $x_{\rm CO_2,max}=1\%$. All threshold values for $\alpha_{\rm at}^{\rm XUV}$ are listed in Table~\ref{tab:AppEtaStar} of Section~\ref{app:inputTable}.

It has to be noted, however, that Figure~\ref{fig:LLplot} only illustrates the 50th percentile of the stellar distribution, i.e., so-called `moderate rotators'. Stars with same mass but at a lower percentile of the rotational distribution will rotate slower (such as the so-called `slow rotator' at the 5th percentile) and, therefore, reach any surface flux threshold earlier than the 50th percentile. Conversely, any star that rotates faster will fall below such value later. To account for this effect, we will implement the entire stellar rotational distribution, as it can be calculated with Mors, and distribute all stars evenly over all percentiles.

\begin{figure*}
\centering
\includegraphics[width = 0.7\linewidth, page=1]{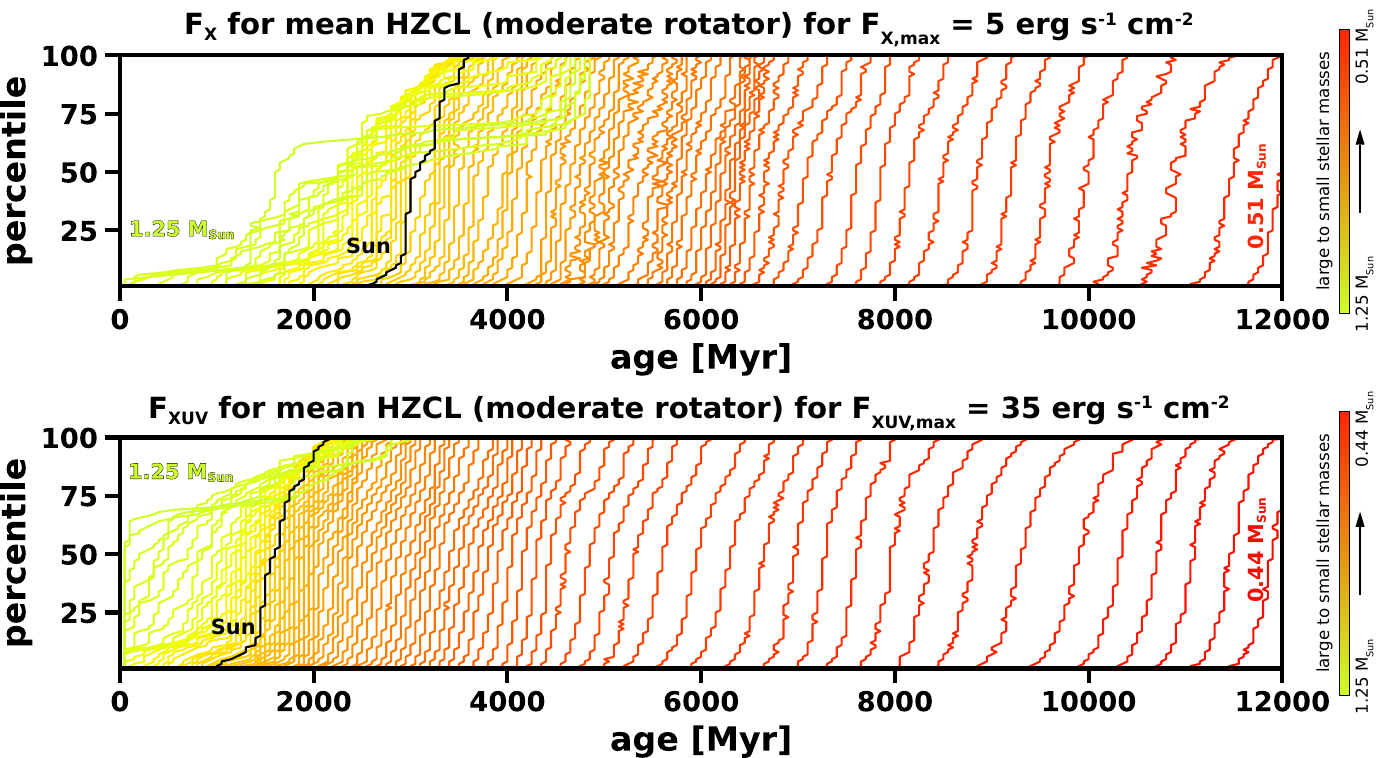}
\caption{The stellar ages at which the X-ray (upper panel) and XUV surface flux (lower panel) at $d_{\rm \langle HZCL \rangle}$ will fall below threshold values of  $F_{\rm X, max} = 5.0\,\rm erg\,s^{-1}\,cm^{-2}$ (upper panel) and $F_{\rm XUV,max} = 35.0 \,\rm erg\,s^{-1}\,cm^{-2}$ (lower panel), respectively, for each percentile (from 0th to 100th) of the stellar rotational distribution and stellar masses of $M_{\star}\,$=\,0.1-1.25\,M$_{\odot}$. The greenish to reddish lines indicate the different stellar masses (from 1.25\,M$_{\odot}$) with masses declining towards the right side of the plot with a step size of $\Delta M_{\star} = 0.01\,$M$_{\odot}$. For $F_{\rm X, max} = 5.0\,\rm erg\,s^{-1}\,cm^{-2}$, the threshold value can only be met within 12\,Gyr for stars that have a mass of $M_{\star} \geq 0.51$\,M$_{\odot}$ while $F_{\rm XUV,max} = 35.0 \,\rm erg\,s^{-1}\,cm^{-2}$ can be met already for $M_{\star} \geq 0.44$\,M$_{\odot}$.}
\label{fig:percentiles}
\end{figure*}

Figure~\ref{fig:percentiles} illustrates the age at which each percentile (from 0th to 100th) of each stellar mass (with $\Delta M_{\star} = 0.01\,$M$_{\odot}$) will decline below the threshold values of both our nominal and minimum cases for $x_{\rm CO_2,max}=10\%$ with $F_{\rm XUV, max} = 35.0\,\rm erg\,s^{-1}\,cm^{-2}$ (lower panel), and $F_{\rm X, max} = 5.0\,\rm erg\,s^{-1}\,cm^{-2}$ (upper panel), respectively. Here, two features are particularly noteworthy.  First, earlier percentiles, i.e., slower rotating and therefore less active stars of the same stellar mass, fall below the given threshold values earlier. However, this effect is more pronounced for higher mass stars (see also Section~\ref{sec:StellarEvo} and Figure~\ref{fig:Lx}). Second, not all stars will meet such criteria during the Hubble time, $t_0$, but $F_{\rm XUV, max}$ will be met slightly faster and by a higher number of stellar masses than $F_{\rm X, max}$. For the given thresholds in Figure~\ref{fig:percentiles}, stars with masses of $M_{\star} \geq 0.44$\,M$_{\odot}$ and $M_{\star} \geq 0.51$\,M$_{\odot}$ will fall below $F_{\rm XUV, max}$ and $F_{\rm X, max}$, respectively, within the first 12\,Gyr of their lifetimes. This already illustrates that most M dwarfs will never meet the criteria for our nominal and minimum cases since their short wavelength radiation declines far too slowly to provide habitable conditions within the current lifetime of the universe.

With the \textit{Runaway Greenhouse} limit from \citet{Kopparapu2014} and the $p$CO$_2$\,=\,0.1\,bar limit from \citet{Ramirez2020} as the inner and outer HZCL boundaries, respectively, and by taking account of the entire stellar rotational distribution, we find that $2.44 \times 10^{9}$ stars out of the initial $1.64 \times 10^{10}$ GHZ stars fulfill the XUV requirement, $\alpha_{\rm at}^{\rm XUV}$, for our nominal case with $x_{\rm CO_2,max} = 10\%$ and $F_{\rm XUV, max} = 35.0\,\rm erg\,s^{-1}\,cm^{-2}$. This is only $\sim15.4$\% and 3.1\% of all remaining stars, and all disk stars, respectively.  By only changing the X-ray or XUV surface flux to $F_{\rm X, max} = 5.0\,\rm erg\,s^{-1}\,cm^{-2}$ (minimum case) and $F_{\rm XUV, max} = 56.0\,\rm erg\,s^{-1}\,cm^{-2}$ (maximum case) we obtain $1.65 \times 10^{9}$ and $4.54 \times 10^{9}$ stars, respectively. By varying all input parameters, including the inner HZCL boundary, which we change to the the \textit{Runaway Greenhouse} limit by \citet{Kopparapu2013} for the minimum case, we obtain an entire range of $2.44(+3.93/-1.79) \times 10^{9}$ stars. For our minimum case, this indicates that only 0.79\% of all disk stars (i.e., $0.65 \times 10^{9}$ stars) are still fulfilling all implemented necessary requirements needed to host an EH.

As a side note, if we change the outer HZCL boundary of our nominal case from \citet{Ramirez2020} to \citet{Schwieterman2019}, the number of suitable stars decreases from $2.44 \times 10^{9}$ to $2.20 \times 10^{9}$ stars due to the HZCL being smaller and $d_{\rm \langle HZCL \rangle}$ being closer to the star. If we take the \textit{Moist Greenhouse} limit from \citet{Kopparapu2013} instead of the \textit{Runaway Greenhouse} limit from \citet{Kopparapu2014}, but keep everything else within our nominal case the same, we obtain a slightly larger fraction of suitable stars, i.e., $2.52 \times 10^{9}$.

\begin{figure}
\centering
\includegraphics[width = 1.0\linewidth, page=1]{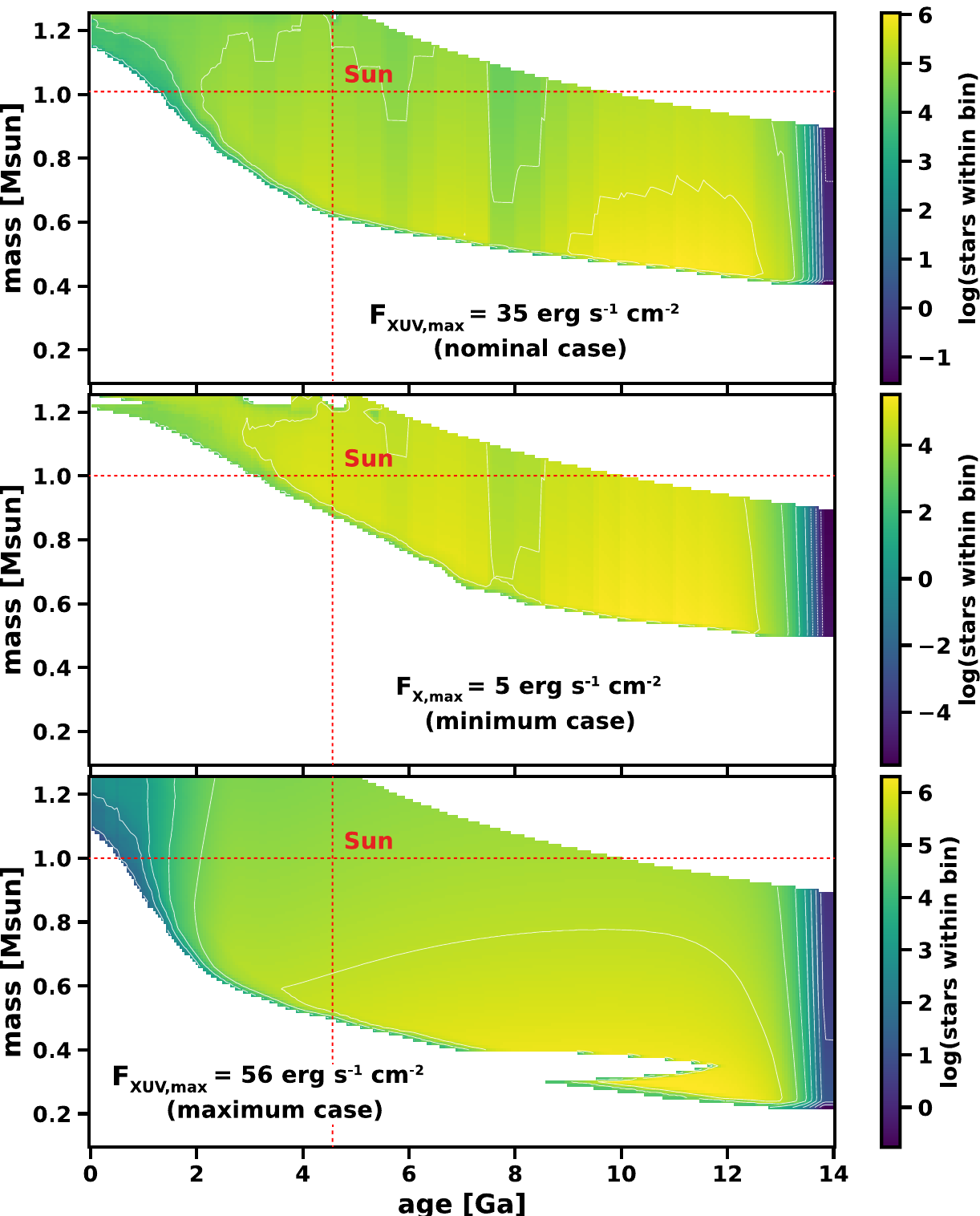}
\caption{The number of remaining GHZ stars that are above the lower limit, $\alpha_{\rm at}^{\rm XUV}$, as a function of age and stellar mass within a bin of $\Delta M_{\star} = 0.01$\,M$_{\odot}$ and $\Delta t = 50\,$Myr. The upper panel shows our nominal case with $F_{\rm XUV, max} = 35.0\,\rm erg\,s^{-1}\,cm^{-2}$, and the outer HZCL boundary from \citet{Ramirez2020} with our standard values for the IMF, SFR, and GHZ. The middle and lower panels illustrate our minimum and maximum cases with $F_{\rm X, max} = 5.0\,\rm erg\,s^{-1}\,cm^{-2}$ and $F_{\rm XUV, max} = 56.0\,\rm erg\,s^{-1}\,cm^{-2}$, respectively, combined with the minimum and maximum values of all other input parameters.}
\label{fig:LLstars}
\end{figure}

Figure~\ref{fig:LLstars} shows the number of remaining GHZ stars that are above the lower limit, $\alpha_{\rm at}^{\rm XUV}$, for our nominal (upper), minimum (middle) and maximum (lower panel) cases. Even though these cases do not only have varying values for $\alpha_{\rm at}^{\rm XUV}$ but also minimum and maximum values for all other input parameters, the effect of the lower limit can easily be seen. The lower the threshold values for $F_{\rm XUV, max}$ or $F_{\rm X, max}$ are, the higher the stellar mass must be to still meet this requirement. While for our nominal case, all stars below $M_{\star} = 0.41$\,M$_{\odot}$ fail to meet this necessary criterion, this threshold changes toward $M_{\star} = 0.22$\,M$_{\odot}$ and $M_{\star} = 0.50$\,M$_{\odot}$ for the maximum and minimum case, respectively. Whether M dwarfs are still part of the remaining stellar distribution, therefore, strongly depends on the lower limit. Stellar masses below $M_{\star} \sim 0.20$\,M$_{\odot}$, however, will likely never meet this requirement, at least as long as their XUV flux cannot be considered exceptionally weak. {This also exemplifies why a priori excluding stars with stellar masses below $M_{\star}=0.1$\,M$_{\odot}$ can at most have negligible effects on our results.}

The lower limit therefore strongly affects the relative fractions of the different spectral classes, much more so than the metallicity threshold. Only 0-22.36\% of the remaining stars will be M dwarfs, while K dwarfs suddenly hold the majority with 53.38-68.66\%. The relative fraction of G and F stars also rise significantly, i.e., from 4.26-4.76\% and 1.12-1.49\% to 18.09-35.54\% and 4.36-11.08\%, respectively.

Figure~\ref{fig:LLmasses} illustrates this distribution. Here, the upper panel depicts our nominal case and shows the number of stars by spectral class that still meet all requirements as a function of birth age. At maximum, only very old M dwarfs remain part of the sample (for the minimum case, no M dwarfs remain, not even very old ones), while the distribution shifts toward younger ages for earlier spectral classes. This is particularly interesting since any N$_2$-dominated atmosphere must be outgassed at an age when the star already allows for its thermal stability. A geologically active planet must therefore be a prerequisite for allowing the build-up of a stable atmosphere. As already briefly discussed in Section~\ref{sec:starProps}, however, this may not be the case for planets older than about 6\,Gyr \citep{Frank2014,Mojzsis2021}. Any M dwarf still part of the distribution has an age much older than the Sun and even older than $\sim$6\,Gyr, indicating that any geological activity on planets orbiting these old stars may already have vanished millions or even billions of years ago, at least for cosmochemically Earth-like planets.

The lower panel of Figure~\ref{fig:LLmasses} illustrates nominal, minimum, and maximum cases for $x_{\rm CO_2,max} = 10\%$ and shows the number of remaining stars as a function of stellar mass. As can be seen, the lower limit predominantly affects low-mass stars.

\begin{figure}
\includegraphics[width = 1.0\linewidth, page=1]{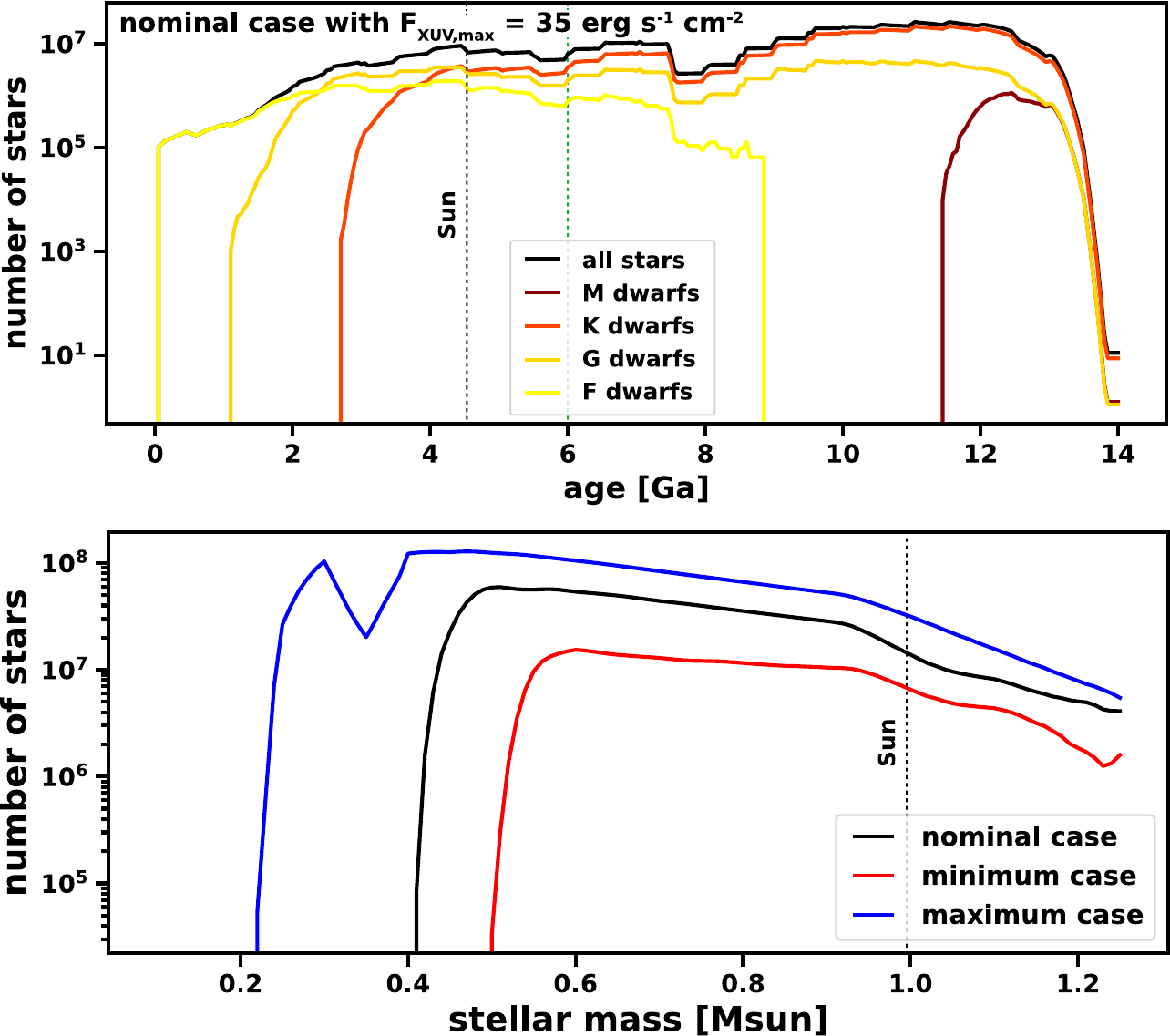}
\caption{{Upper panel: The number of remaining GHZ stars that are above the lower limit, $\alpha_{\rm at}^{\rm XUV}$, for our nominal case with $F_{\rm XUV, max} = 35.0\,\rm erg\,s^{-1}\,cm^{-2}$ as a function of birth age and separated into spectral classes. Here, cutoff birth ages can be seen for each spectral class (i.e., 0.05, 1.1, 2.7, and 11.5 Gyr for F, G, K, and M stars, respectively\textsuperscript{a}). The green dashed line shows the approximate age when geological activity at cosmochemically Earth-like planets will cease -- no M dwarfs are younger than that age. Lower panel: The number of remaining GHZ stars for our nominal, minimum, and maximum cases as a function of stellar mass. Cutoff masses can be seen for each case.}}
\small\textsuperscript{{a}}{Note that the temporal resolution of our model is 0.05\,Gyr}.
\label{fig:LLmasses}
\end{figure}

The numbers presented above, however, are for N$_2$-O$_2$-dominated atmospheres with $x_{\rm CO_2,max} = 10\%$. If we set the mixing ratio to $x_{\rm CO_2,max} = 1\%$ instead, and implement the corresponding outer HZCL boundary from \citet{Schwieterman2019} for $p$CO$_2$\,=\,0.01\,bar, our numbers will decrease significantly. For this, we assume $F_{\rm X}$ and $F_{\rm XUV}$ threshold values in agreement with the upper atmosphere simulations by \citet{Johnstone2021Stars} for N$_2$-dominated atmospheres with $x_{\rm CO_2,max} = 1\%$. As can be seen in Figure~\ref{fig:LxLimit}, N$_2$-atmospheres with 1\% CO$_2$ will strongly expand already for $F_{\rm X} \sim 2.5\,\rm erg\,s^{-1}\,cm^{-2}$. We may, therefore, define $F_{\rm XUV, max} = 15.0\,\rm erg\,s^{-1}\,cm^{-2}$ for our nominal and $F_{\rm X, max} = 2.5\,\rm erg\,s^{-1}\,cm^{-2}$ for our minimum cases. For the maximum case, we take $F_{\rm XUV, max} = 35.0\,\rm erg\,s^{-1}\,cm^{-2}$ to again account for any potential parameters that may enhance the stability of the atmosphere. However, since we have to assume a smaller HZCL with its outer boundary being defined through $p$CO$_2 = 0.01\,$bar for all our cases, the number of remaining stars will be significantly smaller even for our maximum case. For the Sun at 5\,Gyr, for instance, the mean HZCL distance of the maximum case will be located at only $d_{\rm \langle HZCL \rangle} = 1.05$\,AU, which clearly illustrates its small HZCL width.

By applying this criterion, only $0.73(+2.54/-0.58) \times 10^{9}$ stars still remain that can host EHs covered by N$_2$-O$_2$-dominated atmospheres with $x_{\rm CO_2,max} = 1\%$, a stellar fraction that accounts to 0.90(+3.13/-0.76)\% of all stars in the galactic disk. The respective minimum stellar masses decrease toward $M_{\star} = 0.65(+0.09/-0.18)\,$M$_{\odot}$, which implies that no M dwarfs remain within the sample. Of all remaining stars, 2.90-58.32\% are K dwarfs, while 33.49-79.17\% and 8.19-17.92\% are G and F stars, respectively. Figure~\ref{fig:pCO2_001} presents the nominal case for $x_{\rm CO_2,max} = 1\%$.

\begin{figure}
\centering
\includegraphics[width = 1.0\linewidth, page=1]{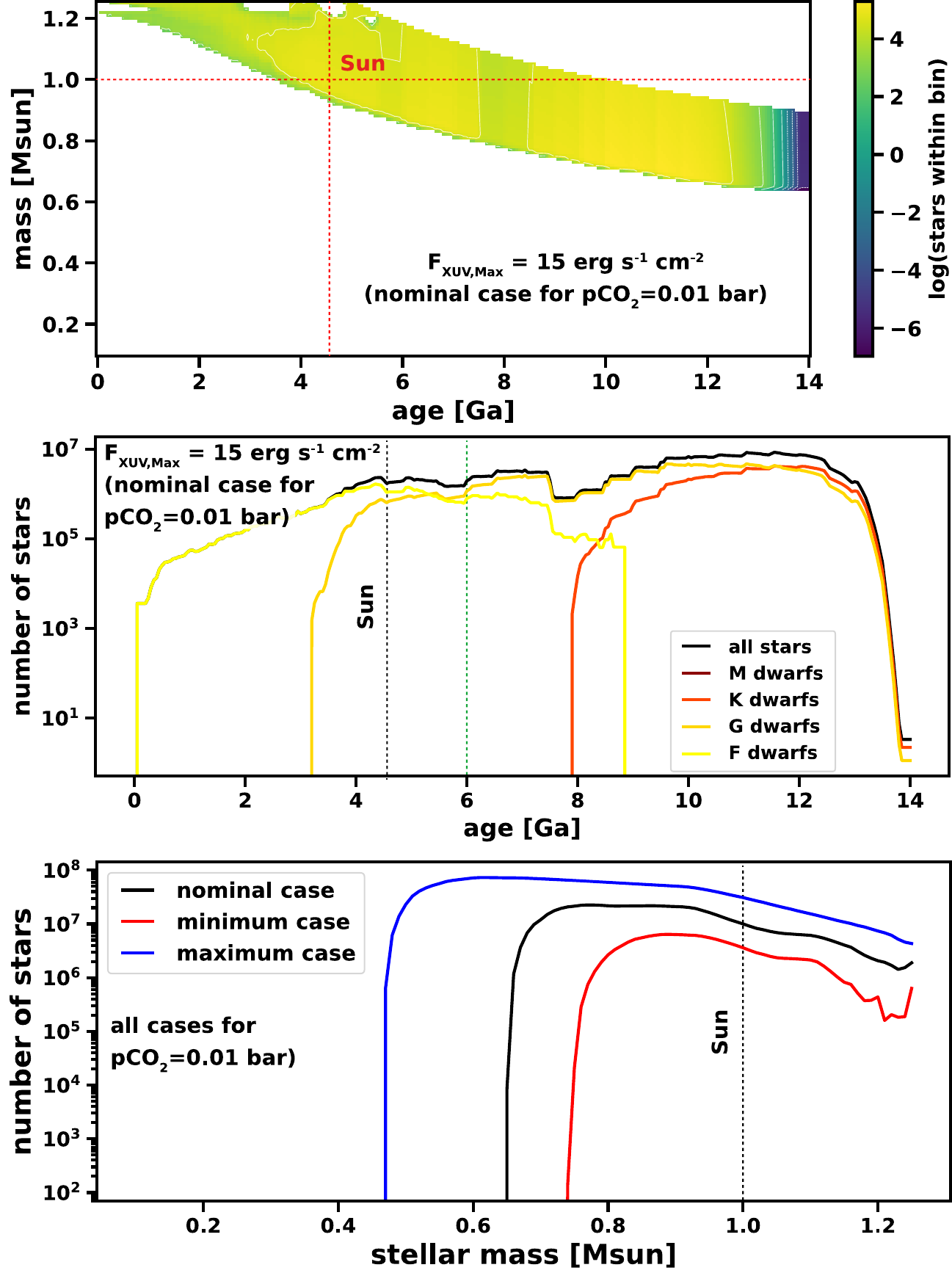}
\caption{Our nominal case for an N$_2$-O$_2$-dominated atmosphere with $x_{\rm CO_2,max} = 1\%$ and for $F_{\rm XUV, max} = 15.0\,\rm erg\,s^{-1}\,cm^{-2}$ (upper and middle panels). The upper panel shows the distribution of remaining stars as a function of birth age and stellar mass, while the middle panel illustrates the distribution of stars within the different spectral classes as a function of birth age. {The green dashed line again illustrates the approximate age at which geological activity will cease at cosmochemically Earth-like planets}. The lower panel shows the number of remaining stars for all cases with $x_{\rm CO_2,max} = 1\%$ as a function of stellar mass. No M dwarfs are part of the remaining stellar distributions.}
\label{fig:pCO2_001}
\end{figure}

Besides, we emphasize that the actual number of stars suitable to host planets with N$_2$-O$_2$-dominated atmospheres will likely be lower, independent of the maximum level of $p$CO$_2$ chosen. Our calculation of the lower limit neither includes the effect of superflares, nor non-thermal escape, the latter of which is triggered through a star's stellar wind and CMEs. Since these processes are important particularly around M to late K dwarfs, these will further reduce the number of small-mass stars being part of the remaining stellar distribution. EHs around M and late K-type dwarfs should, therefore, be exceedingly rare.

\subsubsection{The upper stellar limit, $\alpha_{\rm at}^{S_{\rm eff}}$}\label{sec:ul}

\paragraph{Implementing $\alpha_{\rm at}^{S_{\rm eff}}$:}\label{sec:ulImp}

For the upper limit, $\alpha_{\rm at}^{\rm ul}$, as a necessary requirement, we will make use of the rise of a star's bolometric luminosity, $L_{\rm bol}$. A habitable planet will certainly be rendered uninhabitable when the \textit{Runaway Greenhouse} limit sets in. For complex life, however, even \textit{Moist Greenhouse} conditions may pose serious stress. A planet with such climate conditions can be considered uninhabitable for humans since its global mean temperature of $T_{\rm pl} \gtrsim 320$\, \citep{Wolf2015} is above the {hyperthermia} limit, which sets in when the temperature exceeds 310\,K in a water-soaked climate over an extended period \citep{Sherwood2010}. Generally, exposures to temperatures above $\sim$310\,K and $\sim$315\,K in combination with high and low humidity, respectively, can be lethal for {Earth's} animal life \citep{Asseng2021}, and the metabolism of multicellular poikilotherms (i.e., animals whose body temperature reacts directly to the ambient temperature) may halt above $T_{\rm pl} \sim$320\,K \citep{Silva2017}. The photosynthetic production of oxygen seems to stop above the same temperature \citep{Silva2017} while no eukaryote on Earth can complete its entire life cycle at temperatures above $\sim$330\,K \citep{Clarke2014}. The halt of photosynthesis may particularly present a severe limitation to the stability and evolution of N$_2$-O$_2$-dominated atmospheres since the biological replenishment of oxygen may completely stop at such planets.

We will subsequently denote the upper limit simply as $\alpha_{\rm at}^{S_{\rm eff}}$, where $S_{\rm eff}$ stands for the effective stellar surface flux at the respective HZ planet. {As soon as $S_{\rm eff}$ equals the stellar surface flux at the chosen inner HZ boundary, a planet can be considered to be uninhabitable.}

While \citet{Kopparapu2013} calculated \textit{Moist Greenhouse} and \textit{Runaway Greenhouse} limits to be at $S_{\rm eff,\odot} = 1.014\,$S$_{\rm eff,\oplus}$ and $S_{\rm eff,\odot} = 1.0512\,$S$_{\rm eff,\oplus}$, respectively, a follow-up study by \citet{Kopparapu2014} investigated how the \textit{Runaway Greenhouse} limit changes with planetary mass, $M_{\rm pl}$. They found that the limit for planets with $M_{\rm pl}$\,=\,1\,M$_{\oplus}$ was slightly higher than previously thought, i.e., $S_{\rm eff,\odot} = 1.107$\,S$_{\rm eff,\oplus}$, with no significant difference in $S_{\rm eff,\odot}$ between \textit{Runaway Greenhouse} and \textit{Moist Greenhouse} limits. For the entire investigated mass range of $M_{\rm pl} = 0.1-5.0\,$M$_{\oplus}$, the limit further varied between $S_{\rm eff,\odot}= 0.99$\,S$_{\rm eff,\oplus}$ and $S_{\rm eff,\odot}=1.188$\,S$_{\rm eff,\oplus}$, respectively. Another study by \citet{Leconte2013} investigated climate thresholds with a three-dimensional climate model but found no stable solution for the \textit{Moist Greenhouse} regime. Instead, these authors discovered that the climate directly shifts to the \textit{Runaway Greenhouse} state, which will occur at an incident stellar flux onto the planet's atmosphere of $S_{\rm top} = 375\,\rm W/m^2$ which corresponds to $S_{\rm eff,\odot} \sim 1.10$\,S$_{\rm eff,\oplus}$\footnote{This is for a mean solar effective flux at the Earth's orbit of S$_{\rm eff,\oplus} = 1362\,\rm W/m^2$ with the incident solar flux on the Earth's atmosphere being defined as {S$_{\rm top,\oplus} = \mathrm{S}_{\rm eff,\oplus}/4$ due to the absorption area of the Earth being four times its cross-sectional area \citep[e.g.,][]{Bauer2004}.}}. \citet{Wolf2015} further investigated \textit{Moist Greenhouse} and \textit{Runaway Greenhouse} limits with a modified version of the Community Earth System Model of the National Center for Atmospheric Research\footnote{See \url{https://www.cesm.ucar.edu/}.}. Their research, in contradiction to \citet{Leconte2013}, resulted in a stable \textit{Moist Greenhouse} state that is achieved for $S_{\rm eff,\odot} = 1.125$\,S$_{\rm eff,\oplus}$, while \textit{Runaway Greenhouse} sets in at $S_{\rm eff,\odot} =1.21$\,S$_{\rm eff,\oplus}$. Similarly, \citet{Popp2016} found the moist greenhouse state accompanied by significant water loss to be stable and emerging for surface fluxes as low as $S_{\rm eff,\odot} =1.03$\,S$_{\rm eff,\oplus}$ but for the present day's CO$_2$ mixing ratio.

We will investigate $S_{\rm eff}$ for the mean value $d_{\rm \langle HZCL \rangle}$ of the HZCL. However, here we need to assume a fixed value for the HZCL boundaries since otherwise, the stellar surface flux within the habitable zone would stay constant due to the HZCL boundaries evolving together with the bolometric luminosity. We will, therefore, take fixed HZCL boundaries that correspond to the specific age at which the lower limit, $\alpha_{\rm at}^{\rm XUV}$ is for the first time going to decrease below the respective thresholds for $F_{\rm XUV,max}$ and $F_{\rm X,max}$, respectively, for each stellar mass and percentile. This makes logical sense since this is the earliest age at which an N$_2$-O$_2$-dominated atmosphere becomes stable within the respective HZCL, and the accompanying planet will consequently be rendered uninhabitable (the latest) at the moment it leaves its HZCL boundaries). We can, therefore, define the stable HZCL boundaries of a star as the ones that coincide with the age an Earth-like atmosphere will for the first time be stable against loss to space on a planet that orbits at the respective distance of $d_{\rm \langle HZCL \rangle}$.

\begin{figure*}
\centering
\includegraphics[width = 0.7\linewidth, page=1]{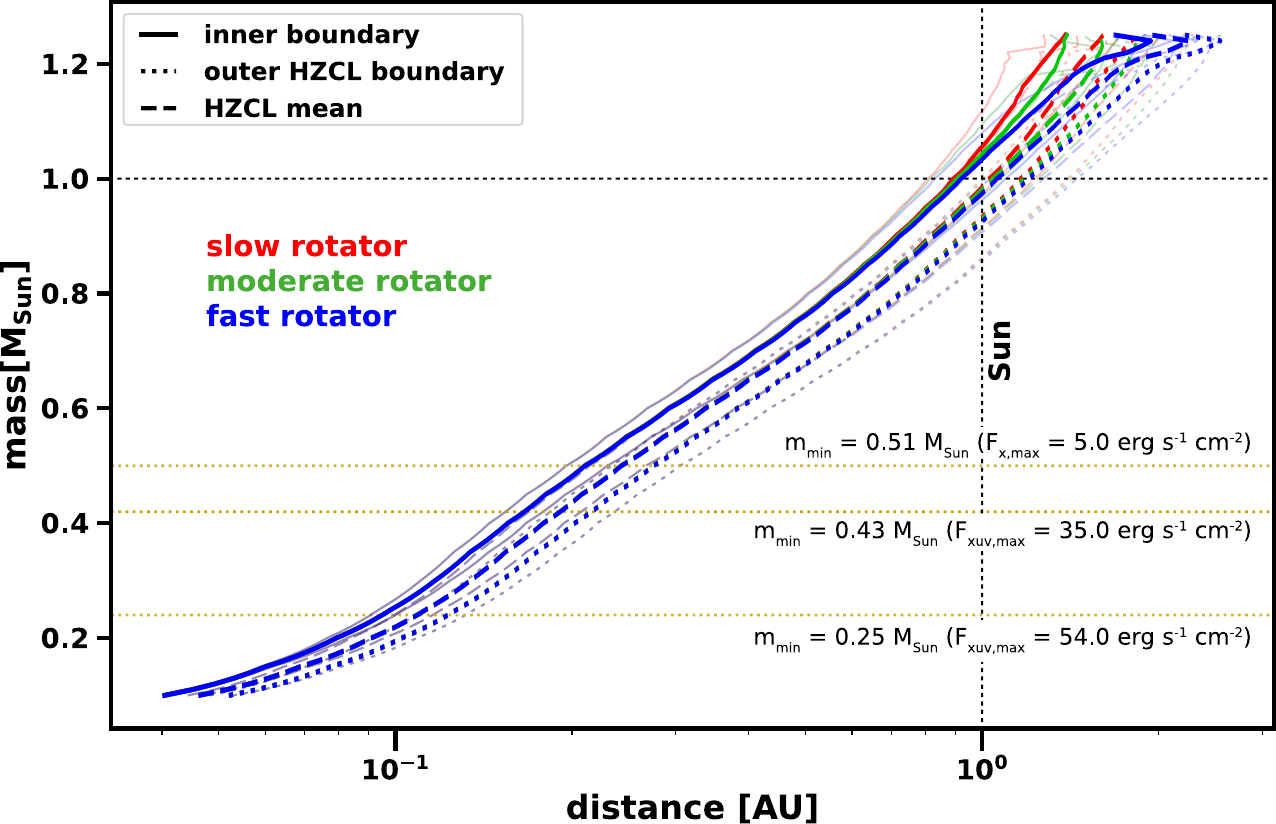}
\caption{The stable HZCL boundaries for slow, moderate, and fast rotators with  $M_{\star}=0.1-1.25$\,M$_{\odot}$, as well as for our nominal (non-transparent lines), minimum (transparent, shifted to the right) and maximum (transparent, shifted to the left) cases. These boundaries define the HZCL boundaries at the age when the lower limit, $\alpha_{\rm at}^{\rm XUV}$, falls below the respective threshold values (i.e., $F_{\rm XUV,max} = 35.0\,\rm erg\,s^{-1}\,cm^{-2}$ for the nominal, $F_{\rm XUV,max} = 54.0\,\rm erg\,s^{-1}\,cm^{-2}$ for the maximum, and $F_{\rm X,max} = 5.0\,\rm erg\,s^{-1}\,cm^{-2}$ for the minimum case). The vertical orange dotted lines illustrate the stellar masses where the lower limit decreases below the threshold value during the current age of the Universe. Note that for the outer HZCL boundary, the values from \citet{Ramirez2020} were used for nominal and maximum cases, while the ones from \citet{Schwieterman2019} for $p$CO$_2 = 0.1$\,bar were used for the minimum case.}
\label{fig:LbolHZR}
\end{figure*}

Figure~\ref{fig:LbolHZR} illustrates the stable HZCL boundaries for a time when slow, moderate, and fast rotators fall below the threshold values for the lower limit, $\alpha_{\rm at}^{\rm XUV}$, for our nominal, maximum and minimum cases, respectively. Here, the non-transparent lines illustrate the nominal case and the transparent lines the minimum (shifted to the left) and maximum cases (shifted to the right). One can see that the HZCL boundaries of slow, moderate, and fast rotators will slowly diverge for increasing stellar masses. Similarly, maximum and minimum cases are shifted towards the left and right. This behavior stems from the different ages at which the various stellar masses and rotational percentiles reach the threshold values of nominal, minimum, and maximum cases. Slow rotators fall below certain X-ray and XUV flux thresholds earlier than fast rotators, so the respective stable HZCL boundaries will be closer to the star because the bolometric luminosity increases over time.


\begin{figure}
\centering
\includegraphics[width = 1.0\linewidth, page=1]{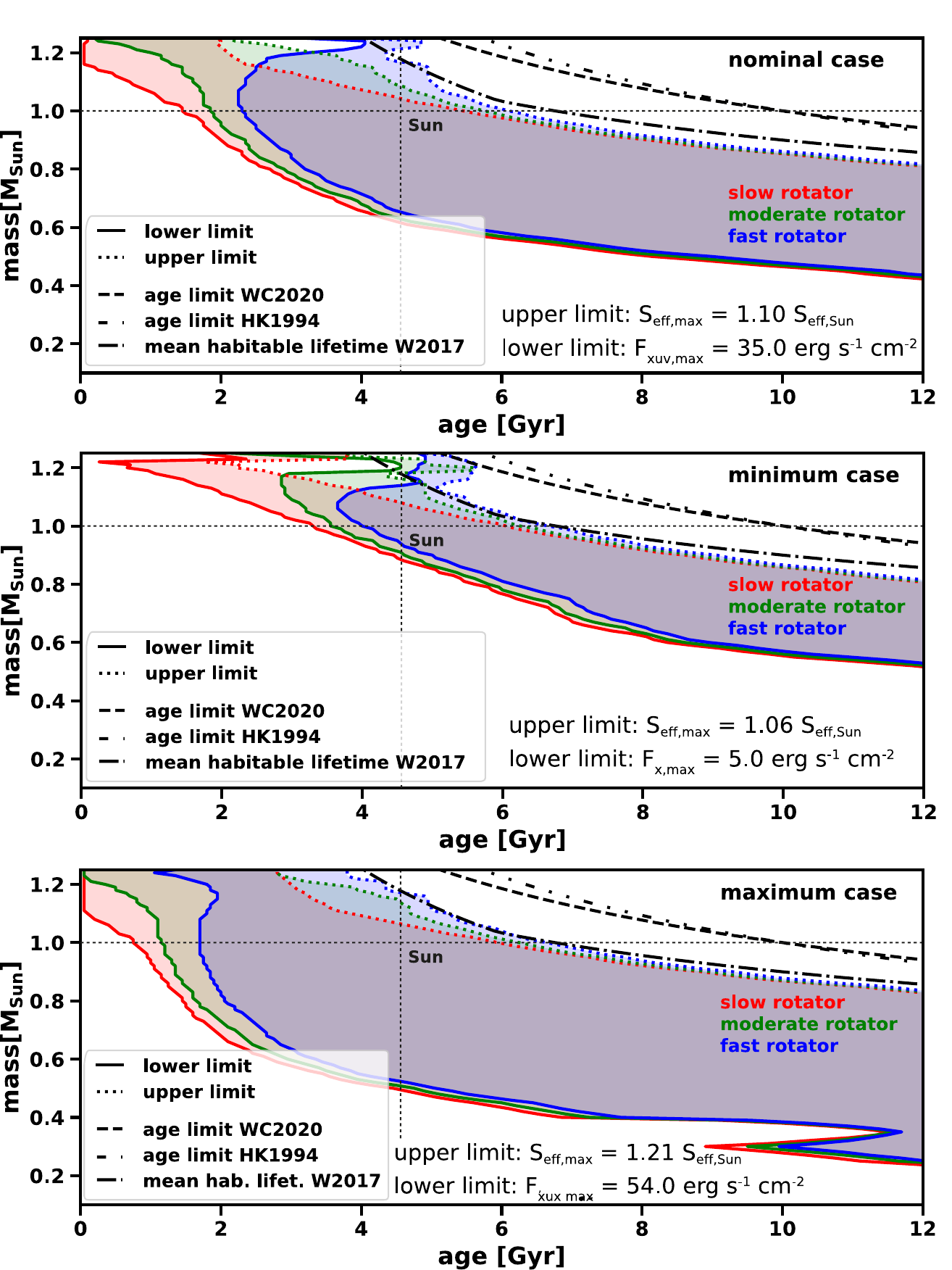}
\caption{The ages when slow, moderate and fast rotators fall below the lower and surpass the upper limits, i.e., $\alpha_{\rm at}^{\rm XUV}$, and $\alpha_{\rm at}^{S_{\rm eff}}$, respectively, for our nominal (upper), minimum (middle) and maximum (lower panel) cases. The shaded areas depict the mass-age parameter space where a stable N$_2$-O$_2$-dominated atmosphere with $x_{\rm CO_2,max} = 10\%$ can in principle exist. The dashed and dash-dot-dotted black lines illustrate the stellar main-sequence lifetimes calculated with the equations from \citet{Westby2020} and \citet{Hansen1994} while the dash-dotted line shows the mean habitable lifetime for planets that {can} possess intelligent observers from \citet{Waltham2017}.}
\label{fig:llul}
\end{figure}

Similar to the lower limit variation, we also change the respective thresholds for the upper limit, $\alpha_{\rm at}^{S_{\rm eff}}$. For our nominal case, we take a value of $S_{\rm eff,max} = 1.107\,S_{\rm eff,\oplus}$ from \citet{Kopparapu2014}, which is also in agreement with $S_{\rm eff,\odot} \sim 1.10$\,S$_{\rm eff,\oplus}$ from \citet{Leconte2013}. For our minimum and maximum cases, we further implement $S_{\rm eff,max} = 1.0512\,S_{\rm eff,\oplus}$ from \citet{Kopparapu2013} and $S_{\rm eff,max} = 1.21\,S_{\rm eff,\oplus}$ from \citet{Wolf2015}, respectively. 

\paragraph{The effect of $\alpha_{\rm at}^{S_{\rm eff}}$:}\label{sec:ulEff}

Figure~\ref{fig:llul} illustrates the ages at which slow, moderate, and fast rotators fall below {the lower limits} and surpass the upper limits, i.e., $\alpha_{\rm at}^{\rm XUV}$, and $\alpha_{\rm at}^{S_{\rm eff}}$, respectively, for our nominal (upper), minimum (middle) and maximum (lower panel) cases. Any stars in the colored areas between these limits can in principle provide habitable conditions and allow for the evolution of an Earth-like Habitat. Interestingly, one can see that depending on the chosen assumptions for the lower and upper thresholds, these may already start to intersect for some of the stellar masses within the F-type spectral class, although mostly for different stellar rotational percentiles. A star where the limits for the same percentiles intersect can never provide the necessary requirements for an EH to evolve, at least not within such boundary conditions. This could pose a serious threat to the habitability of F stars above a mass of $M_{\star} \gtrsim 1.10$\,M$_{\odot}$. It further indicates that stars beyond $M_{\star}\sim1.25$\,M$_{\odot}$ may never provide a time window for habitability.

However, for our nominal case with $x_{\rm CO_2,max}=10\%$, no intersection takes place and stars can provide habitable conditions, at least for a certain time, up to our maximum mass of $M_{\star} = 1.25\,$M$_{\odot}$. In total, $2.10 \times 10^{9}$ stars meet all the necessary requirements implemented so far with our nominal threshold value of $S_{\rm eff, max} = 1.10$\,S$_{\rm eff,\oplus}$ for the lower limit, $\alpha_{\rm at}^{S_{\rm eff}}$. These are 85.88\% of all the stars that meet the XUV limit and only 2.07\% of all the stars in the disk. By keeping the \textit{Runaway Greenhouse} limit from \citet{Kopparapu2014} as inner boundary and only varying the upper limit between a maximum of $S_{\rm eff, max} = 1.21$\,S$_{\rm eff,\odot}$ and a minimum of $S_{\rm eff, max} = 1.0512$\,S$_{\rm eff,\odot}$, we obtain $2.19 \times 10^{9}$ to $2.03 \times 10^{9}$ stars, respectively. By varying all input parameters from minimum to maximum, we obtain an entire range of $2.10(+3.70/-1.59) \times 10^{9}$ stars that still meet all implemented necessary requirements to host an Earth-like Habitat with an N$_2$-O$_2$-dominated atmosphere with $x_{\rm CO_2,max} = 10\%$. The distribution of these stars can be seen in Figure~\ref{fig:Lbolstars} for our nominal (upper), minimum (middle), and maximum (lower panel) cases.

\begin{figure}
\centering
\includegraphics[width = 1.0\linewidth, page=1]{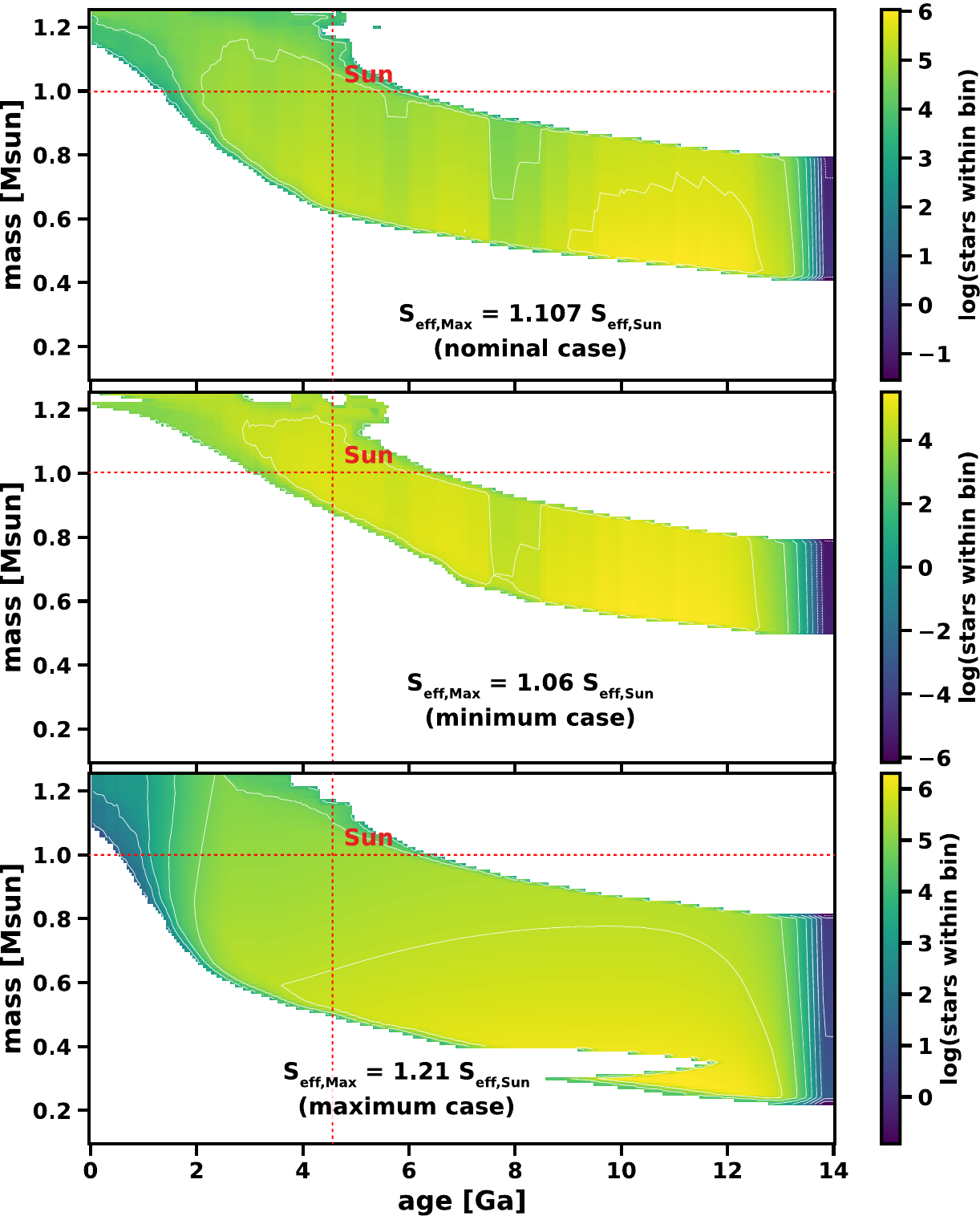}
\caption{The distribution of remaining stars as a function of stellar mass and age, including $\alpha_{\rm at}^{S_{\rm eff}}$ as the upper limit, for our nominal (upper), minimum (middle) and maximum (lower panel) cases. All plots are for N$_2$-O$_2$-dominated atmospheres with $x_{\rm CO_2,max}=10\%$.}
\label{fig:Lbolstars}
\end{figure}

It can also be seen in Figure~\ref{fig:llul} that the specific type of calculation for the main-sequence lifetime, as already introduced in Section~\ref{sec:MSL}, does not influence our final results since the habitability lifetime of EHs will always be shorter than the main-sequence lifetime of the respective star. It happens, however, that the habitable area might intersect with the mean habitable lifetime for intelligent observers from \citet{Waltham2017}. This is not surprising as this author's power law fits define a \textit{typical} habitable lifetime that is highly dependent on the mean habitable lifetime of various locations in the HZ. The further outwards our actual HZCL shifts because of a decrease in the lower threshold value, $\alpha_{\rm at}^{\rm XUV}$, the closer it will get to intersect with the mean habitable lifetime from \citet{Waltham2017}.

This can specifically be observed if we restrict the maximum atmospheric mixing ratio to $x_{\rm CO_2,max} = 1\%$. With $0.45(+2.12/-0.36) \times 10^{9}$ stars for this scenario, the effect of the upper limit becomes much more significant than for $x_{\rm CO_2,max} = 10\%$. Here, only 61.73\% and 57.34\% of the remaining stars adhere to $S_{\rm eff, max} = 1.10$\,S$_{\rm eff,\oplus}$ for our nominal, and $S_{\rm eff, max} = 1.0512$\,S$_{\rm eff,\oplus}$ for our minimum case, respectively. The latter is a particularly interesting scenario. As can be seen in the upper panel of Figure~\ref{fig:llul_0p01}, the HZCL of the higher-mass stars shifts significantly outwards (non-transparent lines) for our minimum case with $x_{\rm CO_2}=1\%$, especially when compared to our nominal case with $x_{\rm CO_2}=10\%$ (transparent lines in the background). For those stars, the lower limit's threshold value of $F_{\rm X,min} = 2.5\,\rm erg\,s^{-1}\,cm^{-2}$ is reached at a time when the bolometric luminosity is already significantly increased. This means that the lower and upper limits of the same percentiles begin to intersect for stellar masses above $M_{\star} \sim 1.10$\,M$_{\odot}$ (middle panel), thereby rendering them uninhabitable as visible in the lower panel. Furthermore, the stable HZCL begins to overlap with the mean habitable lifetime from \citet[][dash-dotted line]{Waltham2017} while remaining below the main-sequence lifetimes of \citet[][dashed line]{Westby2020} and \citet[][dash-dot-dotted line]{Hansen1994}.

An additional problem that may occur for stars with $M_{\star} \gtrsim 1.30$\,M$_{\odot}$ refers to their high rotation rates, which are mostly close to their breakup speeds. As was recently found by \citet{Ahlers2022}, such high rotation rates change a star's luminosity and spectral energy distribution, leading to its HZ shifting closer toward the host star. Since the lower and upper limits of stars with $M_{\star} \gtrsim 1.10$\,M$_{\odot}$ may start to intersect, such a shift in HZ distance might potentially severe this issue further.

\begin{figure*}
\centering
\includegraphics[width = 0.6\linewidth, page=1]{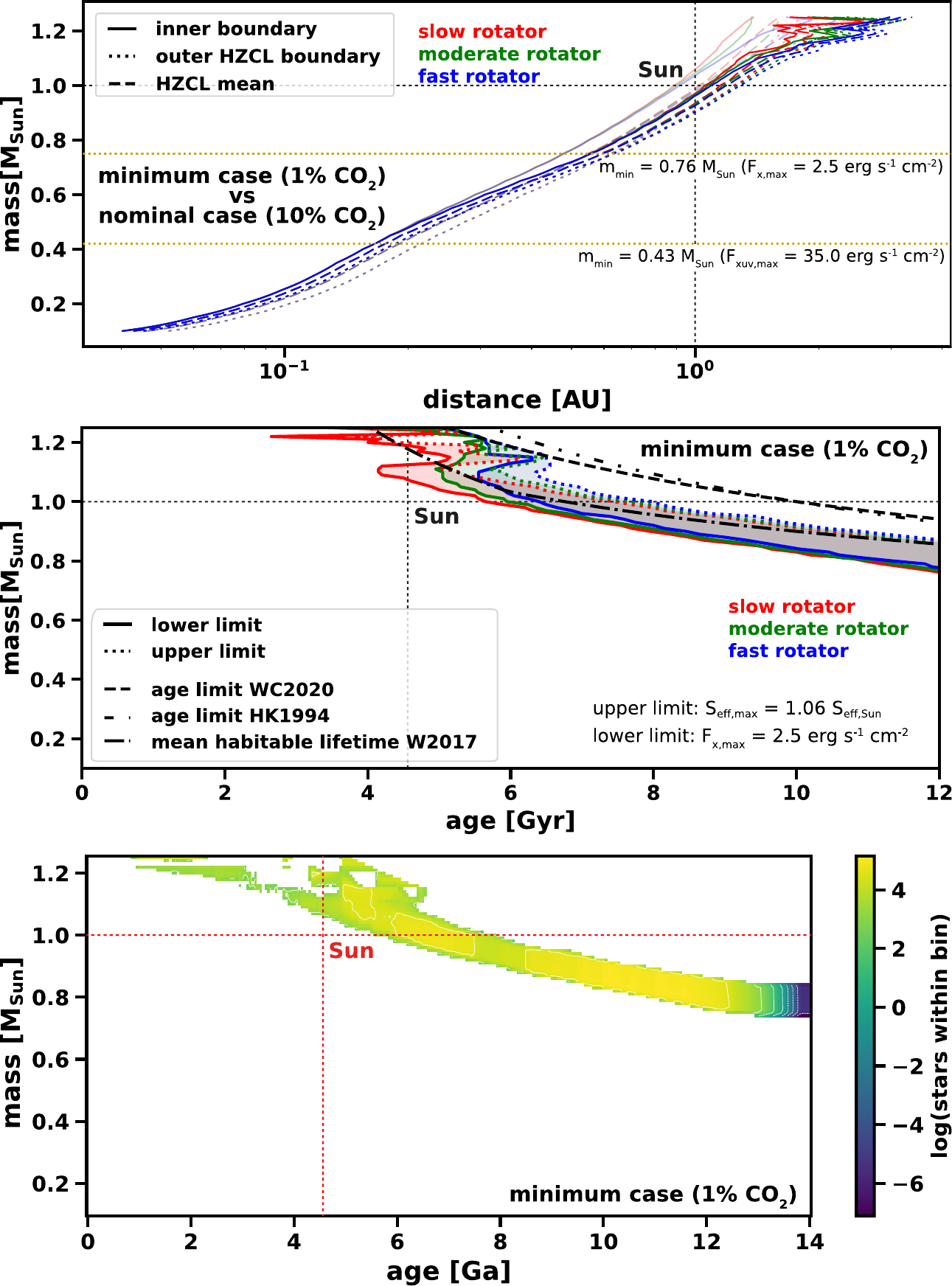}
\caption{The minimum case for an N$_2$-O$_2$-dominated atmosphere with $x_{\rm CO_2}=1\%$. The upper panel shows the same as Figure~\ref{fig:LbolHZR} but for this specific scenario (non-transparent lines) and compares it to our nominal case for $x_{\rm CO_2}=10\%$. The outer HZCL boundary for $p$CO$_2 = 0.01$\,bar by \citet{Schwieterman2019} was implemented for the minimum case and the outer HZCL boundary from \citet{Ramirez2020} for the nominal case. The middle panel shows the same as Figure~\ref{fig:llul} but again for this specific scenario with $x_{\rm CO_2}=1\%$. Here, lower and upper limits start to intersect for stellar masses above $M_{\star} \sim 1.10$\,M$_{\odot}$, implying that no habitable time window can be found for some of the rotational percentiles. This is also visible in the lower panel, which illustrates the distribution of remaining stars as a function of stellar mass and age for the same minimum case.}
\label{fig:llul_0p01}
\end{figure*}

Finally, we emphasize that the actual upper limit may be determined not by the bolometric luminosity of the parent star, but by the properties of the planet itself. Even though such an upper limit would then be part of $\eta_{\rm EH}$ (see Section~\ref{sec:max}) instead of $\eta_{\star}$, it is worth to be briefly mentioned here. Besides the already discussed cessation of geological activity and plate tectonics after $\sim 6$\,Gyr \citep[e.g.,][]{Frank2014,Cheng2018Tectonics,Mojzsis2021}, another severe limitation can occur on planets with active photosynthesis -- at least the Earth may eventually experience such a fate. The increase in bolometric luminosity leads to increased weathering rates, which applies not only to Earth but to every geologically and hydrologically active planet. Besides strong swings to the up and down, this effect steadily decreases the mean CO$_2$ concentration within the terrestrial atmosphere over geological timescales \citep[e.g.,][]{Walker1981,Caldeira1992,Franck2002}. However, photosynthesis requires a certain level of concentration to ensure it works, as was already pointed out by \citet{Lovelock1982}. At low CO$_2$-levels (i.e., around 180-200\,ppm), plants that use the so-called C$_3$ pathway for photosynthesis begin to show a significant decrease in biomass production and none survived at or around the so-called CO$_2$ compensation point of C$_3$ plants at $\sim$50\,ppm, which is the CO$_2$ concentration at which a plant's rate of photosynthesis equals its rate of respiration \citep[see, e.g.,][for a comprehensive review]{Gerhart2010}. Several studies that investigate the lifespan of the biosphere generally assume the lower limit for C$_3$ plants to be $\sim$150\,ppm \citep[e.g.,][]{Lovelock1982,Caldeira1992,Lenton2001,DeSousaMello2020} whereas the limit for plants using the relatively rare C$_4$ pathway is assumed to be around 10\,ppm \citep[e.g.,][]{Caldeira1992,Franck2000,Franck2006,DeSousaMello2020}. As summarized by \citet{DeSousaMello2020}, the C$_3$ plant limit is estimated to be reached in Earth's future at some point between about 0.1\,Gyr \citep{Lovelock1982} and 1\,Gyr \citep{Lenton2001} with a simulation by \citet{Caldeira1992} being in the middle at around 0.5\,Gyr. For C$_4$ plants, the corresponding limit of $\sim$10\,ppm is reached later in about 0.5\,Gyr \citep{Franck1999} to 1.6\,Gyr \citep{Franck2006}. \citet{DeSousaMello2020} themselves calculate that the respective thresholds for C$_3$ and C$_4$ plants are reached in 0.17\,Gyr and 0.84\,Gyr, respectively. Such timescales are mostly below the threshold limits for the stellar surface flux, $S_{\rm eff,max}$, which may be reached in $\gtrsim$0.9\,Gyr \citep[e.g.,][]{Kasting1993,Wolf2014,Waltham2019}. However, whether life capable of photosynthesis here on Earth, or on any other EH, could evolve more efficient pathways to use CO$_2$ to circumvent such an upper limit remains speculation.

However, CO$_2$ levels might not be the only atmospheric threshold that could be reached before the \textit{Moist Greenhouse} limit. As \citet{Ozaki2021}, who couple carbon, oxygen, phosphorus, sulfur, and methane cycles in their model together with the evolution of the Sun, point out, the Earth's atmosphere will be deoxygenated to an O$_2$ content of less than 1\% of the present atmospheric level (PAL) probably before the \textit{Moist Greenhouse} limit is reached. Within their simulations, 1\% PAL and 10\% PAL are achieved in $1.08 \pm 0.14$\,Gyr and $1.05 \pm 0.16$\,Gyr, respectively, largely due to an increase in surface temperature and a prompt halt of photosynthesis, the latter related to the CO$_2$-limits mentioned above \citep{Ozaki2021}. {A similar outcome has been found by \citet{Lingam2021Aquatic}, who calculate that the net primary productivity, i.e., the net production of organic carbon within a biosphere, will decline steeply towards $<$10\% of the modern value at the time the \textit{Runaway Greenhouse} state is reached, which is due to the related increase in the average surface temperature. Importantly, this will be accompanied by a strong decline in atmospheric O$_2$ since the oxygen sinks will outcompete its sources, thereby leading to the extinction of advanced animal life well before the ocean begins to evaporate.}

One should keep such limits in mind and be aware that the \textit{Moist Greenhouse} or even the \textit{Runaway Greenhouse} limit will not be the limiting factor that sets an end to Earth's complex life biosphere. {This will be extinct well before these limits are reached.}

\subsection{Final values for $\eta_{\star}$}\label{sec:etaStar}

As derived step by step in the preceding subsections, not every star with $M_{\star} = 0.1-1.25$\,M$_{\odot}$ in today's galactic disk provides stable conditions for the evolution of EHs. Even though a total of $N_{\star} = 1.015(+0.061/-0.077) \times 10^{11}$ stars are located in the disk between $r = 2-21$\,kpc and $z = 0-2.5$\,kpc, only a relatively small number meet all the necessary requirements needed for EHs to evolve. Besides the predefined stellar mass range, the requirements we considered are:
\begin{description}
  \item[\textbf{Avoiding sterilization by nearby SNs (Section~\ref{sec:SN}):}] Depending on our specific case, $\sim$22-30\% of all disk stars (and $\sim$27-30\% of all disk stars in the mass range) fulfill such requirement. Note that this criterion is particularly important for complex and/or surface life, while a microbial or sub-surface habitat may require less strict conditions. {If organisms on other worlds evolve to adapt to harsh radiation environments and/or complex life requires less time to develop, this requirement could be met by more stellar systems.} Supernovae are, however, not the only high-energy events. GRBs or sudden outbursts of the central supermassive black hole can further reduce the fraction of stars around which a complex biosphere may survive.
  \item[\textbf{Surpassing the metallicity threshold (Section~\ref{sec:metallicity}):}] Only stars surpassing a certain metallicity threshold allow for the accretion of rocky exoplanets. Depending on the chosen threshold, $\sim$39-86\% of all remaining stars meet this requirement. However, this does not cover the potential upper metallicity threshold above which planets may grow too quickly to allow EHs to form.
  \item[\textbf{The stellar lower limit (Section~\ref{sec:XUVEffect}):}] The short-wavelength radiation of young and/or lower-mass stars may be too high to allow for the thermal stability of N$_2$-O$_2$-dominated atmospheres. Depending on the chosen threshold, this lower limit is met either by $\sim$7-27\% or $\sim$2-13\% of the remaining sample of stars, depending on whether the atmospheres have maximum CO$_2$ mixing ratios of 10\% or 1\% CO$_2$, respectively. However, energetic flares and non-thermal escape through stellar winds and CMEs -- effects we do not consider -- will further reduce the number of stars surpassing this critical and often overseen requirement. Importantly, the number of suitable stars is lower for atmospheres with lower CO$_2$ mixing ratios because their thermal stability is also lower.
  \item[\textbf{The stellar upper limit (Section~\ref{sec:ul}):}] This necessary requirement considers the increase of bolometric luminosity, a fate that will ultimately render any planet uninhabitable. Between $\sim$79-91\% and $\sim$57-79\% of all remaining main-sequence stars are below the chosen stellar surface fluxes for atmospheres with 10\% and 1\% CO$_2$, respectively, thereby meeting this requirement. {Be aware, however, that some planetary parameters (e.g., the halt of photosynthesis) can render an EH uninhabitable even prior to its heat death.}
\end{description}
Besides, we must consider the specific stellar mass range that can provide habitable conditions. Even though we predefined a range of $M_{\star} = 0.1-1.25$\,M$_{\odot}$ that holds $\sim$80-99\% of all of today's stars, a number that mostly depends on the chosen lower mass limit, $M_{\star \rm low}$, the actual number of appropriate stars will likely not depend on this predefined range as natural cutoffs may occur for the lower and upper stellar mass limits. The lower stellar mass range is certainly cutoff by the stellar lower limit, $\alpha_{\rm at}^{\rm ll}$, depending on the chosen thresholds somewhere between $M_{\star}$\,=\,0.22-0.50\,M$_{\odot}$ and $M_{\star}$\,=\,0.47-0.74\,M$_{\odot}$ for N$_2$-O$_2$-dominated atmospheres with a maximum of 10\% CO$_2$ and 1\% CO$_2$, respectively. The upper stellar mass range may become uninhabitable above $M_{\star}\sim$1.10-1.25\,M$_{\odot}$ due to an intersection between $\alpha_{\rm at}^{\rm ll}$ and the upper limit, $\alpha_{\rm at}^{\rm ul}$. Our results, therefore, depend only slightly, if at all, on the chosen stellar mass range.

\begin{figure}
\centering
\includegraphics[width = 1.0\linewidth, page=1]{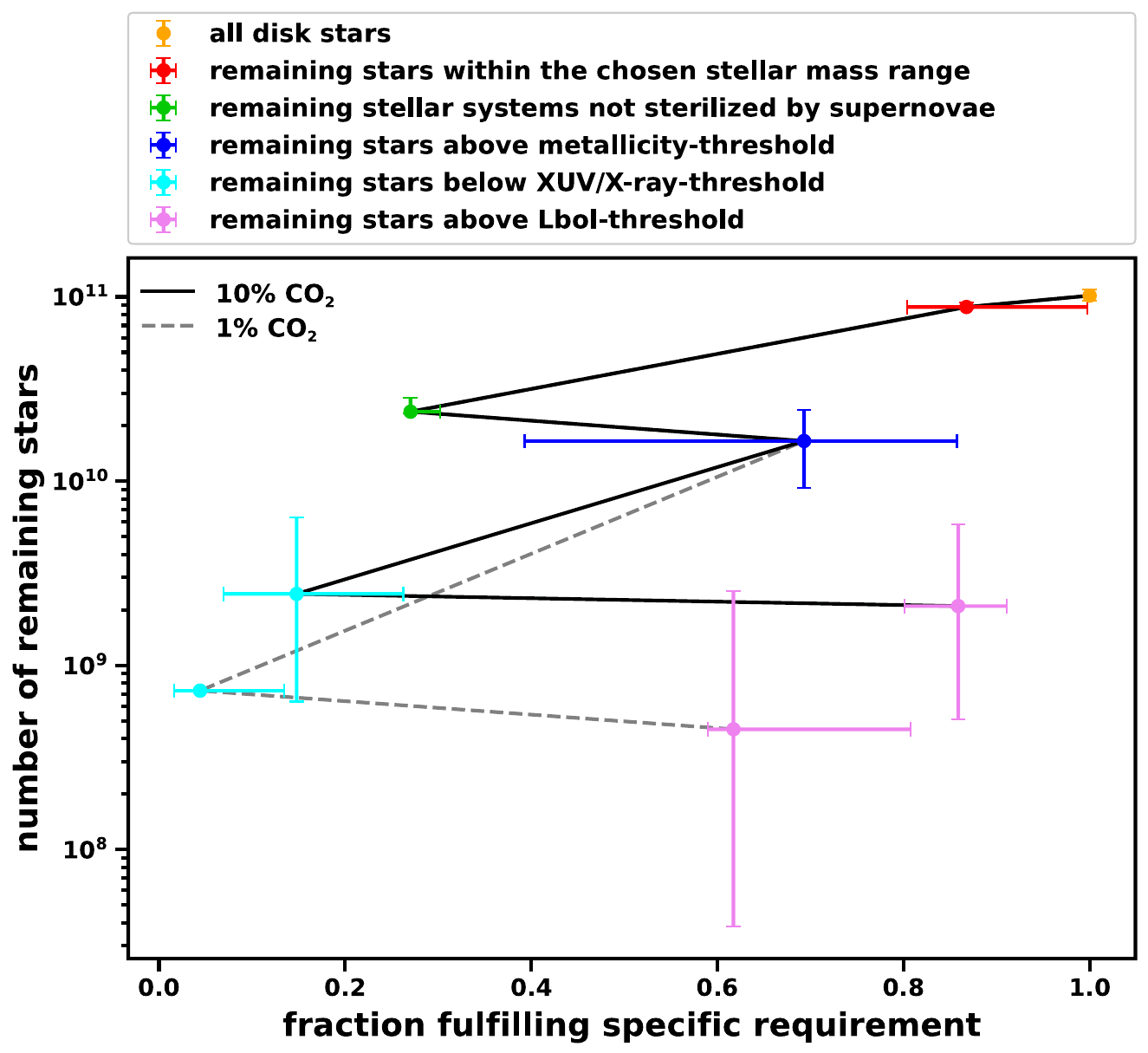}
\caption{The number of remaining stars as a function of necessary requirements feeding into $\eta_{\star}$. The x-axis shows the stellar fraction (including error bars) remaining after implementing each criterion. The figure starts at the top with the entire number of disk stars (orange) and shall be read from the top to the bottom. The y-axis therefore shows the number of stars remaining after implementing each criterion. The fractions listed are not cumulative but only belong to one requirement. As can be seen, the lower limit{, i.e., the XUV/X-ray threshold,} provides the most important implemented limit, thereby removing the highest number of stars. {The calculation of the respective requirements can be found in the following sections: all disk stars and stars within the chosen stellar mass range, Section~\ref{sec:NStar}; stellar systems not sterilized by supernovae, Section~\ref{sec:SN}; stars above the metallicity-threshold, Section~\ref{sec:metallicity}; stars below XUV/X-ray-threshold, Section~\ref{sec:XUVEffect}; stars below $L_{\rm bol}$-threshold, Section~\ref{sec:ul}. Also see Tables~\ref{tab:AppNStar} and \ref{tab:AppEtaStar} for the exact parameter values feeding into our model, and the associated Appendix~\ref{sec:appStellar} for a discussion on stellar evolution and its fundamental effect on Earth-like atmospheres.}}
\label{fig:etaStar}
\end{figure}

By combining all criteria, a maximum of 2.07(+4.11/-1.60)\% of all disk stars provide an appropriate environment for hosting a stable N$_2$-O$_2$-dominated atmosphere with $x_{\rm CO_2}\leq10\%$. Similarly, only 0.44(+2.30/-0.36)\% of all disk stars meet the stricter thresholds for atmospheres with only $x_{\rm CO_2}\leq1\%$. Table~\ref{tab:etaStar} and Figure~\ref{fig:etaStar} display the different requirements feeding into $\eta_{\star}$ together with their outcomes. As can be seen, the limiting factor with the highest effect is the lower limit, $\alpha_{\rm at}^{\rm XUV}$, no matter whether we assume EHs with a maximum of 1\% or 10\% CO$_2$. The host star's radiation environment is the most crucial factor within $\eta_{\star}$ that has to be considered for deriving a maximum number of EHs in the galactic disk, at least among the parameters implemented in our model. Other factors feeding into $\eta_{\star}$ are discussed in the appended Sections~\ref{sec:XUV} and \ref{sec:otherAat}.

\begin{table*}\footnotesize
  \begin{center}
    \caption{The effect of the necessary requirements onto $\eta_{\star}$.}
    \label{tab:etaStar}
    \resizebox{0.9\textwidth}{!}{%
\begin{tabular}{l|c|c|c|c|c|c}
  \hline
   & disk stars $N_{\star}$ & mass range  & supernovae & metallicity & lower limit & upper limit \\
    & $N_{\star}$ & $m=0.1-1.25\,$M$_{\odot}$  & $\alpha_{\rm GHZ}^{\rm SN}$ & $\alpha_{\rm GHZ}^{\rm met}$ & $\alpha_{\rm GHZ}^{\rm XUV}$ & $\alpha_{\rm GHZ}^{S_{\rm eff}}$ \\
   \hline
   & & & & & & \\
   remaining fraction  & 1 & $0.8677^{+0.1296}_{-0.0633}$ & $0.2703^{+0.0319}_{-0.00003}$ & $0.6934^{+0.2240}_{-0.3001}$ & $0.1528^{+0.1332}_{-0.0835}$ & $0.8696^{+0.0538}_{-0.0683}$ \\
   (individual) & & & & & $0.0464^{+0.0954}_{-0.0303}$ & $0.6481^{+0.1596}_{-0.0585}$ \\
   & & & & & & \\
   remaining fraction & 1 & $0.8677^{+0.1296}_{-0.0633}$ & $0.2345^{+0.0669}_{-0.0171}$ & $0.1626^{+0.1139}_{-0.0771}$ & $0.0249^{+0.0542}_{-0.0189}$$^a$ & $0.0216^{+0.0514}_{-0.0169}$$^a$ \\
    (cumulative) & & & & & $0.0076^{+0.0316}_{-0.0062}$$^b$ & $0.0049^{+0.0268}_{-0.0041}$$^b$ \\
   & & & & & & \\
   remaining stars  & $10.149^{+0.591}_{-0.765}$ & $8.806^{+0.550}_{-0.167}$ & $2.380^{+0.478}_{-0.045}$ & $0.165^{+0.944}_{-0.732}$ & $0.252^{+0.490}_{-0.189}$$^a$ & $0.219^{+0.466}_{-0.168}$$^a$ \\
    ($10^{10}$ stars) & & & & & $0.077^{+0.291}_{-0.062}$$^b$ & $0.050^{+0.247}_{-0.041}$$^b$ \\
   \hline
\end{tabular}}
\end{center}\footnotesize
$^a$for an N$_2$-O$_2$-dominated atmosphere with $x_{\rm CO_2,max}=10\%$$^b$same, but with $x_{\rm CO_2,max}=1\%$
\end{table*}

So, although we obtain a total of $N_{\star} = 1.02(+0.06/-0.08) \times 10^{11}$ disk stars, we only find maximum occurrence rates of disk stars that can host EHs of $\eta_{\star,10\%} \leq 0.021(+0.041/-0.016)$ and $\eta_{\star,1\%} \leq 0.0044(+0.0240/-0.0036)$ for atmospheres with maximum CO$_2$ mixing ratios of 10\% and 1\% CO$_2$, respectively. If we, for the time being, set $\eta_{\rm EH} = 1$ by ignoring any factors that may feed into it, Equation~\ref{eq:NEH} results in $N_{\rm EH,10\%} \leq 2.10(+3.70/-1.59) \times 10^{9}$ stars and $N_{\rm EH,1\%} \leq 0.45(+2.12/-0.36) \times 10^{9}$ stars for both atmospheric cases, respectively. However, we discuss and evaluate some of the requirements potentially feeding into $\eta_{\rm EH}$ in the following section and its related appended Sections~\ref{sec:etaHZ}, \ref{sec:h2oMain} and \ref{sec:appMoon}. There, we show that $\eta_{\rm EH}$ must be significantly smaller than the values typically assumed for $\eta_{\oplus}$.

\begin{figure}
\centering
\includegraphics[width = 1.0\linewidth, page=1]{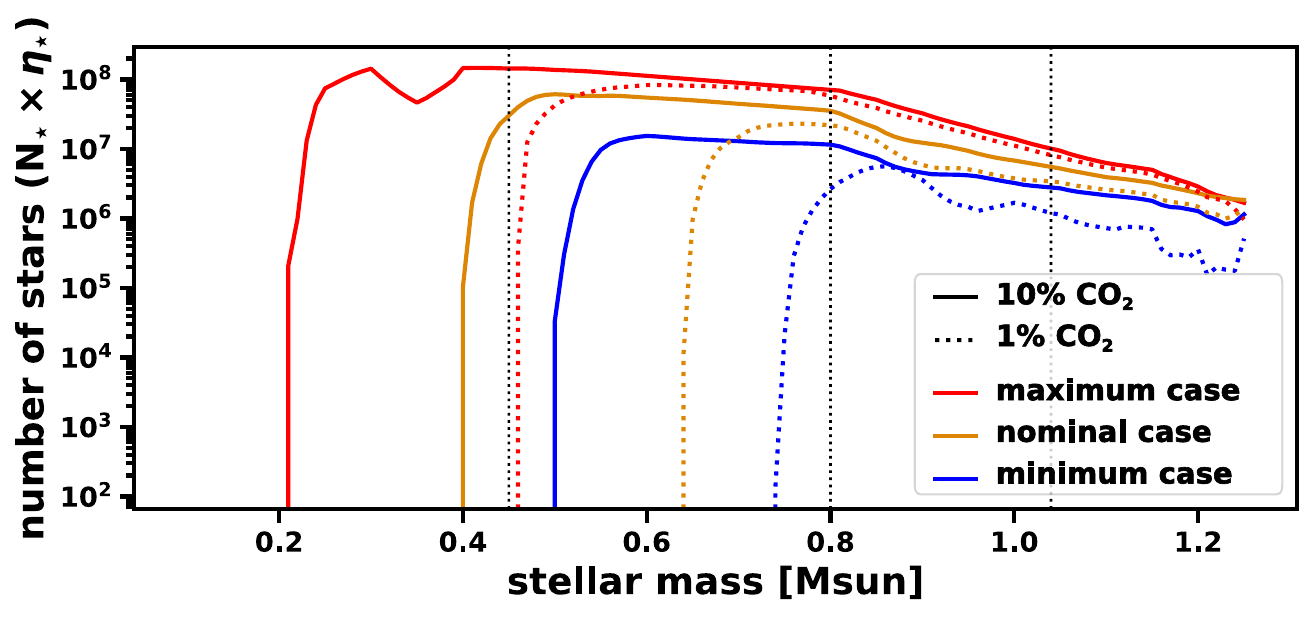}
\caption{The distribution of the remaining stars that are potentially able to host EHs, i.e., of $N_{\star} \times \eta_{\star}$, as a function of stellar mass for N$_2$-O$_2$-dominated atmospheres with $x_{\rm CO_2}\leq10\%$ (solid lines) and $x_{\rm CO_2}\leq1\%$ (dotted lines) for our nominal (orange), maximum (green) and minimum (blue) cases. The vertical dotted lines illustrate the borders of the different stellar spectral classes.}
\label{fig:compNStar}
\end{figure}

Finally, Figure~\ref{fig:compNStar} illustrates the distribution of {the} remaining stars that are potentially able to host EHs, i.e., of $N_{\star} \times \eta_{\star}$, for both types of N$_2$-O$_2$-dominated atmospheres. As can be seen, most stars are located somewhere in the middle of the stellar mass range. However, the fraction of M stars varies greatly. For an atmosphere with $x_{\rm CO_2}\leq10\%$, no suitable M dwarfs exist for the minimum case but over the entire parameter range their fraction may hold 1.05(+23.50/-1.05)\% of the entire stellar distribution. K dwarfs, on the other hand, contain the highest number of stars in every of our three model cases with 67.17(+12.78/-6.57)\%, whereas the fractions of G and F stars vary in a comparatively narrow range of 15.67(+9.60/-2.58)\% and 3.33(+4.23/-1.56)\%, respectively. For an atmosphere with only $x_{\rm CO_2}\leq1\%$, the stellar distribution contains no M dwarfs at all and the fraction of K dwarfs varies greatly with 48.30(+25.76/-43.23)\%. Furthermore, G and F stars hold fractions of 41.94(+38.30/-19.32)\% and 9.77(+4.94/-6.44)\%, respectively. If one compares these fractions with the initial values (i.e., 77.8-81.3\%, 14.5-16.4\%, 3.1-4.5\%, and 0.8-1.2\% for M, K, G, and F stars, respectively), one can recognize that the stellar distribution shifted significantly toward heavier stellar masses. Although M stars dominate the distribution of all disk stars, ultimately few if any of them provide habitable conditions for \textit{life as we know it}, and EHs may indeed be very rare or even absent around M dwarfs. K stars, on the other hand, seem to be the most promising place to search for biosignatures that result from complex life, at least by only considering $\eta_{\star}$. In the following section, we will deduce whether this assessment remains valid after considering $\eta_{\rm EH}$ - based on our current, rather restrictive scientific knowledge.

\section{The Fraction and Number of Suitable Planets, $\eta_{\rm EH}$ and $N_{\rm EH}$}\label{sec:max}

\subsection{The planet occurrence rate of rocky exoplanets within the HZCL, $\beta_{\rm HZCL}$}\label{sec:bHZCLMain}

The occurrence rate, $\beta_{\rm HZCL}$, of rocky exoplanets in the HZCL is the first important criterion that feeds into $\eta_{\rm EH}$. It is the factor that most similarly resembles the typical parameter Eta-Earth, $\eta_{\oplus}$, which is broadly defined as the mean number of rocky exoplanets within the HZ of solar-like stars. We discuss $\eta_{\oplus}$ and its potential value in more detail in the appended Section~\ref{sec:etaHZ}. This also includes Table~\ref{tab:etaEarth} with previous estimates on Eta-Earth and their chosen boundary conditions from the corresponding studies. In this appended Section, we further discuss the role of atmospheric erosion in estimating the occurrence rate of rocky exoplanets in the HZ(CL). Studies that account for this important effect typically receive much lower values for $\eta_{\oplus}$.

\subsubsection{Calculating and implementing $\beta_{\rm HZCL}$}\label{sec:betaHZCL}

Values for $\eta_{\oplus}$ and $\zeta_{\oplus}$\footnote{{The parameter $\zeta_{\oplus}$ takes the lower and higher boundary conditions for the orbital period and planetary radius to be within 20\% of the Earth's values.}} are generally not estimated for the HZCL (see Table~\ref{tab:etaEarth}). To derive occurrence rates for the HZCL, but also for a specific range of planetary radii, we need to convert different $\eta_{\oplus}$ and $\zeta_{\oplus}$-values to fit within the parameter range we choose. To do this, we follow the basic method of \citet{Mulders2018} and \citet{Pascucci2019} as described in the next paragraphs.

To fit the distribution of Kepler planets, several studies \citep[e.g.,][]{Mulders2018,Pascucci2019} assume that the occurrence rate distribution can be described by a separable function in orbital period, $P_{\rm pl}$, and planetary radius, $R_{\rm pl}$, such that its integral can be given as \citep[e.g.,][]{Mulders2018,Pascucci2019}
\begin{equation}\label{eq:planetDist}
  N_{\rm pl} = \int_{P_{\rm pl,min}}^{P_{\rm pl,max}}\int_{R_{\rm pl,min}}^{R_{\rm pl,max}}A f(P_{\rm pl})f(R_{\rm pl})d\log P_{\rm pl}d\log R_{\rm pl}.
\end{equation}
Here, $N_{\rm pl}$ is the number of planets per star over a specific range in period (with $P_{\rm pl,min}$ and $P_{\rm pl,max}$ as lower and upper boundary conditions) and radius (with $R_{\rm pl,min}$ and $R_{\rm pl,max}$ as lower and upper boundary conditions). The parameter $A$ is a normalization factor such that $N_{\rm pl}$ equals the predefined number of planets per star within the respective boundaries. The planet orbital period function $f(P_{\rm pl})$ can be described by a broken power law \citep[e.g.,][]{Mulders2018}, i.e.,
\begin{eqnarray}\label{eq:bplP}
  f(P_{\rm pl}) =
  \begin{cases}
    (P_{\rm pl}/P_{\rm pl,break})^{a_p}, & P_{\rm pl} < P_{\rm pl,break} \\
    (P_{\rm pl}/P_{\rm pl,break})^{b_p}, & \rm else,
  \end{cases}
\end{eqnarray}
where $a_p$ and $b_p$ are fitting parameters and $P_{\rm pl,break}$ is the orbital period reflecting the flattening in sub-Neptune occurrence rates at around ten days \citep{Youdin2011,Howard2012,Mulders2015}. Similarly, a second independent broken power law \citep{Mulders2018} can be applied to the planet radius function $f(R_{\rm pl})$ such that
\begin{eqnarray}\label{eq:bplP2}
  f(R_{\rm pl}) =
  \begin{cases}
    (R_{\rm pl}/R_{\rm pl,break})^{a_r}, & R_{\rm pl} < R_{\rm pl,break} \\
    (R_{\rm pl}/R_{\rm pl,break})^{b_r}, & \rm else,
  \end{cases}
\end{eqnarray}
which reflects (i) the increase of planets toward smaller mass above a break of $R_{\rm pl,break} \sim 2-3$\,R$_{\oplus}$ \citep[e.g.,][]{Howard2012,Petigura2013}  and (ii) a departure from such power law below $R_{\rm pl,break}$.

\citet{Mulders2018} found the respective breaks to be at $P_{\rm pl,break} = 12(+5/-3)$\,days and $R_{\rm pl,break} = 3.3(+0.3/-0.4)$\,R$_{\oplus}$ together with $a_p=1.5(+0.5/-0.3)$, $b_p=0.3(+0.1/-0.2)$, $a_r=-0.5(+0.2/-0.2)$, and $b_r=-6(+2/-3)$ as best fit solutions to the Kepler DR25 planet population by using their newly developed Exoplanet Population Observation Simulator \verb"epos"\footnote{See \url{https://epos.readthedocs.io/en/latest/}.}. Based on Kepler DR25, but also on the then newly published Gaia catalog, \citet{Pascucci2019} further used a single power law for the period and a broken one for the radius for their `model\#4' by fitting orbital periods of $P_{\rm pl}=12-400$\,days and planetary radii of $R_{\rm pl}=1-6$\,R$_{\oplus}$ with \verb"epos", thereby obtaining $R_{\rm pl,break} = 3.2(+0.2/-0.3)$ and $b_p=0.14(+0.07/-0.07)$, $a_r=1.0(+0.5/-0.5)$, and $b_r=-6(+2/-2)$.

To recalculate $\eta_{\oplus}$-values for solar-like stars, we implement these power laws together with the fitting parameters derived by \citet{Pascucci2019} for their `model\#4', as this does not include close-in planets (see appended Section~\ref{sec:etaHZ}). However, since we are only interested in planets significantly below $R_{\rm break} \sim 3.2$, the power laws further simplify toward two single power laws for period and radius with $f(R_{\rm pl})$ even reduced to a linear distribution since we will simply assume $a_r=1.0$ for our nominal case, as found by \citet{Pascucci2019} for their `model\#4' (if neglecting the error bars). For our estimate of $\beta_{\rm HZCL}$, at least for a solar-like star, we can therefore take
\begin{eqnarray}\label{eq:splAll}
    f(P_{\rm pl}) &=& P_{\rm pl}^{0.14}, \\
    f(R_{\rm pl}) &=& R_{\rm pl}/3.2.
\end{eqnarray}

By normalizing Equation~\ref{eq:planetDist} such that $N_{\rm pl}$ will equal specific $\eta_{\oplus}$-values from the literature within their respective boundary conditions, i.e., $A= \eta_{\oplus}/N_{\rm pl}$, we can finally calculate $\beta_{\rm HZCL}$ by taking our specific boundary conditions for the HZCL and the planetary radii as input into the renormalized equation for $N_{\rm pl}$.


For the orbital period distribution function, $f(P_{\rm pl})$, the lower and upper boundaries, $P_{\rm pl,min}$, and $P_{\rm pl,max}$, are defined by the respective inner and outer HZCL boundaries implemented in our nominal, minimum and maximum cases, respectively. For the radius distribution function, $f(R_{\rm pl})$, {we describe the physical constraints that define the adopted lower and upper boundaries i.e., the minimum radius, $R_{\rm pl,min}$, and the maximum radius, $R_{\rm pl,min}$, for our different cases in detail in the following paragraphs\footnote{In addition, Table~\ref{tab:AppBetaHZCL} contains a summary of all relevant constraints and inputs for deriving $R_{\rm pl, min}$ and $R_{\rm pl, min}$.}}. The minimum radius, $R_{\rm pl,min}$, for our nominal and minimum cases {(for both $x_{\rm CO_2}\leq10\%$ and $x_{\rm CO_2}\leq1\%$)}, we calculate through the relation used for the exoplanet yield estimates of the HabEx \citep{Kopparapu2018,Gaudi2020}, Luvoir \citep{Kopparapu2018,Luvoir2019}, and LIFE \citep{Quanz2022} mission concepts. This is derived from an empirical atmospheric loss relationship by \citet{Zahnle2017} and calculates the minimum planetary radius that may hold on to its secondary atmosphere, i.e.,
\begin{equation}\label{eq:rMin}
  R_{\rm pl,min} = 0.8 \times a^{-0.5},
\end{equation}
where $a$ defines the semi-major axis, which we substitute by the fixed mean HZCL distance, $d_{\langle \rm HZCL \rangle}$, at the time a star for the first time allows for the existence of an N$_2$-O$_2$-dominated atmosphere, as defined in Section~\ref{sec:ul}. For simplicity, we assume just one $R_{\rm pl,min}$ for all stellar masses and percentiles, that is, $R_{\rm pl,min}$ calculated for a 50$^{\rm th}$ percentile of a solar-like star with $M_{\star}=1.0\,$M$_{\odot}$. If we were to substitute semi-major axis by insulation, i.e., $R_{\rm pl,min} = 0.8 \times S_{\rm eff}^{0.25}$  \citep[see,][]{Quanz2022}, and recalculate the minimum radius for all stellar masses, $R_{\rm pl,min}$ would decrease slightly toward low-mass stars, which seems unreasonable given $F_{\rm XUV}$ increases for the same stars. Keeping $R_{\rm pl,min}$ constant seems hence preferable. For our nominal case with $x_{\rm CO_2}\leq10\%$, the fixed mean HZCL distance can be found at $d_{\langle \rm HZCL \rangle} = 1.032$\,AU, which gives a minimum radius of $R_{\rm pl,min} = 0.79$\,R$_{\oplus}$ corresponding to a minimum mass of $M_{\rm pl,min} = 0.42$\,M$_{\oplus}$ (see Equation~\ref{eq:MRrel} below). For our minimum case with $x_{\rm CO_2}\leq10\%$, $d_{\langle \rm HZCL \rangle}$ shifts slightly outwards toward $d_{\langle \rm HZCL \rangle} = 1.079$\,AU as $\alpha^{\rm XUV}_{\rm at}$ is reached later. This shift implies a tiny decrease in minimum radius and mass toward $R_{\rm pl,min} = 0.77$\,R$_{\oplus}$ and $M_{\rm pl,min} = 0.38$\,M$_{\oplus}$, respectively.

The maximum radius for our nominal and minimum cases {(again for both $x_{\rm CO_2}\leq10\%$ and $x_{\rm CO_2}\leq10\%$)} is set to be $R_{\rm pl,max} = 1.23$\,R$_{\oplus}$. \citet{Chen2017} derived a probabilistic mass-radius relation based on a sample of 316 well-constrained astronomical objects and found a transition from rocky exoplanets to Neptune-like worlds at a planetary mass of $M_{\rm pl} = 2.04(+0.7/-06)$\,M$_{\oplus}$, which relates to aforementioned radius of $R_{\rm pl} = 1.23$\,R$_{\oplus}$ by considering the mass-radius relation derived by \citet{Chen2017}. This can be written as \citep[see also,][]{Zink2019}
\begin{eqnarray}\label{eq:MRrel}
  M_{\rm pl} =
  \begin{cases}
  0.972 \times \left(\frac{R_{\rm pl}}{R_{\oplus}}\right)^{3.584}, & R_{\rm pl} < 1.23\,R_{\oplus} \\
  1.436 \times \left(\frac{R_{\rm pl}}{R_{\oplus}}\right)^{1.698}, & 1.23\,R_{\oplus} < R_{\rm pl} < 14.31\,R_{\oplus}.
  \end{cases}
\end{eqnarray}
As \citet{Chen2017} point out, their derived transition from rocky to Neptunian at $R_{\rm pl} = 1.23(+0.44/-0.22)$\,R$_{\oplus}$ is fully compatible with the result by \citet{Rogers2015}, who found a transition occurring at $R_{\rm pl} = 1.48(+0.08/-0.04)$\,R$_{\oplus}$. We further note, however, that other studies found different transitions between rocky planets and those with primordial atmospheres, even as high as 2\,R$_{\oplus}$ \citep{Otegi2020}.

Nevertheless, it seems reasonable to use the value of \citet{Chen2017} for $R_{\rm pl,max}$ to be assumed for our nominal {and minimum cases}, since studies of the transition between rocky and Neptunian planets are mainly based on short-period planets that circumvent their stars on highly irradiated orbits. Planets further out in the HZCL of FGK stars should, on average, tend to maintain their primordial atmosphere longer than short-period planets because the XUV surface flux at their orbits is significantly lower early in their systems' evolution. This implies that the radius valley likely shifts to smaller radii for longer orbits, consistent with photoevaporation and core-powered mass-loss models, and already empirically supported by spectroscopic analysis of the California Kepler Survey sample \citep{Martinez2019,Petigura2022}. A recent theoretical study by \citet{Kimura2022} further found that planets above $R_{\rm pl}\sim1.3$\,R$_{\oplus}$ may provide uninhabitable conditions, as they are likely to host a thick primordial atmosphere with an underlying supercritical water layer (see also Section~\ref{sec:h2o}). In addition, \citet{Nakajima2022} suggested that planets with $R_{\rm pl}\gtrsim 1.3-1.6$\,R$_{\oplus}$ may not be able to form large moons, as the building blocks in the satellite's accretion disk would preferentially tend to fall onto the relatively massive planet (see also Section~\ref{sec:moon1}).

However, for our {two} maximum cases, we implement {a maximum radius of} $R_{\rm pl,max} = 1.4$\,R$_{\oplus}$, which is based on a `conservative interpretation' \citep[e.g.,][]{Gaudi2020} of the empirically derived transition from rocky to Neptune-like exoplanets by aforementioned \citet{Rogers2015}, and which was also taken by the different exoplanet yield calculations for the space mission concepts HabeX, Luvoir and LIFE. By applying Equation~\ref{eq:MRrel}, this radius limit would translate to a planetary mass of $M_{\rm pl} \sim 2.55$\,R$_{\rm \oplus}$, if the planet already resided above the transition from rocky to Neptune-like, for which the second part of Equation~\ref{eq:MRrel} would be valid. However, by still assuming such a planet to be rocky, the first part of the same equation may lead to a more reasonable estimate of $M_{\rm pl} \sim 3.25$\,M$_{\rm \oplus}$\footnote{By alternatively taking the mass-radius relation from \citet{Otegi2020}, it gives $\sim 2.9$\,M$_{\rm \oplus}$.}. Such a mass range is comparable to the transition mass of 2-3\,$M_{\oplus}$ above which volcanic degassing starts to significantly decrease because of the increasing pressure gradient in the planet's interior \citep[e.g.,][]{Dorn2018}. Bodies with a higher planetary mass might therefore pose significant problems degassing and maintaining a secondary atmosphere.

The minimum radius, $R_{\rm pl,min}$, for our {two} maximum cases, we define following \citet{Zink2019}, who set $R_{\rm pl,min} = 0.72$\,R$_{\oplus}$ corresponding to a minimum mass of $M_{\rm pl,min} = 0.3$\,M$_{\oplus}$. This value derives from a study by \citet{Raymond2007}, who applied the radiogenic heat flux model by \citet{Williams1997} to show that planets with $M_{\rm pl} < 0.3$\,M$_{\oplus}$ will fail to maintain a long-term carbon-silicate cycle\footnote{The radiogenic heat flux model by \citet{Williams1997} is based on an Earth-like radiogenic inventory per unit mass. However, one should keep in mind that the related inventory, and hence the maintenance of a carbon-silicate cycle, could be vastly different. This value could, for instance, be much higher for planets that formed in closer proximity to r-process events such as neutron star mergers \citep{Lingam2020Radiogenic}, vary in dependence of stellar birth age \citep{ONeill2020Tectonics}, stellar composition \citep[e.g.,][]{Wang2022PlanetComp}, and the amount of early devolatilization \citep[e.g.,][]{Erkaev2022}. {However, a vastly different heat budget would also have significant impact on a planet's tectonic regime \citep[e.g.,][]{ONeill2020Review}.}}.

{To summarize, the minimum and maximum radii for our nominal, minimum, and maximum cases with $x_{\rm CO_2}\leq10\%$ are $0.79\,\mathrm{R}_{\oplus} < R_{\rm pl} < 1.23\,\mathrm{R}_{\oplus}$, $0.77\,\mathrm{R}_{\oplus} < R_{\rm pl} < 1.23\,\mathrm{R}_{\oplus}$, and $0.72\,\mathrm{R}_{\oplus} < R_{\rm pl} < 1.4\,\mathrm{R}_{\oplus}$, respectively. Based on Equation~\ref{eq:MRrel}, this translates} to $0.42\,\mathrm{M}_{\oplus} < M_{\rm pl} < 2.04\,\mathrm{M}_{\oplus}$, $0.38\,\mathrm{M}_{\oplus} < M_{\rm pl} < 2.04\,\mathrm{M}_{\oplus}$, and $0.3\,M_{\rm pl,min} < M_{\rm pl} < 3.25\,\,M_{\rm pl,max}$  {for the same three cases. For our nominal and minimum cases with $x_{\rm CO_2}\leq1\%$, $R_{\rm pl, min}$  changes slightly to $R_{\rm pl, min} = 0.78\,$R$_{\oplus}$ and $R_{\rm pl, min} = 0.74\,$R$_{\oplus}$, respectively, based on the considerations described above. These values further relate to a minimum planetary mass of $M_{\rm pl, min} = 0.4\,$M$_{\oplus}$ and $M_{\rm pl, min} = 0.33\,$M$_{\oplus}$, respectively\footnote{It is worth noting that our minimum radii are relatively close to $R_{\rm pl}\sim 0.7\,$R$_{\oplus}$, the planetary radius that, according to \citet{Chin2024}, exhibits the highest non-thermal escape rates for constant stellar radiation and wind conditions. This is therefore in support of our chosen values, as this behavior provides a potential lower radius limit for a habitable planet (see also Section~\ref{sec:XUVEffect})}.}


\begin{figure*}
\centering
\includegraphics[width = 0.7\linewidth, page=1]{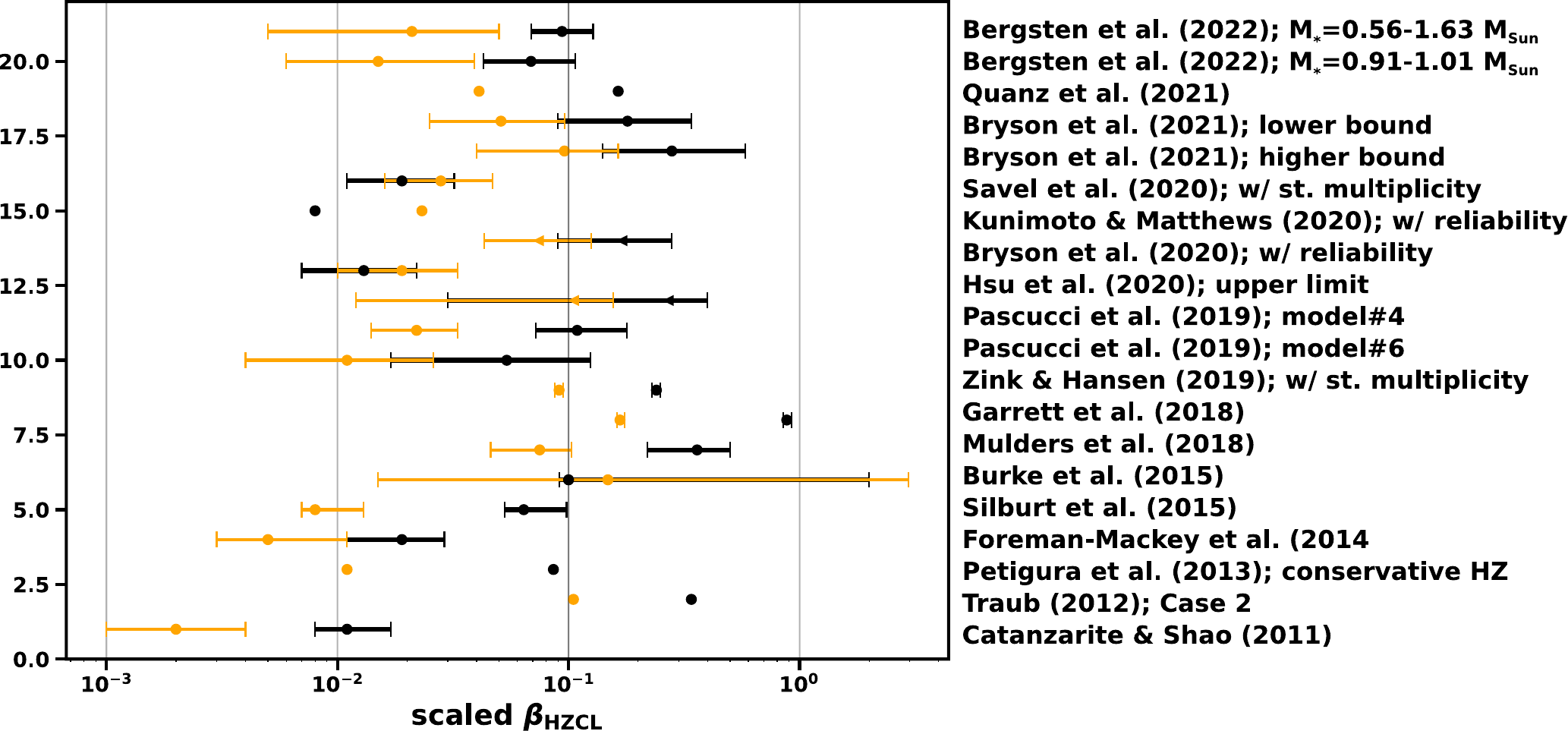}
\caption{Different values for planet occurrence rates around solar-like stars from the literature (black points and error bars) for the same studies and boundary conditions as listed in Table~\ref{tab:etaEarth}. The orange points and error bars display the calculated values for $\beta_{\rm HZCL}$ for a solar-like star based on the corresponding $\eta_{\oplus}$ and $\zeta_{\oplus}$ values. To obtain $\beta_{\rm HZCL}$ for this figure, $f(R_{\rm pl})$ was integrated within $R_{\rm pl,min} = 0.79$\,R$_{\oplus}$ and $R_{\rm pl,max} = 1.23$\,R$_{\oplus}$, while $f(P_{\rm pl})$ was calculated for the \textit{Runaway Greenhouse} limit by \citet{Kopparapu2014} as the inner and the \citet{Ramirez2020}-HZCL as the outer boundary condition. Compare also with Fig.~14 from \citet{Kunimoto2020}, which gives {the differential planet occurrence rate of Earth-analogs}, $\Gamma_{\oplus}$, for a similar sample of studies.}
\label{fig:betaHZCL}
\end{figure*}

Figure~\ref{fig:betaHZCL} and the right column of Table~\ref{tab:etaEarth} compare different $\eta_{\oplus}$ (and $\zeta_{\oplus}$) literature values with the corresponding $\beta_{\rm HZCL}$-occurrence rates for a solar-like star. The latter are for our nominal case and are derived by scaling the respective literature values to our orbital period and planetary radius range via Equation~\ref{eq:splAll}, i.e., the broken power law relation adopted from \citet{Pascucci2019}. Depending on the initial parameter range, most of the initial $\eta_{\oplus}$-values decrease significantly due to the relatively small HZCL and a tightened range for the planetary radius. In most cases, $\beta_{\rm HZCL}$ will therefore be below 10\%. The $\eta_{\oplus}$-values from `model\#4' and `model\#6' by \citet{Pascucci2019}, for instance, decrease toward $\beta_{\rm HZCL} = 0.022$ and $\beta_{\rm HZCL} = 0.011$, respectively, whereas the one from `model~4' by \citet{Neil2020} slightly increases toward $\beta_{\rm HZCL} = 0.023$ as their initial parameter range is more restrictive than the one from our nominal case. By taking the values from \citet{Zink2019} and \citet{Savel2020}, who took into account planetary and stellar multiplicity, respectively, but no atmospheric escape, we obtain $\beta_{\rm HZCL} = 0.091$ and $\beta_{\rm HZCL} = 0.028$, respectively.

For our two nominal cases, we take the $\beta_{\rm HZCL}$-values derived from `model\#4' of \citet{Pascucci2019} since this model (at least partially) excludes evaporating cores and provides the power laws used in our estimate. In addition, it fits very well with the HZCL occurrence rates derived from `model 4' of \citet{Neil2020} and from \citet{Bergsten2022} for FGK stars. For our two minimum cases, we implement $\beta_{\rm HZCL}$ derived from `model\#6' of \citet{Pascucci2019} as this variant excludes evaporating cores even more efficiently (by using orbital periods of 25-400 days vs 12-400 days in their `model\#4') and fits relatively well with the HZCL value derived from \citet{Bergsten2022} for stars with stellar masses between $M_{\star}=0.91-1.01$\,M$_{\odot}$. However, we decided to take this specific result as our minimum and `model\#4' as our nominal cases since the simulations by \citet{Pascucci2019} do not include stellar and/or planetary multiplicity, which both may slightly increase occurrence rates as was shown by \citet{Zink2019} and \citet{Savel2020}, respectively. Although the $\eta_{\oplus}$-value of \citet{Savel2020} does not exclude evaporating cores, it is still almost as low as `model\#4' of \citet{Pascucci2019}, suggesting that our nominal case could be quite optimistic. In contrast, the HZCL occurrence rate derived from \citet{Zink2019} is significantly higher and close to the value derived from the `higher bound' for $\zeta_{\oplus}$ of \citet{Bryson2021}. Since the latter may again not fully account for evaporating cores, similar to \citet{Zink2019}, we instead derive $\beta_{\rm HZCL}$ for our maximum cases from the `lower bound' for $\zeta_{\oplus}$ of \citet{Bryson2021}. All values discussed in this paragraph can be found in the appended Table~\ref{tab:etaEarth}, where $\beta_{\rm HZCL}$ is scaled to our nominal case with $x_{\rm CO_2}\leq10\%$.

Before implementing $\beta_{\rm HZCL}$ into our model, we need to extrapolate it to each stellar mass. Since a higher number of planets can likely be found around low-mass stars \citep[e.g.,][]{Yang2020,He2021}, as described in Section~\ref{sec:etaHZ}, we take a different occurrence rate for M dwarfs and scale it to both the HZCL and the same planetary radius range as for the initial $\eta_{\oplus}$-value used for solar-like stars by making the simplification that the planet orbit and radius distributions behave the same for M as for FGK dwarfs. For this, we implement M dwarf planet occurrence rates from \citet{Dressing2013}, who found $\eta = 0.16(+0.17/-0.07)$ for the conservative HZ and $R_{\rm pl} = 1.0-1.5$\,$R_{\oplus}$, from which we take $\eta = 0.16$ for our nominal and $\eta = 0.09$ for our minimum cases. For the maximum cases, we take $\eta = 0.312$ from \citet{Quanz2022}, again for the conservative HZ and $R_{\rm pl} = 0.82/0.62-1.4$\,$R_{\oplus}$, where $R_{\rm pl} = 0.82$  and $R_{\rm pl} = 0.62$ are the minimum radii at the outer and inner HZ boundaries.

We then let the occurrence rate increase linearly as a function of stellar effective temperature, $T_{\rm eff}$, by calculating the slope from the planet occurrence rates around a G star with $M_{\star} = 1.0$\,M$_\odot$ and an M dwarf with a mean mass of $M_{\star} = 0.26$\,M$_\odot$. Simulated with the stellar evolution model Mors \citep{Johnstone2021Stars}, these stellar masses correspond to mean stellar surface temperatures of $T_{\rm eff} \sim 3400$\,K and $T_{\rm eff} \sim 5780$\,K, respectively. Assuming such a linear relation, {with stellar temperature}, {i.e.,}
\begin{equation}\label{eq:corrHZCL}
  \beta_{\rm HZCL}(T_{\rm eff}) = a_{\rm HZCL} T_{\rm eff} + b_{\rm HZCL},
\end{equation}
{where the coefficients $a_{\rm HZCL}$ and $b_{\rm HZCL}$ are derived from the occurrence rates around a solar-like star and a mean-mass M dwarf, respectively,} is relatively reasonable since several studies (see Section~\ref{sec:etaHZ}) found a linear increase of the number of planets per star as a function of $T_{\rm eff}$ toward later spectral types, albeit the parameter ranges of these studies were mostly set to be much broader than for rocky HZ planets. {In addition, we should note that planetary population synthesis models \citep{Burn2021} recently found the occurrence rate of Earth-sized planets in the HZ to peak around early M dwarfs at $M_{\star}\sim0.3-0.5$\,M$_{\odot}$), which potentially indicates that we overestimate planet occurrence rates for late-type M-dwarfs.} 


If we implement $\beta_{\rm HZCL}$ simply as described above, however, we will likely underestimate the occurrence rate of planets around our remaining stellar sample. In Section~\ref{sec:metallicity}, we implemented a metallicity threshold into our model, which takes account of the distribution of metals in the galactic disk and excludes stars that are below our metallicity threshold, $Z_{\rm min}$. However, planet occurrence rates, mostly based on the Kepler sample, cover the entire distribution of stars, including very metal-poor dwarfs. All Kepler planets are additionally located toward only one direction in the Galaxy (i.e., the Kepler stellar field) and, besides few exceptions, most of the discovered planets are within $\sim$1000\,pc\footnote{See, e.g., NASA exoplanet archive, \url{https://exoplanetarchive.ipac.caltech.edu/}.}, for which the galactic metallicity distribution does not change significantly. It makes, therefore, sense to weight the planet occurrence rate with the metallicity of the Solar Neighborhood by re-scaling $\beta_{\rm HZCL}$ to only account for stars above $Z_{\rm min}$. We would otherwise underestimate occurrence rates in galactic regions with lower metallicities and vice versa if assuming $\beta_{\rm HZCL}$ to be the same in the entire disk. To achieve such a weighted occurrence rate, we have to divide $\beta_{\rm HZCL}$ by the fraction of stars in the solar region that meet our metallicity threshold, which is equivalent to a star's probability, $\mathcal{P}(Z_{\rm min,SN})$, to be above $Z_{\rm min}$ in the Solar Neighborhood, i.e.,
\begin{equation}\label{eq:mProbE}
  \beta_{\mathrm{HZCL},Z_{\rm min}} = \frac{\beta_{\rm HZCL}}{\mathcal{P}(Z_{\rm min,SN})}.
\end{equation}
which gives the weighted HZCL planet occurrence rate for stars with metallicities above $Z_{\rm min}$.

By only taking account of stars with sufficient metallicity, the corrected $\beta_{\mathrm{HZCL},Z_{\rm min}}$ will be slightly increased than the initial value for $\beta_{\rm HZCL}$ as the same number of planets will be distributed among fewer stars. However, for our nominal and maximum cases, the effect of this weighting is minuscule since for $Z_{\rm min} = 0.3$ and $Z_{\rm min} = 0.1$, the probability of being above this threshold in the Solar Neighborhood is $\mathcal{P}(Z_{\rm min,SN}) = 0.9923$ and $\mathcal{P}(Z_{\rm min,SN}) > 0.999$, respectively. However, the weighting factor becomes significant for our minimum case with $Z_{\rm min} = 0.75$ for which $\mathcal{P}(Z_{\rm min,SN}) = 0.624$.

\subsubsection{The effect of $\beta_{\rm HZCL}$}\label{sec:betaHZCL2}

As exemplified in Figure~\ref{fig:betaHZCL} and Table~\ref{tab:etaEarth}, converting $\eta_{\oplus}$ to $\beta_{\rm HZCL}$ significantly decreases the occurrence rate of potential EHs. This is also illustrated in Figures~\ref{fig:bHZCL} and \ref{fig:bHZCL2p50}, which show $\beta_{\rm HZCL}$ (orange line; only visible for the minimum case as it otherwise overlaps with the blue lines) and its metallicity weighted equivalent, $\beta_{\mathrm{HZCL},Z_{\rm min}}$, (blue lines) as a function of stellar mass for our atmospheric cases with $x_{\rm CO_2}\leq10\%$ and $x_{\rm CO_2}\leq1\%$, respectively. The red and orange crosses show the corresponding occurrence rates for an M dwarf with a mean mass of 0.26\,M$_{\odot}$ (nominal and minimum cases; 0.27\,M$_{\odot}$ for the maximum case) and a solar-like star with 1.0\,M$_{\odot}$ (all cases). The ones surrounded by a black outline represent the initial $\eta_{\oplus}$ (or $\zeta_{\oplus}$) literature values for M and G dwarfs, respectively. The black lines further illustrate the initial occurrence rates scaled to our planetary radius range and the conservative HZ by \citet{Kopparapu2014} as a linear function of stellar temperature. As can be seen, scaling to the HZCL leads to a significantly lower occurrence rate in all displayed cases. The additional weighting of $\beta_{\rm HZCL}$ with the Solar Neighborhood's average metallicity, $\mathcal{P}(Z_{\rm min,SN})$, significantly affects only the minimum case, since $Z_{\rm min} = 0.75$ is the only metallicity threshold that a significant number of stars in the Solar Neighborhood do not meet.

\begin{figure}
\centering
\includegraphics[width = 1.0\linewidth, page=1]{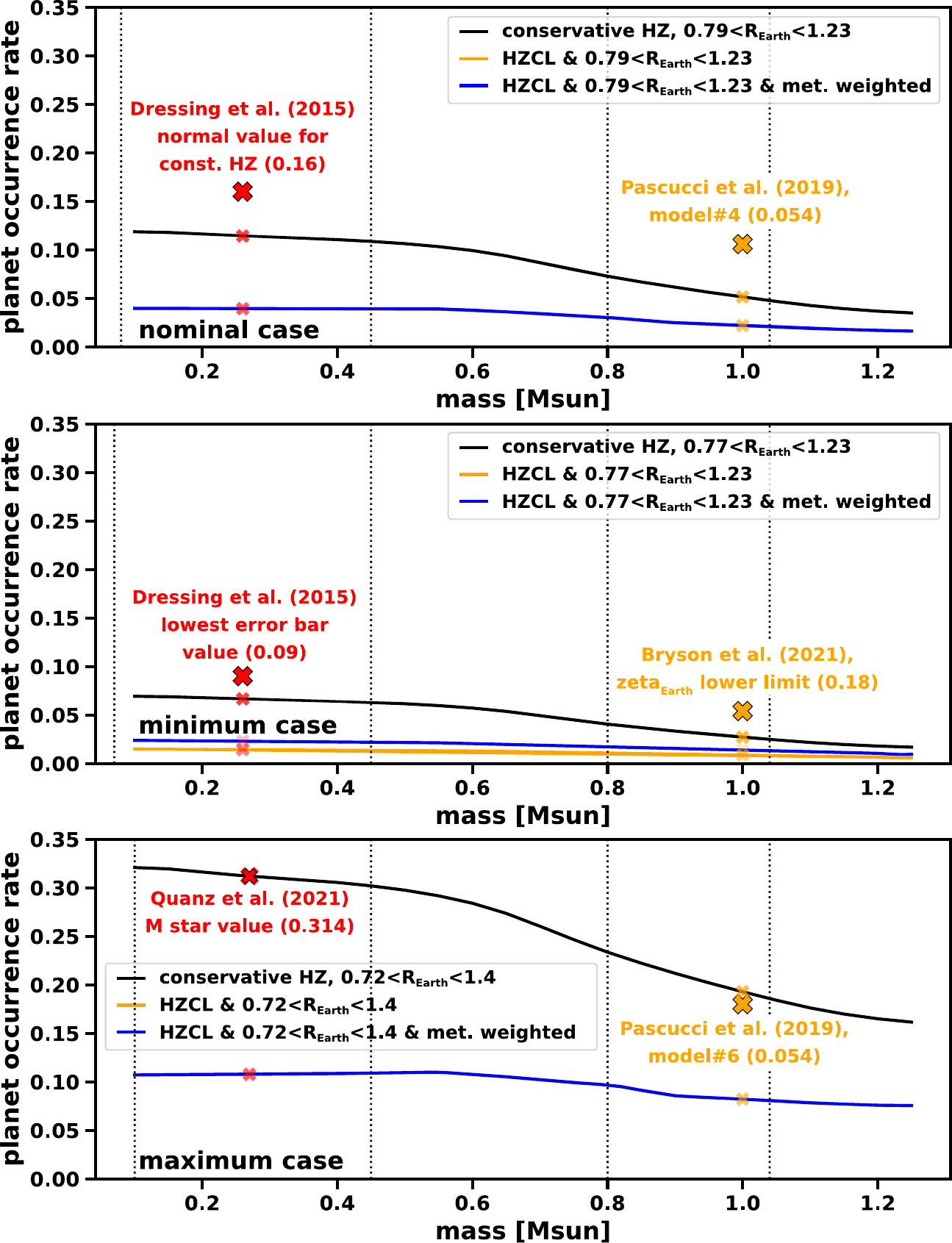}
\caption{Planet occurrence rates for our nominal (upper), minimum (middle) and maximum (lower panel) cases for an N$_2$-O$_2$-dominated atmosphere with $x_{\rm CO_2}\leq10\%$. The big red and orange crosses with the black outlines show the initial planet occurrence rates from the literature for M and G stars, respectively, from which $\beta_{\rm HZCL}$ is calculated. The black lines show the occurrence rates (i) scaled to the conservative HZ and the specific planetary radius ranges used for the various cases, and (ii) extrapolated to the entire stellar mass range by assuming a linear dependence on the stellar effective temperature. These mass-dependent and recalculated $\eta_{\oplus}$-values are then scaled to the HZCL of each stellar mass, which gives $\beta_{\rm HZCL}$ (orange lines, below blue lines for nominal and maximum cases). Weighting $\beta_{\rm HZCL}$ by metallicity gives $\beta_{\mathrm{HZCL},Z_{\rm min}}$ (blue lines). The smaller crosses show the corresponding occurrence rates for a mean-mass M dwarf and a solar-like star. For the nominal and maximum cases, $\beta_{\mathrm{HZCL},Z_{\rm min}}$ is indistinguishable from $\beta_{\rm HZCL}$. All scalings are performed with the power laws of `model\#4' from \citet{Pascucci2019}.}
\label{fig:bHZCL}
\end{figure}

For an atmosphere with $x_{\rm CO_2}\leq10\%$, the recalculated rocky planet occurrence rate {in the HZCL of solar-like stars} for our nominal case results in $\beta_{\rm HZCL} = 0.0219$ and a weighted value of $\beta_{\mathrm{HZCL},Z_{\rm min}} = 0.021$. This corresponds to an almost fivefold drop compared to an initial rate of $\eta_{\oplus} = 0.109$ as taken from `model\#4' of \citet{Pascucci2019}. For our maximum and minimum cases, we further find $\beta_{\rm HZCL} = 0.0822$ and $\beta_{\rm HZCL} = 0.0088$, respectively. The latter increases toward $\beta_{\mathrm{HZCL},Z_{\rm min}} = 0.014$ due to the significant metallicity weighting factor of $1/\mathcal{P}(Z_{\rm min,SN})=1.603$, whereas this is negligible for our maximum case. The final weighted HZCL planet occurrence rate averaged over all stellar masses is further found to be $\beta_{\rm HZCL,Z_{\rm min}}\,=\,$0.035(+0.070/-0.019) for atmospheres with $x_{\rm CO_2}\leq10\%$.

All these values were scaled with the broken power law coefficients from `model\#4' of \citet{Pascucci2019}. If we take the same broken power laws but with the coefficients published by \citet{Mulders2018}, the rocky planet occurrence rate in the HZCL of a solar-like star will change slightly from $\beta_{\rm HZCL}$\,=\,0.021 to $\beta_{\rm HZCL}$\,=\,0.023 for our nominal case. If varying the coefficient $a_{r}$ from \citet{Pascucci2019} within its error bars, $\beta_{\rm HZCL}$ can increase up to 0.034. This shows that varying the initial values of $\eta_{\oplus}$ usually has a larger impact on the final value of $\beta_{\rm HZCL}$ (see Table~\ref{tab:etaEarth} and Figure~\ref{fig:etaStar} for an extensive collation).

\begin{figure}
\centering
\includegraphics[width = 1.0\linewidth, page=1]{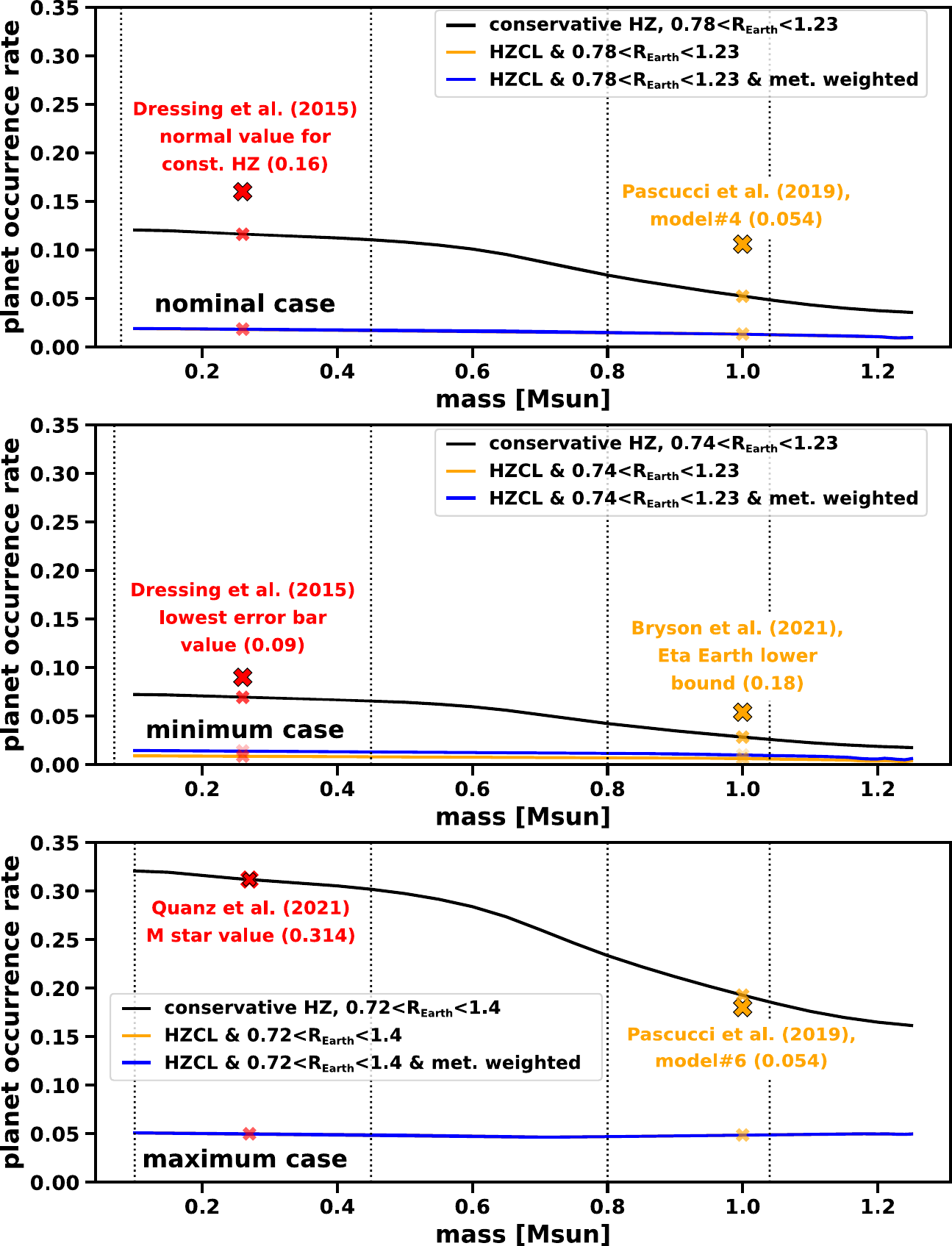}
\caption{The same as Figure~\ref{fig:bHZCL}, but for an N$_2$-O$_2$-dominated atmosphere with $x_{\rm CO_2}\leq1\%$. Here, the values for $\beta_{\rm HZCL}$ are smaller than for an atmosphere with $x_{\rm CO_2}\leq1\%$, since the HZCL is smaller to.}
\label{fig:bHZCL2p50}
\end{figure}

Another obvious factor that affects $\beta_{\rm HZCL}$ is the HZCL size. Because of implementing the outer HZCL boundary for $p$CO$_2 = 0.01$\,bar from \citet{Schwieterman2019} into our nominal case for atmospheres with $x_{\rm CO_2}\leq1\%$, the occurrence rate in the HZCL of solar-like stars decreases from $\beta_{\rm HZCL,10\%} = 0.0219$ to $\beta_{\rm HZCL,1\%} = 0.0132(+0.0355/-0.0069)$. The mean weighted value for the entire stellar mass range finally changes from $\beta_{\rm HZCL,Z_{\rm min},10\%}=0.035(+0.070/-0.019)$ to $\beta_{\mathrm{HZCL},Z_{\rm min},1\%} = 0.015(+0.0326/-0.0055)$.

If we apply these occurrence rates to the sample of stars, $N_{\star} \times \eta_{\star}$, capable of hosting EHs with N$_2$-O$_2$-dominated atmospheres containing $x_{\rm CO_2}\leq10\%$, we find a maximum number of $7.25(+53.07/-6.34) \times 10^{7}$ planets for the entire parameter range, i.e., from our minimum to maximum case. By only varying the initial planet occurrence rates from minimum to maximum (i.e., $\eta_{\oplus}\,=\,0.054-0.18$ for a solar-like star and $\eta=0.09-0.312$ for a mean-mass M dwarf), but keeping all other parameters the same, the number of planets varies between $3.99 \times 10^{7}$ and $13.37 \times 10^{7}$ planets, i.e., by almost an order of magnitude.

For N$_2$-O$_2$-dominated atmospheres with $x_{\rm CO_2}\leq1\%$, we obtain a maximum number of $0.66(+11.48/-0.57)\times 10^{7}$ planets, a number that theoretically can be as low as half a million planets. However, by now these numbers do not consider additional planetary requirements that must be met for a planet to evolve into an EH, suggesting that their actual number in the galactic disk will most likely be significantly lower.

\subsection{The appropriate planetary compositional/mineralogical set-up, $B_{\rm pc}^i$}\label{sec:pc}

The requirement for having an appropriate planetary compositional and mineralogical set-up, $B_{\rm pc}(\prod_{i=1}^n \beta_{\rm pc}^i)$ {from Equation~\ref{eq:etaEH}}, confines any necessary requirements that are related to the right environmental elemental abundances that a planet needs for \textit{life as we know it}. Besides the CHNOPS elements, this may also include radioactive heat-producing isotopes, specific planetary Mg/Si and C/O ratios, important ion donors such as Na, K, Mg, and Ca, or the availability of transition metals like Fe, Cu, Mo, or Ni \citep[see, e.g.,][]{Moore2017,Covone2022}. This requirement certainly also involves the necessity to hold an appropriate amount of a specific solvent for biological reactions, which - at least for aerobic complex life - is equivalent with H$_2$O as this is a very effective solvent for ions and polar molecules, but not for organic compounds \citep[e.g.,][]{SchulzeMakuch2008}.

Since our knowledge about exoplanet characteristics, their geophysical parameters, and statistics, as well as their relation to the origin and evolution of life, is still at a basic level, the importance of only a few necessary requirements is at present quantifiable to at some certain extent. In contrast, many are debated, or may not even be known yet. Within Sections~\ref{sec:pc} and \ref{sec:env}, we therefore focus on the chosen few that can already be implemented via {relatively} reasonable and quantifiable scientific arguments, and which are further refinable, or potentially refutable, by future observations}.

Further necessary requirements specifically feeding into $B_{\rm pc}(\prod_{i=1}^n \beta_{\rm pc}^i )$ are discussed in the {appended} Section~\ref{sec:otherBCHNOPS}. Even though these may not be quantifiable at present, it is obvious that at least some of them are crucial for the existence of EHs. We therefore strongly encourage the reader to consult this appended section, as well as  Sections~\ref{sec:otherEtaStar}, \ref{sec:otherBenv} and \ref{sec:otherBlife}, which discuss additional potential requirements that may feed into $\eta_{\star}$ {from Equation~\ref{eq:etaStar}}, and $B_{\rm env}(\prod_{i=1}^n \beta_{\rm env}^i )$ and $B_{\rm life}(\prod_{i=1}^n \beta_{\rm life}^i )$ {from Equation~\ref{eq:etaEH}}, respectively.

For $B_{\rm pc}(\prod_{i=1}^n \beta_{\rm pc}^i )$, we focus on  $\beta_{\rm pc}^{\rm H_2O}$, a planet's necessary requirement to possess the right abundance of water. While we present the effect of $\beta_{\rm pc}^{\rm H_2O}$ onto the distribution and maximum number of EHs in the subsequent section, we discuss the actual importance and potential amount of water needed to allow for the evolution of EHs -- and complex life -- in more detail in the appended Section~\ref{sec:h2oMain}. There, Section~\ref{sec:h2o} presents several arguments as to why it is very likely important for the evolution of an EH to possess both subaerial land and oceans to be present on its surface. In that section, we hence argue for the likely need for the so-called `habitable trinity' \citep[see,][]{Dohm2014}, i.e., the simultaneous presence of oceans, subaerial land, and a secondary atmosphere as necessary requirements for an EH to evolve  \citep[thereby also excluding sub-surface ocean worlds covered by a global ice layer such as Europa and Enceladus; see, e.g.,][for this class of habitats]{Lammer2009,Lammer2013,Hand2020}. In the related Section~\ref{sec:h2ofreq} we further discuss the frequency of planets that indeed host both subaerial land and oceans, based on a wide range of scientific studies.

Here, we highlight the caveat that it cannot be precisely answered at present whether an EH, and as a next step complex life, can also evolve on water worlds without any subaerial land, or even on subsurface-ocean worlds (which are excluded from our definition of EHs). A potential habitat fully covered by an ocean certainly provides very different conditions than a world with subaerial land. However, future models and observations of exoplanets can give further insights and potentially correct and refine the parameter space of our study. Next, we, therefore, implement $\beta_{\rm pc}^{\rm H_2O}$ into our model based on (i) the assumption that the simultaneous presence of subaerial land and oceans is a necessary requirement for the evolution of EHs, and (ii) on a wide range of potential occurrence rates of such planets, based on the discussion in the appended Section~\ref{sec:h2oMain}.

\subsubsection{The H$_2$O requirement, $\beta_{\rm pc}^{\rm H_2O}$}\label{sec:h2oEffect}

\paragraph{Calculating and implementing $\beta_{\rm pc}^{\rm H_2O}$:}\label{sec:h2oImp}

For implementing $\beta_{\rm pc}^{\rm H_2O}$ into our model, we need to extrapolate the frequency of planets that possess an appropriate water mass fraction for simultaneously hosting oceans and subaerial land, to the entire stellar mass range, as we did for $\beta_{\rm HZCL}$ (Section~\ref{sec:betaHZCL}). As discussed in the appended Section~\ref{sec:h2ofreq}, planet formation models suggest that $\beta_{\rm pc}^{\rm H_2O}$ could depend to some extent on the stellar spectral class, as rocky exoplanets around M dwarfs are more often either completely dry or completely covered by oceans compared to planets around K and G stars \citep[e.g.,][]{Mulders2015,Ciesla2015}. There could be a corresponding correlation between a planet's water abundance and the evolution of a star's short-wavelength flux, $L_{\rm X}$ and/or $L_{\rm XUV}$, as the studies by \citet{Tian2015} and \citet{Kimura2022} suggest. In \citet{Tian2015}, the bimodal distribution of planets is strongest for M dwarfs with $M_{\star}\,=\,$0.3\,M$_{\rm \odot}$, but generally increases for higher $L_{\rm XUV}/L_{\rm bol}$ values, i.e., for higher XUV surface fluxes, $F_{\rm XUV}$, in the HZ. High values of $F_{\rm XUV}$ and $F_{\rm X}$, in particular when the radiation remains high for long periods (as is the case with late-type M dwarfs), can lead to extensive loss of water, which can desiccate planets with a comparably small initial amount of H$_2$O. On the other hand, at certain water mass fractions much higher than Earth's, atmospheric escape may not be sufficient to deplete a planet's H$_2$O to the point where the simultaneous existence of oceans and subaerial land would be possible (see also Section~\ref{sec:h2ofreq}).

\citet{Tian2015} calculated the loss of water via the well-known energy-limited escape formalism, a relatively simple prescription of thermal atmospheric escape that is limited by a star's XUV luminosity \citep[e.g.,][]{Watson1981,Salz2016}, and which is known to often underestimate volatile loss for small and/or highly irradiated planets \citep[e.g.,][]{Kubyshkina2018,Krenn2021}. \citet{Kimura2022}, on the other hand, simulated atmospheric escape through a hydro-based approximation (HBA) equation found by \citet{Kubyshkina2018} through fitting a vast range of hydrodynamic upper atmosphere simulations. Both approaches are linearly dependent on the incident XUV flux from the host star, although the HBA approach can deviate from linear dependence for highly irradiated close-in planets where the atmospheric escape scales with $F_{\rm XUV}^{0.4222}$ -- but in the HZ this should mostly be linear \citep[see][]{Kubyshkina2018}. The  energy-limited equation can thus be written as \citep[e.g.,][]{Watson1981,Tian2015,Salz2016,Kubyshkina2018}
\begin{equation}\label{eq:elc}
  \dot{M_{\rm en}} = \frac{\epsilon R_{\rm XUV}^2R_{\rm pl}F_{\rm XUV}}{G M_{\rm pl}K},
\end{equation}
where $\dot{M_{\rm en}}$ is a planet's atmospheric mass-loss rate of, e.g., hydrogen, $\epsilon$ is the heating efficiency, $R_{\rm XUV}$ is the absorption radius of the XUV flux in the planetary atmosphere, $G$ is the gravitational constant, and $K$ is a factor that accounts for the so-called Roche-lobe effects \citep[see,][]{Erkaev2007}, which can be assumed $K = 1$ for planets in the HZ(CL). As can be seen, atmospheric escape described by the energy-limited formulation is linearly correlated with the XUV surface flux, $F_{\rm XUV}$, and hence also with a star's XUV luminosity via $F_{\rm XUV} \sim L_{\rm XUV}/d_{\rm pl}^2$, where $d_{\rm pl}$ is the orbital distance of the planet.

\citet{Tian2015} provide results for stars with $M_{\star} = \{0.3, 0.5, 1.0\}$\,M$_{\odot}$ based on planet formation, migration and atmospheric escape through the first $\sim$100\,Myr. Similarly, \citet{Kimura2022} present simulations for stellar masses of $M_{\star} = \{0.1, 0.3, 0.5, 1.0\}$\,M$_{\odot}$, but additionally include water enrichment {in the primordial atmosphere through the oxidation of atmospheric hydrogen by incoming planetesimals and the underlying magma ocean, resulting in an atmospheric water mass fraction} of $X_{\rm H_2O} = 0.8$. As these are the only studies that consider these mechanisms in one framework and provide results for different stellar masses, we use the values from \citet{Tian2015} and \citet{Kimura2022} as the main inputs to our model, but limit ourselves to planets with a water mass fraction of $w_{\rm H_2O}=0.003-0.2\%$, as explained in detail in the appended Section~\ref{sec:h2ofreq}.  From the simulations published by \citet{Tian2015} we follow their ``green'' scenario, which is based on a heating efficiency of $\epsilon=10\%$ and $L_{\rm XUV}/L_{\rm bol} = 10^{-4}$. In Mors, the stellar evolution model we implemented, $L_{\rm XUV}/L_{\rm bol}$ is also found to be on the order of $10^{-4}$ over the entire stellar mass range for the relevant stellar ages up to $~\sim$100\,Myr. From \citet{Kimura2022} we make use of both main scenarios, where (i) water enrichment from the primordial atmosphere is excluded (i.e., $X_{\rm H_2O} = 0$), and (ii) is included ($X_{\rm H_2O} = 0.8)$.

Fig.~2 in \citet{Tian2015} provides a detailed breakdown of the final number of planets within certain water mass fractions.
Since this figure only shows ratios in orders of magnitude, it is not entirely clear how many of the planets displayed fall within our range. For the G-type star we, therefore, take 74 out of 407 planets that are between a water mass fraction of $10^{-5}$ to $10^{-3}$ in Fig.~2 by \citet{Tian2015} for both our nominal cases. For our two maximum cases, we take 104 out of 407 planets, which additionally includes planets between a water mass fraction of $10^{-3}$ and $10^{-2}$. For our two minimum cases, we assume that 8\% out of all planets in the HZ of a solar-mass star have an appropriate water mass fraction. This is the fraction of planets that are not entirely covered by oceans, as obtained by \citet{Simpson2017} through Bayesian reasoning. However, this value might be somewhat optimistic since this fraction also includes desert planets.

For K dwarfs with $M_{\star}\,=\,$0.5\,M$_{\odot}$, we again use the values from \citet{Tian2015} by taking 5 out of 292 planets for our minimum cases, again corresponding to a water mass fraction between $10^{-5}$ and $10^{-3}$, as shown in their Fig.~2. For our nominal and maximum cases, we further take the values from \citet{Kimura2022} for $X_{\rm H_2O}=0$ and $X_{\rm H_2O}=0.8$, which give occurrence rates of $\sim0.0215$ and $\sim0.043$, respectively (see their Fig.~3).

For M dwarfs with $M_{\star}\,=\,$0.3\,M$_{\odot}$, we take 1 out of 55 planets for our two maximum cases, which corresponds to planets between water mass fraction of $10^{-5}$ and $10^{-2}$ in \citet{Tian2015}. For the nominal and minimum cases, we again implement the results from \citet{Kimura2022}, where fractions of $\sim0.013$ and $\sim 0.001$ end up between water contents of $w_{\rm H_2O}=0.003- 0.2\%$ for $X_{\rm H_2O}=0.8$ and $X_{\rm H_2O}=0$, respectively. For M dwarfs with $M_{\star}\,=\,$0.1\,M$_{\rm \odot}$, we finally implement the values provided from the same study for $X_{\rm H_2O}=0.8$ as our maximum, and for $X_{\rm H_2O}=0$ as our nominal and minimum cases, respectively. As can be calculated from Fig.~3 in \citet{Kimura2022}, fractions of $\sim0.003$ for $X_{\rm H_2O}=0.8$ and $\sim0.0015$ for $X_{\rm H_2O}=0$ are between our chosen water content range of $w_{\rm H_2O}=0.003-0.2\%$.

To further obtain occurrence rates of planets with the correct water mass fraction over the entire stellar mass spectrum, we simply assume that the occurrence rates between stars with stellar masses of  $M_{\star} = \{0.1, 0.3, 0.5, 1.0\}$\,M$_{\odot}$ are linearly correlated with the incident XUV surface flux for our nominal and maximum cases, and with the incident X-ray surface flux for our minimum cases. Since stellar spectral class also appears to have some influence on the occurrence rate, we additionally correlate it with the reciprocal of the stellar effective temperature, $T_{\rm eff}$, in the same way as we applied it to $\beta_{\rm HZCL}$. For each stellar mass, $F_{\rm XUV}(M_{\star})$, $F_{\rm X}(M_{\star})$, and $T_{\rm eff}(M_{\star})$ are averaged over the first 100\,Myr, the timescale for which \citet{Tian2015} carried out their simulations. We scale the frequency of planets with the right amount of water therefore via
\begin{equation}\label{eq:corrH2O}
  \beta_{\rm pc}^{\rm H_2O}(F_{\rm XUV},T_{\rm eff}) = a_{\mathrm{H_2O},i} \frac{F_{\rm XUV}}{T_{\rm eff}} + b_{\mathrm{H_2O},i},
\end{equation}
where the coefficients $a_{\mathrm{H_2O},i}$ and $b_{\mathrm{H_2O},i}$ {for the stellar mass ranges $M_{\star,1}\,=\,0.1-0.3$\,M$_{\odot}$, $M_{\star,2}\,=\,0.3-0.5$\,M$_{\odot}$, $M_{\star,3}\,=\,0.5-1.0$\,M$_{\odot}$, and $M_{\star,4}\,=\,1.0-1.25$\,M$_{\odot}$,} are derived from the corresponding occurrence rates at $M_{\star} = \{0.1, 0.3, 0.5, 1.0\}$\,M$_{\odot}$, respectively. {Note that this monotonic increase with $F_{\rm XUV}$ might not be entirely realistic at all values, since extremely high XUV fluxes could swiftly remove (nearly) all of the water from the planet.}

\paragraph{The effect of $\beta_{\rm pc}^{\rm H_2O}$:}\label{sec:h2oEff2}

\begin{figure*}
\centering
\includegraphics[width = 0.7\linewidth, page=1]{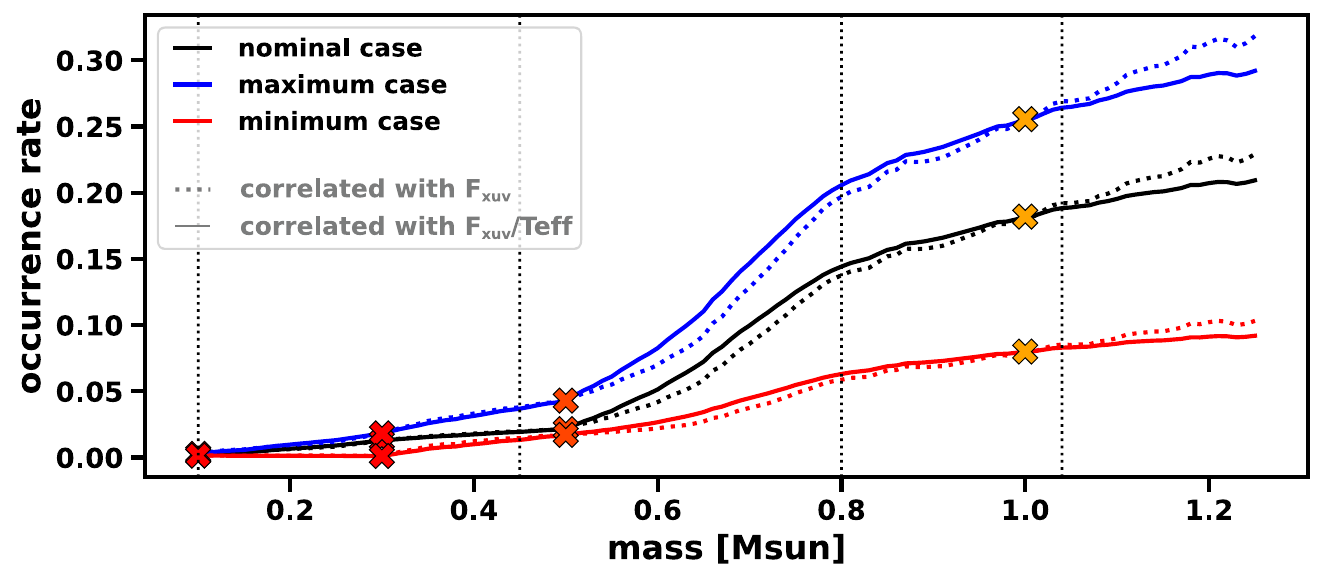}
\caption{The frequency of planets with an appropriate water mass fraction as a function of stellar mass for our nominal (black), maximum (blue) and minimum (red) cases. The dotted lines show occurrence rates for which we assumed a simple linear correlation with $F_{\rm XUV}$ (nominal and maximum cases) or $F_{\rm X}$ (minimum case) between $M_{\star}=\{0.3, 0.5, 1.0\}$\,M$_{\odot}$, whereas the solid lines show the same but correlated via $F_{\rm XUV}/T_{\rm eff}$ or $F_{\rm X}/T_{\rm eff}$, respectively. The crosses illustrate the implemented occurrence rates for $M_{\star}=\{0.3, 0.5, 1.0\}$\,M$_{\odot}$ with water contents of $w_{\rm H_2O} = 0.003-0.02\%$ as taken from \citet{Tian2015}, \citet{Kimura2022}, and \citet{Simpson2017}.}
\label{fig:fWater}
\end{figure*}

Figure~\ref{fig:fWater} shows the derived occurrence rates for our nominal, maximum, and minimum cases as a function of stellar mass. The different crosses display the planet frequencies with appropriate water mass fractions based on the results of \citet{Tian2015}, \citet{Simpson2017} and \citet{Kimura2022}, as described above. The dotted lines display occurrence rates derived from linear correlations with stellar $F_{\rm XUV}$ (nominal and maximum cases) and $F_{\rm X}$ (minimum cases), whereas the solid lines are additionally correlated to the inverse of the stellar effective temperature, i.e., to $F_{\rm XUV}/T_{\rm eff}$ and $F_{\rm X}/T_{\rm eff}$, respectively. The latter leads to lower occurrence rates for stars with $M_{\star}>1.0$\,M$_{\rm \odot}$, but to higher rates for $M_{\star}\,=\,0.5-1.0$\,M$_{\rm \odot}$, a behavior that seems more realistic than correlating the occurrence rate only with $F_{\rm XUV}$ or $F_{\rm X}$. Correlating $\beta_{\rm pc}^{\rm H_2O}$ with both slightly increases the total number of planets that remain in our sample compared to the simpler correlation.

\begin{figure}
\centering
\includegraphics[width = 1.0\linewidth, page=1]{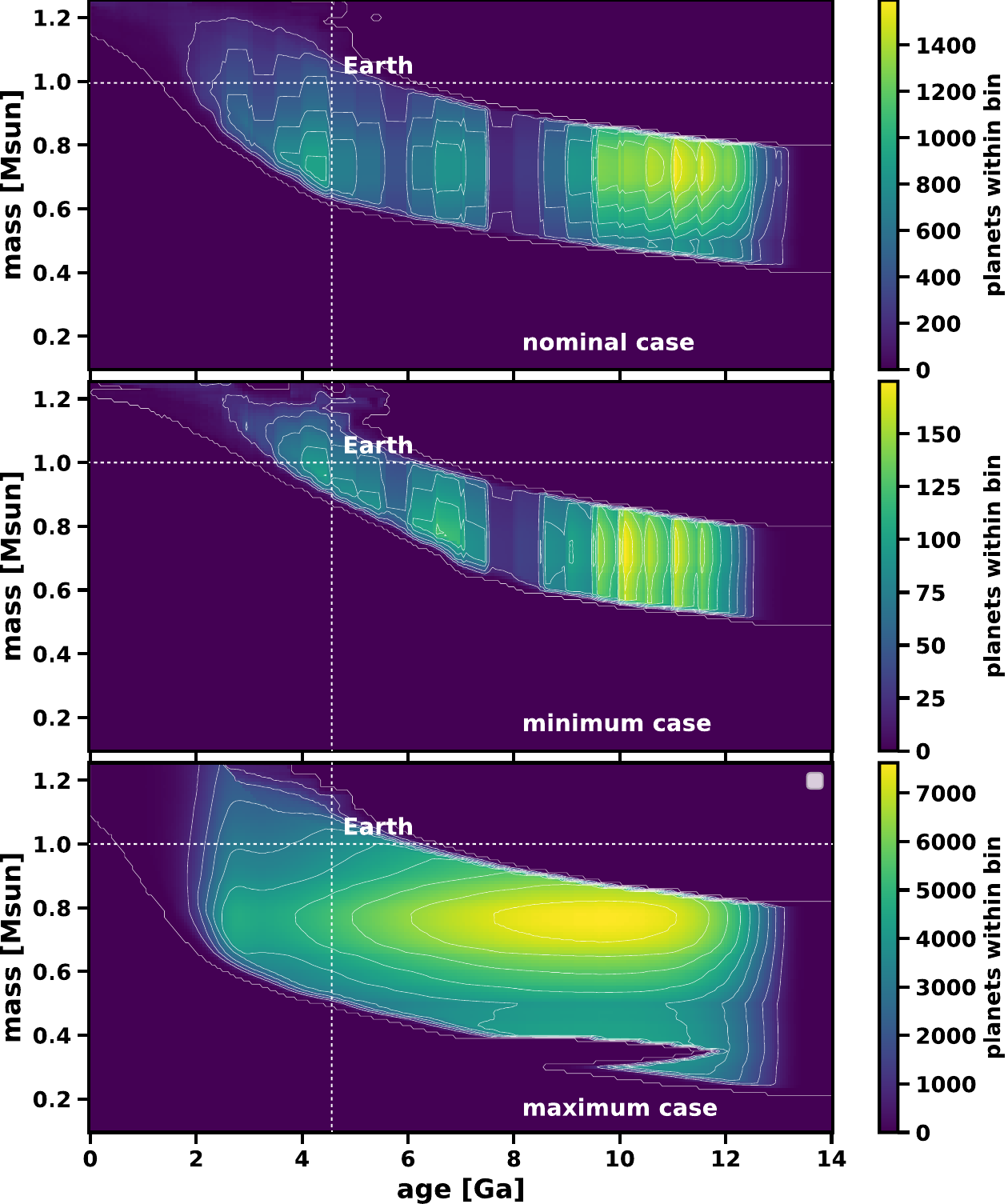}
\caption{The distribution of remaining planets with the right amount of water that can in principal host an N$_2$-O$_2$-dominated atmosphere with $x_{\rm CO_2,max}=10\%$ as a function of stellar mass and age for our nominal (upper), minimum (middle) and maximum (lower panel) cases.}
\label{fig:H2OPlanets}
\end{figure}

For our nominal case with $x_{\rm CO_2,max}=10\%$, a total of $5.44\times10^{6}$ planets meet the water requirement. This amounts to 7.51\% of all rocky exoplanets in the HZCL. If we instead correlate $\beta_{\rm pc}^{\rm H_2O}$ with the XUV surface flux only, the number of remaining planets will decrease slightly toward $4.95\times10^{6}$ worlds. By varying $\beta_{\rm pc}^{\rm H_2O}$ only from our maximum to minimum values and keeping all other parameters constant, the numbers range from $2.75\times10^{6}$ to $8.25\times10^{6}$ planets. Correlating our minimum case with $F_{\rm X}$ instead of $F_{\rm XUV}$ to stay consistent with the implementation of $\alpha_{\rm at}^{\rm ll}$, the lower value decreases slightly from $2.75\times10^{6}$ to $2.65\times10^{6}$ planets. By changing all parameters from minimum to maximum, we obtain an entire range of $5.44(+50.55/-5.01)\times10^{6}$ worlds, for which the host stars and planets still meet each of the as of yet implemented requirements. Figure~\ref{fig:H2OPlanets} shows the distribution of remaining planets for our nominal, minimum and maximum cases with atmospheric CO$_2$ mixing ratios of $x_{\rm CO_2,max}=10\%$ as a function of stellar mass and birth age. Be aware, however, that this figure's panels show the planet density per mass and age bin in a non-logarithmic scale, in contrast to all earlier distribution plots.

So, by only including $\beta_{\rm HZCL}$ and $\beta_{\rm pc}^{\rm H_2O}$ into our model, we obtain a value of $\eta_{\rm EH,10\%} < 0.0026(+0.0071/-0.0017)$ for atmospheres with $x_{\rm CO_2,max}=10\%$. For N$_2$-O$_2$-dominated atmospheres with $x_{\rm CO_2,max}=1\%$, we further obtain $\eta_{\rm EH,1\%} < 0.0021(+0.0042/-0.0003)$, which amounts to $0.94(+16.67/-0.79)\times10^{6}$ remaining planets\footnote{The reason for this significantly lower number of suitable planets compared to our 10\% CO$_2$ case, even though $\eta_{\rm EH}$ is almost identical in both scenarios, is mostly due to $\eta_{\star}$ being much lower for $x_{\rm CO_2,max}=1\%$ than for $x_{\rm CO_2,max}=10\%$.}.
{Notably, our obtained average fraction of planets having subaerial land and oceans at the same time for all our cases and the entire stellar mass range is $\beta_{\rm pc}^{\rm H_2O} = 0.093(+0.055/-0.018)$, which is in very good agreement with the recent estimate by \citet{Lingam2019WaterFrac}, who found this fraction to be $\sim$0.1. {A very recent study by \citet{Stern2024} estimates this fraction to be $\beta_{\rm pc}^{\rm H_2O}=0.0002-0.01$, i.e., by one to three orders of magnitude lower than ours.}}


\subsection{Long-term environmental stability, $B_{\rm env}^i$}\label{sec:env}

Several necessary parameters feed into the required long-term environmental stability term, $B_{\rm env}(\prod_{i=1}^n \beta_{\rm env}^i)$, such as the particularly important $\beta_{\rm env}^{\rm cycle}$, the necessary requirement of functioning carbon-silicate and nitrogen cycles. Although most of them are difficult to evaluate scientifically at present (see specifically appended Section~\ref{sec:other}), there is a factor, albeit a controversial one \citep[see, e.g.,][]{Waltham2019}, that could play a crucial role in the development and stability of EHs and the frequency of which can already be quantified to some certain extent. This is $\beta_{\rm env}^{\rm moon}$, the potential need of a planet to host a large satellite, as the Earth does with the Moon.

We know that the importance of a large moon for the evolution of an Earth-like biosphere may not be entirely clear. However, we include it into our model (i) since its potential prevalence can be discussed and estimated, and (ii) to illustrate one of the potential requirements feeding into $B_{\rm env}$. Even if large moons turn out to be negligible, our derived estimates will most likely remain absolute maximum values (also see Section~\ref{sec:discussion}). This is because other factors not considered in our model such as $\beta_{\rm env}^{\rm cycle}$ and the origin of life play a crucial role for EHs to emerge and evolve. Even though their frequency cannot be estimated accurately with our present knowledge, it is reasonable to assume that both these requirements may restrict the number of EHs to a large extent. However, there are several reasonable arguments as to why a large moon is indeed a requirement for the development of an EH and complex life. We therefore address these various strands of argument in detail in the appended Section~\ref{sec:moon1} and discuss the corresponding occurrence rate based on the scientific literature in the appended Section~\ref{sec:moonFreq}. Building on these arguments and occurrence rates, we will next implement $\beta_{\rm env}^{\rm moon}$ into our model.

\subsubsection{The large moon requirement, $\beta_{\rm env}^{\rm moon}$}\label{sec:moonEff}

\paragraph{Implementing $\beta_{\rm env}^{\rm moon}$:}\label{sec:moonImp}

For implementing the large moon requirement, we start with occurrence rate simulations based on solar-like stars. As outlined in the appended Section~\ref{sec:moonFreq}, \citet{Elser2011} simulated the frequency of large moons around Earth-like planets. We will implement their results for our nominal, minimum, and maximum cases for stars with a stellar mass of $M_{\star}=1.0$\,M$_{\odot}$. Even though the results by \citet{Elser2011} are the only simulations that specifically modeled the frequency of such moons, it was, e.g., shown by \citet{Brasser2013} that not all of these systems will turn out to evolve habitable conditions. Satellites with the right characteristics to aid the evolution and stability of EHs may hence be rarer still since these can either collide with the planet or even enhance obliquity instabilities. Taking the occurrence rates from \citet{Elser2011} may therefore be a relatively optimistic way of implementing $\beta_{\rm env}^{\rm moon}$ into our model but leaves the possibility of moon-less planets to nevertheless evolve to an EH by resolving all the various arguments given in Section~\ref{sec:moon1} through other means.

For M dwarfs, we can apply the results by \citet{MartinezRodriguez2019}, who specifically studied the orbital stability of moons around the by-then-known planets in the HZ of M stars. As these authors have calculated, 4 out of 33 planets are in principle capable of providing stable conditions for timescales longer than the Hubble time, $t_0$. To obtain an estimate on the occurrence rate of large satellites around M dwarfs, we can simply multiply the fraction of stable moons found by \citet{MartinezRodriguez2019} with the fraction of systems that may form a large moon in the first place, as simulated by \citet{Elser2011}. This implicitly assumes that the general formation of large satellites around M dwarfs occurs with a frequency similar to that of G stars and that each of these moons indeed allows habitable conditions over large timescales (a potential overestimate, as explained above). The obtained value will then be applied to the mean stellar mass of the M-type spectral class.

To obtain a continuous frequency distribution over the entire range of stellar masses, we correlate the frequency of large moons with the host planet's Hill radius, $r_{\rm H}$, i.e.,
\begin{equation}\label{eq:rH}
  r_{\rm H} = d_{\rm \langle HZCL \rangle} \sqrt[3]{\frac{M_{\rm pl}}{3 M_{\star}}},
\end{equation}
via the linear function
\begin{equation}\label{eq:betaMoon}
\beta_{\rm env}^{\rm moon}(r_{\rm H}) = a_{\rm moon} r_{\rm H} + b_{\rm moon},
\end{equation}
where the coefficients $a_{\rm moon}$ and $b_{\rm moon}$ are again derived through our predefined occurrence rates at $M_{\star} = 0.26$\,M$_{\odot}$ (for nominal and minimum cases; $M_{\star} = 0.27$\,M$_{\odot}$ for the maximum case), and $M_{\star} = 1.0$\,M$_{\odot}$ (for all cases), respectively. In Equation~\ref{eq:rH}, we further set the planetary mass $M_{\rm pl} = 1\,$M$_{\oplus}$, since we assume an Earth-mass as the average planetary mass within our model (see Section~\ref{sec:betaHZCL}). Scaling the frequency of long-lived large satellites between our predefined frequencies with $r_{\rm H}$ is a reasonable choice since various studies have shown that the long-term stability of a large moon is strongly dependent on the expansion of its host planet's Hill sphere with no stable circumplanetary satellite orbits existing outside such a sphere \citep[e.g.,][]{Domingos2006,Sasaki2012,Sasaki2014,Piro2018,Martinez2019,Tokadjian2020,Dobos2021,Hansen2022}.

\begin{figure}
\centering
\includegraphics[width = 1.0\linewidth, page=1]{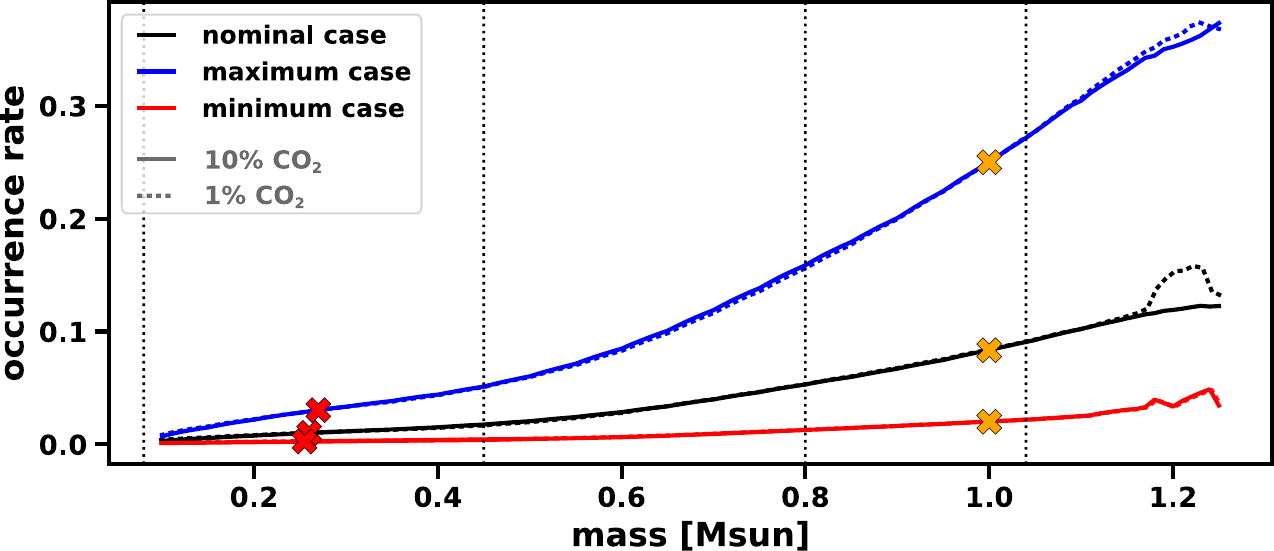}
\caption{The frequency of planets with a large moon as a function of stellar mass for our nominal (black), maximum (blue) and minimum (red) cases, {and for the corresponding atmospheric scenarios with $x_{\rm CO_2,max}=10\%$ (solid lines) and $x_{\rm CO_2,max}=1\%$ (dotted lines)}. The crosses show the assumed occurrence rates for $M_{\star} = 1.0$\,M$_{\odot}$ (orange) taken from \citet{Elser2011} and for the mean stellar mass of M dwarfs (red) through multiplying the fraction of stable planet-moon systems around M stars from \citet{MartinezRodriguez2019} with the values of \citet{Elser2011}. {There is a small difference between the two atmospheric scenarios, since $d_{\langle \rm HZCL \rangle}$ shows slight differences between the scenarios with $x_{\rm CO_2,max}=10\%$ and $x_{\rm CO_2,max}=1\%$.}}
\label{fig:mOf}
\end{figure}

\paragraph{The effect of $\beta_{\rm env}^{\rm moon}$:}\label{sec:moonEff2}
Figure~\ref{fig:mOf} shows the occurrence rate distributions for our nominal (black), maximum (blue), and minimum cases (red), {and for our corresponding atmospheric scenarios with $x_{\rm CO_2,max}=10\%$ (solid lines) and $x_{\rm CO_2,max}=1\%$ (dotted lines),} as derived through the methodology described above. {The slight differences between the two atmospheric scenarios stem from the fact that the mean HZCL distance, $d_{\langle \rm HZCL \rangle}$, which feeds into $r_{\rm H}$, is slightly different in both scenarios since their stable HZCL is not entirely similar (see Section~\ref{sec:ul}).} The orange crosses indicate the frequency of planets with a long-lived large moon at $M_{\star}\,=\,$1\,M$_{\odot}$, which correspond to 0.08, 0.25, and 0.02 for our nominal, maximum, and minimum cases, i.e., to the mean, maximum, and minimum values for the occurrence rate of large moons derived by \citet{Elser2011}. For the mean stellar mass of M dwarfs (red crosses at $M_\star = \{0.255, 0.26, 0.27\}$\,M$_{\odot}$), the occurrence rates are 0.01, 0.03, and 0.002 for our nominal, maximum, and minimum cases, respectively, derived by multiplying the values from \citet{Elser2011} with the findings by \citet{MartinezRodriguez2019}. Even though these values seem to be relatively low, particularly for M dwarfs, we point out that they may nonetheless overestimate the frequency of large moons around low-mass stars since, as outlined in appended Section~\ref{sec:moonFreq}, various studies find that large satellites may never be stable around these stars, potentially even for stellar masses as large as $M_{\star}=0.64$\,M$_{\odot}$ \citep{Sasaki2014}. Furthermore, we have neglected that not all the satellites formed will ultimately lead to the planet becoming more habitable, as shown by \citet{Brasser2013}.

To put our occurrence rates into further context, we can compare them with some of the lunar survival rates calculated by \citet{Dobos2021}, even though there are not many HZ planets currently known to have a radius comparable to the Earth. However, the planet Kepler-186f with a radius of $R_{\rm pl}\sim$1.15\,R$_{\oplus}$ orbits a host star with a mass of $M_{\star}=0.54$\,M$_{\odot}$ toward the outer reaches of its respective HZ. For this body, \citet{Dobos2021} find a survival rate of 0.09 for a large moon hypothetically existing in this system. If one multiplies this value with the frequency of large satellites forming in the first place of $\sim$0.02-0.25 from \citet{Elser2011}, this gives a rate of $\sim$0.002-0.023. With the same stellar mass, we get a higher value of $\sim$0.005-0.07, even though our assumed average planet orbits closer to the star and has a lower mass. For all the Trappist-1 planets and for Proxima Centauri~B, \citet{Dobos2021} find a survival rate of 0, whereas we obtain $\sim$0.001-0,007 and $\sim$0.0015-0.015 for $M_{\star}=0.08$\,M$_{\odot}$ and $M_{\star}=0.12$\,M$_{\odot}$, respectively. For higher stellar masses, Kepler-452b with $R_{\rm pl}\sim$1.6\,R$_{\oplus}$, a planet orbiting within (or slightly inside) the HZ of a star with $M_{\star}=1.04$\,M$_{\odot}$, gives a survival rate of 0.41 \citep{Dobos2021}, which amounts to an occurrence rate of $\sim$0.008-0.1. This is again lower than our derived frequency of $\sim$0.02-0.27 for the same stellar mass. The last planet that somehow makes sense to compare with is Kepler-1638b at the inner HZ boundary of a star with $M_{\star}=0.97$\,M$_{\odot}$, although its radius is already estimated to be $R_{\rm pl}=1.83$\,M$_{\oplus}$. According to \citet{Dobos2021}, this system shows a survival rate of 0.33, which corresponds to an inferred occurrence rate of $\sim0.007-0.08$. Our value for this stellar mass, on the other hand, is between $\sim$0.02-0.27. These examples illustrate that our inferred occurrence rate distribution may be optimistic over the entire range of stellar masses.

\begin{figure}
\centering
\includegraphics[width = 1.0\linewidth, page=1]{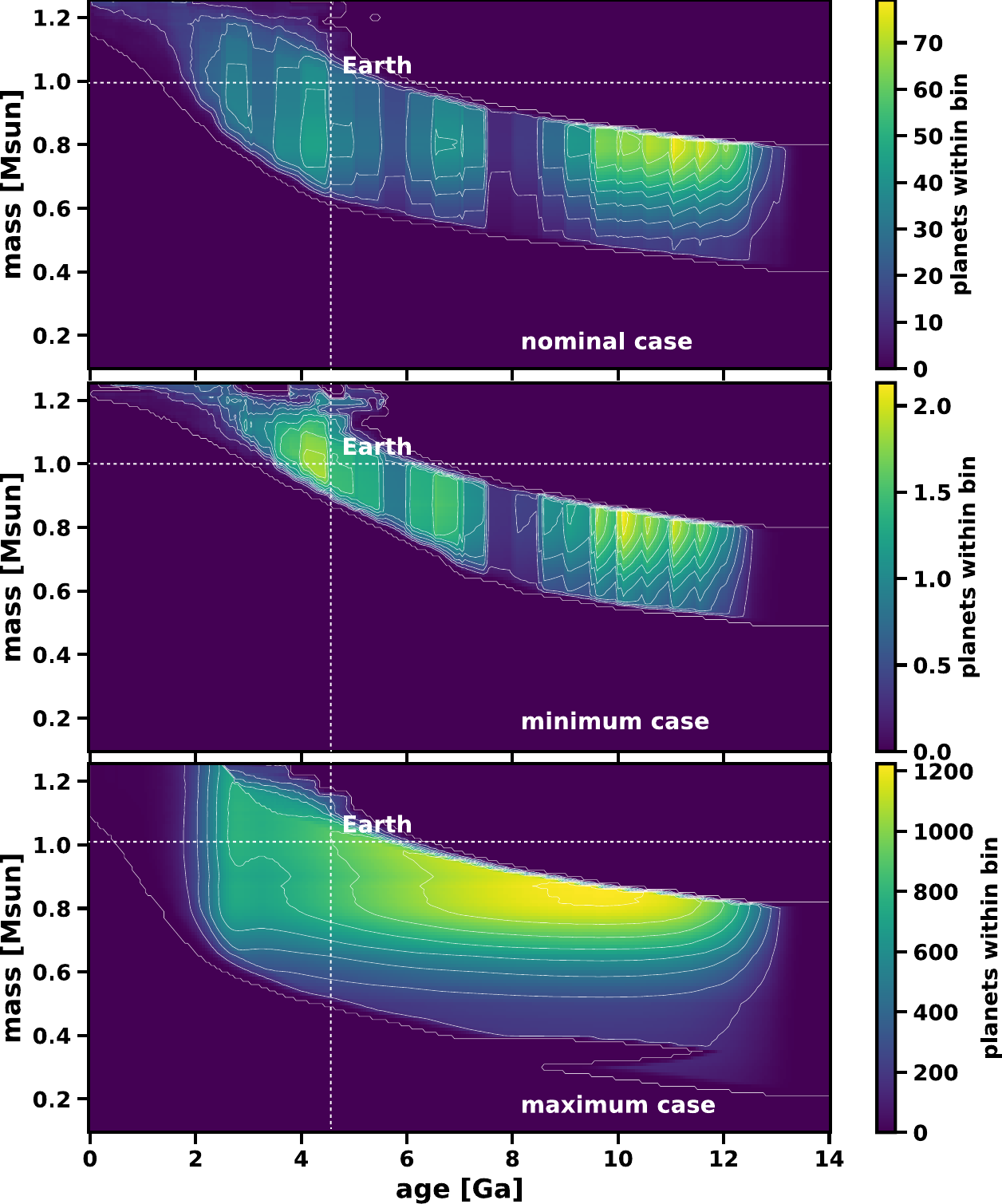}
\caption{The distribution of planets with N$_2$-O$_2$-dominated atmosphere containing $x_{\rm CO_2,max}=10\%$ that meet all implemented requirements, including $\beta_{\rm env}^{\rm moon}$, for our nominal (upper panel), minimum (middle panel) and maximum cases (lower panel).}
\label{fig:mO}
\end{figure}

Based on our derived distribution, we get a total range of $2.54(+71.64/-2.48)\times10^{5}$ planets from our minimum to the maximum case for atmospheres with $x_{\rm CO_2,max}=10\%$ (illustrated in Figure~\ref{fig:mO}). For $x_{\rm CO_2,max}=1\%$, we further obtain $0.57(+27.07/-0.56)\times10^{5}$ planets that still meet all the implemented requirements. As a comparison, if we implement an occurrence rate distribution based on the calculated survival rates from \citet{Dobos2021} for Proxima Centauri~B, Kepler-186f, and Kepler-452b by additionally assuming that no stable moons can exist in the HZ of stars with $M_{\star}\leq0.2$\,M$_{\odot}$, our maximum number of EHs drops two to three times to $9.08(+255.42/-8.87)\times10^{4}$ and $2.31(+105.88/-2.26)\times10^{4}$ planets for $x_{\rm CO_2,max}=10\%$ and $x_{\rm CO_2,max}=1\%$, respectively. {If large moons turn out to be irrelevant for the emergence of Earth-like Habitats, their maximum number will remain at $5.44(+50.55/-5.01)\times10^{6}$ and $0.94(+16.67/-0.79)\times10^{6}$ planets for $x_{\rm CO_2,max}=10\%$ and $x_{\rm CO_2,max}=1\%$, respectively, i.e., at the values we obtained after implementing $\beta_{\rm pc}^{\rm H_2O}$.}

By including the large moon requirement, however, $\eta_{\rm EH}$ can be found to be $\eta_{\rm EH,10\%} = 1.21(+11.58/-1.10)\times 10^{-4}$ for atmospheres with $x_{\rm CO_2,max}=10\%$ and $\eta_{\rm EH,1\%} = 1.28(+9.48/-1.15)\times 10^{-4}$ for $x_{\rm CO_2,max}=1\%$. As can be seen, $\eta_{\rm EH}$ is slightly higher for atmospheres with $x_{\rm CO_2,max}=1\%$, at least for our nominal case. This effect can also be seen by comparing the average moon-hosting frequency among the planets remaining in our distribution. This is $0.047(+0.086/-0.034)$ for the 10\%\,CO$_2$ and $0.061(+0.093/-0.044$ for the 1\%\,CO$_2$ case, respectively. If we take the occurrence rate distribution based on the results by \citet{Dobos2021} these values decline towards $0.017(+0.031/-0.012$ and $0.025(+0.036/-0.018$, respectively.


\subsection{Final values for $\eta_{\rm EH}$ and $N_{\rm EH}$}\label{sec:NEH}

Our scientific knowledge about the planetary conditions to be met for a system to develop into an EH is relatively poor. Accordingly, only three requirements feeding into $\eta_{\rm EH}$ were discussed, evaluated, and implemented in our model in the last sections. Not even the importance of all of them, such as the role of a large moon, is very well established. However, there are a number of other necessary and potential requirements that we have not yet considered in $\eta_{\rm EH}$. This makes it unlikely that the actual number of EHs in the galactic disk will be higher than our inferred maximum range, even if the role of a large moon or the need for oceans and subaerial land is overemphasized. The three implemented requirements feeding into $\eta_{\rm EH}$ are

\begin{description}
  \item[\textbf{The frequency of rocky exoplanets in the HZCL:}] By scaling different estimates of $\eta_{\oplus}$ to the HZCL and a certain planetary radius range, we find an occurrence rate of $\beta_{\rm HZCL} \sim 0.02 - 0.10 $ for N$_2$-O$_2$-dominated atmospheres with $x_{\rm CO_2,max}=10\%$, whereas we find $\beta_{\rm HZCL} \sim 0.01 - 0.05$ for atmospheres with $x_{\rm CO_2,max}=1\%$. The lower the upper CO$_2$ threshold is set, the lower the number of rocky exoplanets in the correspondingly reduced HZCL will be. Earth-like atmospheres with about 400\,ppm (i.e., 0.04\%) CO$_2$ will therefore be rarer than our derived values because of the corresponding decrease in the size of the HZCL and, accordingly so, the rate of planets contained therein, but also due to a decrease in atmospheric thermal stability against high XUV surface fluxes.
\item[\textbf{The appropriate water mass fraction:}] Based on different geophysical and planet formation studies, we find that $\sim 5-9\%$ (for $x_{\rm CO_2,max}=10\%$) and $\sim 7-15\%$ (for $x_{\rm CO_2,max}=1\%$) of the rocky exoplanets in the HZCL could have adequate water mass fractions to support surface water and subaerial land simultaneously. Here, the actual occurrence rate depends on the remaining sample of stars. By considering N$_2$-O$_2$-dominated atmospheres with $x_{\rm CO_2,max}=1\%$, there are simply no more low-mass stars in the remaining distribution for which $\beta_ {\rm pc}^{\rm H_2O}$ could be particularly low.
  \item[\textbf{The importance of a large moon:}] Several arguments point to the potential importance of a large moon to allow for the evolution of EHs. However, to account for the possibility that habitable worlds may also emerge by resolving the arguments listed in appended Section~\ref{sec:moon1} without the need of a large satellite, we find that $\sim 1-13\%$ (for $x_{\rm CO_2,max}=10\%$) and $\sim 2-15\%$ (for $x_{\rm CO_2,max}=1\%$) of the remaining planets can meet this requirement. A stricter interpretation of $\beta_{\rm env}^{\rm moon}$, by taking into account different survival and formation rates based on both \citet{Dobos2021} and  \citet{Elser2011} leads to values of $\sim$0.4-4\% and $\sim$0.6-6\% for our two atmospheric scenarios, respectively.
\end{description}

\begin{figure}
\centering
\includegraphics[width = 1.0\linewidth, page=1]{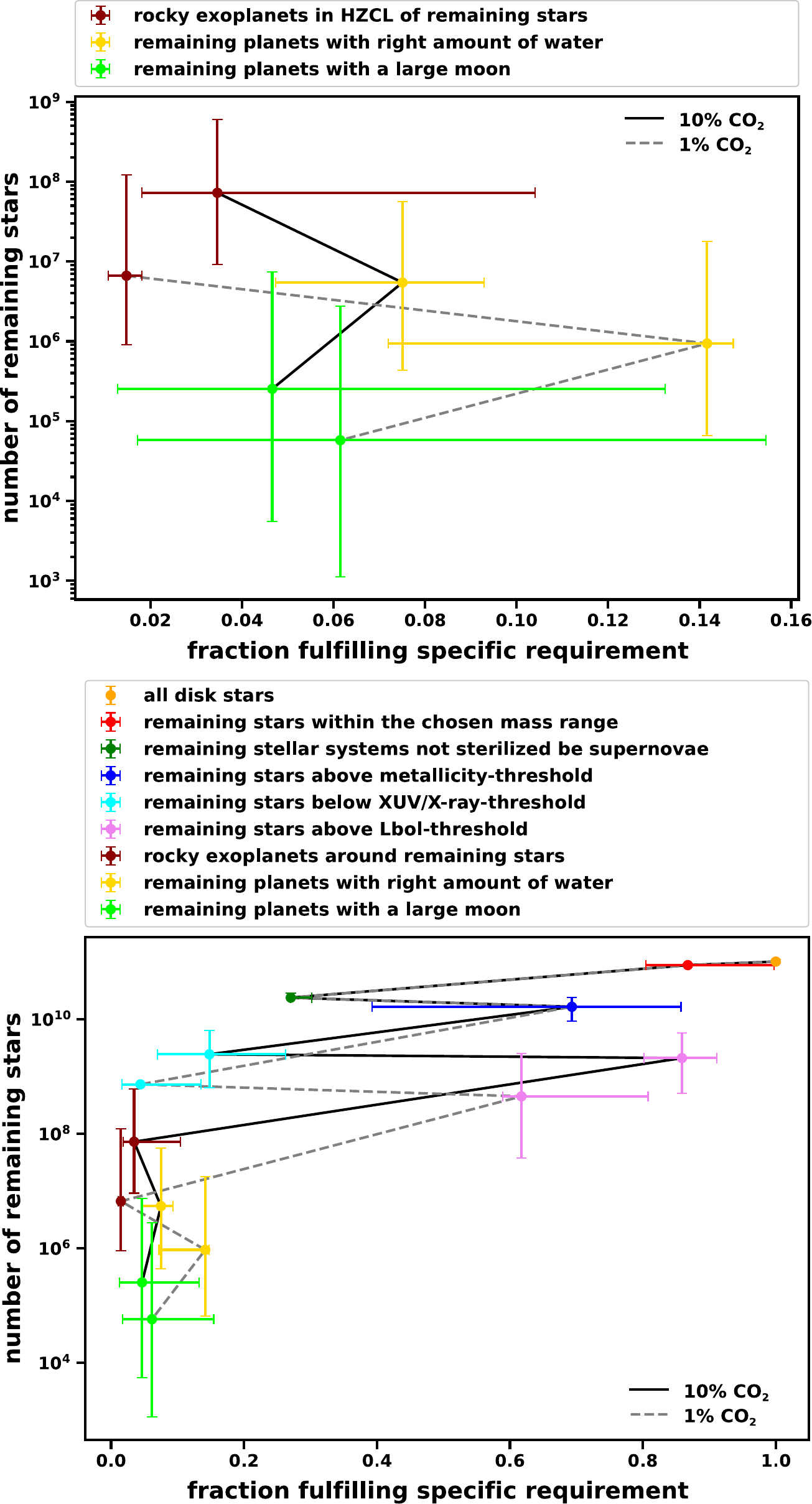}
>
\caption{Same as Figure~\ref{fig:etaStar} but for necessary requirements only related to $\eta_{\rm EH}$ (upper panel) and for all implemented requirements combining $\eta_{\rm \star}$ and $\eta_{\rm EH}$ (lower panel). {The planetary requirements were calculated in the following sections: occurrence rate of rocky exoplanets in the HZCL, Section~\ref{sec:bHZCLMain}; planets with an appropriate water mass fraction, Section~\ref{sec:h2oEffect}; planets with a large moon, Section~\ref{sec:moonEff}. See also Tables~\ref{tab:AppBetaHZCL} and \ref{tab:AppEtaEH} for the exact parameter values feeding into the planetary parameters.}}
\label{fig:etaEH}
\end{figure}


\begin{table*}\footnotesize
  \begin{center}
    \caption{The effect of the necessary requirements onto $\eta_{\rm EH}$.}
    \label{tab:etaEH}
    \resizebox{0.8\textwidth}{!}{%
\begin{tabular}{l|c|c|c|c}
  \hline
   & suitable stars & planet frequency  & water & moon \\
    & $N_{\star}\times \eta_{\star}$ & $\beta_{\rm HZCL}$  & $\beta_{\rm pc}^{\rm H_2O}$ & $\beta_{\rm env}^{\rm moon}$  \\
   \hline
   & & & & \\
   remaining fraction  & 1$^a$ & $0.0346^{+0.0694}_{-0.0165}$$^a$ & $0.0750^{+0.0178}_{-0.0277}$$^a$ & $0.0466^{+0.0858}_{-0.0339}$$^a$ \\
   (individual) & 1$^b$ & $0.0147^{+0.0326}_{-0.0040}$$^b$ & $0.1416^{+0.0057}_{-0.0696}$$^b$ & $0.0614^{+0.0930}_{-0.0443}$$^b$ \\
   & & & & \\
   remaining fraction & 1$^a$ & $0.0346^{+0.0694}_{-0.0165}$$^a$ & $0.0026^{+0.0071}_{-0.0017}$$^a$ & $0.00012^{+0.0012}_{-0.00011}$$^a$ \\
    ($\eta_{\rm EH}$ cumulative) & 1$^b$ & $0.0147^{+0.0326}_{-0.0040}$$^b$ & $0.0021^{+0.0049}_{-0.0013}$$^b$ & $0.00013^{+0.00095}_{-0.00012}$$^b$ \\
   & & & & \\
   remaining fraction & $206.51^{+411.55}_{-159.58}$$^a$ & $7.1459^{+57.1509}_{-6.2942}$$^a$ & $0.5363^{+5.4324}_{-0.4960}$$^a$ & $0.0250^{+0.7656}_{-0.0245}$$^a$ \\
    $(10^{-4}\eta_{\star}\eta_{\rm EH}$ cumulative) & $44.301^{+229.37}_{-36.411}$$^b$ & $0.6532^{+12.292}_{-0.5685}$$^b$ & $0.0925^{+1.8151}_{-0.0864}$$^b$ & $0.0057^{+0.2889}_{-0.0056}$$^b$ \\
   & & & & \\
   remaining stars/planets$^c$  & $2095.9^{+3702.6}_{-1590.9}$$^a$ & $72.5260^{+530.69}_{-63.36}$$^a$ & $5.4440^{+50.5548}_{-5.0096}$$^a$ & $0.2538^{+7.1637}_{-0.2483}$$^a$  \\
    ($10^{6}$ stars/planets) & $449.62^{+2117.86}_{-364.73}$$^b$& $6.630^{+114.82}_{-5.7184}$$^b$ & $0.939^{+16.958}_{-0.8734}$$^b$ & $0.0577^{+2.7063}_{-0.0566}$$^b$  \\
   \hline
\end{tabular}}
\end{center}\footnotesize
\raggedright
$^a$for an N$_2$-O$_2$-dominated atmosphere with a maximum of 10\% CO$_2$; $^b$same, but with a maximum of 1\% CO$_2$; $^c$all planets, except for `suitable stars'
\end{table*}

Interestingly, all three implemented planetary requirements show broadly similar importance, as can also be seen in Table~\ref{tab:etaEH} and Figure~\ref{fig:etaEH}. The upper panel of the latter shows the requirements implemented into $\eta_{\rm EH}$, whereas the lower panel illustrates all considered criteria including the ones for $\eta_{\star}$. The only necessary requirement feeding into $\eta_{\star}$ that is similarly crucial than the planetary requirements is the stellar lower limit, $\alpha_{\rm at}^{\rm ll}$. The importance of the planetary requirements strongly suggests that not every rocky exoplanet in the HZCL of a suitable star will be habitable, at least not for \textit{life as we know it}.

Whereas for our two atmospheric cases the maximum frequency of suitable stars was found to be $\eta_{\star,10\%} \leq 2.07(+4.12/-1.60)\times 10^{-2}$ and $\eta_{\star,1\%} \leq 0.44(+2.29/-0.36) \times 10^{-2}$, respectively, the maximum frequency of EHs is  lower by roughly 2 orders of magnitude, that is, $\eta_{\rm EH,10\%} \leq 1.21(+11.58/-1.10)\times 10^{-4}$ and $\eta_{\rm EH,1\%} \leq 1.28(+9.48/-1.15)\times 10^{-4}$, respectively. The values of $\eta_{\rm EH}$ are additionally smaller by around 2 to 3 orders of magnitude than Eta-Earth values published since 2019 (see Table~\ref{tab:etaEarth}), i.e., $\eta_{\oplus} \sim 0.01-0.3$. That $\eta_{\rm EH}$ is generally smaller than $\eta_{\oplus}$, however, is logical since the latter only relates to rocky planets in the HZ without considering any additional habitability criteria.

Combining both, $\eta_{\rm EH}$ and $\eta_{\star}$, illustrates that only a small (maximum) fraction of stellar systems may indeed host rocky exoplanets that are or have the potential to evolve into an Earth-like Habitat. Combining both occurrence rates gives
\begin{eqnarray}
\begin{split}\label{eq:etaAll}
  \eta_{\star}\eta_{\rm EH} \leq
  \begin{cases}
    2.50^{+76.56}_{-2.45}\times 10^{-6}, & x_{\rm CO_2,max} = 10\%\\
    0.57^{+28.89}_{-0.56}\times 10^{-6}, & x_{\rm CO_2,max} = 1\%,
  \end{cases}
\end{split}
\end{eqnarray}
for N$_2$-O$_2$-dominated atmospheres with $x_{\rm CO_2,max} = 10\%$ and $x_{\rm CO_2,max} = 1\%$, respectively. Substituting these values into Equation~\ref{eq:NEH} finally results in
\begin{eqnarray}
\begin{split}\label{eq:etaAll}
  N_{\rm EH} \leq
  \begin{cases}
    2.538^{+71.637}_{-2.483}\times 10^{5}, & x_{\rm CO_2,max} = 10\%\\
    0.577^{+27.063}_{-0.566}\times 10^{5}, & x_{\rm CO_2,max} = 1\%,
  \end{cases}
\end{split}
\end{eqnarray}
which gives a plausible range for the maximum number of EHs in the galactic disk. We, however, caution to take the lower and upper bounds as realistic numbers, but to better regard them as statistically improbable limits (see Section~\ref{sec:caveats}.).

\begin{figure}
\centering
\includegraphics[width = 1.0\linewidth, page=1]{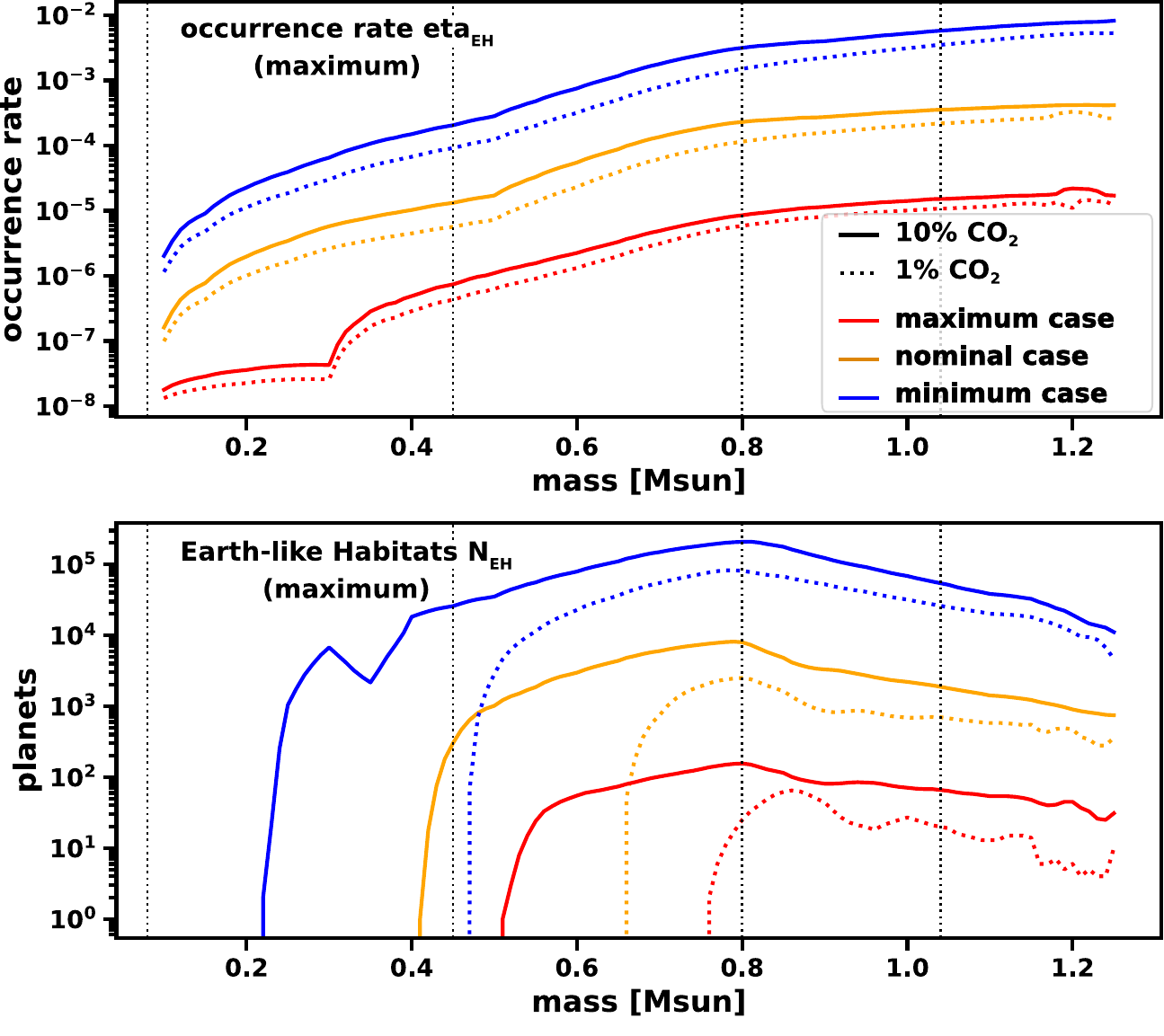}
\caption{Maximum values for $\eta_{\rm EH}$ (upper panel) and $N_{\rm EH}$ (lower panel) as a function of stellar mass for both N$_2$-O$_2$-dominated atmospheres with $x_{\rm CO_2,max}=10\%$ (solid lines) and $x_{\rm CO_2,max}=1\%$ (dotted lines), respectively. While the upper panel shows the fraction of rocky exoplanets that have the potential to evolve into EHs around suitable stars, the lower panel directly gives the maximum number of EHs, i.e., $N_{\rm EH} \leq N_{\star}\eta_{\star}\eta_{\rm EH}$, per stellar mass.}
\label{fig:compNEH}
\end{figure}

Our maximum values for $\eta_{\rm EH}$ (upper panel) and $N_{\rm EH}$ (lower panel) are illustrated in Figure~\ref{fig:compNEH} as a function of stellar mass. In the upper panel, the occurrence rate drops significantly towards lower stellar masses for all simulated cases. For solar-like stars, the maximum occurrence rate of EHs around suitable stars can be found to be $\eta_{\rm EH} \leq 3.32(+49.20/-3.18)\times 10^{-4}$, whereas it significantly drops by two orders of magnitude for a typical M dwarf with a mass of $M_{\star}=0.25$\,M$_{\odot}$, for which we find $\eta_{\rm EH} \leq 3.44(+35.85/-3.40)\times 10^{-6}$. The actual occurrence rate of EHs around low-mass stars, however, will even be rarer than suggested by $\eta_{\rm EH}$, as $\eta_{\star}$ declines in a similar manner to lower stellar masses (see Section~\ref{sec:NStar}).

The largest number of EHs can be expected to exist in the K-type spectral class for all our cases. This is in very good agreement with a study by \citet{Cuntz2016}, who also found early-type K dwarfs to be the most promising targets for finding advanced lifeforms by implementing the frequency of stellar types, the speed of stellar evolution, the size of the respective habitable zones, and the stellar radiation environment into their model. {It also fits very well with analysis by \citet{Lingam2021Life}, who found that different models of stellar habitability, e.g., the maximum habitability interval, the propensity of a planet to host life based on major evolutionary events, and the likelihood of attaining post-GOE oxygen levels, converge toward early K-type stars as the most viable targets for life.

In our model, a stellar mass of $M_{\star} \sim 0.8$\,M$_{\odot}$ seems to be a specifically favorable spot for providing habitable conditions, as can be seen in the lower panel of Figure~\ref{fig:compNEH}. The occurrence rate of EHs at $M_{\star}=0.8$\,M$_{\odot}$ can be found to be $(\eta_{\star}\eta_{\rm EH})_{10\%} \leq 3.25(+77.67/-3.19)\times 10^{-5}$ and $(\eta_{\star}\eta_{\rm EH})_{1\%} \leq 1.02(+30.73/-1.01)\times 10^{-5}$ for $x_{\rm CO_2,max}=10\%$ and $x_{\rm CO_2,max}=1\%$, respectively. Starting from this stellar mass, the maximum number of EHs decreases in both directions of the stellar mass spectrum and it reaches an occurrence rate for solar-like stars with $M_{\star}=1.0$\,M$_{\odot}$ of $(\eta_{\star}\eta_{\rm EH})_{\rm 10\%} \leq 1.51(+39.61/-1.47)\times 10^{-5}$ and $(\eta_{\star}\eta_{\rm EH})_{1\%} \leq 0.47(+18.69/-0.46)\times 10^{-5}$, which is lower by roughly a factor of 2 than for $M_{\star}=0.8$\,M$_{\odot}$. At lower stellar masses, clear cutoffs occur below which no EHs may be found. These cutoff stellar masses are $M_{\star,\rm min,10\%} \geq 0.41(+0.09/-0.19)$\,M$_{\odot}$ and $M_{\star,\rm min,1\%} \geq 0.65(+0.09/-0.18)$\,M$_{\odot}$. The occurrence rates directly at the respective cutoff stellar masses for the nominal cases, i.e., $M_{\star,\rm min,10\%}=0.41$\,M$_{\odot}$ and $M_{\star,\rm min,10\%}=0.65$\,M$_{\odot}$, are $(\eta_{\star}\eta_{\rm EH})_{10\%} \leq 9.76\times 10^{-10}$ and $(\eta_{\star}\eta_{\rm EH})_{1\%} \leq 8.14\times 10^{-10}$, respectively. This equates to a maximum probability of finding EHs at these specific stellar masses within the entire galactic disk of below unity, i.e., $\mathcal{P}(N_{\rm EH,10\%})\lesssim0.92$ and $\mathcal{P}(N_{\rm EH,1\%})\lesssim0.33$, respectively. This means that an entirety of 3 Milky Way disks is needed to statistically find just one EH with an N$_2$-O$_2$-dominated atmosphere containing $x_{\rm CO_2,max}=1\%$ around a star with a stellar mass of $M_{\star}=0.65$\,M$_{\odot}$. A rare occasion indeed.

\begin{figure}
\centering
\includegraphics[width = 1.0\linewidth, page=1]{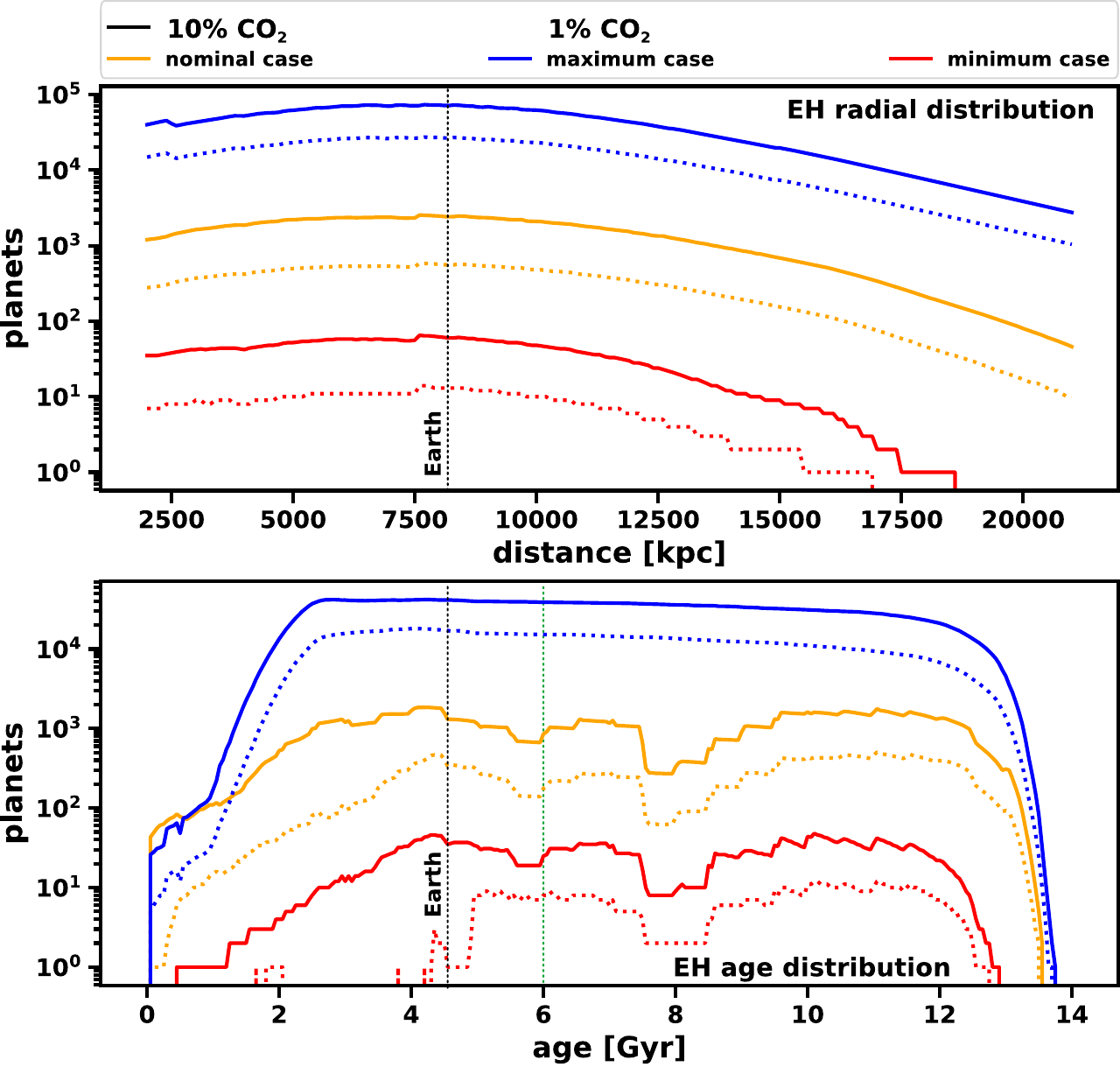}
\caption{Maximum numbers for $N_{\rm EH}$ as a function of galactic distance (upper panel) and stellar birth age (lower panel), for both N$_2$-O$_2$-dominated atmospheres with $x_{\rm CO_2,max}=10\%$ (solid lines) and $x_{\rm CO_2,max}=1\%$ (dotted lines), respectively. {The vertical dashed line indicates the approximate age at which geological activity on cosmochemically Earth-like planets may cease.}}
\label{fig:compNEHra}
\end{figure}

Figure~\ref{fig:compNEHra} further illustrates the distribution of $N_{\rm EH}$ as a function of galactic distance (upper panel) and stellar birth age (lower panel). From the upper panel, one can see that the maximum number of EHs slightly increases from the outskirts of the galaxy towards its center until about $r\sim7.5$\,kpc, after which it starts to decrease toward the galactic center, where sterilization becomes more important than metallicity and stellar frequency. As can be seen, the position of the Sun at $r\sim8.2$\,kpc is only slightly offset from the maximum. Interestingly, almost exactly 50\% of all potential EHs will be inside the Sun's galactic orbit for our nominal and maximum cases for both atmospheric scenarios. For our minimum cases, the ratio slightly shifts toward the inner disk, with $\sim$55\% located inside the solar orbit.

As can further be seen in the lower panel of Figure~\ref{fig:compNEHra}, most of the planets in our final sample are older than the Earth (see also Figure~\ref{fig:mO}). Many of these planets may already be geologically inactive and, therefore, likely uninhabitable. This again illustrates that the actual number of EHs within the galactic disk will be significantly lower than our maximum number. To illustrate this point, for atmospheres with $x_{\rm CO_2,max}=10\%$ and $x_{\rm CO_2,max}=1\%$ about 72-80\% and 73-97\% of all planets in our remaining sample, respectively, are older than the Earth with an average age of 7.25(+0.1/-0.5)\,Gyr. Still, $\sim$52-60\%  and $\sim$50-75\% of the planets are older than 6\,Gyr. If we assume that cosmochemically Earth-like planets older than this threshold will be geologically inactive \citep{Frank2014,Mojzsis2021}, this reduces our maximum number of EHs by 50-75\%, depending on the specific scenario. For our nominal cases, this gives decreased maximum numbers of $N_{\rm EH,10\%} \lesssim 1.5\times 10^{5}$ EHs and $N_{\rm EH,1\%} \lesssim 0.4\times 10^{5}$ EHs, respectively. {If we only take planets around G-type stars, however, the average age of the remaining planets in our sample reduces to 5.65(+0.55/-0.8)\,Gyr.}

\begin{figure}
\centering
\includegraphics[width = 1.0\linewidth, page=1]{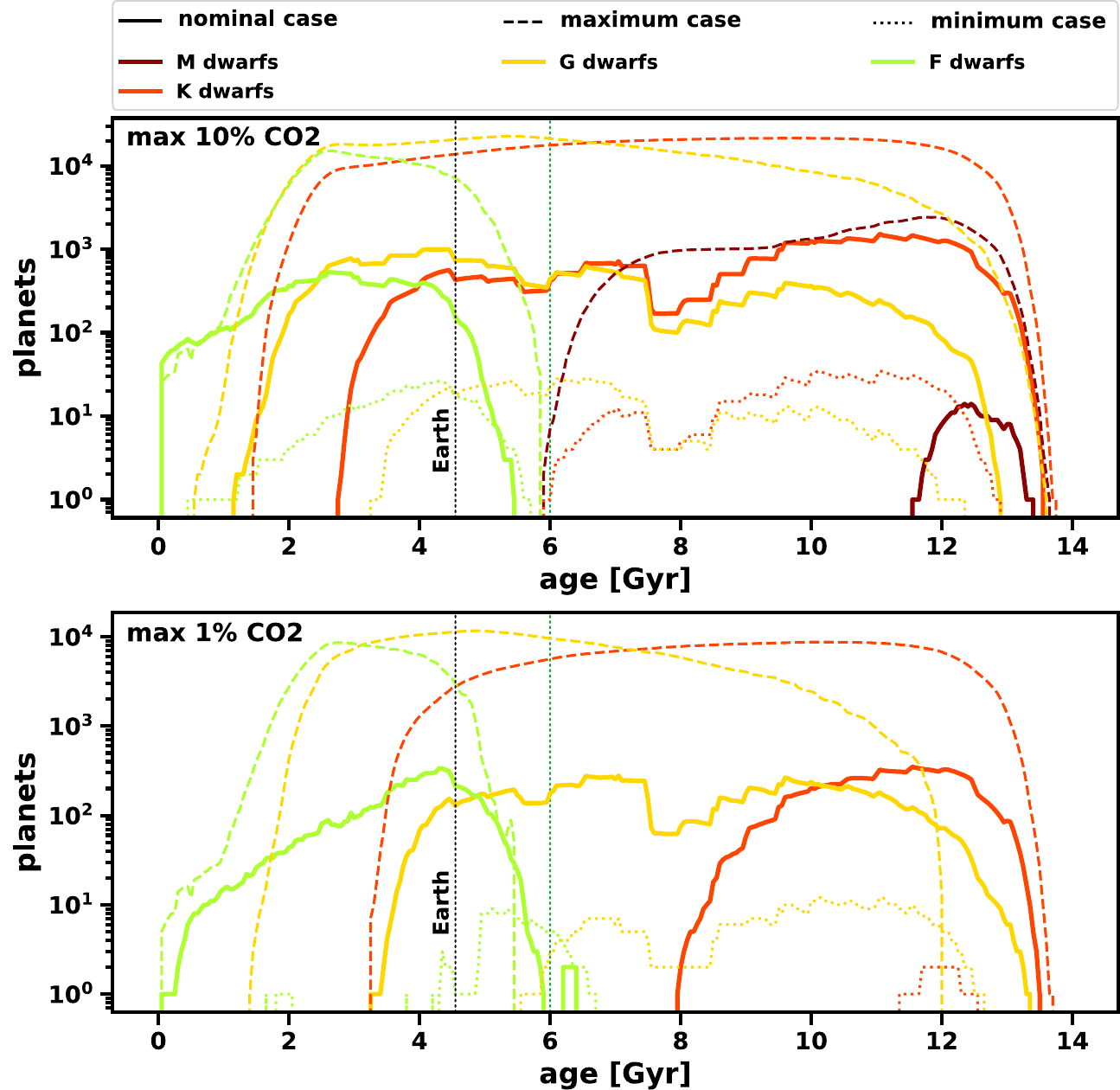}
\caption{The birth age of all remaining planets within our sample by spectral class for an N$_2$-O$_2$-dominated atmosphere with a maximum of 10\%\,CO$_2$ (upper) and 1\%\,CO$_2$ (lower panel), respectively. Here, the different colors illustrate the various spectral classes (with black as the sum of all) and solid, dashed, and dotted lines illustrating nominal, maximum, and minimum cases. {The vertical dashed line indicates the approximate age at which geological activity on planets cosmochemically similar to the Earth may cease (see text).}}
\label{fig:planetsSpec}
\end{figure}

Finally, Figure~\ref{fig:planetsSpec} illustrates the sample of remaining planets as a function of birth age and spectral class for both of our atmospheric cases, whereas Table~\ref{tab:NEHspec} gives the fractions of stars within the different spectral classes. The relative distribution between the spectral classes changed significantly from the initial distribution of $N_{\star}$, as found via the IMF. It also deviates from the distribution we obtained after implementing $\eta_{\star}$ by shifting even further towards higher-mass stars. Even though initially $79.80(+1.51/-10.46)$\% of all stars are M stars, the majority of planets, $N_{\rm EH}$, that meet all the implemented requirements for evolving into an EH can probably be found around either K or G dwarfs, while in most of our cases none or only a very small number around the highly abundant M star population can be found. The relative fractions further vary between our two atmospheric scenarios, with the planet distribution again shifting farther toward higher-mass stars for lower CO$_2$ mixing ratios. If we consider $x_{\rm CO_2,max}=10\%$, $48.45(+5.63/-8.11)$\% of suitable planets can be found around K dwarfs, while this value decreases toward $32.33(+10.43/-29.03)$\% for $x_{\rm CO_2,max}=1\%$. Conversely, the fraction of planets increases with decreasing CO$_2$ mixing ratio for G and F stars, that is, from $39.93(+0.90/-4.84)$\% and $10.72(+8.10/-7.65)$\% to $48.22(+27.37/-4.32)$\% and $21.11(+1.67/-6.20)$\% for $x_{\rm CO_2,max}=10\%$ and $x_{\rm CO_2,max}=1\%$, respectively.

\begin{table*}\footnotesize
  \begin{center}
    \caption{The distribution of potentially habitable planets within the different stellar spectral classes.}
    \label{tab:NEHspec}
    \resizebox{\textwidth}{!}{%
\begin{tabular}{l|l|c|c|c|c|c}
  \hline
   case & & total [number] & M [\%] & K [\%]  & G [\%] & F [\%] \\
   \hline
   \multirow{5}{*}{max 10\%\,CO$_2$:}& initial $N_{\star}$ (all masses)  & $10.15^{+0.61}_{-0.77}\times10^{10}$ & $79.80^{+1.51}_{-10.46}$ & $14.50^{+1.17}_{-0.13}$ & $3.38^{+1.21}_{-0.25}$ & $0.84^{+2.90}_{-0.06}$ \\
   & initial $N_{\star}$ (M$_\star =0.1-1.25$\,$M_{\odot}$)  & $8.81^{+0.55}_{-0.17}\times10^{10}$ & $77.25^{+0.78}_{-0.003}$ & $18.06^{+3e-3}_{-1.59}$ & $3.90^{+0.62}_{-1e-4}$ & $0.75^{+0.23}_{-3e-5}$ \\
   & $N_{\star}\eta_{\star}$ & $2.09^{+3.70}_{-1.59}\times10^{9}$ & $1.05^{+23.50}_{-1.05}$ & $67.17^{+12.78}_{-6.57}$ & $15.67^{+9.60}_{-2.58}$ & $3.33^{+4.23}_{-1.56}$  \\
   & $N_{\rm EH}$ & $2.54^{+71.64}_{-2.48}\times10^{5}$ & $0.11^{+2.23}_{-0.11}$ & $48.45^{+5.63}_{-8.11}$ & $39.93^{+0.90}_{-4.84}$ & $10.72^{+8.10}_{-1.45}$ \\
   & younger than 6\,Gyr & $1.00^{+34.39}_{-0.98}\times10^{5}$ & $0^{+0.12}_{-0}$ & $25.57^{+6.10}_{-23.94}$ & $51.83^{+5.42}_{-2.95}$ & $27.09^{+19.46}_{-7.65}$ \\
    \hline
   \multirow{3}{*}{max 1\%\,CO$_2$:$^a$} & $N_{\star}\eta_{\star}$ & $0.45^{+2.12}_{-0.36}\times10^{9}$ & 0 & $48.30^{+25.76}_{-43.23}$ & $41.94^{+38.30}_{-19.32}$ & $9.77^{+4.94}_{-6.44}$ \\
   & $N_{\rm EH}$ & $0.58^{+27.06}_{-0.57}\times10^{5}$ & 0 & $32.33^{+10.43}_{-29.03}$ & $48.22^{+27.37}_{-4.23}$ & $21.11^{+1.67}_{-6.20}$ \\
   & younger than 6\,Gyr & $0.17^{+13.51}_{-0.17}\times10^{5}$ & 0 & $0^{+15.48}_{-0}$ & $47.27^{+10.51}_{-28.70}$ & $64.96^{+16.47}_{-38.21}$ \\
   \hline
\end{tabular}}
\end{center}\footnotesize
$^a$the first two rows (i.e.,initial $N_{\star}$ for the entire and for the restricted mass range) are equivalent with the 10\%\,CO$_2$ case
\end{table*}

As discussed earlier and as illustrated in Figure~\ref{fig:planetsSpec}, any remaining planets around M dwarfs, but even many around K and to a lesser extent G dwarfs, belong to old stellar systems. If we, consequently, remove any stars that are older than 6\,Gyr by assuming that bodies older than this age have, {on average,} stopped being geologically active, the distribution of remaining planets further shifts toward higher-mass stars (see also Table~\ref{tab:NEHspec}). For N$_2$-O$_2$-dominated atmospheres with $x_{\rm CO_2,max}=10\%$, suitable planets are then most often found around G dwarfs with $51.83(+5.42/-2.95)$\%, whereas F stars may be the most promising targets for finding Earth-like atmospheres with $x_{\rm CO_2,max}=1\%$. In the latter case, a majority of $64.96(+16.47/-38.21)$\% of all suitable planets can indeed be found around F dwarfs with the rest mostly located around G dwarfs. Even though F stars are the least abundant stars among the studied (M, K, G, and F) spectral classes, EHs with CO$_2$ mixing ratios similar to the Earth's, may preferentially be found around this type of stars, at least by only assuming planets younger than an age of 6\,Gyr. The Earth with an age of 4.56\,Gyr may therefore be somewhat atypical.

This also indicates that most of the planets theoretically suitable to evolve into EHs may be relatively short-lived due to the limited lifetime of F stars. Whether enough time remains on such planets to reach atmospheric oxygenation (see Section~\ref{sec:oxy}) similar in abundance to the Earth's for allowing the subsequent evolution of complex life remains unanswered by our study. However, if a lifetime comparable to that of today's Earth turns out to be crucial \citep[see, e.g.,][]{Catling2005}, complex life could actually be favored around late-type G and early-type K dwarfs, potentially making Earth special among all habitable planets, but not atypical among Earth-like Habitats able to host complex life.

\section{Discussion \& Summary}\label{sec:discussion}

Our final outcome for $N_{\star}$, $\eta_{\star}$, and $\eta_{\rm EH}$ together with $N_{\rm EH}$ were already discussed in detail in Sections~\ref{sec:NStar}, \ref{sec:etaStar}, and \ref{sec:NEH}, respectively, and their final values can be found in Table~\ref{tab:NEHspec}. Here, we will {extend} our discussions and present further conclusions and implications that can be drawn from our model outcome in the following {subsections}.


\subsection{Compared to N$_{\star}$, EHs will be rare}\label{sec:rare}

{We find relatively low values for the fraction of stars, $\eta_{\star}$, that can presently host EH, and for the frequency of EHs, $\eta_{\rm EH}$, around these stars. These are $\eta_{\star,10\%} < 2.07^{+4.12}_{-1.60}\times 10^{-2}$ and $\eta_{\rm EH,10\%} < 1.21^{+11.58}_{-1.10}\times 10^{-4}$, and $\eta_{\star,1\%} < 0.44^{+2.29}_{-0.36}\times 10^{-2}$ and $\eta_{\rm EH,1\%} < 1.28^{+9.48}_{-1.15}\times 10^{-4}$  for N$_2$-O$_2$-dominated atmospheres with $x_{\rm CO_2,max}=10\%$ and $x_{\rm CO_2,max}=1\%$, respectively.} Based on these rates, we can conclude that only a low fraction of all presently existing stars have the potential to host an EH. By applying these maximum occurrence rates to the number of stars, $N_\star = 10.15(+0.61/-0.77)\times 10^{10}$, in the galactic disk between a galactocentric distance of $r=2.0-21.0$\,kpc and a stellar mass range of $M_{\star} = 0.1$\,M$_{\odot} - 1.25$\,M$_{\odot}$, one can see that EHs might indeed be rare, at least compared to $N_{\star}$, with $N_{\rm EH,10\%}\leq2.538^{+71.637}_{-2.483}\times10^{5}$ and $N_{\rm EH,1\%}\leq0.577^{+27.063}_{-0.566}\times10^{5}$, respectively. 

{In addition, we emphasize that $\eta_{\rm EH}$ will also be rare compared to $\beta_{\rm HZCL}$, the occurrence rate of rocky exoplanets in the HZCL. In Sections~\ref{sec:betaHZCL} and \ref{sec:betaHZCL2}, we estimated this value to be in the order of $\beta_{\rm HZCL}\sim 0.01 - 0.1$, which is in itself slightly lower than current estimates of $\eta_{\oplus}\sim 0.02 - 0.3$ (see Section~\ref{sec:etaHZ}). If we even take the product $\eta_{\star}\times\eta_{\rm EH}$ to indicate the maximum occurrence rate of EHs around any star in the galactic disk, the discrepancy between this occurrence rate and  $\beta_{\rm HZCL}$ or $\eta_{\oplus}$ becomes even more apparent. In Section~\ref{sec:copernicus}, we further compare $\eta_{\star}$ and $\eta_{\rm EH}$  with $\beta_{\rm HZCL}$ and $\eta_{\oplus}$ in the context of the Copernican Principle of Mediocrity.}

\subsection{Our derived numbers are a plausible maximum range}\label{sec:absMax}

As already pointed in Section~\ref{sec:caveats}, {we first caution to take our values from the minimum and maximum cases as realistic maximum numbers}, since we always implemented minimum and maximum values from the scientific literature for our minimum and maximum cases, respectively. It can therefore be assumed that the actual maximum number of EHs is closer to our nominal case. {We also remind the reader that our minimum and maximum cases are, strictly speaking, the minimum and maximum among all the possible maximum values, thereby displaying the variation on the maximum number of Earth-like Habitats. The `true' minimum of $N_{\rm EH}$ in the galactic disk can in principle be as low as $N_{\rm EH} = 1$ (the Earth as the only EH), whereas the maximum values from our maximum cases should in principle be `real' maxima.

That said, the reason why our estimate is a plausible upper bound for the prevalence of EHs in the galactic disk is simple. As already discussed, several necessary requirements feed into $N_{\rm EH}$, $\eta_{\rm EH}$, and $\eta_{\star}$ that were not included into our model since these requirements are, at present, not sufficiently quantifiable to make a scientifically sound estimate on their prevalence. Similarly, it might not be clear how different fctors could be {correlated} with each other. We therefore set the prevalence of all of them equal to unity, even though at least some of them certainly have much lower occurrence rates.

For the sake of the argument, we can simply assume that the combined requirements of having long-term working carbon-silicate and nitrogen cycles have a prevalence among our remaining sample of planets of $\beta_{\rm env}^{\rm cycle}=0.1$\footnote{The real number might be completely different -- albeit even $<$0.1 seems entirely reasonable if one considers that only about one-third of all exoplanets may have a proper composition to evolve plate tectonics in the first place; see \citet{Unterborn2017}.}. This would reduce our maximum number of EHs to a hypothetical maximum, $\tilde{N}_{\rm EH}$, of $\tilde{N}_{\rm EH,10\%}\leq2.538^{+71.637}_{-2.483}\times10^{4}$ and $\tilde{N}_{\rm EH,1\%}\leq0.577^{+27.063}_{-0.566}\times10^{4}$, respectively. Nevertheless, these numbers would be still maxima since various crucial factors are still missing. If the origin of life, another essential requirement, also occurred with a frequency of $\beta_{\rm life}^{\rm origin}=0.1$, we would be left with $\tilde{N}_{\rm EH,10\%}\leq2.538^{+71.637}_{-2.483}\times10^{3}$ and $\tilde{N}_{\rm EH,1\%}\leq0.577^{+27.063}_{-0.566}\times10^{3}$, i.e., a maximum of $\sim 2\,500$ and $\sim$500 EHs for nominal cases in the entire galactic disk.

This brief exercise illustrates that EHs might indeed be rare in the Galaxy and that their actual number can be much lower than our maximum estimate, potentially by orders of magnitude. Such a reasoning is also illustrated by calculating the minimum mean occurrence rate that each of the requirements not implemented into our model can have for $N_{\rm EH}\geq 1$ to be still valid, meaning that the number of EHs in the galactic disk remains equal or greater than unity. In Section~\ref{sec:other}, we discuss a non-exhaustive list of 22 potential requirements that may feed into $N_{\rm EH}$ but are neglected by our study. Some of them will be less important and some will be correlated with each other. However, on average each of these, by now neglected requirements can have a mean rate of $\langle\theta_{\rm n.r,10\%}\rangle\gtrsim0.57^{+0.11}_{-0.08}$ and $\langle\theta_{\rm n.r,1\%}\rangle\gtrsim0.61^{+0.12/-0.10}$ for nitrogen-dominated atmospheres with $x_{\rm CO_2,max}=10\%$ and $x_{\rm CO_2,max}=1\%$, respectively, to still achieve $N_{\rm EH}\geq 1$. If one would further assume above mentioned hypothetical occurrence rates of 0.1 for working carbon-silicate and nitrogen cycles as well as for the origin of life, it would even give $\langle\theta_{\rm n.r,10\%}\rangle\gtrsim0.68^{+0.13}_{-0.11}$ and $\langle\theta_{\rm n.r,1\%}\rangle\gtrsim0.73^{+0.16}_{-0.13}$, respectively. As a comparison, the mean occurrence rates over all our implemented requirements for a stellar mass range of $M_{\star}=0.1-1.25$\,M$_{\odot}$ are significantly lower with $\langle\theta_{\rm i.r,10\%}\rangle = 0.16^{+0.10}_{-0.07}$ and $\langle\theta_{\rm i.r,1\%}\rangle=0.13^{+0.09}_{-0.06}$, respectively.


\subsection{Could $N_{\rm EH}$ be an underestimate?}\label{sec:increase}

There are certain arguments and reasons for our maximum values to potentially underestimate EHs within the galactic disk, mostly based on the following two arguments:

\paragraph*{Underestimated occurrence rates}
we could theoretically underestimate the occurrence rate of one of our implemented requirements significantly, a possibility that cannot be excluded. It may, for instance, turn out that N$_2$-O$_2$-atmospheres are to a higher extent thermally stable than currently expected through several different studies. {It may also turn out that the sterilizing effect of supernovae is too conservative, if the evolution of complex life on other planets on average happens faster and/or any putative lifeforms on these planets evolved a better resilience against harsh radiation environments compared to Earth. In addition, large moons could turn out to be non-essential for EHs to evolve with any argument in favor of their importance being resolved by other means.} However, we tried to implement each requirement to the best of our scientific knowledge and attempted to cover a wide parameter range for each factor as found in the scientific literature. {If we took the much narrower range for the appropriate water mass fraction from \citet{Stern2024} for our nominal case with $x_{\rm CO_2,max}=10\%$, for instance, our maximum number of EH would decline by about an order of magnitude from $N_{\rm EH}\sim 5\times10^5$ to $N_{\rm EH}\sim 4\times10^4$ planets, which illustrates the relatively wide parameter range chosen for our model.}

As another example, if it were to turn out that Earth-like atmospheres are thermally stable for much higher XUV fluxes than currently expected, as suggested by one recent study \citep[i.e.,][see also discussion in Section~\ref{sec:XUV}]{Nakayama2022}, it would only marginally change our results. Besides the XUV requirement, several other implemented requirements significantly affect planets around low-mass stars. So, even if we assume that such atmospheres are still stable for XUV fluxes as high as 100\,F$_{\rm XUV, \oplus}$ at the Earth, our results will only slightly increase by less than a factor of 2  -- from $N_{\rm EH,10\%}\leq2.538^{+71.637}_{-2.483}\times10^{5}$ towards $\tilde{N}_{\rm EH,10\%}\leq3.370^{+74.289}_{-3.266}\times10^{5}$ for an atmosphere with $x_{\rm CO_2,max}=10\%$ -- because of the other requirements taking their toll on low-mass stars as well. In addition, we did not include non-thermal escape into our model, which would also affect M stars significantly, illustrated by several different studies, and specifically pointed out by \citet{Nakayama2022}.

So, even if Earth-like atmospheres were to be stable for higher XUV fluxes, they would likely still be unstable around M dwarfs due to non-thermal escape being highly detrimental. The non-thermal losses will even be further enhanced in such a case since the stable HZCL boundaries moves closer to the star by increasing the XUV threshold (because the XUV threshold is reached earlier, the HZCL boundaries will be closer to the host star at this particular age, implying that the stellar wind ram pressure and hence non-thermal escape will be higher than at larger orbital distances).

\paragraph*{Factors that may increase $N_{\rm EH}$}
Second, there could be factors that might actually increase the prevalence of EHs. Besides some benefits that one or the other factor discussed in appended Section~\ref{sec:other} may have, the most obvious example are EHs that form not on exoplanets but on exomoons. Exomoons, however, will only increase our results negligibly, if at all -- and certainly far within any reasonable error bars -- because of the following reasons.

First, any gas giant with an exomoon that could host an Earth-like atmosphere has to be within the HZCL as well. \citet{Fernandes2019}, for instance, found the occurrence rate of giant planets between 1 and 20 Jupiter masses, M$_{\rm Jup}$, at orbital distance between 0 and 100\,AU to be $\eta_{\rm gp} = 0.062^{+0.015}_{-0.012}$ with a peak around the snow line. So, the occurrence rate of gas giants with $M_{\rm pl}\geq 1$\,M$_{\rm jup}$ in the HZCL will be much smaller. If we (optimistically) assumed a giant planet frequency within the HZCL in the order of $\eta_{\rm gp,HZCL}\sim0.01$ averaged over all stellar masses, this would already be below the average HZCL occurrence rates of rocky exoplanets within our model, i.e., $\eta_{\rm EH,10\%}=0.035^{+0.070}_{-0.019}$ and $\eta_{\rm EH,10\%}=0.015^{+0.0326}_{-0.0055}$, respectively.

These gas giants, however, would also need to have exomoons and these exomoons would have to be very large. Otherwise, any N$_2$-O$_2$-dominated atmosphere would not be stable around such a moon. If we put Titan today at the position of the Earth, its N$_2$-dominated atmosphere would not be stable \citep[see, e.g.,][]{Spross2021}, even for the Earth's presently low XUV surface flux, which is nonetheless $\sim$100 times higher compared to Saturn's orbit at 10\,AU. So, the mass of such an exomoon would have to be significantly higher than Titan's, possibly comparable to the mass of the Earth. But how high might the occurrence rate of such exomoons be? These will be very rare and could potentially exist only around gas giants much more massive than Jupiter. They will logically have a frequency that is much lower than mentioned above.

But are such massive exomoons even possible? \citet{Canup2006} found a common mass scaling of $M_{\rm sat}/M_{\rm gp} \lesssim10^{-4}$ between the gas giant's and their entire satellites' mass, $M_{\rm gp}$ and $M_{\rm sat}$, respectively, based on the scalings in the Solar System and on N-body simulations of moon formation around giant planets. This ratio was further confirmed as universal by \citet{Sasaki2010} and \citet{Heller2015a} and should therefore apply to any extrasolar gas giant. Indeed, \citet{Benisty2021} report the discovery of a circumplanetary disk around the gas giant PDS70c and found a mass for its accretion disk that is below $M_{\rm acc}=10^{-4}$ times the mass of PDS70c. By optimistically assuming that the entire mass of such a circumplanetary disk leads to the formation of just one massive exomoon, a gas giant would have to be about $\sim$3.4 times as massive as Jupiter to host a Mars-mass exomoon and $\sim$31.5 times as massive to host an Earth-mass exomoon. The mass limit for deuterium burning and the formation of brown dwarfs, however, is around $\sim$13\,M$_{\rm Jup}$ \citep[e.g.,][]{Spiegel2010}.

It is not hard to see that Earth-mass or even Mars-mass exomoons in the HZCL of suitable stars will be exceedingly rare. And this does not even consider any potential additional habitability issues of exomoons around massive gas giants. These, for example, include the highly energetic particle environment within their extended magnetospheres through which such exomoons have to pass at least when orbiting through their tails \citep[e.g.,][]{Heller2013} -- in case that their orbits are not entirely within the magnetosphere anyway. Tidal forces would also be critical for the habitability of exomoons around low-mass stars. As was found by \citet{Zollinger2017}, no habitable moon would be possible for stellar masses of $M_{\star}\leq0.2$\,M$_{\odot}$ due to extreme tidal heating for any configuration, and detrimental effects could remain up to a stellar mass of $M_{\star}\sim0.5$\,M$_{\odot}$ for increasing masses of the host planet and depending on the location within the HZ. However, \citet{Zollinger2017} only simulated host masses up to $M_{\rm pl} = 1.0$\,M$_{\rm Jup}$. For higher host masses, tidal problems could therefore potentially remain even for stellar masses of $M_{\star}>0.5$\,M$_{\odot}$. However, if we assumed an optimistic occurrence rate of habitable Earth-like exomoons of $\eta_{\rm EM}\sim0.001$ for FGK stars that decreases towards 0 as a function of the Hill radius somewhere between $M_{\star}\sim0.2-0.5$\,M$_{\odot}$, our nominal case for an atmosphere with $x_{\rm CO_2,max}=10\%$ would increase by far less than 1 permille.

Also, habitable planets around brown and white dwarfs \citep{Barnes2013}, other potential environments to consider, are unlikely because of their dim and steadily decreasing habitable zones \citep{Barnes2013,Lingam2020BrownDwarfs} and, at least in the case of white dwarfs, efficient tidal heating and high UV surface fluxes \citep{Barnes2013}. One should note, however, that a recent study by \citet{Becker2023} found specific scenarios for which tidal heating may keep a planet's surface around a white dwarf above freezing for up to 10 Gyr. However, as the authors point out, they did not consider how tidal heating and the UV flux from the host dwarf might affect such planet's atmosphere.

Here, one could further note the hypothetical habitability of planets around neutron stars. To evaluate such habitability, \citet{Patruno2017} calculated HZ boundaries and atmospheric escape for various neutron star environments and planetary masses. According to their estimates, super-Earths could retain their atmospheres for up to several hundred million years if their initial atmospheric mass fraction was a substantial part of the planet's mass. If a moderately strong planetary magnetic field were additionally present and/or the pulsar winds were non-isotropic, habitable conditions would even be preserved for several billion years. To allow for temperatures above freezing, however, \citet{Patruno2017} argue that the atmospheres of these planets will have to be heated via X-ray and gamma-ray radiation. For this to happen, the XUV surface fluxes in the respective `habitable zones' would have to reach values that are about six orders of magnitude higher than at today's Earth orbit. Any atmosphere would therefore be lost significantly faster than estimated by \citet{Patruno2017}, who assumed hydrostatic atmospheres with no XUV-induced expansion at all. This can be regarded to be highly unrealistic. Moreover, planets around neutron stars would experience significant tidal heating and obliquity variations, the letter as much as $\sim 50-100^{\circ}$ within 10 days to 3\,Myr \citep{Iorio2021}.

The movie \textit{Interstellar}\footnote{See \citet{Thorne2014} for a description on the science behind the movie Interstellar.} also spurred a whole bunch of scientific publications on whether planets around black holes can exist and even host life. Although these may indeed exist around stellar black holes and could potentially even form around SMBHs \citep{Wada2019,Wada2021,Giang2022}, the emergence of life, and specifically complex life, will be highly restrictive due to several astrophysical and relativistic reasons as discussed in several studies \citep{Opatrny2017,Schnittman2019,Bakala2020,Iorio2020SMBHHab,Veysi2023}. Even if life may evolve on some of these rare habitats though, it would be entirely different from EHs with very narrow timescales for evolution due to the relativistic phenomenon of time dilation \citep[e.g.,][]{Veysi2023}. Besides, although habitats other than EHs may potentially exist, these will not increase our maximum number of EHs, as these are by definition no EHs. This is true for any other kind of hypothetical habitat.

\subsection{EHs around M dwarfs could be exceedingly rare}\label{sec:mdwarfs}

As our results show, EHs around M dwarfs may only exist under very rare circumstances. Our model found cutoff masses of $M_{\star,\rm min,10\%} \geq 0.41^{+0.09}_{-0.19}$\,M$_{\odot}$ and $M_{\star,\rm min,1\%} \geq 0.65^{+0.09}_{-0.18}$\,M$_{\odot}$, below which Earth-like Habitats cannot exist. In reality, one might expect that EHs at lower stellar masses can occasionally exist for outliers in stellar activity and under very beneficial circumstances with system parameters far away from average so that these do not fit within our model parameter range. The Sun, for instance, is very weakly active \citep[see, e.g.,][and the discussion in Section~\ref{sec:XUV}]{Reinhold2020,Johnstone2021Stars}, clearly below average and also below the XUV flux range covered by the stellar evolution model from \citet{Johnstone2021Stars}. However, some stars will be active on levels above the same model distribution. Outliers will therefore exist at both extremes and a small number of EHs around M dwarfs cannot be excluded because of extreme variations.

{Our lower stellar cutoff masses are in very good agreement with \citet{Lingam2021Life} and \citet{Cuntz2016}, who both discuss lower stellar mass thresholds below which the existence of complex life becomes unlikely. For most of their investigated criteria, \citet{Lingam2021Life} observe a sharp decrease in their likelihood below a stellar mass range of $M_{\star}\sim$0.4-0.6\,M$_{\odot}$. In their model, the likelihood of abiogenesis decreases strongly for $M_{\star}\lesssim$0.4\,M$_{\odot}$, whereas the probability of obtaining the n$^{\rm th}$ critical step in evolution decreases sharply at $M_{\star}\sim$0.5\,M$_{\odot}$. Also, the atmospheric build-up of oxygen, similar to the Earth's after the GOE, may not occur on planets around stars with $M_{\star}\lesssim$0.57\,M$_{\odot}$. Based on the results from \citet{Lingam2021Life}, one could therefore expect a lower stellar cutoff mass for the evolution of EHs that is located in the late K dwarf regime, thereby basically excluding M stars as well. Similarly, \citet{Cuntz2016} argue that the evolution of complex life is less likely even for spectral types later than about K3, i.e., for stellar masses below $M_{\star}\sim$0.75\,M$_{\odot}$. This corresponds well with the upper range of our stellar cutoff masses for EHs with $x_{\rm CO_2,max}=1\%$.}

Our result that EHs around M dwarfs are exceedingly rare is also quite robust. Many crucial criteria disfavor M dwarfs as hosts for surficial \textit{life as we know it} and there are several more crucial criteria than implemented into our model (see Section~\ref{sec:other}). So, even if some requirements turned out to be less important, other important factors would remain. {We discuss the implications of this reasoning on the falsifiability of our model further down below in Section~\ref{sec:quantification}.}

Criteria disfavoring EHs at M dwarfs -- potentially non-exhaustively -- are:
\begin{itemize}
\item The XUV surface flux in the HZCL of M dwarfs will remain high for billions of years \citep[e.g.,][]{McDonald2019,Fleming2020,Birky2021,Johnstone2021Stars,Engle2024XUV}, thereby leading to thermal atmospheric escape and the loss of water \citep[e.g.,][]{Tian2008,Tian2015,Tian2015O2,Johnstone2019,France2020,VanLooveren2024Trappist}.
\item High rates of flaring \citep{Howard2019,Vida2019,Guenther2020} will increase atmospheric escape \citep[e.g.,][]{France2020} and potentially sterilize the surface \citep[e.g.,][]{Howard2019,Tilley2019}. But it was also suggested to increase UV levels to allow prebiotic chemistry \citep[e.g.,][]{Mullan2018} -- a very thin line indeed between erosion and sterilization on the one hand and prebiotic chemistry on the other.
\item The strong stellar winds in the HZCL \citep[e.g.,][]{Garaffo2016} will lead to very high rates of non-thermal losses \citep[e.g.,][]{Airapetian2017,Dong2020TOI,France2020}, in addition to thermal escape, even if a magnetosphere might be present \citep[e.g.,][]{GarciaSage2017,RodriguezMozos2019}. Non-thermal losses alone can lead to a very fast loss of the atmosphere \citep[e.g.,][]{Lichtenegger2010}; taken both, thermal and non-thermal escape, together, even CO$_2$-dominated atmospheres might not be stable around most low-mass M dwarfs.
\item {The strong and rapidly varying stellar magnetic field at the orbit of a rocky exoplanet around an M dwarf will lead to immense Joule Heating in the upper atmosphere comparable or even larger than the heating induced by the incident XUV surface flux, as was recently shown by \citet{Cohen2024JouleHeating}. This process alone will likely be sufficient to completely erode a secondary atmosphere.}
\item The very slow decrease in stellar luminosity will lead to a strong shift of the HZCL so that planets initially within will later be outside \citep[e.g.,][]{Ramirez2014}. So, even if such planets finally end up in the HZCL, they might have already lost their atmosphere and water \citep[e.g.,][]{Lammer2011,Luger2015,Tian2015} or entered the \textit{Runaway Greenhouse} state early-on \citep[e.g.,][]{Ramirez2014}.
\item The probability of possessing an appropriate water mass fraction for simultaneously having surface water and subaerial land, might be significantly lower than for planets around FGK stars \citep[e.g.,][]{Tian2015,Kimura2022}. These planets could either completely desiccate due {to} the high water loss into space, end up as water worlds, or even with steam atmospheres \citep[e.g.,][]{Marounina2020}.
\item Planets around M dwarfs have a low chance of forming and keeping a large moon due to stronger tidal interaction with the host star \citep[e.g.,][]{Piro2018,MartinezRodriguez2019,Tokadjian2020}. For low-mass M dwarfs, it may even be impossible. However, tidal interaction with the host star could at least partially substitute for some of a large moon's positive effects -- but it could also overheat the planet, thereby rendering it uninhabitable \citep[e.g.,][]{McIntyre2022}.
\item Induction heating around M dwarfs with strong magnetic fields could lead to extreme volcanism \citep[][]{Kislyakova2018,Kislyakova2020Induction}. This could be favorable for the carbon-silicate cycle. It could, however, also overheat the planet or desiccate it from all volatiles during the active phase of the host star. The same is true for flare-CME-induced interior heating, another relevant process recently discussed by \citet{Grayver2022}.
\item After the long, active phase of the host star decreases below the stability threshold level for N$_2$-O$_2$-dominated atmospheres, the planets may already be depleted by all volatiles and/or be geologically dead. Such planets cannot build up Earth-like atmospheres, thereby inhibiting the emergence of EHs.
\item The UV flux availability will likely be too low to allow for substantial prebiotic chemistry \citep[e.g.,][]{Buccino2006,Buccino2007}.  The chance for an origin of life might therefore be highly reduced. The effectiveness of photosynthesis might similarly be limited {and the oxygenation time around M dwarfs will take too long for complex life to evolve \citep[e.g.,][]{Lingam2021Life}.} In general, the available total energy received at such planet over the entire Hubble time may likely be too low to allow for complex life \citep[e.g.,][]{HaqqMisra2019}. However, flares could increase such energy, but this might be a double-edged sword.
\item Abiotic build-up of O$_2$ could reduce the chance for the origin of life and/or limit its early evolution due to oxidized conditions disfavoring the origin and evolution of early \textit{life as we know it} \citep[e.g.,][]{Lingam2020Oxygen}. It could further decrease the UV availability at a planet's surface due to ozone formation.
\item {Photochemical conditions in the atmospheres of planets orbiting M dwarfs can lead to relatively high atmospheric levels of the highly toxic CO \citep{Schwieterman2019,Schwieterman2019CO}, providing an additional obstacle for complex life \citep{Schwieterman2019}. {The same seems to be the case for ozone, as \citet{Cooke2024} found surface O$_3$ concentrations to substantially surpass 40\,ppb in several of their atmospheric simulations for Trappist-1e and Proxima Centauri~b, with maximum concentrations reaching up to 2200 ppb -- values lethal for life on Earth.}}
\item Planets around M dwarfs might not receive enough reduced atmospheric compounds -- an important source of prebiotic molecules -- to kick-start prebiotic chemistry and the origin of life, as late comet and meteorite impacts will be highly reduced or even absent  \citep[e.g.,][]{Lichtenberg2022,Anslow2023}. {In general, asteroid belt analogs for the delivery of volatiles and prebiotic molecules may be rare around M dwarfs. But still, M dwarfs will suffer significantly more high-velocity impacts than planets around larger host stars, a substantial challenge for the evolution of complex life \citep{Anslow2023}.}
\item Planets around M dwarfs are likely tidally locked \citep[e.g.,][]{Barnes2017}. Warm climates can evolve at such worlds with low CO$_2$ partial pressures as long as the planet is not too dry \citep[e.g.,][]{Turbet2016}. However, it remains to be seen whether EHs could indeed evolve on such planets.
\end{itemize}
The listed criteria further increase the likelihood of EHs being exceedingly rare around M dwarfs. Such rareness will also resolve the so-called `red sky paradox' as posited by \citet{Kipping2021}, in-line with this author’s supposed 'resolution IV', which states that 'M dwarfs have fewer habitable worlds'.

\subsection{The highest rates of EHs}\label{sec:highEH}

The highest probability to find EHs could be within the K-type spectral class, specifically around $\sim$0.8\,M$_{\odot}$, in agreement with other studies \citep[e.g.,][]{Cuntz2015,Lingam2018LifeGStars,HaqqMisra2018,HaqqMisra2019,Lingam2021Life}. For Earth-like atmospheres with $x_{\rm CO_2,max}=1\%$ the highest frequency of EHs will still remain around $M_{\star}\sim$0.8\,M$_{\odot}$, but summed over the entire stellar mass range, most of them will be found around G-type stars. Interestingly, if we exclude planets older than $\sim$6.0\,Gyr, most EHs with $x_{\rm CO_2,max}=1\%$ may even be orbiting F stars, and almost none will be left around K dwarfs. If we also exclude planets younger than $\sim$4\,Gyr to account for an average oxygenation time similar to the Earth's, {about half of the remaining EHs will be around G and F dwarfs for $x_{\rm CO_2,max}=1\%$, but with strong variations between the different cases (see Tables~\ref{tab:AppSpec0p1} and \ref{tab:AppSpec0p01}). For the emergence of complex life, this suggests that the Earth's location around a G-type star is indeed relatively typical.} In this case, however, only $N_{\rm EH,10\%}\leq 4.98^{+160.59}_{-4.85}\times10^{4}$ and $N_{\rm EH,1\%}\leq 1.18^{+66.01}_{-1.16}\times10^{4}$ EHs remain, {leaving relatively few habitats for the potential evolution of complex life.} On a galactic scale, almost exactly 50\% of EHs are inside the galactic orbit of the Sun with a peak density slightly inside of it. {Again, this makes the location of the Sun quite a typical place for \textit{life as we know it} to emerge.}

\subsection{The rate {and distance} of EHs in the Solar Neighborhood}\label{sec:rateEH}

By only considering G stars and the Solar Neighborhood, with a sphere of 0.5\,kpc around the Sun, the occurrence rate of EHs will be higher than over both the entire stellar mass range and galactic disk. For this, we find occurrence rates of $(\eta_{\star}\eta_{\rm EH})_{\rm SN,10\%}\leq 6.76^{+146.069}_{-6.57}\times10^{-5}$ and $(\eta_{\star}\eta_{\rm EH})_{\rm SN,1\%}\leq 1.97^{+66.66}_{-1.92}\times10^{-5}$, respectively. This is more than an order of magnitude larger than the values of the entire galactic disk and stellar mass range. Nevertheless, it is still roughly three to four orders of magnitude lower than the value range of $\eta_{\oplus}\sim 0.01-0.3$. This indicates that it will be difficult to find an EH relatively close to the Earth with present and upcoming exoplanet missions. However, it would be exciting if one or more EHs could be found in the Solar Neighborhood, as this would suggest (if not a serendipitous coincidence) that some parameters in our model are fundamentally wrong. Proposed missions such as LIFE \citep[e.g.,][]{Alei2022,Dannert2022,Kammerer2022,Quanz2022,Angerhausen2023} are therefore of paramount importance for better understanding habitability and life in the Universe.

{We can also calculate the average minimum distance, $d_{\rm EH,SN}$, between EHs in the Solar Neighborhood since we know the stellar distribution and occurrence rate of EHs in the vicinity of the Sun. By assuming that each of them occupies an average spherical volume of $4\pi r_{\rm EH,SN}^3/3$ \citep[see, e.g.,][]{Lingam2021Life}, we can derive the radius, $r_{\rm EH,SN}$, of such volume via the relation}
\begin{equation}\label{eq:distEH}
  \frac{4\pi r_{\rm EH,SN}^3}{3}\times N_{\rm EH,SN} \sim V_{\rm SN},
\end{equation}
{where $N_{\rm EH,SN}$ is the maximum number of EHs in the Solar Neighborhood and $V_{\rm SN}$ is a predefined volume of the Solar Neighborhood (e.g., one spatial bin within our simulation); the average minimum distance, $d_{\rm EH,SN}$, between two EHs can then simply be calculated through $d_{\rm EH,SN} = 2 r_{\rm EH,SN}$. However, we have to ensure that the maximum number of EHs within $V_{\rm SN}$ is at least 2 or higher for obtaining reasonable values for $d_{\rm EH,SN}$. For one spatial bin (i.e., with $100\times100\times100$\,pc) within our model this is only valid for both our maximum cases. For all other scenarios we need to take larger volumes, i.e., cubes with side lengths of 200\,pc (nominal case with $x_{\rm CO_2,max}=10\%$), 300\,pc (nominal case with $x_{\rm CO_2,max}=1\%$), 600\,pc (minimum case with $x_{\rm CO_2,max}=10\%$), and 1000\,pc (minimum case with $x_{\rm CO_2,max}=1\%$), respectively, for which the maximum number of EHs is always $>3$. This is important since $d_{\rm EH,SN}$ increases for larger $V_{\rm SN}$ due to the exponential stellar distribution chosen for our galactic disk model.

This behavior can be seen in Figure~\ref{fig:distances}, which compares the average distance between EHs in the galactic disk, $d_{\rm EH,d}$, (thick dashed grey lines) with $d_{\rm EH,SN}$ from our different model cases (upper panel for $x_{\rm CO_2,max}=10\%$; lower panel for $x_{\rm CO_2,max}=1\%$). For the Solar Neighborhood, we obtain average distances of $d_{\rm EH,SN,10\%}\sim155_{-110}^{+450}$\,pc and $d_{\rm EH,SN,1\%}\sim265^{+900}_{-200}$\,pc, whereas these increase to $d_{\rm EH,d,10\%}\sim370^{+960}_{-250}$\,pc and $d_{\rm EH,d,1\%}\sim610^{+1660}_{-440}$\,pc, respectively, by averaging over the entire galactic disk. We further note that our average minimum distances, $d_{\rm EH,SN}$, are by roughly an order of magnitude larger than the closest, `life-harboring world' estimated by \citet{Madau2023}. These authors found this distance to be $\lesssim$20\,pc by simply assuming that microbial life arose as soon as it did on Earth on $\gtrsim$1\% of all rocky exoplanets in the HZ of K dwarfs.}

\begin{figure}
\centering
\includegraphics[width = 1.0\linewidth, page=1]{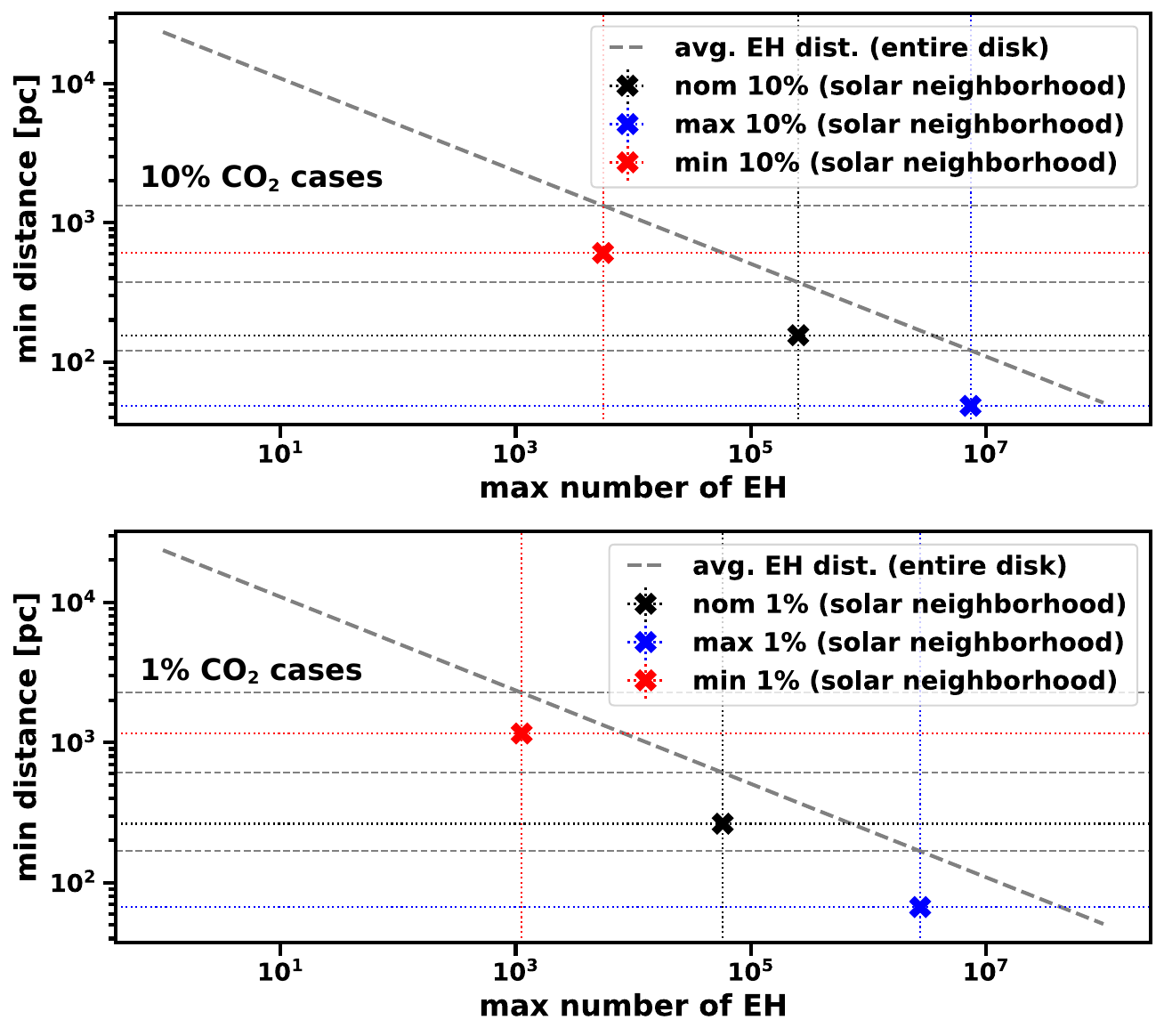}
\caption{{ Upper panel: The minimum average distance, $d_{\rm EH,SN,10\%}$, between EHs in the Solar Neighborhood for our nominal (black cross), maximum (blue cross) and minimum (red cross) cases with $x_{\rm CO_2,max}=10\%$, as well as the average distance, $d_{\rm EH,d,10\%}$, for the entire galactic disk (thick dashed grey line). One can see that $d_{\rm EH,SN,10\%}$ remains always smaller than {$d_{\rm EH,d,10\%}$} as long as the minimum number of EHs in the galaxy is $>2$. Lower panel: The same but for $x_{\rm CO_2,max}=1\%$.}}
\label{fig:distances}
\end{figure}

\subsection{Quantification of further requirements and our results}\label{sec:quantification}

As already discussed, some additional requirements will likely become scientifically quantifiable in the next decades. Current and future exoplanet missions are therefore highly important. With the ground-based facilities ELT and TMT, space missions JWST, Ariel and PLATO, and their potential successors LIFE and the Habitable Worlds Observatory (HWO), further criteria can be quantified to such an extent that these can be included in future assessments on the prevalence of EHs. Exoplanet atmosphere characterization might relatively soon be able to tell us, whether rocky exoplanets accrete relatively fast and mostly within the disk, and how many of them will host primordial atmospheres. This could also tell us something about the prevalence of working carbon-silicate and nitrogen cycles, possibly even about the frequency of \textit{life as we know it}, by statistically characterizing the compositions of exoplanetary atmospheres. Upcoming missions will at some time be able to give us estimates on the occurrence rates of CO$_2$-dominated and O$_2$-rich atmospheres, planets without thick H$_2$- and He-envelopes, and maybe even on N$_2$-dominated atmospheres. Future missions such as PLATO \citep{Rauer2014} will also refine our knowledge on $\eta_{\oplus}$ and probably on the frequency of large moons.

All these factors taken together will make our results testable in the relatively near future. Discovering scientific hints on $N_{\rm EH,SN}>1$ in the Solar Neighborhood, in addition to the Earth, will clearly falsify our outcome. Finding none, on the other hand, can lead to a robust refinement of our maximum numbers, and further scientific results, theoretically and observationally, will shed light on additional requirements that can be thoroughly addressed in the future.

{This also relates to our reasoning that EHs around M dwarfs are exceedingly rare. This conclusion implies that the search for signs of life at planets orbiting low-mass stars will likely return a lack of robust atmospheric biosignatures\footnote{See, e.g., \citet{Schwieterman2018} for a recent review on atmospheric biosignatures.}. This endeavour has already recently started to be partly on its way, most prominently with JWST, which can already perform detections of atmospheric molecules that are often assumed to be related to life such as O$_3$, O$_2$, CH$_4$ and H$_2$O. JWST can further give hints on whether or not a rocky exoplanet is hosting a secondary atmosphere. This was already tentatively illustrated for the rocky M dwarf planets Trappist-1b \citep{Greene2023}, Trappist-1c \citep{Zieba2023}\footnote{Note that this study is debated. Another study, for instance, found that the JWST data for Trappist-1c could also be in agreement with thicker $\sim$0.1\,bar CO$_2$-O$_2$-atmospheres \citep{Lincowski2023}.}{, GJ\,341b \citep{Kirk2024} and LHS\,475b \citep{LustigYaeger2023}, whose observations are all consistent with having no or only very tenuous atmospheres.}}

{It is hence worth pointing out that robust discoveries of biosignatures and/or N$_2$-dominated atmospheres on rocky exoplanets in the HZCL of M dwarfs would serve as a means of falsifying our model's predictions, thereby indicating that at least certain arguments within our model must be wrong. From a scientific point of view this is good, as it meets the Popperian standard of science.}

{Finally, our work can help guiding the design of next-generation telescopes, potentially beyond LIFE and HWO, in case that our results will not be falsified by finding EHs in the Solar Neighborhood. Our average minimum distance, $d_{\rm EH,SN}\gtrsim 150-250$\,pc, to the closest EH constrains the minimum aperture of ground- and space-based telescopes that are required for detecting atmospheric biosignatures from these planets.}

\subsection{Habitats other than EHs}\label{sec:exohab}

Our study tells nothing about any habitats other than EHs, whether these may be sub-surface ocean worlds or water worlds, Titan-like habitats, planets with completely different biochemistry, or even worlds with anaerobic microbial life still able to evolve into EHs.

One may, however, simply extend our formulation to take account of other habitats, $N_i$, i.e.,
\begin{equation}\label{eq:NHab}
N_{\rm Hab}(N_i) = \sum_{i=1}^{n}N_{\star}\times N_i = \sum_{i=1}^{n}N_{\star}\times\eta_{\star,i}\times\eta_{i}.
\end{equation}
By including hypothetical Titan-like habitats, $N_{\rm TH}$, and sub-surface ocean worlds, $N_{\rm SSOW}$, for instance, this equation can be further written as
\begin{equation}\begin{split}\label{eq:NHab2}
N_{\rm Hab}(N_{\rm EH},N_{\rm TH},N_{\rm SSOW}) & = N_{\star}\times\eta_{\star,\rm EH}\times\eta_{\rm EH}\\ & + N_{\star}\times\eta_{\star,\rm TH}\times\eta_{\rm TH}\\ & + N_{\star}\times\eta_{\star,\rm TH}\times\eta_{\rm TH},
\end{split}\end{equation}
where the different terms for $\eta_{\star,i}$ and $\eta_{i}$ can again be populated with various necessary galactic, stellar, and planetary environments that are specific for the respective $N_i$.

These and other potential habitats may be common in the Galaxy, but our present study is completely agnostic about them and whether life evolves on these more exotic worlds. Here, we should hence note that, while the existence of an EH may already imply the presence of {aerobic denitrifying microbes, anaerobic anammox species, and/or any other microbes that recycle N$_2$ back into the atmosphere}, any estimated numbers for, e.g., $N_{\rm TH}$ and $N_{\rm SSOW}$ provide no information about whether life actually exists on these bodies, nor about the frequency of life in the Galaxy, as it is unlikely that life itself is a prerequisite for the evolution of these habitats. However, if some form of non-technological life can indeed evolve on such worlds, even if its origin at such bodies would be less likely, exotic biospheres could turn out to be the most common ones in the Galaxy and the entire Universe. They could even outnumber 2$^{\rm nd}$ Earth's, as we are keen to find them, by orders of magnitude. Possibly.

\subsection{ETIs in the Galaxy}\label{sec:eti}

Our result that Earth-like Habitats are relatively rare, also implies that {the emergence of} (aerobic) communicating extraterrestrial intelligence (ETI) in the Galaxy  -- the main target of SETI \citep[e.g.,][]{Drake1961,Drake1965,Tarter2001} -- will {likely be rare as well}. In fact, it will be significantly rarer than EHs per se since additional requirements must be met for ETIs to evolve, specifically, if these should be technologically advanced species that intend to send radio messages into space and who are able to listen to what others are sending. That no definitive signals from ETIs were hence detected by SETI comes by relatively little surprise.

{However, that the emergence of aerobic ETIs will be rare due to a small number of viable sites for their initial evolution, does not necessarily mean that once an ETI indeed evolved, it, or its postbiological successors, could not become more widespread in the Milky Way than complex life itself \citep[e.g.,][]{Lingam2019ETI,Wright2022ETI}. While the latter will evolve within a specific biosphere with a finite lifetime, a technological civilization emerging from an EH can outlive its initial biosphere \citep[e.g.,][]{Balbi2021ETI}. In principle, it can even colonize planets other than EHs, and hence spread efficiently\footnote{However, recently \citet{Lingam2019ETI} estimated that the likelihood of detecting intelligent life might nevertheless be at least 2 orders of magnitude less probable than the detection of primitive life.} throughout the galactic disk and even beyond \citep[e.g.,][]{Walters1980ETI}. One should keep such uncertainties and caveats in mind for the following discussion.}

That said, if (i) EHs indeed are rare, but (ii) some other aerobic ETIs nevertheless exist in today's Galaxy, intentionally sending signals into space to communicate with whoever might be out there, can be an unwise action to take. In this case, the potential rareness of EHs implies that the Earth itself, with its N$_2$-O$_2$-dominated atmosphere, is a very valuable resource for any aerobic ETI currently existing in the Milky Way. This is even more true if these ETIs indeed possess the relevant technology to spread from their original Earth-like Habitat into interstellar space and toward other valuable planets. It would thus come by no surprise if technologically advanced species would have a certain interest in our `Pale Blue Dot'. {This reasoning would be strongly supported if some of the controversially discussed unidentified aerial phenomena (UAP)\footnote{For further details on UAPs, see the NASA independent study team report on unidentified aerial phenomena \citep{NASA2023UAP}} had actually originated on an EH other than ours, as our results would then not only indicate that these objects must have travelled a considerable distance, but could also explain the actual reason for their visit: the Earth as a rare and valuable resource.} Even if these ETIs were benevolent, however, first contact could have negative impact on humanity \citep[e.g.,][]{Schetsche2018}.

Intentionally signaling that we, and the Earth specifically, are actually here, as is the aim of METI \citep[e.g.,][]{Baum2011,HaqqMisra2013METI,Vakoch2011,Vakoch2016} and as was recently proposed by \citet{Jiang2022}, can therefore add some kind of unnecessary, diffuse and incalculable danger that, at least with our present knowledge, cannot be properly assessed and may therefore neither be neglected nor evoked. Here, it is important to note that radio leakage and other techno- and atmospheric signatures can in principle be detected from close-by ETIs, but an intentional, strong, and focused signal such as the Arecibo message in 1974 will reach much farther into space \citep[see, e.g.,][]{HaqqMisra2013METI}.

Sending messages into space must be discussed broadly, specifically by also considering the argument that EHs might be rare. Even though some of the past METI initiatives have been more transient than other terrestrial radio sources \citep[e.g.,][]{Sullivan1978,Baum2011,HaqqMisra2013METI}, this was not the case for the Arecibo message and will not be the case for certain future attempts. It should also be emphasized that justifying METI by pointing to the detectability of unintended radio leaks from Earth appears to be a relatively poor argument for intentional signaling to ETIs, as this can also be seen as an argument for minimizing the leakage of technosignatures in the future. Besides, \citet{Baum2011} stressed that intentional messages to ETIs should not contain specific biological information about humans, as we do not know whether these will be received by a malicious ETI. These authors further argue that we should avoid being recognized as an expanding civilization to minimize the possibility of being perceived as a threat. This argument is further strengthened by our finding that EHs are rare. The destruction of one of these rare habitats indeed leaves quite an alarming impression on any intelligent observer.

But how many {separately emerged and evolved} aerobic ETIs could indeed exist in the Milky Way at present? For some simple estimate, we can easily extend our formulation, i.e.,
\begin{equation}\label{eq:NETI}
N_{\rm ETI}(N_{\rm Hab}\equiv N_{\rm EH}) = N_{\rm EH}\times \eta_{\rm ETI},
\end{equation}
where $\eta_{\rm ETI}$ is the present occurrence rate of communicating, aerobic ETIs in the Galaxy. This can further be written as
\begin{equation}\begin{split}\label{eq:etaETI}
\eta_{\rm ETI} = \Pi_{\rm bio}\left(\prod_{i=1}^n \pi_{\rm bio}^i\right)\times\Pi_{\rm civ}\left(\prod_{i=1}^n \pi_{\rm civ}^i\right)\times \frac{\tau_{\rm ETI}}{\tau_{\rm EH}}.
\end{split}\end{equation}
Here, $\Pi_{\rm bio}$\footnote{In this case, $\Pi$ and $\pi$ were chosen based on the Greek translation of civilisation, i.e., $\uppi \mathrm{o}\uplambda \upiota\uptau\upiota\upsigma\upmu\acute{\mathrm{o}}\mathrm{\varsigma}$.} defines any biological requirements, $\pi_{\rm bio}^i$, that have to be met for ETIs to evolve and are not yet part of $B_{\rm life}$. These could, among others, include the origin of complex and intelligent life (i.e., $f_{\rm in}$ in the Drake Equation; see Equation~\ref{eq:drake}). The other term, $\Pi_{\rm civ}$, further depicts any societal and technological requirements such as the origin of a technically communicating civilization (i.e., $f_{\rm civ}$ in the Drake Equation). Finally, the parameter $\tau_{\rm ETI}$ is equivalent to $L$ within the Drake Equation and signifies the average lifetime of ETIs. Since our formulation is based on the present number of stars in the Galaxy instead of the SFR, however, we have to divide $\tau_{\rm ETI}$ by the average lifetime, $\tau_{\rm EH}$, of an EH.

If we assume $\tau_{\rm EH}$ to be in the order of $\tau_{\rm EH}\sim10^9$ years and take a commonly used ETI lifetime of $\tau_{\rm ETI}\sim 10^4$\,years \citep[see, e.g.,][Table~2]{HaqqMisra2013METI}\footnote{{In reality, there will likely be a relatively large spread for $\tau_{\rm ETI}$, including relatively long-lived ETIs \citep[e.g.,][]{Kipping2020ETI,Balbi2021ETI}; see discussion in \citet{Wright2022ETI}.}}, $N_{\rm ETI}$ will already be lower than $N_{\rm EH}$ by 5 orders of magnitude just by taking into account $\tau_{\rm ETI}$. This specific estimate therefore results in $N_{\rm ETI}\lesssim 3^{+72}_{-3}$ for $x_{\rm CO_2,max}=10\%$, a maximum number that is around unity or lower for our nominal and minimum cases - but likely it will be lower than that.

This rather simple estimate does not include any other of the potential requirements needed to properly estimate $N_{\rm ETI}(N_{\rm Hab}\equiv N_{\rm EH})$, such {as the occurrence rate of tighter $p$O$_2$-limits needed for the emergence of technology \citep[e.g.,][see also Paper I]{Balbi2023}} or complex and intelligent life. It does not include criteria that were already argued to be rare, such as an unlikely emergence of eukaryotic cells \citep[see, e.g.,][]{Lane2017}, the timing of evolutionary transitions \citep[e.g.,][]{SnyderBeattie2021}, or the emergence of intelligence itself \citep[see, e.g.,][]{Lineweaver2022}. But still, this simple estimate already indicates that the Fermi paradox \citep[e.g.,][]{Hart1975Fermi,Tipler1980Fermi,Cirkovic2018,Forgan2019} could theoretically be resolved through the rareness of EHs, specifically if we further consider potential settlement dynamics of technological civilizations within the Milky Way that can result in large unsettled galactic regions \citep[e.g.,][]{ Landis1998,CarrollNellenback2019}. It even suggests that {the emergence of} intelligent life could be extremely rare and that recent estimates on ETIs in the Galaxy obtained numbers that are clearly too high \citep[e.g.,][]{Cai2021,Westby2020,Song2022}.

{If one wants to calculate the maximum number of planets inhabited by ETIs or visited by self-replicating probes \citep[e.g.,][]{Bracewell1960,Tipler1980Fermi,Freitas1980} instead of the maximum number of separately emerged and evolved (aerobic) ETIs, however, this formalism would have to be extended for covering interstellar travel within the Galaxy, by including an additional settling factor $S$ \citep[e.g.,][]{Lingam2021Life}, as was already suggested to be added to the Drake Equation \citep[e.g.,][]{Walters1980ETI,Brin1983}.} {In addition}, the above formulation does not include non-biological, AI-based civilizations that descended from biological, aerobic ETIs or any other intelligent extraterrestrials (biological and non-biological) that originated on habitats other than EHs. For this, we can simply extend Equation~\ref{eq:NETI} as follows:
\begin{equation}\label{eq:NETI2}
N_{\rm ETI}(N_{\rm Hab}=N_i) = \sum_i^n N_{\star}\times N_{\rm i}\times \eta_{\rm{ETI},i},
\end{equation}
where $i$ again denotes the different kinds of habitats that may exist in the Milky Way, or more broadly, in the Universe. Whether an extension of this formalism indeed increases the number of ETIs in the Galaxy significantly remains speculative, even though it seems reasonable to assume that the origin of technological civilizations on bodies such as sub-surface ocean worlds or even Titan-like habitats could be highly unlikely due to physical limitations \citep[see also][for a recent discussion on technological intelligence on land- and ocean-based habitats]{Lingam2023}. This also suggests that, {at least if ignoring the prospects of interstellar travel and colonization,} the next ETI will likely be located at a great distance from Earth (if any within the Milky Way), much farther than any leaked signature has traveled so far. {This reasoning can easily be derived from the average minimum distance, $d_{\rm EH,SN}$, between the Earth an the next EH in the Solar Neighborhood, which we calculated to be at least $d_{\rm EH,SN}\gtrsim155_{-110}^{+450}$\,pc in Subsection~\ref{sec:rateEH}. This distance is well beyond the sphere of space through which human-generated electromagnetic transmissions have yet travelled, a sphere with an approximate lookback time of $t_{\rm lb}\sim100$\,years corresponding to $r_{\rm lb}\sim30$\,pc. The average distance to the next ETI emerging from an EH, however, must logically be $d_{\rm EH,SN}\gtrsim155_{-110}^{+450}$\,pc, at least as long as we ignore interstellar travel.} Any messages, intentional as well as unintentional, might therefore remain unheard for quite some time yet.

{And it is, in turn, very likely that SETI observations will have to wait quite a while to detect technological emissions from ETIs in the Galaxy, if they exist at all. This conclusion is well supported by \citet{Grimaldi2023}, who, based on Bayesian reasoning, finds `optimistic waiting times' for such a signal to be 60-1800 years with a probability of 50\%. If EHs, and hence ETIs, are very rare, these waiting times may indeed be quite optimistic.}

\subsection{The Copernican Principle of Mediocrity}\label{sec:copernicus}

Finally, we emphasize that the Principle of Mediocrity, or more specifically, the Copernican Principle {of Mediocrity }\citep[see, e.g.,][]{Gott1993,Cirkovic2012,Scharf2014,Westby2020} in the sense of the Earth not being special and (complex or even intelligent) life consequently being common in the Universe \citep[e.g.,][]{Sagan1994}, cannot be regarded to be valid on the specific level of individual planets, as was recently also pointed out by \citet{Balbi2023Mediocrity}. The Earth is certainly not special in the sense of being central to the Galaxy or even the Universe. {However, the occurrence rate of rocky exoplanets, $\beta_{\rm HZCL}$, in the HZCL diverges significantly from $\eta_{\rm EH}$, the maximum occurrence rate of EHs around suitable stars, or even more so from the product $\eta_{\rm EH}\times \eta_{\star}$, the maximum occurrence rate of EHs around any star. Their ratios can simply be illustrated through}
\begin{equation}\label{eq:EEvsEEH}
  f_{\rm EH} = \frac{\eta_{\rm EH}}{\beta_{\rm HZCL}},
\end{equation}
and
\begin{equation}\label{eq:EEStarvsEEH}
  f_{\rm EH,\star} = \frac{\eta_{\rm EH}\eta_{\star}}{\beta_{\rm HZCL}},
\end{equation}
{respectively. If we, for instance, assume our nominal case for N$_2$-dominated atmospheres with $x_{\rm CO_2,max}=10\%$ with obtained occurrence rates of $\eta_{\star,10\%} < 2.07\times 10^{-2}$, $\eta_{\rm EH,10\%} < 1.21\times 10^{-4}$, and $\beta_{\rm HZCL,10\%} = 0.022$, we receive $f_{\rm EH,10\%}< 5.5\times 10^{-3}$ and $f_{\rm EH,\star}<6.25\times10^{-5}$, respectively. Taken all our cases together, these ratios can be calculated to vary between $f_{\rm EH}< \{1.1\times 10^{-2}, 9.3\times 10^{-4}\}$ and $f_{\rm EH,\star}<\{7.3\times10^{-4}, 7.4\times10^{-7}\}$, respectively, which comprises a mismatch of up to almost 7 orders of magnitude and implies. {The latter of these values further imply that, on average, a minimum of $1.4\times 10^3$ to $1.4\times 10^6$ planets in the HZCL are needed for one EH to evolve.} These numbers nicely illustrate why the average distance to the closest, `life-harboring world' in the Solar Neighborhood, as recently estimated by \citet{Madau2023}, is closer by roughly an order of magnitude than $d_{\rm EH,SN}$ in our study.}

It is therefore also not surprising that our maximum estimate on ETIs in the galactic disk is much lower than the one provided by \citet{Westby2020}, who defined and applied a specific form of the Copernican Principle called `astrobiological Copernican Principle'.{ They define it as follows: Any sufficiently Earth-like planet (i.e., `suitable planet') in the HZ of a `suitable star' will form (intelligent) life over a time frame of $\sim$5\,Gyr in a similar fashion than the Earth. \citet{Westby2020} further define `suitable star' as any star with an age greater than 5\,Gyr and a sufficiently high metallicity to allow for the evolution of advanced biology and ETIs. The term `suitable planet' further refers to any rocky exoplanet within the habitable zones of such stars. Based on this astrobiological Copernican Principle, \citet{Westby2020} derive the following modification of the Drake Equation, i.e.,}
\begin{equation}\label{eq:ACPEq}
  N = N_{\star}\times f_{\rm L} \times f_{\rm HZ}\times f_{\rm M} \left(\frac{\tau_{\rm ETI}}{\tau'}\right),
\end{equation}
{where $N_{\star}$ and $\tau_{\rm ETI}$ are the already discussed number of stars in the Galaxy and the average lifetime of ETIs, respectively. The other variables comprise the fraction of stars that are older than 5\,Gyr, $f_{\rm L}$, the fraction of rocky exoplanets in the HZ, $f_{\rm HZ}$, and the fraction of stars with sufficient metallicity, $f_{\rm M}$. Finally, the variable $\tau'$ is defined as the average age of the stars in the Galaxy minus 5\,Gyr, i.e., the average time intelligent life needs to evolve, as set in the their model.}

{\citet{Westby2020} further split the astrobiological Copernican Principle into a `Weak' and `Strong' scenario, of which the first one presumes that intelligent life needs at least 5\,Gyr to evolve, while the second assumes that intelligent life must form between 4.5 and 5.5\,Gyr.} If we take their `Strong' scenario with $\tau_{\rm ETI}=100$~years (i.e., their scenario with the lowest number of ETIs), our maximum number of ETIs in the galactic disk would be $N_{\rm ETI}\lesssim 10^{-3}$ instead of their $N_{\rm ETI}=36^{+175}_{-32}$, even though many necessary requirements for $N_{\rm EH}$, and even more so for $N_{\rm ETI}$, are not implemented in our model. {If we take the value of their `Strong' scenario by ignoring the error bars, this can be simply converted to a ratio of $f_{\rm ETI} < 10^{-3}/36 \sim 2.8\times 10^{-5}$, while applying the `Weak' scenario by \citet{Westby2020} with $N_{\rm ETI}=928^{+1980}_{-818}$ ETIs\footnote{To be precise, this is the result of the `Weak' scenario in \citet{Westby2020} with an assumed metallicity threshold of $Z_{\rm min} = 0.1$\,$Z_{\odot}$} leads to a decreased ratio of even $f_{\rm ETI} < 10^{-3}/928 \sim 1.1 \times 10^{-6}$.}

{These exemplary discrepancies} clearly illustrate that not every rocky planet in the HZ(CL) can simply be considered habitable. Such an assumption {can }even be wrong by orders of magnitude. If planetary parameters are very similar to each other, then it is not of great importance whether we live on this or on any other planet that fulfills certain parameters. In this tautological sense, the Copernican Principle logically holds. However, there are these specific requirements that shall be met for making a planet habitable for \textit{life as we know} it and only a small fraction of planets will meet these criteria. In such a stricter sense, the Copernican Principle cannot be applied and the Earth is indeed special or even rare (at least in relation to the total number of stars and the planets orbiting in their HZCLs) and, based on such principle, it is a fallacy to assume that (complex) life is common in the Universe, even if one sets the unknown frequency of life originating to be equal to unity.

One could, however, argue for some kind of combined Anthropic-Copernican Principle that states that certain special conditions have to be met for \textit{life as we know} it to evolve. As long as these conditions can evolve at some place, the Copernican Principle suffices, and life as we know it might be common on such worlds \citep[see also][]{Gott1993}. However, these combined conditions will be rare on a galactic scale, and in that sense the Anthropic Principle \citep[e.g.,][]{Carter1983} holds as well. It might thus not be a coincidence that we live somewhere in the middle of the Galaxy on an Earth-mass planet below an Earth-like atmosphere that orbits within the HZCL of a mid-aged, anomalously weakly active G-type star, that has surface water, subaerial land, and a large moon on which intelligent species can set foot on for embarking onto their journey into space.

\section{Tables of input parameters and model results}\label{app:inputTable}

{Here, we provide a comprehensive overview on all the input parameters} that feed into our different simulations (i.e., into nominal, minima and maxima cases for N$_2$-O$_2$-dominated atmospheres with $x_{\rm CO_2,max}=10\%$ and $x_{\rm CO_2,max}=1\%$, respectively) {together with our obtained model results}. If the input parameters are the same for different cases these are only displayed once, distributed over the columns of the respective cases. Each {of the first three} tables covers all 6 cases. However, there are no differences in the parameters for the 10\%\,CO$_2$ and 1\%\,CO$_2$ scenarios for our calculation of $N_{\star}$, as listed in Table~\ref{tab:AppNStar}. We have split the parameters into four separate tables, i.e., into tables for $N_{\star}$ (Table~\ref{tab:AppNStar}), $\eta_{\star}$ (Table~\ref{tab:AppEtaStar}), $\beta_{\rm HZCL}$ (Table~\ref{tab:AppBetaHZCL}), and $\eta_{\rm EH}$ (Table~\ref{tab:AppEtaEH}). These tables also include the number of stars/planets that resulted for each step, as well as the different fractions and cumulative fractions of the different requirements.

{In addition, Tables~\ref{tab:AppSpec0p1} and \ref{tab:AppSpec0p01} provide similar information than the smaller Table~\ref{tab:NEHspec} but lists the distribution of remaining stars/planets within the different stellar spectral classes for each of the implemented requirements, {and additionally provides numbers for stars that are both older than 6\,Gyr and younger than 4\,Gyr to account for (i) an Earth-like oxygenation time of $T_{\rm O_2}=4\,$Gyr and (ii) the potential cessation of geological activity at cosmochemically Earth-like planets}. Table~\ref{tab:AppSpec0p1} lists these values for $x_{\rm CO_2,max}=10\%$, while Table~\ref{tab:AppSpec0p01} gives the same for $x_{\rm CO_2,max}=1\%$. Note that both tables are identical prior to the upper limit, $\alpha_{\rm at}^{\rm XUV}$.}

\begin{table*}\footnotesize
  \begin{center}
    \caption{Input parameters for simulating the number of stars in the disk, $N_{\star}$, and the resulting stellar properties. Note that parameters that are the same for several cases are only listed once per row and distributed over the respective cases they refer to.}
    \label{tab:AppNStar}\vspace{-3mm}
    \resizebox{\textwidth}{!}{%
\begin{tabular}{l|l|l|c|c|c|c|c|c}
  \hline
   & & & \multicolumn{6}{c}{model cases}  \\
    & & & \multicolumn{2}{c}{nominal} & \multicolumn{2}{c}{maximum} & \multicolumn{2}{c}{minimum} \\
    & & & 10\% CO$_2$ & 1\% CO$_2$ & 10\% CO$_2$ & 1\% CO$_2$ & 10\% CO$_2$ & 1\% CO$_2$ \\

   \hline
   \multirow{10}{*}{\parbox{2cm}{IMF \& SFR (Sect.~\ref{sec:NStarImp})}} & \multicolumn{2}{l}{implemented IMF} & \multicolumn{2}{c}{\citet{Kroupa2001}} & \multicolumn{2}{c}{\citet{Chabrier2003}} & \multicolumn{2}{c}{\citet{Kroupa2001}} \\
    & \multicolumn{2}{l}{implemented SFR} & \multicolumn{2}{c}{\citet{Snaith2015}} & \multicolumn{2}{c}{\citet{Naab2006}} & \multicolumn{2}{c}{\citet{Snaith2015}} \\
    & \multicolumn{2}{l}{max. main-sequence age} & \multicolumn{6}{c}{\citet{Westby2020}} \\
    & \multicolumn{2}{l}{IMF mass range [$M_{\odot}$]} & \multicolumn{2}{c}{$0.08-100.00$} & \multicolumn{2}{c}{$0.1-100.00$} & \multicolumn{2}{c}{$0.07-100.00$}\\
    & \multicolumn{2}{l}{implemented stellar range [$M_{\odot}$]} & \multicolumn{6}{c}{$0.10-1.25$} \\
   \cline{2-9}
    & \multirow{7}{*}{\parbox{2cm}{mass range\\spectral classes [M$_{\odot}$]\\\citep[e.g.,][]{Habets1981}}} & M & \multicolumn{2}{c}{$0.08 - 0.44$} & \multicolumn{2}{c}{$0.10 - 0.44$} & \multicolumn{2}{c}{$0.07 - 0.44$} \\
    & & K & \multicolumn{6}{c}{$0.45 - 0.79$} \\
    & & G & \multicolumn{6}{c}{$0.80 - 1.03$} \\
    & & F & \multicolumn{6}{c}{$1.04 - 1.39$} \\
    & & A & \multicolumn{6}{c}{$1.40 - 2.09$} \\
    & & B & \multicolumn{6}{c}{$2.10 - 15.99$} \\
    & & O & \multicolumn{6}{c}{$\geq 16.00$} \\
    \hline
    \multirow{9}{*}{\parbox{2cm}{galactic\\parameters (Sect.~\ref{sec:NStar})}} & \multicolumn{2}{l}{galactic mass model} &\multicolumn{6}{c}{\citet{McMillan2017}} \\
        & \multirow{3}{*}{thin disk} & $\Sigma_0$ [M$_{\oplus}$\,pc$^{-2}$] & \multicolumn{6}{c}{896.0} \\
    & & R$_{\rm d}$ [kpc] & \multicolumn{6}{c}{2.50} \\
    & & Z$_{\rm d}$ [kpc] & \multicolumn{6}{c}{0.30} \\
    & \multirow{3}{*}{thick disk} & $\Sigma_0$ [M$_{\oplus}$\,pc$^{-2}$] & \multicolumn{6}{c}{183.0} \\
    & & R$_{\rm d}$ [kpc] & \multicolumn{6}{c}{3.02} \\
    & & Z$_{\rm d}$ [kpc] & \multicolumn{6}{c}{0.90} \\
    \cline{2-9}
    & \multirow{2}{*}{galactic disk region} & $r$ range [kpc] & \multicolumn{6}{c}{2.0 - 21.0} \\
    & & $z$ range [kpc] & \multicolumn{6}{c}{0 - 2.5} \\
   \hline
    \multirow{13}{*}{disk mass (Sect.~\ref{sec:NStar})} & \multicolumn{2}{l}{stellar remnant fraction} & \multicolumn{6}{c}{20\% assumed \citep[e.g.,][]{Xiang2018}} \\
    \cline{2-9}
    & \multirow{3}{*}{\parbox{2cm}{with remnants (entire disk)}} & thin disk [M$_{\odot}$] & \multicolumn{6}{c}{$3.52\times10^{10}$} \\
    & & thick disk [M$_{\odot}$] & \multicolumn{6}{c}{$1.05\times10^{10}$} \\
    & & total [M$_{\odot}$] & \multicolumn{6}{c}{$5.57\times10^{10}$} \\
    \cline{2-9}
    &  \multirow{3}{*}{\parbox{2cm}{without remnants (entire disk)}} & thin disk [M$_{\odot}$] & \multicolumn{6}{c}{$2.82\times10^{10}$} \\
    & & thick disk [M$_{\odot}$] & \multicolumn{6}{c}{$0.84\times10^{10}$} \\
    & & total [M$_{\odot}$] & \multicolumn{6}{c}{$3.65\times10^{10}$} \\
    \cline{2-9}
    & \multirow{3}{*}{\parbox{2cm}{with remnants ($2.0-21.0$\,kpc)}} & thin disk [M$_{\odot}$] & \multicolumn{6}{c}{$2.86\times10^{10}$} \\
    & & thick disk [M$_{\odot}$] & \multicolumn{6}{c}{$0.85\times10^{10}$} \\
    & & total [M$_{\odot}$] & \multicolumn{6}{c}{$3.71\times10^{10}$} \\
    \cline{2-9}
    &  \multirow{3}{*}{\parbox{2cm}{without remnants ($2.0-21.0$\,kpc)}} & thin disk [M$_{\odot}$] & \multicolumn{6}{c}{$2.29\times10^{10}$} \\
    & & thick disk [M$_{\odot}$] & \multicolumn{6}{c}{$0.68\times10^{10}$} \\
    & & total [M$_{\odot}$] & \multicolumn{6}{c}{$2.97\times10^{10}$} \\
    \hline
    \multirow{19}{*}{disk stars (Sect.~\ref{sec:NStar})} & \multicolumn{2}{l}{mean stellar mass (total)  [$M_{\odot}$]} & \multicolumn{2}{c}{0.57} & \multicolumn{2}{c}{0.65} & \multicolumn{2}{c}{0.53} \\
    & \multicolumn{2}{l}{mean stellar mass (present)  [$M_{\odot}$]} & \multicolumn{2}{c}{0.29} & \multicolumn{2}{c}{0.32} & \multicolumn{2}{c}{0.28} \\
     & \multicolumn{2}{l}{stellar mass ever existing [$M_{\odot}$]} & \multicolumn{2}{c}{$6.33\times10^{10}$} & \multicolumn{2}{c}{$6.82\times10^{10}$} & \multicolumn{2}{c}{$6.27\times10^{10}$} \\
     & \multicolumn{2}{l}{stars ever existing} & \multicolumn{2}{c}{$11.13\times10^{10}$} & \multicolumn{2}{c}{$10.53\times10^{10}$} & \multicolumn{2}{c}{$11.73\times10^{10}$} \\
     & \multicolumn{2}{l}{stellar mass still existing [\%]} & \multicolumn{2}{c}{46.86} & \multicolumn{2}{c}{43.53} & \multicolumn{2}{c}{47.34} \\
     & \multicolumn{2}{l}{stars yet existing (entire mass range) [\%]} & \multicolumn{2}{c}{91.15} & \multicolumn{2}{c}{89.11} & \multicolumn{2}{c}{91.75} \\
     & \multicolumn{2}{l}{stars yet existing (within 0.1-1.25\,M$_{\odot}$) [\%]} & \multicolumn{2}{c}{97.39} & \multicolumn{2}{c}{97.49} & \multicolumn{2}{c}{97.39} \\
     \cline{2-9}
     & \multirow{3}{*}{\parbox{2cm}{present stars (entire $r$\,\&\,$M_{\star}$)}} & thin disk & \multicolumn{2}{c}{$9.60\times10^{10}$} & \multicolumn{2}{c}{$8.44\times10^{10}$} & \multicolumn{2}{c}{$10.18\times10^{10}$} \\
    & & thick disk & \multicolumn{2}{c}{$2.86\times10^{10}$} & \multicolumn{2}{c}{$2.52\times10^{10}$} & \multicolumn{2}{c}{$3.03\times10^{10}$} \\
    & & total & \multicolumn{2}{c}{$12.46\times10^{10}$} & \multicolumn{2}{c}{$10.96\times10^{10}$} & \multicolumn{2}{c}{$13.21\times10^{10}$} \\
    \cline{2-9}
    & \multirow{3}{*}{\parbox{2cm}{present stars (entire $M_{\star}$; $2.0-21.0$\,kpc)$^a$}} & thin disk & \multicolumn{2}{c}{$7.83\times10^{10}$} & \multicolumn{2}{c}{$6.27\times10^{10}$} & \multicolumn{2}{c}{$8.31\times10^{10}$} \\
    & & thick disk & \multicolumn{2}{c}{$2.32\times10^{10}$} & \multicolumn{2}{c}{$1.85\times10^{10}$} & \multicolumn{2}{c}{$2.45\times10^{10}$} \\
    & & total & \multicolumn{2}{c}{$10.15\times10^{10}$} & \multicolumn{2}{c}{$9.38\times10^{10}$} & \multicolumn{2}{c}{$10.76\times10^{10}$} \\
    \cline{2-9}
    & \multirow{3}{*}{\parbox{2cm}{present stars (0.1-1.25\,M$_{\odot}$; $2.0-21.0$\,kpc)}} & thin disk & \multicolumn{2}{c}{$6.80\times10^{10}$} & \multicolumn{2}{c}{$7.22\times10^{10}$} & \multicolumn{2}{c}{$6.67\times10^{10}$} \\
    & & thick disk & \multicolumn{2}{c}{$2.01\times10^{10}$} & \multicolumn{2}{c}{$2.14\times10^{10}$} & \multicolumn{2}{c}{$1.97\times10^{10}$} \\
    & & total & \multicolumn{2}{c}{$8.81\times10^{10}$} & \multicolumn{2}{c}{$9.36\times10^{10}$} & \multicolumn{2}{c}{$8.64\times10^{10}$} \\
    & \multicolumn{2}{l}{fraction within mass range [\%]} & \multicolumn{2}{c}{86.77} & \multicolumn{2}{c}{99.73} & \multicolumn{2}{c}{80.29} \\
    & \multicolumn{2}{l}{fraction below mass range [\%]} & \multicolumn{2}{c}{12.76} & \multicolumn{2}{c}{0} & \multicolumn{2}{c}{19.28} \\
    & \multicolumn{2}{l}{fraction above mass range [\%]} & \multicolumn{2}{c}{0.47} & \multicolumn{2}{c}{0.27} & \multicolumn{2}{c}{0.43} \\
    \hline
    \multirow{2}{*}{\parbox{2cm}{stellar black\\{}holes in disk$^b$}} & \multicolumn{2}{l}{for threshold mass $\geq15$\,M$_{\odot}$} & \multicolumn{2}{c}{$2.93\times10^{8}$} & \multicolumn{2}{c}{$3.19\times10^{8}$} & \multicolumn{2}{c}{$2.87\times10^{8}$} \\
    & \multicolumn{2}{l}{for threshold mass $\geq20$\,M$_{\odot}$} & \multicolumn{2}{c}{$1.93\times10^{8}$} & \multicolumn{2}{c}{$2.09\times10^{8}$} & \multicolumn{2}{c}{$1.89\times10^{8}$} \\
    \hline
\end{tabular}}
\end{center}\footnotesize
\raggedright
$^a$note that these values are equivalent with $N_{\star}$ in our model; $^b$derived in Section~\ref{sec:NStar}
\end{table*}

\begin{table*}\footnotesize
  \begin{center}
    \caption{Input parameters feeding into simulating $\eta_{\star}$ and the resulting numbers of stars \& stellar fractions.  Note that parameters that are the same for several cases are only listed once per row and distributed over the respective cases they refer to.}
    \label{tab:AppEtaStar}
    \resizebox{\textwidth}{!}{%
\begin{tabular}{l|l|l|c|c|c|c|c|c}
  \hline
   & & & \multicolumn{6}{c}{model cases}  \\
    & & & \multicolumn{2}{c}{nominal} & \multicolumn{2}{c}{maximum} & \multicolumn{2}{c}{minimum} \\
    & & & 10\% CO$_2$ & 1\% CO$_2$ & 10\% CO$_2$ & 1\% CO$_2$ & 10\% CO$_2$ & 1\% CO$_2$ \\
   \hline
   \multirow{3}{*}{\parbox{2cm}{SNe (Sect.~\ref{sec:SN})}} & \multicolumn{2}{l}{implemented probability} & \multicolumn{2}{c}{model~4$^a$} & \multicolumn{2}{c}{model~2$^b$} & \multicolumn{2}{c}{model~4$^a$} \\
    & \multicolumn{2}{l}{non-sterilized stars$^c$} & \multicolumn{2}{c}{$2.38\times10^{10}$} & \multicolumn{2}{c}{$2.83\times10^{10}$} & \multicolumn{2}{c}{$2.34\times10^{10}$} \\
    & \multicolumn{2}{l}{fraction [\%]$^d$} & \multicolumn{2}{c}{27.04} & \multicolumn{2}{c}{30.22} & \multicolumn{2}{c}{27.04} \\
    & \multicolumn{2}{l}{cumulative fraction [\%]$^e$} & \multicolumn{2}{c}{23.45} & \multicolumn{2}{c}{30.14} & \multicolumn{2}{c}{21.70} \\
   \hline
   \multirow{6}{*}{\parbox{2cm}{metallicity (Sect.~\ref{sec:metallicity})}} & \multicolumn{2}{l}{implemented present [Fe/H] dist.} & \multicolumn{6}{c}{\citet{Hayden2015}} \\
    & \multicolumn{2}{l}{implemented galactic [Fe/H] evol.$^f$} & \multicolumn{6}{c}{\citet{Snaith2015}; wd$^g$} \\
    \cline{2-9}
    & \multicolumn{2}{l}{metallicity threshold $Z_{\rm min}$$^h$ [$Z_{\odot}$]}  & \multicolumn{2}{c}{0.3} & \multicolumn{2}{c}{0.1} & \multicolumn{2}{c}{0.75} \\
    \cline{2-9}
    & \multicolumn{2}{l}{remaining stars} & \multicolumn{2}{c}{$1.64\times10^{10}$} & \multicolumn{2}{c}{$2.42\times10^{10}$} & \multicolumn{2}{c}{$0.92\times10^{10}$} \\
    & \multicolumn{2}{l}{fraction [\%]} & \multicolumn{2}{c}{69.31\%} & \multicolumn{2}{c}{85.74} & \multicolumn{2}{c}{39.29} \\
    & \multicolumn{2}{l}{cumulative fraction [\%]} & \multicolumn{2}{c}{16.25\%} & \multicolumn{2}{c}{25.84} & \multicolumn{2}{c}{8.53} \\
   \hline
   \multirow{2}{*}{\parbox{2cm}{HZ boundaries (Sect.~\ref{sec:XUVImp})}} & \multicolumn{2}{l}{implemented inner HZ boundary} & \multicolumn{4}{c}{K14/RG$^i$} & \multicolumn{2}{c}{K13/RG$^j$} \\
   & \multicolumn{2}{l}{implemented outer HZ boundary} & R20/HZCL$^k$ & S19/HZCL$^l$ & R20/HZCL$^k$ & S19/HZCL$^l$  & R20/HZCL$^k$ & S19/HZCL$^l$ \\
   \hline
   \multirow{5}{*}{\parbox{2cm}{lower limit (XUV limit; Sect.~\ref{sec:XUVEffect})}} & \multicolumn{2}{l}{implemented lower limit$^m$} & \multicolumn{2}{c}{$F_{\rm XUV}$} & \multicolumn{2}{c}{$F_{\rm XUV}$} & \multicolumn{2}{c}{$F_{\rm X}$} \\
   & \multicolumn{2}{l}{surface flux [$\rm erg\,s^{-1}\,cm^{-2}$]} & 35.0 & 17.5 & 56.0 & 35.0 & 5.0 & 2.5 \\
   \cline{2-9}
   & \multicolumn{2}{l}{remaining stars} & $2.44\times10^9$ & $0.73\times10^9$ & $6.37\times10^9$ & $3.27\times10^9$ & $0.65\times10^9$ & $0.15\times10^9$ \\
   & \multicolumn{2}{l}{fraction [\%]} & 14.80 & 4.42 & 26.26 & 13.48 & 6.93 & 1.61 \\
   & \multicolumn{2}{l}{cumulative fraction [\%]} & 3.01 & 0.90 & 7.84 & 4.03 & 0.59 & 0.14 \\
   \hline
   \multirow{5}{*}{\parbox{2cm}{upper limit ($S_{\rm eff}$ limit; Sect.~\ref{sec:ul})}} & \multicolumn{2}{l}{implemented limit [$S_{\rm eff,\odot}$]$^n$} & \multicolumn{2}{c}{1.107$^o$} & \multicolumn{2}{c}{1.21$^p$} & \multicolumn{2}{c}{1.0512$^q$} \\
   \cline{2-9}
   & \multicolumn{2}{l}{remaining stars$^r$} & $2.10\times10^9$ & $0.45\times10^9$ & $5.80\times10^9$ & $2.57\times10^9$ & $0.50\times10^9$ & $0.08\times10^9$ \\
   & \multicolumn{2}{l}{fraction [\%]} & 85.88 & 61.73 & 91.08& 78.55 & 79.46 & 57.33 \\
   & \multicolumn{2}{l}{cumulative fraction [\%]$^s$} & 2.07 & 0.44 & 6.18 & 2.74 & 0.47 & 0.08 \\
   \hline
\end{tabular}}
\end{center}\footnotesize
\raggedright
$^a$model~4 by \citet{Gowanlock2011}, taken from the upper panel of their Fig.~6; $^b$model~2 by \citet{Gowanlock2011}, again from the upper panel of their Fig.~6; $^c$from here onwards we only consider stars within our chosen mass range, i.e. $0.10-1.25$\,M$_{\odot}$; $^d$from here onwards `fraction' denotes the fraction of stars that remains after implementing the specific requirement onto the \textit{remaining sample} of stars; $^e$from here onwards `cumulative fraction' denotes the fraction of stars that remains from the \textit{initial sample} of disk stars as are found within the entire mass range; $^f$metallicity evolution normalized to fit present-day [Fe/H]-value of \citet{Hayden2015} for each galactic spatial bin in our model; $^g$for the outer disk ($r>7.5$\,kpc), we took the metallicity evolution of \citet{Snaith2015} `with dilution'; $^h$to calculate the percentage of stars per galactic spatial bin with a certain [Fe/H] value to be above $Z_{\rm min}$, we followed \citet{Westby2020} -- see Sect.~\ref{sec:MetImp}; $^i$`runaway greenhouse' limit as calculated through \citet{Kopparapu2014}; $^j$runaway greenhouse limit calculated through \citet{Kopparapu2013}; $^k$outer HZCL boundary for $p$CO$_2 = 0.1$\,bar from \citet{Ramirez2020}, their Fig.~2; $^l$outer HZCL boundary for $p$CO$_2 = 0.01$\,bar calculated through \citet{Schwieterman2019}; 
$^n$in units of effective stellar surface flux at Earth; $^o$runaway greenhouse limit from \citet{Kopparapu2014}; $^p$runaway greenhouse limit from \citet{Leconte2013}; $^q$runaway greenhouse limit from \citet{Kopparapu2013}; $^r$this equals the maximum amount of stars suitable for EHs in our model, i.e., $N_{\star}\eta_{\eta_{\star}}$; $^s$this, finally, equals the maximum value for $\eta_{\star}$ in our model.
\end{table*}

\begin{table*}\footnotesize
  \begin{center}
    \caption{Input parameters feeding into simulating $\beta_{\rm HZCL}$ and the resulting numbers of planets \& planet fractions.  Note that parameters that are the same for several cases are only listed once per row and distributed over the respective cases they refer to.}
    \label{tab:AppBetaHZCL}
    \resizebox{\textwidth}{!}{%
\begin{tabular}{l|l|l|c|c|c|c|c|c}
  \hline
   & & & \multicolumn{6}{c}{model cases}  \\
    & & & \multicolumn{2}{c}{nominal} & \multicolumn{2}{c}{maximum} & \multicolumn{2}{c}{minimum} \\
    & & & 10\% CO$_2$ & 1\% CO$_2$ & 10\% CO$_2$ & 1\% CO$_2$ & 10\% CO$_2$ & 1\% CO$_2$ \\
   \hline
   \multirow{14}{*}{\parbox{2cm}{planets\\in HZCL\\(Sect.~\ref{sec:bHZCLMain})}} & \multirow{2}{*}{\parbox{2cm}{initial $\eta_{\oplus}$\\{}(in HZ)$^a$}} & FGK stars$^b$ & \multicolumn{2}{c}{0.106$^c$} & \multicolumn{2}{c}{0.18$^d$} & \multicolumn{2}{c}{0.054$^e$} \\
    & & M stars$^f$ & \multicolumn{2}{c}{0.16$^g$ (0.26\,M$_{\odot}$)} & \multicolumn{2}{c}{0.312$^h$ (0.27\,M$_{\odot}$)} & \multicolumn{2}{c}{0.09$^i$ (0.26\,M$_{\odot}$)}\\
    \cline{2-9}
   & \multirow{6}{*}{\parbox{2cm}{$\beta_{\rm HZCL}$$^{j,k}$}} & power law$^l$ &  \multicolumn{6}{c}{1D broken power law from model\#4 by \citet{Pascucci2019}$^m$} \\
   & & $R_{\rm pl,min}$$^n$ [R$_{\oplus}$] & 0.79$^o$ & 0.78$^o$ & \multicolumn{2}{c}{0.72$^p$} & 0.77$^o$ & 0.74$^o$ \\
   & & $R_{\rm pl,max}$$^q$ [R$_{\oplus}$] & \multicolumn{2}{c}{1.23$^r$} & \multicolumn{2}{c}{1.4$^s$} & \multicolumn{2}{c}{1.23$^r$} \\
   \cline{3-9}
   & & FGK stars$^b$ & 0.022 & 0.013 & 0.082 & 0.049 & 0.009 & 0.006 \\
   & & M stars$^t$ & 0.039 & 0.018 & 0.108 & 0.050 & 0.014 & 0.009 \\
   & & lin. extrap.$^u$ & \multicolumn{6}{c}{stellar effective surface temperature ($T_{\rm eff}$)} \\
   \cline{2-9}
   & \multirow{3}{*}{\parbox{2cm}{$\beta_{\mathrm{HZCL},Z_{\rm min}}$}} & $\mathcal{P}(Z_{\rm min,SN})$$^v$ & \multicolumn{2}{c}{1.008} & \multicolumn{2}{c}{1.0} & \multicolumn{2}{c}{1.603}  \\
   & & FGK stars$^o$ & 0.022 & 0.013 & 0.082 & 0.049 & 0.014 & 0.010 \\
   & & M stars$^t$ & 0.040 & 0.018 & 0.108 & 0.050 & 0.023 & 0.014 \\
   \cline{2-9}
   & \multicolumn{2}{l}{planets in HZCL} & $7.25\times10^7$ & $0.66\times10^7$ & $60.32\times10^7$ & $12.15\times10^7$ & $0.92\times10^7$ & $0.09\times10^7$ \\
   & \multicolumn{2}{l}{fraction in HZCL [\%]$^w$} & 3.46 & 1.15 & 10.40& 4.73 & 1.81 & 1.07 \\
   & \multicolumn{2}{l}{cumulative fraction (all) [\%]} & $7.15\times10^{-2}$ & $0.65\times10^{-2}$ & $64.30\times10^{-2}$ & $12.95\times10^{-4}$ & $0.85\times10^{-2}$ & $0.08\times10^{-2}$ \\
   \hline
\end{tabular}}
\end{center}\footnotesize
\raggedright
    $^a$the initial values for the used radius and orbital period ranges can be found in the cited studies; $^b$occurrence rate always taken for a stellar mass of 1.0\,M$_{\oplus}$; $^c$from \citet{Pascucci2019}, model\#4; $^d$from \citet{Bryson2021}, lower bound; $^e$from \citet{Pascucci2019}, model\#6; $^f$occurrence rate always taken for the mean stellar mass of the M dwarf spectral class in the respective case (values are different for the different cases and the corresponding masses are written in brackets); $^g$from \citet{Dressing2013}; $^h$from \citet{Quanz2022}; $^i$from \citet{Dressing2013}, lower value of the uncertainty range; $^j$the respective orbital period ranges are taken from the HZCL boundaries as defined in Table~\ref{tab:AppEtaStar}, whereas the radius range are displayed in this table; $^k$the derived $\beta_{\rm HZCL}$ occurrence rates for FGK and M stars are then calculated with Eq.~\ref{eq:planetDist} (and subsequent equations) as described in Sect.~\ref{sec:betaHZCL}; $^l$the power law used for scaling $\eta_{\rm EH}$ to the respective HZCL boundaries and radius range; $^m$the power law reduces to $f(P_{\rm pl})= P^{0.14}_{\rm pl}$ and $f(R_{\rm pl})= R_{\rm pl}/3.2$, see Sect.~\ref{sec:betaHZCL} and Eq.~\ref{eq:splAll}; $^n$planetary minimum radius used for scaling $\eta_{\oplus}$ with power law; $^o$calculated with the relation used for the exoplanet yield estimates of space missions HabEx, Luvoir and LIFE, i.e., $R_{\rm pl,min}=0.8\times a^{-0.5}$ where $a = d_{\rm \langle HZCL\rangle}$, see Eq.~\ref{eq:rMin} and Sect.~\ref{sec:betaHZCL}; $^p$this radius corresponds to a minimum planetary mass of 0.3\,M$_{\oplus}$, see again Sect.~\ref{sec:betaHZCL} and \citet{Zink2019}; $^q$planetary maximum radius for scaling $\eta_{\oplus}$; $^r$the derived transition radius between rocky and Neptunian planets according to \citet{Chen2017}, see also Sect.~\ref{sec:betaHZCL}; $^s$based on a `conservative interpretation' \citep{Gaudi2020} of the derived transition between rocky and Neptunian planets from \citet{Rogers2015}; $^t$for the same average stellar masses as for the M star $\eta_{\oplus}$-values; $^u$stellar variable through which the displayed occurrence rates for FGK (i.e., for 1\,M$_{\oplus}$) and M mean stellar masses are linearly scaled to all stellar masses (see Eq.~\ref{eq:corrH2O} and discussion in Sect.~\ref{sec:h2oImp}); $^v$factor through which $\beta_{\rm HZCL}$ is weighted to account for the galactic metallicity distribution, which gives the weighted occurrence rate $\beta_{\mathrm{HZCL},Z_{\rm min}}$, see Eq.~\ref{eq:mProbE} and discussion in Sect.~\ref{sec:betaHZCL}; $^w$equivalent with the mean occurrence rate, $\langle\beta_{\rm HZCL}\rangle$ of rocky exoplanets in the HZCL over the entire stellar mass range of $M_{\rm star}=0.1-1.25$\,M$_{\odot}$);
\end{table*}


\begin{table*}\footnotesize
  \begin{center}
    \caption{Input parameters feeding into simulating $\eta_{\rm EH}$ (H$_2$O and large moon requirements) and the resulting numbers of EHs \& planet fractions.  Note that parameters that are the same for several cases are only listed once per row and distributed over the respective cases they refer to.}
    \label{tab:AppEtaEH}
    \resizebox{\textwidth}{!}{%
\begin{tabular}{l|l|l|c|c|c|c|c|c}
  \hline
   & & & \multicolumn{6}{c}{model cases}  \\
    & & & \multicolumn{2}{c}{nominal} & \multicolumn{2}{c}{maximum} & \multicolumn{2}{c}{minimum} \\
    & & & 10\% CO$_2$ & 1\% CO$_2$ & 10\% CO$_2$ & 1\% CO$_2$ & 10\% CO$_2$ & 1\% CO$_2$ \\
   \hline
   \multirow{9}{*}{\parbox{2cm}{planets with\\right amount\\of H$_2$O$^a$\\(Sect.~\ref{sec:h2oEffect})}} & \multirow{5}{*}{\parbox{2cm}{occurrence rates}} & G star (1.0\,M$_{\odot}$) & \multicolumn{2}{c}{0.182$^b$} & \multicolumn{2}{c}{0.256$^c$} & \multicolumn{2}{c}{0.080$^d$} \\
   & & K star (0.5\,M$_{\odot}$) & \multicolumn{2}{c}{0.021$^e$} & \multicolumn{2}{c}{0.043$^f$} & \multicolumn{2}{c}{0.017$^b$} \\
   & & M star (0.3\,M$_{\odot}$) & \multicolumn{2}{c}{0.013$^f$} & \multicolumn{2}{c}{0.018$^c$} & \multicolumn{2}{c}{0.001$^e$} \\
   & & M star (0.1\,M$_{\odot}$) & \multicolumn{2}{c}{0.001$^f$} & \multicolumn{2}{c}{0.003$^e$} & \multicolumn{2}{c}{0.001$^f$} \\
   & & lin. extrap.$^g$ & \multicolumn{4}{c}{$F_{\rm XUV}T_{\rm eff}^{-1}$} & \multicolumn{2}{c}{$F_{\rm X}T_{\rm eff}^{-1}$}\\
   \cline{2-9}
   & \multicolumn{2}{l}{remaining planets} & $5.44\times10^6$ & $0.94\times10^6$ & $60.00\times10^6$ & $17.90\times10^6$ & $0.43\times10^6$ & $0.07\times10^6$ \\
   & \multicolumn{2}{l}{fraction [\%]} & 7.51 & 14.16 & 9.28 & 14.73 & 4.74 & 7.20 \\
   & \multicolumn{2}{l}{cumulative fraction (planets) [\%]} & $2.60\times10^{-1}$ & $2.09\times10^{-1}$ & $9.66\times10^{-1}$ & $0.86\times10^{-1}$ & $0.77\times10^{-1}$ & $0.08\times10^{-1}$ \\
      & \multicolumn{2}{l}{cumulative fraction (all) [\%]} & $5.36\times10^{-3}$ & $0.93\times10^{-3}$ & $59.69\times10^{-3}$ & $12.07\times10^{-3}$ & $0.40\times10^{-3}$ & $0.06\times10^{-3}$ \\
   \hline
   \multirow{9}{*}{\parbox{2cm}{planets with large moon\\(Sect.~\ref{sec:moonEff})}} & \multirow{3}{*}{\parbox{2cm}{occurrence rates}} & G star (1.0\,M$_{\odot}$) & \multicolumn{2}{c}{0.083$^h$} & \multicolumn{2}{c}{0.250$^i$} & \multicolumn{2}{c}{0.020$^{j}$} \\
   & & M star$^{k}$ & \multicolumn{2}{c}{0.010$^{l}$} & \multicolumn{2}{c}{0.030$^{l}$} & \multicolumn{2}{c}{0.002$^{l}$} \\
   & & lin. extrap.$^{m}$ & \multicolumn{6}{c}{with planetary Hill radius, i.e. $r_{\rm H} = \sqrt{m_{\rm pl}/(3M_{\star})}$}\\
   \cline{2-9}
   & \multicolumn{2}{l}{remaining planets$^n$} & $2.54\times10^5$ & $0.58\times10^5$ & $74.18\times10^5$ & $27.64\times10^5$ & $0.06\times10^5$ & $0.01\times10^5$ \\
   & \multicolumn{2}{l}{fraction [\%]} & 4.66 & 6.15 & 13.25 & 15.44 & 1.28 & 1.71 \\
   & \multicolumn{2}{l}{cumulative fraction (planets) [\%]$^{o}$} & $1.21\times10^{-2}$ & $1.28\times10^{-2}$ & $12.79\times10^{-2}$ & $10.77\times10^{-2}$ & $0.11\times10^{-2}$ & $0.13\times10^{-2}$ \\
      & \multicolumn{2}{l}{cumulative fraction (all) [\%]$^{p}$} & $2.50\times10^{-4}$ & $0.57\times10^{-4}$ & $79.06\times10^{-4}$ & $29.46\times10^{-4}$ & $0.05\times10^{-4}$ & $0.01\times10^{-4}$ \\
   \hline
\end{tabular}}
\end{center}\footnotesize
\raggedright
    $^a$the appropriate planetary water mass fraction for simultaneously hosting subaerial land and ocean was always assumed to be $w_{\rm H_2O}$=0.003-0.2\% of the planetary mass; $^b$from \citet{Tian2015}, planets in Fig.~2 between water mass fractions of $w_{\rm H_2O}=10^{-5}-10^{-3}$; $^c$from \citet{Tian2015}, planets in Fig.~2 between water mass fractions of $w_{\rm H_2O}10^{-5}-10^{-2}$; $^d$from \citet{Simpson2017}, fraction not entirely covered by oceans from the right panel of their Fig.~5 (solid line); $^e$from \citet{Kimura2022} for $X_{\rm H_2O}$ = 0.8 from their Fig.~3; $^f$from \citet{Kimura2022} for $X_{\rm H_2O}$ = 0 from their Fig.~3; $^g$stellar variable through which the displayed occurrence rates for the specific predefined G, K, and M stellar masses are linearly scaled to all stellar masses; $^h$from \citet{Elser2011}, mean value; $^i$from \citet{Elser2011}, max value; $^{j}$from \citet{Elser2011}, min value; $^{k}$same mean M star masses as for the planet occurrence rate, i.e., 0.26, 0.27, and 0.26\,M$_{\odot}$ for nominal, maximum and minimum cases, respectively; $^{l}$fraction of stable moons around M stars (i.e., 4/33 moons) from \citet{MartinezRodriguez2019} multiplied with min, max and mean values from \citet{Elser2011} for G stars (i.e., our corresponding occurrence rates for G stars); $^{m}$variable through which the displayed occurrence rates for G and M mean stellar masses are linearly scaled to all stellar masses - in this case, the planetary Hill radius (see Eq.~\ref{eq:rH} \& \ref{eq:betaMoon} and discussion in Sect.~\ref{sec:moonImp}); $^n$this equals the final maximum number of EHs found in our model, i.e., $N_{\rm EH}\leq\eta_{\star}\eta_{\rm EH}$; $^{o}$this equals the final maximum occurrence rate, $\eta_{\rm EH}$, in our model; $^{p}$this finally equals the maximum occurrence rate $(\eta_{\star}\eta_{\rm EH})$.
\end{table*}

\begin{table*}\footnotesize
  \begin{center}
    \caption{The distribution of potentially habitable planets/stars that remain after implementing the respective requirement as a function of different stellar spectral classes (for a maximum of 10\% CO$_2$)$^a$.}
    \label{tab:AppSpec0p1}
    \resizebox{0.8\textwidth}{!}{%
\begin{tabular}{l|c|c|c|c|c}
  \hline
   requirement & total [number] & M [\%] & K [\%] & G [\%] & F [\%] \\
   \hline
   initial $N_{\star}$ (all masses)  & $10.15^{+0.61}_{-0.77}\times10^{10}$ & $79.80^{+1.51}_{-10.46}$ & $14.50^{+1.17}_{-0.13}$ & $3.38^{+1.21}_{-0.25}$ & $0.84^{+2.90}_{-0.06}$ \\
   initial $N_{\star}$ (M$_\star =0.1-1.25$\,$M_{\odot}$)  & $8.81^{+0.55}_{-0.17}\times10^{10}$ & $77.25^{+0.78}_{-0.003}$ & $18.06^{+3e-3}_{-1.59}$ & $3.90^{+0.62}_{-1e-4}$ & $0.75^{+0.23}_{-3e-5}$ \\
   stars in GHZ & $2.38^{+0.45}_{-0.05}\times10^{10}$ & $77.27^{+0.77}_{-0}$ & $18.06^{+0}_{-1.60}$ & $3.87^{+0.64}_{-1e-5}$ & $0.80^{+0.19}_{-1e-5}$ \\
   stars above met. threshold & $1.65^{+0.77}_{-0.73}\times10^{10}$ & $76.70^{+1.00}_{-0.44}$ & $17.93^{+0.10}_{-1.43}$ & $4.43^{+0.33}_{-0.16}$ & $1.15^{+0.37}_{-0.03}$ \\
   \hline
   lower ($F_{\rm XUV/X}$) limit & $2.44^{+3.93}_{-1.81}\times10^{9}$ & $0.90^{+21.46}_{-0.90}$ & $55.20^{+13.47}_{-1.82}$ & $24.39^{+11.15}_{-6.31}$ & $6.04^{+5.04}_{-1.68}$ \\
   upper (L$_{\rm bol}$) limit (i.e., $N_{\star}\eta_{\star}$) & $2.09^{+3.70}_{-1.59}\times10^{9}$ & $1.05^{+23.50}_{-1.05}$ & $67.17^{+12.78}_{-6.57}$ & $15.67^{+9.60}_{-2.58}$ & $3.33^{+4.23}_{-1.56}$  \\
   \hline
   planets in HZCL & $7.25^{+53.07}_{-6.34}\times10^{7}$ & $1.20^{+24.35}_{-1.20}$ & $72.75^{+12.19}_{-10.90}$ & $12.06^{+10.29}_{-0.79}$ & $1.80^{+3.10}_{-0.47}$  \\
   planets with right amount of H$_2$O & $5.44^{+50.56}_{-5.01}\times10^{6}$ & $0.30^{+7.02}_{-0.30}$ & $61.32^{+8.13}_{-3.54}$ & $27.41^{+5.83}_{-1.90}$ & $4.74^{+4.24}_{-0.79}$  \\
   planets with large moon (i.e.,$N_{\rm EH}$) & $2.54^{+71.64}_{-2.48}\times10^{5}$ & $0.11^{+2.23}_{-0.11}$ & $48.45^{+5.63}_{-8.11}$ & $39.93^{+0.90}_{-4.84}$ & $10.72^{+8.10}_{-1.45}$ \\
   \hline
   younger than 6\,Gyr & $1.00^{+34.39}_{-0.98}\times10^{5}$ & $0-0.12$ & $25.57^{+6.10}_{-23.94}$ & $51.83^{+5.42}_{-2.95}$ & $27.09^{+19.46}_{-7.65}$ \\
    \hline
   younger than 6, older than 4\,Gyr & $4.98^{+160.59}_{-4.85}\times10^{4}$ & $0-0.001$ & $35.10^{+2.22}_{-34.84}$ & $55.90^{+7.49}_{-3.10}$ & $9.00^{+0.89}_{-27.35}$ \\
    \hline
\end{tabular}}
\end{center}\footnotesize
\raggedright
$^a$The max. 10\% CO$_2$ scenario is identical with max. 1\% CO$_2$ for all requirements prior to the upper limit.
\end{table*}

\begin{table*}\footnotesize
  \begin{center}
    \caption{The distribution of potentially habitable planets/stars that remain after implementing the respective requirement as a function of different stellar spectral classes (for a maximum of 1\% CO$_2$)$^a$.}
    \label{tab:AppSpec0p01}
    \resizebox{0.8\textwidth}{!}{%
\begin{tabular}{l|c|c|c|c|c}
  \hline
   requirement & total [number] & M [\%] & K [\%] & G [\%] & F [\%] \\
   \hline
   initial $N_{\star}$ (all masses)  & $10.15^{+0.61}_{-0.77}\times10^{10}$ & $79.80^{+1.51}_{-10.46}$ & $14.50^{+1.17}_{-0.13}$ & $3.38^{+1.21}_{-0.25}$ & $0.84^{+2.90}_{-0.06}$ \\
   initial $N_{\star}$ (M$_\star =0.1-1.25$\,$M_{\odot}$)  & $8.81^{+0.55}_{-0.17}\times10^{10}$ & $77.25^{+0.78}_{-0.003}$ & $18.06^{+3e-3}_{-1.59}$ & $3.90^{+0.62}_{-1e-4}$ & $0.75^{+0.23}_{-3e-5}$ \\
   stars in GHZ & $2.38^{+0.45}_{-0.05}\times10^{10}$ & $77.27^{+0.77}_{-0}$ & $18.06^{+0}_{-1.60}$ & $3.87^{+0.64}_{-1e-5}$ & $0.80^{+0.19}_{-1e-5}$ \\
   stars above met. threshold & $1.65^{+0.77}_{-0.73}\times10^{10}$ & $76.70^{+1.00}_{-0.44}$ & $17.93^{+0.10}_{-1.43}$ & $4.43^{+0.33}_{-0.16}$ & $1.15^{+0.37}_{-0.03}$ \\
   \hline
   lower ($F_{\rm XUV/X}$) limit & $7.28^{+25.40}_{-5.80}\times10^{8}$ & 0 & $29.81^{+28.51}_{-26.91}$ & $57.41^{+21.77}_{-23.92}$ & $12.78^{+5.15}_{-4.59}$ \\
   upper (L$_{\rm bol}$) limit (i.e., $N_{\star}\eta_{\star}$) & $0.45^{+2.12}_{-0.36}\times10^{9}$ & 0 & $48.30^{+25.76}_{-43.23}$ & $41.94^{+38.30}_{-19.32}$ & $9.77^{+4.94}_{-6.44}$ \\
   \hline
   planets in HZCL & $6.63^{+114.82}_{-5.72}\times10^{6}$ & 0 & $51.04^{+22.71}_{-45.44}$ & $41.23^{+41.96}_{-18.45}$ & $7.73^{+3.49}_{-4.26}$  \\

   planets with right amount of H$_2$O & $9.93^{+169.58}_{-8.73}\times10^{5}$ & 0 & $43.09^{+15.33}_{-38.39}$ & $46.18^{+35.72}_{-11.07}$ & $10.73^{+2.68}_{-4.26}$  \\
   planets with large moon (i.e., $N_{\rm EH}$) & $0.58^{+27.06}_{-0.57}\times10^{5}$ & 0 & $32.33^{+10.43}_{-29.03}$ & $48.22^{+27.37}_{-4.23}$ & $21.11^{+1.67}_{-6.20}$ \\
   \hline
   younger than 6\,Gyr & $0.17^{+13.51}_{-0.17}\times10^{5}$ & 0 & $0-15.48$ & $47.27^{+10.51}_{-28.70}$ & $64.96^{+16.47}_{-38.21}$ \\
    \hline
   younger than 6, older than 4\,Gyr & $1.18^{+66.01}_{-1.15}\times10^{4}$ & $0$ & $0-22.47$ & $50.84^{+15.96}_{-43.57}$ & $49.16^{+43.47}_{-38.43}$ \\
    \hline
\end{tabular}}
\end{center}\footnotesize
\raggedright
$^a$The max. 1\% CO$_2$ scenario is identical with max. 10\% CO$_2$ for all requirements prior to the upper limit.
\end{table*} 

\section{Conclusions}\label{sec:conclusion}

In this study, we applied our formulation to calculate the maximum number of EHs in the galactic disk, as introduced in Paper~I. Our methodological and scientifically quantifiable approach shows that, {in agreement with the Rare Earth Hypotheses \citep{Ward2000},} EHs in the Galaxy should indeed be rare by deriving maximum numbers of  $N_{\rm EH,10\%}\leq2.54^{+71.64}_{-2.48}\times10^{5}$ and $N_{\rm EH,1\%}\leq0.58^{+27.06}_{-0.57}\times10^{5}$ EHs with N$_2$-O$_2$-dominated atmospheres containing a maximum of 10\% and 1\% CO$_2$, respectively. In subsequent discussions, we further illustrated that these numbers are indeed plausible maximum numbers since a wide variety of necessary requirements are not included in our model as these cannot be scientifically quantified by the present state of research. One can, however, expect that the actual number of EHs in the Milky Way will be significantly below the values derived by our model, potentially by orders of magnitude. Future ground-based and space-based observatories will be able to quantify some of the other requirements {(e.g., the accretion rate of rocky exoplanets inside the protoplanetary gas disk)}, and will give further insights into the distribution of EHs, and, consequently, complex life, in the Galaxy. These missions will also be able to confirm or contradict our results by statistically assessing the atmospheres of rocky exoplanets in the Solar Neighborhood. Upcoming missions such as LIFE are therefore crucially important for scientifically advancing the fields of astrobiology and planetary habitability.

Our results have profound additional implications, as the likely rareness of EHs further implies complex aerobic, and intelligent life to be rarer still. Since a breathable atmosphere will present a very valuable resource for any aerobic ETIs, they might show a certain interest in our planet. Intentionally sending messages into space should therefore be performed with high caution. We also point out that our study is agnostic about life originating on hypothetical habitats other than EHs. Any more exotic habitats (e.g., subsurface ocean worlds) could significantly outnumber planets with Earth-like atmospheres, at least in principle. Finally, we argue that the Copernican Principle of Mediocrity cannot be valid in the sense of the Earth and consequently complex life being common in the Galaxy. Certain requirements must be met to allow for the existence of EHs and only a small fraction of planets indeed meet such criteria. It is therefore unscientific to deduce {complex aerobic} life to be common in the Universe, at least based on the Copernican Principle. Instead, we argue, at maximum, for a combined Anthropic-Copernican Principle stating that life as we know it may be common, as long as certain criteria are met to allow for its existence. {Extremophiles, anaerobic and simple aerobic lifeforms, however, could be more common.}

In a future study, we plan to assess the maximum number of EHs on a statistical basis, and to implement further criteria into our model. We therefore plan to include volatile budgets, volcanic degassing, and possible geological activity timescales, as these can refine our results. Studying the evolution of habitability throughout galactic history is another research topic that we like to address in the future.

\begin{appendix}
    \section{Stellar evolution and its importance for the stability of N$_2$-dominated atmospheres}\label{sec:appStellar}

\subsection{Stellar evolution}\label{sec:StellarEvo}

So-called slow, moderate, and fast rotators correspond to the 5$^{\rm th}$, 50$^{\rm th}$, and 95$^{\rm th}$ percentile of the rotational distribution of main sequence stars \citep[e.g.,][]{Tu2015,Johnstone2021Stars}. {Most of the young stars with an age of $t_{\star}\lesssim150\,$Myr have rotation rates between 1 and 10 days with the peaks for K and G stars slightly shifted toward 10 days and for M stars toward 1 day; see, e.g., Fig.~11 in \citet{Irwin2008}, Fig.~3 in \citet{Johnstone2015WindII}, and Figs.~2 and 3 in \citet{Johnstone2021Stars}.} The notations `slow', `moderate', and `fast' are analogies to stars that are weakly, moderately, and strongly active since the rotational evolution of a star on the main sequence is highly correlated with its stellar mass loss, extreme ultraviolet and X-ray flux evolution \citep[e.g.,][]{Ribas2005,Johnstone2015WindII,Tu2015,Guedel2020,Johnstone2021Stars}.

Based on observations of stellar clusters of different ages, \citet{Tu2015} reconstructed the evolutionary tracks of solar-mass stars through their pre-main sequence and main sequence lifetimes by combining rotational evolution with the stellar X-ray, $L_{\rm X}$, and EUV flux, $L_{\rm EUV}$, evolution (together $L_{\rm XUV}$). Although all solar-like stars start their evolution in a saturation phase with all of {them} having comparable values for $L_{\rm XUV}$, slow rotators desaturate much earlier than moderate and fast rotators. After about 1\,Gyr, however, all these tracks converge towards one, which makes it impossible to simply reconstruct the solar evolution based on observations of its present-day behavior \citep{Tu2015}. Isotope studies, however, indicate that the Sun has likely been a slow rotator \citep[e.g.,][]{Saxena2019,Lammer2020a}, which agrees well with recent simulations indicating that the Earth's N$_2$-dominated atmosphere would not have survived the Archean eon \citep{Johnstone2021}, if the Sun had not been a slow rotator. It is worth noting that isotopic data supports a relatively high partial surface pressure of N$_2$ around 3.0-3.5\,Ga comparable to the present day or lower \citep[e.g.,][]{Marty2013,Avice2018}, whereas raindrop imprints at indicate a total surface pressure of only $0.23\pm0.23$\,bar at 2.7\,Ga \citep{Som2016}. Further geological data support the assumption that a substantial atmosphere was indeed present throughout the Archean eon \citep[see, e.g.,][]{Catling2020}. See the next subsection for further discussions on atmospheric stability.

\citet{Johnstone2021Stars} further reconstructed the XUV flux evolution of stars between stellar masses of $M_{\star}=0.1-1.25$\,M$_{\odot}$ (see Figure~\ref{fig:Lx}) and found a behaviour similar to that of solar-like stars, but with a few noteworthy exceptions relevant to our study. Although, slow, moderate, and fast rotators -- as any percentile of the rotational distribution in general -- diverge early on and converge at a later stage in stellar evolution for all masses, the differences are more modest for smaller mass stars and increase towards higher masses. Their convergence, on the other hand, takes place later on for later-type stars than for earlier-type ones. The most important difference, however, stems from the speed of decline in intensity. While more massive stars, such as the G and F types, decline by an order of magnitude within a few million years, K and particularly M stars, stay active for far longer \citep{Johnstone2021Stars}.

The second important difference is the XUV luminosity, $L_{\rm bol}$, of stars with different masses. Although low-mass stars are much less luminous than higher-mass stars, the ratio of $L_{\rm XUV}$/$L_{\rm bol}$ is significantly elevated for the lower-mass ones. Planets with the same equilibrium temperature in the HZs of later-type stars obtain not only a far higher X-ray and EUV surface flux than planets around earlier-types, but they also obtain it for much longer, as was similarly found by several other studies \citep[e.g.,][]{McDonald2019,Pineda2021,Engle2024XUV}. As \citet{Johnstone2021Stars} point out, stars with a mass below $M_{\star}\sim$0.4\,M$_{\odot}$ may never reach a threshold value for the X-ray surface flux in the HZ of $F_{\rm X} = 10\rm \,erg s^{-1} cm^{-2}$ over their entire lifetime, which is about 15 times the present-day average value for the Earth's orbit of $F_{\rm X,\oplus} = 0.65\rm \,erg s^{-1} cm^{-2}$. This can be seen in the upper panel of Figure~\ref{fig:Lx}, where the X-ray surface flux in the middle of the HZ is shown for slow (red lines), moderate (green lines) and fast rotators (blue lines) for stellar masses of $M_{\star} = \{0.25,0.50,0.75,1.00\}\,$M$_\odot$ as found by \citet{Johnstone2021Stars}. The grey dash-dotted line illustrates the threshold of $F_{\rm X} = 10\rm\, erg s^{-1} cm^{-2}$ as set by these authors. As a comparison, the grey lines in the background illustrate the derived evolution of $F_{\rm X}$ for solar-like stars based on the study by \citet{Tu2015}. Here, it has to be noted that these authors required their power laws for the different evolutionary tracks to fit the Sun at an age of $t_{\odot} = 4570$\,Myr. The evolution derived by \citet{Johnstone2021Stars}, on the other hand, does not include such a requirement.

That small-mass stars potentially never reach such a threshold value for the X-ray and/or EUV flux is also indicated by research that investigated specific M-type stars. \citet{Fleming2020} find that the XUV surface flux (i.e., X-ray and EUV) in the HZ of Trappist-1, a late-type M dwarf with $M_{\star}=0.08$\,M$_{\odot}$ and $t_{\star}\sim8$\,Gyr, is still about $F_{\rm X}\sim100\,F_{\rm XUV, \oplus}$ and that it could have even experienced $F_{\rm X}10^3-10^4\,F_{\rm XUV, \oplus}$ over its first Gyr of existence. That \citet{Birky2021} downsized the total XUV flux over the whole lifetime of the system by $\sim 15$\% does not affect the conclusion that these values are orders of magnitude above the aforementioned threshold value. Similarly, \citet{Duvvuri2021} investigated the EUV flux (for wavelengths between 10-91.2\,nm) of several different M dwarfs, among them Trappist-1, for which these authors found an EUV surface flux at 1\,AU of $F_{\rm EUV} = 0.762^{+1.30}_{-0.744}\,\rm erg\,s^{-1} cm^{-2}$, which scales to $F_{\rm EUV} = 888^{+8720}_{-867}\,\rm erg\,s^{-1} cm^{-2}$ for the orbit of Trappist-1e in the middle of this star's HZ. This is $F_{\rm EUV}=444^{+4360}_{-436}\,F_{\rm EUV,\oplus}$, if one assumes an integrated flux of $F_{\rm EUV, \oplus} = 1.99\,\rm erg\,s^{-1} cm^{-2}$ for the Earth's orbit \citep{Woods2009} for the same wavelength range. Barnard's star with $t_{\star}\sim$10\,Gyr and a mass of $M_{\star}\sim$0.14\,M$_{\odot}$, on the other hand, seems to be relatively weak with a flux of only $F_{\rm EUV} \sim 1-15\,\rm erg\,s^{-1} cm^{-2}$ under quiescent conditions \citep{Duvvuri2021}.

Another study by \citet{Peacock2020} generally finds that even early-type M stars with $M_{\star}\sim 0.4$\,M$_{\odot}$ expose planets at the inner edge of their HZ over the first hundreds of million years to XUV radiation that is up to $\sim$100 times stronger than the Earth experiences over its entire lifetime, with their XUV flux reaching up to $F_{\rm X}\sim10^4\,F_{\rm X, \oplus}$ \citep{Peacock2020}. \citet{Johnstone2021Stars} also find such high initial values, but for stars with masses below $M_{\star}\sim0.25$\,M$_{\odot}$. \citet{Duvvuri2021} further provide the EUV surface flux of the 9.22\,Gyr old Gliese 832 with a mass of $M_{\star}\sim0.45\,$M$_{\odot}$. Its HZ planet, Gliese~832b, still receives a flux of $F_{\rm EUV} \sim 22-110\,\rm erg\,s^{-1} cm^{-2}$, according to their study, compared to $F_{\rm EUV} \sim 18-20\,\rm erg\,s^{-1} cm^{-2}$ for the same stellar mass and age in the model of \citet{Johnstone2021Stars}. For AU\,Mic with $t_{\star}\sim$22\,Myr and $M_{\star}\sim0.5\,$M$_{\odot}$, the values of \citet{Duvvuri2021} result in $F_{\rm EUV} \sim 70-240\,\rm erg\,s^{-1} cm^{-2}$ in the middle of its HZ, which is conversely lower than calculated with the model of \citet{Johnstone2021Stars}, i.e., $F_{\rm EUV} \sim 545-820\,\rm erg\,s^{-1} cm^{-2}$.

\begin{figure*}
\centering
\includegraphics[width = 0.6\linewidth, page=1]{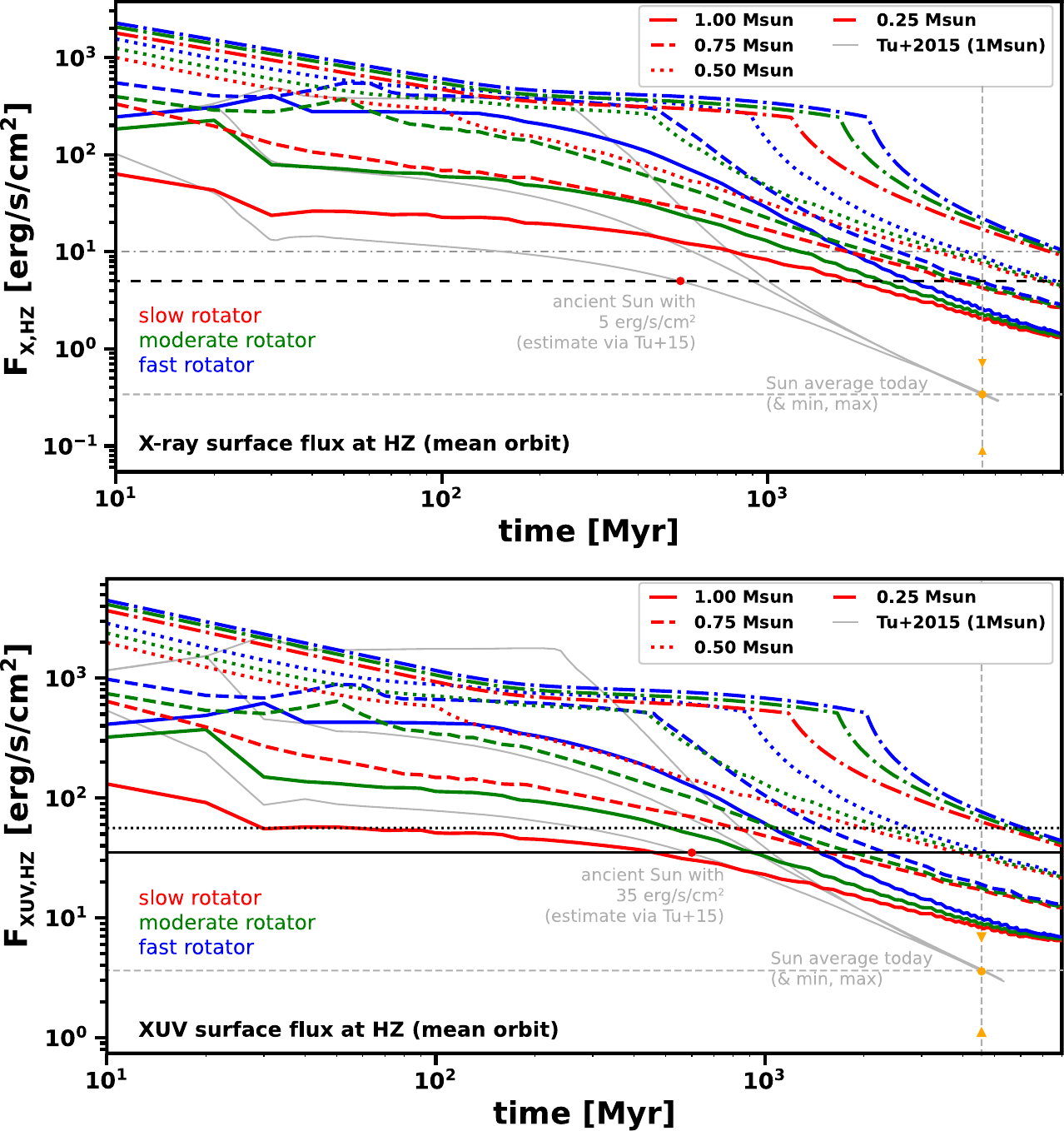}
\caption{Upper panel: The evolution of the stellar X-ray surface flux, $F_{\rm X}$, in the middle of the HZ for slow (red), moderate (green) and fast rotators (blue) for stars with 1.0\,M$_{\odot}$ (solid), 0.75\,M$_{\odot}$ (dashed),0.5\,M$_{\odot}$ (dotted), and 0.25\,M$_{\odot}$ (dashed-dotted lines) according to \citet{Johnstone2021Stars}. Importantly, stars with lower stellar masses (i) clearly shift towards higher $F_{\rm X}$ values, and (ii) remain at high values significantly longer. The range of the present-day Sun is indicated with orange symbols, the grey lines in the back indicate the evolution of $F_{\rm X}$ for solar-like stars according to \citet{Tu2015}. The horizontal solid line indicates the value for our nominal and minimum cases, while the dotted one indicates the value for our maximum case (see text). Lower panel: The same but for the evolution of the XUV surface flux, $F_{\rm XUV}$.}
\label{fig:Lx}
\end{figure*}\vspace{1mm}


Stars also show a wide range of flaring events, with those above an energy of $\sim10^{32}$\,erg typically contributing $\sim$10\% of the entire energy emitted within the X-ray wavelengths \citep{Audard2000}. At young G-type stars, individual superflares were observed that emit a total energy of up to $\sim10^{36}$\,erg \citep[e.g.,][]{Notsu2019,Doyle2020,Guenther2020} while values up to $\sim10^{35}$\,erg were reported for M and K dwarfs \citep[e.g.,][]{Paudel2019,RodriguezMartinez2020,Schmidt2019}. Even though their total emitted energy and frequency might be lower than for higher-mass stars, superflares at low-mass dwarfs can nevertheless have a stronger effect on their HZ planets. These planets do not only receive higher associated surface fluxes but also receive them more frequently. The reason for the latter stems from the fact that a given threshold value for the surface flux will be reached more frequently than for higher-mass stars, since the orbits of HZ planets around low-mass dwarfs are closer to the star \citep{Johnstone2021Stars}. For example, Barnard's star provides an EUV surface flux in the middle of the HZ during flaring conditions of $F_{\rm EUV} \sim 6-30\,\rm erg\,s^{-1} cm^{-2}$ \citep{Duvvuri2021}, which is significantly higher than for quiet conditions. HZ planets around the M dwarf GJ~876, as another example, {were suggested \citep{Youngblood2017} to be impacted} by large Carrington-like flares with a peak X-ray surface flux of $F_{\rm X}\geq1\,\rm erg\,s^{-1} cm^{-2}$ and proton fluxes of $\sim 10^2-10^3$\,proton flux units (pfu) by about 4 orders of magnitude more frequently than modern day Earth. {These fluxes, however, are extrapolated from solar scaling relations, which is questionable. For CMEs, for instance, a study by \citet{Moschou2019} indicates that even though energetic stellar CMEs likely follow the average relation between solar CME mass and X-ray flaring energy, stellar CMEs appear to have less kinetic energy by about two orders of magnitude than predicted by the solar extrapolation.}

Besides a star's radiation environment, its mass loss is another important factor to consider. Like the decrease in XUV luminosity and flaring activity, the stellar mass loss of a solar-like star declines together with the rotation rate of the star. This reduction in mass loss also leads to a decrease of the stellar wind -- of its velocity and its density -- over time \citep{Wood2005,Johnstone2015WindII,Airapetian2016} and potentially to a significant decrease in CME intensity and frequency as well \citep[e.g.,][]{Odert2017,Kay2019}. Both, increased values for the quiet stellar wind and energetic CMEs, will also enhance non-thermal escape at the respective planets, and they might be significantly stronger for low-mass stars. \citet{Garaffo2016}, for instance, estimated that the stellar wind pressure of the M dwarf Proxima Centauri\,B could be about 2000 times higher than the Sun's. Taken all these findings on the evolution of the plasma and radiation environments of low-mass stars together, it comes by little surprise that recent observations with JWST indicate that the inner planets of the Trappist-1 system -- Trappist-1b \citep{Greene2023} and Trappist-1c \citep{Zieba2023} -- are devoid of thick atmospheres {and may even represent airless bodies,} although these planets are about 30-40\% more massive than the Earth.

Finally, it has to be noted that the Sun seems to be exceptionally weakly active for a star of its age and mass, as has been determined through an analysis of 369 solar-like stars based on data gathered by the Gaia space mission \citep{Reinhold2020}. This is in agreement with the empirical stellar model by \citet{Johnstone2021Stars}, who also point out that the Sun is less active than stars with similar age and characteristics. This can be exemplified by the X-ray surface flux, $F_{\rm X, \oplus}$, that is received by the Earth at the present day. While this is on average $F_{\rm X, \oplus}\sim 0.65\, \rm erg\,s^{-1} cm^{-2}$, it is $F_{\rm X}\sim 2.0\, \rm erg\,s^{-1} cm^{-2}$ for a slow rotator with the Sun's age in the model of \citet{Johnstone2021Stars}. However, such an exceptionally low activity seems to be less pronounced in the EUV region of the spectrum. Whereas the solar luminosity in the EUV, $L_{\rm EUV,\odot}$, is typically about 4 to 7 times higher than $L_{\rm X,\odot}$ during high solar activity, the ratio of $L_{\rm EUV}$/$L_{\rm X}$ over the entire the stellar distribution in the model of \citet{Johnstone2021Stars} is ranging between 1.5 to 4 only.

\subsection{XUV flux and its effect on atmospheric stability}\label{sec:XUV}

The host star's XUV surface flux heats the upper atmosphere of a planet, which can lead to significantly expanded thermospheres and, therefore, to extreme thermal escape into space. Although none of the terrestrial planets in today's Solar System experiences such high XUV-driven losses, there are solid indications that this process affected the early planets and is still shaping the distribution of exoplanets in the Galaxy. The famous `Fulton Gap', which is the radius valley between sub-Neptunes and rocky super-Earths \citep[e.g.,][]{Fulton2017,Fulton2018}, is -- together with core-driven mass loss \citep[][]{Gupta2019} -- mostly attributed to hydrodynamic escape of primordial hydrogen-dominated atmospheres \citep[e.g.,][]{Owen2017,King2021,Rogers2021a,Rogers2021b,Kubyshkina2022}.

Although the lightweight hydrogen is the main constituent in primordial atmospheres, heavier elements can also be subject to hydrodynamic escape, as was shown by various studies. The hydrodynamic upper atmosphere model of \citet{Tian2008,Tian2008b}, for instance, successfully reproduces today's upper atmosphere structures of Earth and Mars, and it was applied to an Earth-like N$_2$-dominated atmosphere illuminated by multitudes of today's solar XUV flux. For $F_{\rm XUV}\sim5-7\,F_{\rm XUV,\oplus}$\footnote{The value taken by \citet{Tian2008} for a wavelength range of 0.5-400\,nm was 5.1\,$\rm erg\,s^{-1}cm^{-2}$.}, the upper atmosphere starts to expand adiabatically with its bulk flow velocity exceeding escape velocity. For $F_{\rm XUV} = 20\,F_{\rm XUV,\oplus}$, the exobase, $r_{\rm exo}$, reaches an altitude of $r_{\rm exo}\sim12$\,R$_{\oplus}$, an altitude that is even above the Earth's present-day magnetopause standoff-distance.

Any such expanded atmosphere is lost within a million years by thermal escape alone, as was more recently also shown by, e.g., \citet{Johnstone2019,Johnstone2021}. These authors confirmed the results by \citet{Tian2008,Tian2008b} for N$_2$-dominated atmospheres with an independently developed hydrodynamic upper atmosphere model called \textit{Kompot} \citep{Johnstone2018}. Similarly, \citet{Erkaev2021} showed with a sub-/transonic upper-atmosphere model that the N$_2$-dominated atmosphere of Saturn's moon Titan expands and escapes hydrodynamically for EUV fluxes less than about four times the present-day surface flux at the Earth's orbit. Titan's atmosphere may even experience high loss rates in today's HZ \citep{Spross2021}. This particular result further illustrates that the atmospheres of smaller-mass bodies can experience hydrodynamic escape for XUV fluxes that are significantly lower than the potential threshold values for more massive Earth-like planets.

Molecules that cool the upper atmosphere, i.e., infrared coolers such as CO$_2$ and N$_2$O, absorb and reflect part of the incident XUV flux as infrared radiation back into space, thereby cooling the upper atmosphere and decreasing its expansion, which can be easily exemplified by comparing the structure of Venus' CO$_2$-atmosphere with the Earth's. For Venus, the exobase level can be found to be around $r_{\rm exo} \sim 200$\,km, whereas it is $r_{\rm exo} \sim 500$\,km in Earth's case, even though Venus is slightly less massive and the incident XUV flux at its orbit is about twice as high as at 1\,AU \citep{Way2022}. However, it was already shown by \citet{Tian2009} that even infrared coolers cannot prevent the hydrodynamic escape of a highly irradiated terrestrial atmosphere. In the case of Mars, they have shown that its CO$_2$-atmosphere would be lost within less than 10\,Myr for an XUV flux of $F_{\rm XUV}=20\,F_{\rm XUV,\oplus}$, mostly because of CO$_2$ being dissociated into C and O.

\citet{Johnstone2021} investigated the stability of Earth's N$_2$-dominated atmosphere during the Archean eon as a function of different X-ray surface fluxes and CO$_2$-mixing ratios (without atmospheric O$_2$). To prevent the nitrogen-dominated atmosphere from rapid escape to space, they found that the Sun must have been a slow rotator and that the CO$_2$-mixing ratio during the Archean eon had to be significant, i.e., even sufficiently high to resolve the Faint Young Sun Paradox. Even at Earth, a pure CO$_2$-dominated atmosphere would not have been stable prior to about 3.5\,Ga, if the Sun had been a moderate or even fast rotator \citep{Johnstone2021}. An N$_2$-dominated atmosphere with $x_{\rm CO_2}\,=\,10\%$ starts to adiabatically expand and hydrodynamically escape for a surface flux of $F_{\rm X}\sim 5\, \rm erg s^{-1} cm^{-2}$. Figure~\ref{fig:LxLimit} shows the exobase levels of an N$_2$-atmosphere with different CO$_2$ mixing ratios irradiated by different levels of $F_{\rm X}$ as simulated by \citet{Johnstone2021}. Even though this figure only shows $F_{\rm X}$, \citet{Johnstone2021} used reconstructed synthetic solar spectra between 0.5-400\,nm by \citet{Claire2012} that are enhanced over the entire wavelength range as an input into their simulations.

Even though oxygen and water vapor were not included in the simulations performed by \citet{Johnstone2021}, these authors argue that potential cooling agents such as N$_2$O and H$_2$O would not significantly alter their results since (i) their effect will be minor compared to CO$_2$, particularly for higher values of $p$CO$_2$, (ii) H$_2$O will be refined to the lower atmosphere due to the cold trap, and (iii) high XUV fluxes that already resulted in a strong expansion of the upper atmosphere would also start to dissociate other cooling agents.

However, a recent study by \citet{Nakayama2022} investigated the stability of N$_2$-O$_2$-dominated atmospheres with a newly developed 1D upper atmosphere model that includes atomic radiative line cooling for a higher number of transition lines than the model of \citet{Johnstone2021}. These authors argue that atomic line cooling by oxygen prevents strong hydrodynamic expansion of the thermosphere for fluxes as high as up to $F_{\rm XUV}1000 = F_{\rm XUV,\oplus}$. However, besides implementing atomic line cooling to a higher degree, \citet{Nakayama2022} handle radiative transfer in their code in a relatively simplified manner by assuming that the radiation emitted by recombination processes is directly lost into space without being reabsorbed by atmospheric gases. One may wonder whether such assumption can be physically valid, specifically since the related cooling from radiative recombination in their model \citep[see Fig.~5 in][]{Nakayama2022} starts to significantly increase for $F_{\rm XUV}\sim$4-5\,$F_{\rm XUV,\oplus}$ and even becomes the clearly dominant cooling process for surface fluxes higher than $F_{\rm XUV}\sim$20-30\,$F_{\rm XUV,\oplus}$ by re-emitting almost all the incident radiation back to space. In reality, more than 50\% of the radiation may be reabsorbed by the atmospheric gas below the absorption level and on the way upwards to the exosphere. If an emitted photon moves in a downward direction (which, by chance, happens in $\sim$50\% of the emission cases), it will be absorbed in denser layers and deposits its energy back into the ambient gas, thereby heating the atmosphere. For a photon moving in the upwards direction, its fate will depend on the altitude of its generation and on the zenith angle of its path through the atmosphere. The lower the optical depth at the emission altitude and the smaller the zenith angle, the higher the probability is that the photon will escape and vice versa. The assumption that all radiation is emitted into space may therefore not be physically valid and heating could consequently be underestimated in model of \citet{Nakayama2022}. Even if their recent simulations indeed indicate a higher thermal stability of N$_2$-O$_2$-dominated atmospheres, non-thermal losses to space would increase sharply for these atmospheres, as \citet{Nakayama2022} highlight, specifically because of atomic line cooling and radiative recombination dominating the shape of highly irradiated thermospheres in their model.

\begin{figure}
\centering
\includegraphics[width = 1.0\linewidth, page=1]{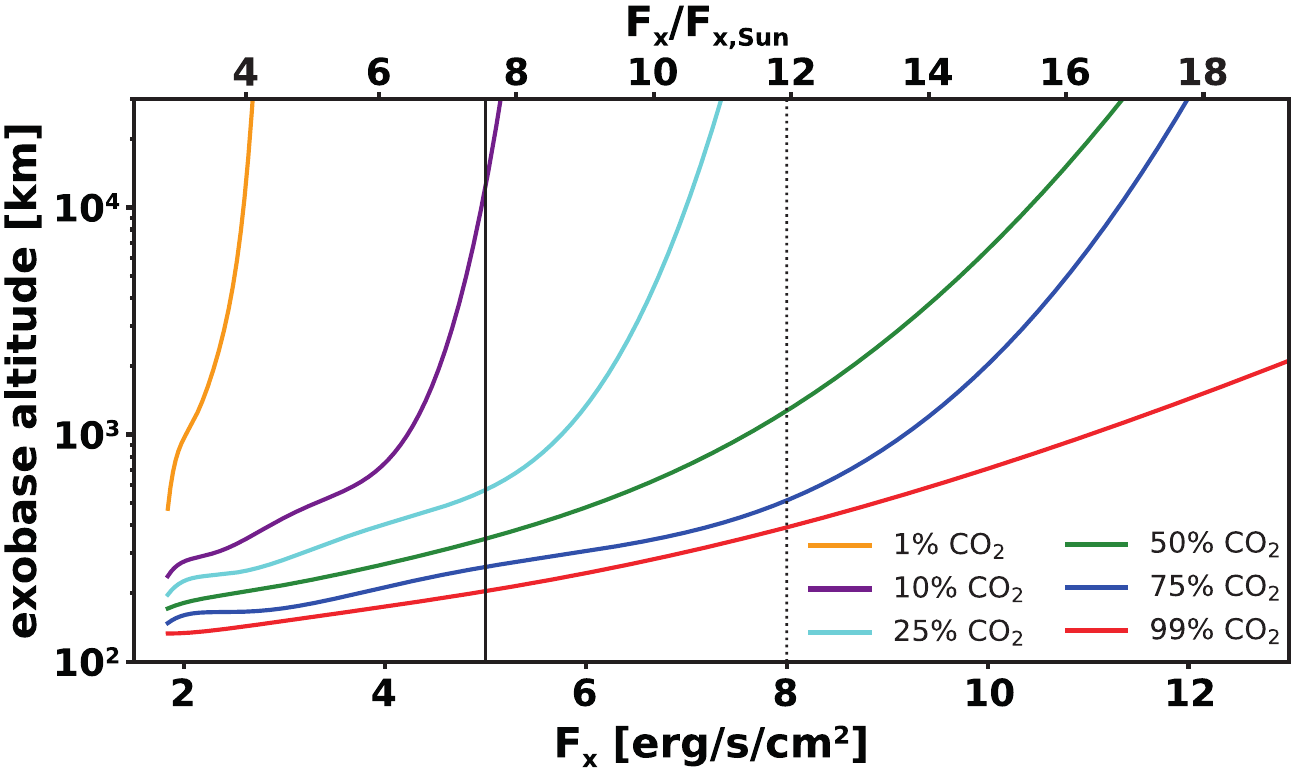}
\caption{Exobase altitudes for N$_2$-atmospheres with different CO$_2$ mixing ratios and for different X-ray surface fluxes, $F_{\rm X}$, simulated with the upper atmosphere model \textit{Kompot} \citep{Johnstone2021}. Even for a mixing ratio of $x_{\rm CO_2}\sim 0.1$ the upper atmosphere expands and escapes hydrodynamically already for $F_{\rm X} \sim 5\, \rm erg\,s^{-1} cm^{-2}$. The vertical solid black line displays our nominal and minimum cases whereas the vertical dotted line illustrates our maximum case (see Section~\ref{sec:fat}). Data from \citet{Johnstone2021}}
\label{fig:LxLimit}
\end{figure}\vspace{1mm}

For Trappist-1, \citet{Peacock2019} found that its HZ planets will have likely lost their potential oceans and atmospheres due to the high irradiation from the host star, a finding that broadly agrees with previous studies such as \citet{Roettenbacher2017} and \citet{Winter2018a}{, and with the recent findings by JWST that the planets Trappist-1b \citep{Greene2023} and c \citep{Zieba2023} may not possess {a dense or any atmosphere at all}}. {This agrees well with recent \textit{Kompot} simulations by \citet{VanLooveren2024Trappist}, who found that none of them can sustain nitrogen-dominated atmospheres because of the high XUV irradiation from their host star}. For the otherwise relatively quiescent M dwarf Barnard's star, on the other hand, flaring alone would still be sufficient to erode 87 Earth atmospheres per billion years \citep{France2020}. The latter finding indicates that flares in addition to quiescent conditions have to be considered, if one wants to study the potential stability of N$_2$-O$_2$-dominated atmospheres and that even exceptionally weak low-mass stars might pose a significant {threat} to Earth-like atmospheres and aerobic complex life within it.

Flares further have a critical impact on any potential ozone layer\footnote{Here, we should note that complex life could in principle hide in the shade or under water to avoid the detrimental effects of ozone destruction. But whether such an environment resembles a fruitful place for a complex biosphere is questionable.}. If an atmosphere at a planet orbiting an M dwarf indeed survived, flares would  be able to destroy $>90\%$ of the entire O$_3$ within just 10 years \citep{Tilley2019}. Similarly, \citet{Vida2017} find that frequent flaring would continuously alter any atmosphere that might potentially be present in the Trappist-1 system, making its planets less favorable for hosting life. This is also in agreement with a study by \citet{Yamashiki2019}, who find that close-in HZ planets such as Trappist-1e and Proxima Centauri\,B receive radiation doses from frequent flaring that can reach fatal levels at least for atmospheres with 10\% of the Earth's atmospheric pressure. From Proxima Centauri, a superflare with an energy of $10^{35.5}$\,erg was detected that irradiated the surface of Proxima Centauri\,B with 65 times the intensity needed to kill simple microbes -- if the present activity rate of the host star proceeds, it will destroy $>$99.9\% of all O$_3$ in a potentially existing atmosphere within 10$^5$ years \citep{Howard2018}. Another study by \citet{Youngblood2017} estimates that the flare activity at GJ~876 completely strips any O$_3$ from an Earth-like atmosphere within $10^2-10^5$\,years. For Trappist-1, however, \citet{Glazier2020} suggests that its superflare rate is insufficient to drive ozone depletion, but at the same time the UV flux provided by these flares would also be too low to trigger prebiotic chemistry.

\citet{Howard2019} further observed 575 flares on 284 out of 4068 cool stars with a median bolometric energy of 10$^{34.0}$\,erg and they found one superflare with sufficient energy to photodissociate Earth's entire ozone. Another 17 stars that may erode the whole atmosphere just via flaring were additionally found by these authors. In total, $\sim15\%$ of the stars observable by TESS exhibit large flares with at least 10\% of the bolometric energy needed to dissociate the entire ozone layer by just one superflare \citep{Howard2019}. Based on results by \citet{Loyd2018}, \citet{Howard2019} estimate such a minimum bolometric energy to be $10^{36.2}$\,erg, {which is in agreement with an older calculation by \citet{Lingam2017Superflare}}.

There are also studies that suggest some organisms might survive frequent flaring. \citet{Abrevaya2020}, for instance, suggest that a fraction of the microbial surface population can survive a typical superflare affecting Proxima Centauri\,B, but these authors do not take into account any atmospheric erosion and assume that there will be sufficient CO$_2$ to prevent escape. \citet{Estrela2020} obtained similar results for the Trappist-1 system, but again without considering escape. However, \citet{DoAmaral2022} investigated the additional effect of the XUV flux received by frequent stellar flaring on the thermal escape of planetary atmospheres. These authors found that flares increase the XUV surface flux by $\sim$10\% compared to quiescent levels, which can result in the removal of up to two additional terrestrial ocean equivalents over the first 1\,Gyr for planets orbiting stars with masses of $M_{\star}=0.2-0.6$\,M$_{\odot}$.

A star's plasma environment can additionally erode planetary atmospheres, particularly in unison with high XUV fluxes. \citet{Lichtenegger2010} showed that an extended nitrogen-dominated atmosphere on early Earth would be eroded by non-thermal escape due to its interaction with the extreme solar wind within a few Myr for an incident XUV flux of $F_{\rm XUV}\sim7\,F_{\rm XUV,\oplus}$. Polar outflow can possess an additional risk for an N$_2$-O$_2$-dominated atmosphere of a magnetized planet, even around a solar-like star. As \citet{Kislyakova2020} found, ion escape could have increased back until the end of the Archean eon by one and more than two orders of magnitude for O$_2$ and N$_2$, respectively. A follow-up study by \citet{Grasser2023} further shows that the Earth's present atmospheric nitrogen inventory in the Archean eon could entirely be eroded via polar outflow even for a nitrogen-dominated atmosphere with $x_{\rm CO_2}=10\%$.

Studies of M stars with stellar winds in the HZ significantly stronger than at Earth today also show that non-thermal escape can render exoplanets uninhabitable. \citet{GarciaSage2017}, for instance, simulated the interaction of a magnetized Earth-twin in the HZ of Proxima Centauri\,B and found that the ionospheric outflow might increase by orders of magnitude compared to present-day Earth's. A nitrogen-dominated atmosphere will, therefore, not survive the extreme XUV flux and stellar wind of Proxima Centauri\,B. Similarly, \citet{Airapetian2017} showed that magnetized Earth-like planets will experience significant ion escape through polar winds around M and active K dwarfs, which would be sufficient to render them inhospitable within a few tens of Myr. This is also in agreement with a study by \citet{RodriguezMozos2019}, who found that magnetized and tidally locked planets will have a close to zero likelihood of keeping their atmospheres due to interactions with their host star's stellar wind. \citet{France2020} further estimated that non-thermal escape at Barnard's star could, besides thermal losses due to flares, remove an additional $\sim$3 Earth atmospheres within a billion years.

A recent study by \citet{Dong2020TOI} investigated non-thermal ion escape with a sophisticated MHD model from the Earth-sized HZ planet TOI-700\,d with a planetary mass of $M_{\rm pl}\sim$1.72\,M$_{\oplus}$ that orbits around an M dwarf with a stellar mass of $M_{\star}\sim$0.42\,M$_{\odot}$. They found that a 1\,bar Earth-like N$_2$-O$_2$-dominated atmosphere around an unmagnetized TOI-700\,d could be lost via ion escape in less than 1\,Gyr due to the strong stellar winds. As \citet{Dong2020TOI} point out, however, they did not take into account the high incident XUV surface flux, but modeled non-thermal escape by implementing the same upper atmosphere structure as for present-day Earth, thereby obtaining lower limits for the erosion timescale. For Venus-like CO$_2$-dominated atmospheres, on the other hand, \citet{Dong2020TOI} recently found, {in agreement with an earlier study by \citet{Lammer2007},} that escape would be lower by almost two orders of magnitude with the only differences between both cases being the higher thermosphere temperature and exobase levels of the Earth-like case. So, even if N$_2$-O$_2$-dominated atmospheres were thermally stable against high XUV fluxes, non-thermal escape would still erode such atmosphere in a geologically short timescale, and, for high XUV levels, likely much faster than in the model runs by \citet{Dong2020TOI}. The latter can reasonably be inferred since thermospheric temperatures and exobase levels of nitrogen-dominated atmospheres increase substantially for highly irradiated planets. Non-thermal escape will additionally be larger for planets with lower masses and around late-type M-dwarfs.

All these studies strongly suggest that, besides a star's quiescent XUV flux, flaring and non-thermal escape induced by stellar winds and CMEs must considered for studying the habitability of planets. If we additionally implemented these parameters into our model, our maximum number of EHs would likely decrease because of strong non-thermal escape induced by a star's plasma environment, enhanced thermal losses due to frequent flaring, and because of the detrimental effects of flares on atmospheric ozone. In addition, one should also consider the evolution of a star's UV flux, as we briefly discuss in the appended Section~\ref{sec:other}.

    \section{$\beta_{\rm HZCL}$: The rocky planet occurrence rate in the HZ}\label{sec:etaHZ}

{As outlined in Section~\ref{sec:bHZCLMain}, the occurrence rate, $\beta_{\rm HZCL}$, of rocky exoplanets in the HZCL is the one parameter in our formulation (see Section~\ref{sec:eq}) that most closely resembles the traditional parameter `Eta-Earth', $\eta_{\oplus}$. As already pointed out,} different studies use very different boundary conditions for the orbital periods and planetary radii feeding into $\eta_{\oplus}$ (see Table~\ref{tab:etaEarth}), which makes a direct comparison difficult.

\begin{table*}\footnotesize
  \begin{center}
    \caption{Literature values of $\eta_{\oplus}$ and their scaled $\beta_{\rm HZCL}$ values$^a$ for $R_{\rm pl,min} = 0.79$\,R$_{\oplus}$ and $R_{\rm pl,max} = 1.23$\,R$_{\oplus}$ and the outer HZCL boundary of \citep{Ramirez2020}.}
    \label{tab:etaEarth}
    \resizebox{\textwidth}{!}{%
\begin{tabular}{l|c|c|c|c|c|c}
  \hline
   & $r_{\rm min}$ [R$_{\oplus}$] & $r_{\rm max}$ [R$_{\oplus}$] & $S_{\rm eff,max}$ [$S_{\rm eff,\odot}$] & $S_{\rm eff,min}$ [$S_{\rm eff,\odot}$] & $\eta_{\oplus}$ or $\zeta_{\oplus}$ & scaled $\beta_{\rm HZCL}$ \\
   \hline
 \citet{Bergsten2022}; for $M_{\star} = 0.56-1.63$\,M$_{\odot}$ & 0.7 & 1.5 & 1.00$^b$ & 0.34$^b$  & 0.094 & 0.021 \\
 \citet{Bergsten2022}; for $M_{\star} = 0.91-1.01$\,M$_{\odot}$ & 0.7 & 1.5 & 0.98$^b$ & 0.34$^b$ & 0.069 & 0.015 \\
 \citet{Quanz2022} & 0.82-0.62$^c$ & 1.4 & 1.11 & 0.36 & 0.164 & 0.041$^d$ \\
 \citet{Neil2020}; model 4 w/ $0.5<M_{\rm pl}<2.0$ & 0.8 & 1.2 & 1.13 & 0.86 & 0.008 & 0.023 \\
 \citet{Bryson2021}; lower bound & 0.82 & 1.4 & 1.11 & 0.36 & 0.18$^{+0.16}_{-0.09}$ & 0.051$^{+0.045}_{-0.026}$ \\
 \citet{Bryson2021}; higher bound & 0.82 & 1.4 & 1.11 & 0.36 & 0.28$^{+0.30}_{-0.14}$ & 0.096$^{+0.068}_{-0.056}$ \\
 \citet{Savel2020}; w/ st. multiplicity & 0.8 & 1.2 & 1.35 & 0.79 & 0.019$^{+0.013}_{-0.008}$ & 0.028$^{+0.019}_{-0.012}$ \\
 \citet{Kunimoto2020}; w/ reliability & 0.75 & 1.5 & 1.02 & 0.35 & $<$0.17$^{+0.11}_{-0.08}$ & $<$0.074$^{+0.051}_{-0.031}$ \\
 \citet{Bryson2020}; w/ reliability & 0.8 & 1.2 & 1.35 & 0.79 & 0.013$^{+0.009}_{-0.006}$ & 0.019$^{+0.014}_{-0.009}$ \\
 \citet{Hsu2019}; upper limit & 0.75 & 1.5 & 1.78 & 0.66 & $<$0.27$^{+0.13}_{-0.24}$ & $<$0.105$^{+0.051}_{-0.093}$ \\
 \citet{Pascucci2019}; model\#4 & 0.7 & 1.5 & 1.15 & 0.35 & 0.109$^{+0.055}_{-0.037}$ & 0.022$^{+0.011}_{-0.008}$ \\
 \citet{Pascucci2019}; model\#6 & 0.7 & 1.5 & 1.15 & 0.35 & 0.054$^{+0.07}_{-0.037}$ & 0.011$^{+0.015}_{-0.007}$ \\
 \citet{Zink2019}; w/ pl. multiplicity & 0.72 & 1.23 & 1.11 & 0.36 & 0.24$^{+0.01}_{-0.01}$ & 0.091$^{+0.004}_{-0.005}$ \\
 \citet{Garrett2018} & 0.5 & 1.5 & 1.11 & 0.36 & 0.88$^{+0.04}_{-0.03}$ & 0.167$^{+0.008}_{-0.005}$ \\
 \citet{Mulders2018} & 0.7 & 1.5 & 1.15 & 0.35 & 0.36$^{+0.14}_{-0.14}$ & 0.075$^{+0.028}_{-0.029}$ \\
 \citet{Burke2015} & 0.8 & 1.2 & 1.35 & 0.79 & 0.1$^{+1.9}_{-0.09}$ & 0.148$^{+2.812}_{-0.133}$ \\
 \citet{Silburt2015} & 1.0 & 2.0 & 1.02 & 0.35 & 0.064$^{+0.034}_{-0.011}$ & 0.008$^{+0.005}_{-0.001}$ \\
 \citet{ForemanMackey2014}& 1.0 & 2.0 & 2.23 & 0.89 & 0.019$^{+0.01}_{-0.008}$ & 0.005$^{+0.006}_{-0.002}$ \\
 \citet{Petigura2013}; conservative HZ & 1.0 & 2.0 & 1.02 & 0.35 & 0.086 & 0.011 \\
 \citet{Traub2012}; Case 2 & 0.5 & 2.0 & 1.56 & 0.31 & 0.34 & 0.105 \\
 \citet{Catanzarite2011}& 0.8 & 2.0 & 1.11 & 0.53 & 0.011$^{+0.06}_{-0.003}$ & 0.002$^{+0.0015}_{-0.0003}$ \\
   \hline
\end{tabular}}
\end{center}\footnotesize
$^a$scaled with the broken power law of model\#4 by \citet{Pascucci2019}; $^b$for solar-like star with $M_{\star} = 1.0\,$M$_{\odot}$; $^c$radius scaled with surface flux $S_{\rm eff}$ through $r_{\rm min} = 0.8 \times S_{\rm eff}^{0.25}$; $^d$a mean radius of 0.72\,R$_{\oplus}$ was taken for calculating $\beta_{\rm HZCL}$
\end{table*}

The first planet occurrence rates were estimated for short-period planets \citep[e.g.,][]{Youdin2011,Howard2012} and based on the earliest Kepler data \citep{Borucki2011}. First values for $\eta_{\oplus}$ derived from the same data set were estimated by \citet{Catanzarite2011}, who found $\eta_{\oplus} = 0.011^{+0.006}_{-0.003}$ for the HZ of \citet{Kasting1993}, and by \citet{Traub2012}, who found $\eta_{\oplus} = 0.34\pm0.14$ (their Case 2 for an average G-type star with $M_{\star}=1.08$\,M$_{\odot}$), respectively. As corresponding boundary conditions, \citet{Catanzarite2011} chose orbital distances and planetary radii between $d_{\rm pl}=0.95-1.37$\,AU and $R_{\rm pl}=0.8-2.0$\,R$_{\oplus}$, whereas \citet{Traub2012} took $d_{\rm pl}=0.8-1.8$\,AU and $R_{\rm pl}=0.5-2.0$\,R$_{\oplus}$, respectively. In addition, both studies had to extrapolate data from close-in rocky exoplanets to wider orbital periods to be able to estimate occurrence rates in the HZ of solar-like stars.

Subsequent studies based on newer planet catalogs found $\eta_{\oplus}$-values mostly ranging between the estimates of \citet{Catanzarite2011} and \citet{Traub2012} with the notable exception of \citet{Garrett2018}, as can be seen in Table~\ref{tab:etaEarth}. Since 2018, estimates for $\eta_{\oplus}$ \citep[e.g.,][]{Mulders2018,Hsu2019,Pascucci2019,Bryson2020,Savel2020,Bryson2021} can make use of Kepler DR25 \citep{Thompson2018}, which significantly enhanced the issues of completeness and reliability \citep[e.g.,][]{Bryson2021}. Nevertheless, the problems of extrapolating to the HZ of G stars and the differing definitions of $\eta_{\oplus}$ remain. However, already \citet{Burke2015} introduced another value called $\zeta_{\oplus}$, which sets its boundary conditions to be within 20\% of the Earth's orbital period and radius. Moreover, the differential occurrence rate for Earth-analogs,
\begin{equation}\label{eq:GammaEarth}
\Gamma_{\oplus} = \frac{d^2N_{\rm pl,tot}}{d(\ln P_{\rm pl}) d (\ln R_{\rm pl})},
\end{equation}
where $N_{\rm pl,tot}$ represents the number of planets per star, and $P_{\rm pl}$ and $R_{\rm pl}$ certain intervals of the orbital period and planetary radius, respectively, was introduced to compare various estimates of $\eta_{\oplus}$ \citep[e.g.,][]{Youdin2011,ForemanMackey2014,Burke2015,Kunimoto2020,Bryson2021}. However, we have to scale any value for $\eta_{\oplus}$, $\zeta_{\oplus}$ or $\Gamma_{\oplus}$ to fit our parameter range for $P_{\rm pl}$ and $R_{\rm pl}$ (as described in Section~\ref{sec:betaHZCL}).

Extrapolating occurrence rates from close-in orbits and low-mass stars to the HZ of G-type stars can lead to significant overestimates of Earth analogs, as first pointed out by \citet{Lopez2018}. The radius distribution of short-period planets is strongly dependent on atmospheric escape, as the XUV surface flux at these orbits can lead to the complete photoevaporation of primordial atmospheres, an effect that, at least partially, explains the well-known Fulton gap \citep[e.g.,][]{Fulton2017,Lopez2018}. A recent study by \citet{Pascucci2019}, which took this bias into account, found that $\eta_{\oplus}$ indeed drops significantly by excluding short-period planets. By not excluding short-period planets, \citet{Pascucci2019} obtained $\eta_{\oplus} = 0.406^{+0.149}_{-0.173}$ through fitting two broken power laws (i.e. `2D broken') onto the distribution of planetary periods and radii (`model\#1' in their study), based on Kepler DR25, stellar properties from Gaia DR2 \citep{Berger2018} and the Exoplanet Population Observation Simulator \verb"epos" \citep{Mulders2018}. They even get $\eta_{\oplus} = 0.536^{+0.297}_{-0.297}$ (`model\#2') by keeping a broken power law for the distribution of planetary periods but fitting a single one onto the respective radii (i.e., `R broken'). In both cases, the boundary conditions for $\eta_{\oplus}$ were chosen to be  $P_{\rm pl}=\{0.95,2.2\}\,$P$_{\oplus}$ and $R_{\rm pl}=\{0.7,1.5\}\,$R$_{\oplus}$.

If, however, \citet{Pascucci2019} excluded planets with periods $P_{\rm pl} < 12$~days (their `model\#4'), the occurrence rate dropped to $\eta_{\oplus} = 0.109^{+0.055}_{-0.037}$ by employing their `R broken' power law formulation. By excluding planets with $P_{\rm pl} < 25$\,days (their `model\#6'), their obtained occurrence rate even decreased to $\eta_{\oplus} = 0.054^{+0.070}_{-0.037}$ (again with `R broken'). This demonstrates that atmospheric escape of primordial atmospheres must be considered since -- in contrast to close-in planets -- these atmospheres may not be lost in the HZ of G stars.

This substantial drop in occurrence rate by excluding evaporated cores was confirmed by \citet{Neil2020}, who took into account three different types of planets for their study, i.e., evaporated cores, planets with gaseous envelopes, and intrinsically rocky planets. By implementing XUV-driven atmospheric escape and employing hierarchical Bayesian modeling onto the Kepler sample, they discovered a drop in $\eta_{\oplus}$ by almost an order of magnitude, i.e., from $\eta_{\oplus} = 0.076$ (their `model 1' without evaporation) to $\eta_{\oplus} = 0.009$ (`model 4', with evaporation), both for $R_{\rm pl} = 0.8-1.2$\,R$_{\oplus}$ and $P_{\rm pl} = 292-438$\,days. By further restricting themselves to planetary masses of $M_{\rm pl} = 0.5-2.0$\,M$_{\rm \oplus}$, the values further dropped to $\eta_{\oplus} = 0.055$ and $\eta_{\oplus} = 0.008$, respectively.

\citet{Bergsten2022} calculated $\eta_{\oplus}$ and $\Gamma_{\oplus}$ for FGK stars between stellar masses of $M_{\star} = 0.56-1.53$\,M$_{\oplus}$ by taking into account reliability \citep[i.e., the likelihood of false alarms due to instrument statistics; see][]{Thompson2018} and by splitting their planetary sample into close-in super-Earths and sub-Neptunes with $R_{\rm pl}\sim1-2$\,R$_{\oplus}$ and $R_{\rm pl}\sim2-3.5$\,R$_{\oplus}$, respectively, to circumvent the effect of atmospheric erosion. In agreement with \citet{Pascucci2019} and \citet{Neil2020}, these authors found that evaporated cores contaminate the population of super-Earths. By excluding them, \citet{Bergsten2022} also estimated comparatively low occurrence rates for FGK stars of $\eta_{\oplus} = 0.0937^{+0.0340}_{-0.0248}$ and $\Gamma_{\oplus}=0.1529^{+0.0555}_{-0.0405}$ for $R_{\rm pl} = 0.7-1.5$\,R$_{\oplus}$ and $P_{\rm pl}=363-811$\,days. For solar-like stars with $M_{\star} = 0.91-1.01$\,M$_{\odot}$, \citet{Bergsten2022} further found $\eta_{\oplus} = 0.0688^{+0.0381}_{-0.0261}$ and $\Gamma_{\oplus}=0.1126^{+0.0624}_{-0.0427}$, respectively, for the same radius range and orbital periods between 370-826 days. Note, however, that \citet{Bergsten2022} had to extrapolate the occurrence rate of planets with $R_{\rm pl} < 1.0$\,R$_{\oplus}$ from bigger planets, as these were not included in the fitting range of their study.

\citet{Bryson2021}, for the first time, derived $\eta_{\oplus}$ by taking into account the stellar surface flux distribution of the Kepler stars. They tried to tackle the bias pointed out by \citet{Lopez2018} and \citet{Pascucci2019}, by restricting themselves to planets within $0.2\,S_{\rm eff,\oplus} \leq S_{\rm eff} \leq 2.2\,S_{\rm eff,\oplus}$ and $0.5\,R_{\oplus} \leq R_{\rm pl} \leq 1.5\,R_{\oplus}$ for stars with $T_{\rm eff}=3900-6300$\,K. For the conservative HZ of \citet{Kopparapu2014} and $R_{\rm pl} = 0.82-1.4$\,R$_{\oplus}$, the radius range used to estimate the exoplanet yields around G stars for the HabEx \citep{Kopparapu2018,Gaudi2020}, Luvoir \citep{Kopparapu2018,Luvoir2019}, and LIFE \citep{Quanz2022} missions, \citet{Bryson2021} retrieved an occurrence rate of $\eta_{\oplus} = 0.18^{+0.16}_{-0.09}$ as a lower and $\eta_{\oplus} = 0.28^{+0.30}_{-0.14}$ as a higher bound. One reason for such high values compared to \citet{Pascucci2019} could be that a surface flux of $S_{\rm eff} \leq 2.2\,S_{\rm eff,\oplus}$ may not entirely account for the early atmospheric escape of primordial atmospheres. The relevant parameter for photoevaporation is $F_{\rm XUV}$ and/or $F_{\rm X}$, but not $S_{\rm eff}$. If one calculates $F_{\rm X}$ for a planet orbiting a star with $T_{\rm eff}=3900$\,K at a distance coinciding with $S_{\rm eff} = 2.2\,$\,S$_{\rm eff,\oplus}$ through the stellar evolution model \textit{Mors} \citep{Johnstone2021Stars}, one obtains $F_{\rm X} \sim 2000\rm\,erg\,s^{-1}\,cm^{-2}$ and $F_{\rm X} \sim 550\rm\,erg\,s^{-1}\,cm^{-2}$ (i.e., $F_{\rm X}\sim3000\,F_{\rm X,\odot}$ and $F_{\rm X}\sim770\,F_{\rm X,\odot}$) for ages of 50 and 500\,Myr, respectively. Such a high insolation will very likely erode any primordial dominated atmosphere around Earth-sized planets \citep[e.g.,][]{Erkaev2022}, thereby still leading to an overestimate of $\eta_{\oplus}$.

Another parameter rarely accounted for when estimating $\eta_{\oplus}$ stems from the potential multiplicity of Earth analogs in the HZ. This effect was first calculated by \citet{Zink2019}, who found that $5.2\pm0.4\%$ of GK dwarfs may have two planets with $R_{\rm pl}=0.72-1.48$\,R$_{\oplus}$ in their conservative HZ. Still, $0.019\pm0.006\%$ of all GK dwarfs host five and an entirety of $30.0\pm1.0\%$ at least one of such planets in their HZ. If they do not take into account multiplicity, the occurrence rate drops to $\eta_{\oplus}\lesssim$21\%. Another study by \citet{Savel2020} evaluated how stellar (not planetary) multiplicity affects Kepler occurrence rates by obtaining adaptive optics observations of 71 Kepler target stars. They detected 3 bound stellar companions and found that accounting for this specific multiplicity may generally increase Kepler-based $\eta_{\oplus}$-estimates by about 6\%. Their estimates correspondingly increased from $\zeta_{\oplus} = 0.11^{+0.08}_{-0.05}$ to $\zeta_{\oplus} = 0.12^{+0.08}_{-0.05}$. Accounting for multiplicities, planetary as well as stellar, therefore increases the number of stars that may indeed host Earth analogs in their respective HZ.

Most of these values, however, are only covering solar-like stars although different stellar masses will host a different number of rocky exoplanets. \citet{Howard2012} first found a linear increase of the occurrence rate with decreasing stellar effective temperature, $T_{\rm eff}$, for planets with orbits less than 50 days. \citet{Mulders2015} later confirmed this result by deriving a planet-radius distribution function for different stellar spectral classes and showed that the mass of a planetary system, but also the individual number of planets, increase linearly with $T_{\rm eff}$ from FGK stars to M dwarfs with the latter having on average 3.5 times more small planets than the other spectral classes. However, \citet{Mulders2015} also stated that they did not find a simple linear or power law correlation between occurrence rate and stellar mass.  \citet{Garrett2018} further recognized that the planet occurrence rate for FGK stars depends on a quadratic function of stellar temperature, while they found no such behavior for M dwarfs.

\citet{HardegreeUllman2019} further investigated planet numbers for a parameter range of $R_{\rm pl} = 0.5-2.5$\,R$_{\oplus}$ for M dwarfs and obtained an increase of planets per star from early to late stellar-types. Similarly, \citet{Yang2020} found that planet occurrence rates and planet multiplicities increase for decreasing stellar temperatures. Whereas only 35\% of early-type stars with $T_{\rm eff} > 6500$\,K host Kepler-like planets with an average multiplicity of $\sim$1.8, this fraction increases to 75\% and $\sim$2.8 for late-type stars with $T_{\rm eff} < 5000$\,K. \citet{He2021} found that the inner planet occurrence rate of FGK dwarfs (with $P_{\rm pl} = 3-300$\,days and $R_{\rm pl} = 0.5-10\,$R$_{\oplus}$) increases linearly from early- to late-type stars in agreement with \citet{Yang2020}.

Consequently, several studies found the occurrence rate of rocky exoplanets in their respective HZ to be higher around M dwarfs than average $\eta_{\oplus}$-values around G-type stars. \citet{Dressing2013}, for instance, calculated the occurrence rate of planets with $R_{\rm pl} = 1.0-1.5$\,R$_{\oplus}$ in the conservative HZ around M dwarfs to be $\eta=0.16^{+0.17}_{-0.07}$. A more recent study by \citet{Quanz2022} estimated this rate to be $\eta=0.312$, again in the conservative HZ but for $R_{\rm pl} = 0.82-1.4$\,R$_{\oplus}$. 

    \section{$B_{\rm pc}^{\rm H_2O}$: The right amount of H$_2$O and its related frequency on rocky exoplanets}\label{sec:h2oMain}

\subsection{The potential simultaneous need for oceans and subaerial land}\label{sec:h2o}

That water is a necessary ingredient for life as we know it is a very well-established scientific fact \citep[see, e.g.,][and references therein]{Westall2018}\footnote{See also \citet{Budisa2014} and \citet{SchulzeMakuch2006} for supercritical \ce{CO2} and other potential solvents, but these are not part of our study.}, although its role at the origin of life may have been ambivalent since hydrolysis counteracts the formation of organic molecules \citep[e.g.,][]{DoNasciementoVeira2020}. It can therefore be stated that an EH must not be a desiccated desert planet, but should contain a certain amount of H$_2$O in stable liquid reservoirs.

However, can a planet that is entirely covered by an ocean with no subaerial land masses evolve into an EH, on which aerobic complex life can emerge? {Although a definitive answer to this question is not possible at the current state of research, several arguments can be made in favor of the simultaneous need for an ocean and subaerial land to be present on a planet. We therefore argue that EHs most likely} need to have (i) oceans, (ii) subaerial land, and (iii) an N$_2$-O$_2$-dominated atmosphere with minor amounts of CO$_2$ on top, which is in direct contact with both water and land. There are several important and independent lines of arguments that this `habitable trinity' \citep[see,][]{Dohm2014} is indeed needed for an EH and complex life to evolve. We list and discuss these arguments in the following.

\vspace{\baselineskip}
\textbf{\textsl{Desiccated and/or desert planets:}}

Desert planets with small amounts of water may still be habitable \citep[e.g.,][]{McKay2014} with an HZ that could even be larger than for planets with Earth-like water inventories \citep[e.g.,][]{Abe2011,Leconte2013b,Zsom2013,Kodama2021}. Defining a lower limit for the water mass fraction that renders a world uninhabitable, however, is difficult, as there are no known minimum levels of \ce{H2O} on a planetary level at which life can still originate. Such a minimum fraction can hence be set by the origin of life itself or by the prevalence of complex life, but not by the lower water limit for the survival of extremophiles. Completely desiccated planets, where the water fraction approaches zero, however, can nevertheless be considered uninhabitable for complex life, as this necessarily needs \ce{H2O} as an essential solvent. Dry worlds may further lack different advantageous characteristics such as tectonic activity \citep[e.g.,][]{Korenaga2013,Korenaga2020}, the formation of continents \citep[e.g.,][]{Campbell1983}, and chemical weathering \citep[e.g.,][]{West2005}. Desiccated planets can therefore be excluded from the sample of planets, as they cannot evolve into EHs.

\vspace{\baselineskip}
\textbf{\textsl{Climate stability, carbon and nitrogen cycling:}}

It was already suggested by \citet{Kite2009} that planets completely covered by oceans may show a weaker climate-stabilizing feedback due to (i) submarine weathering being less dependent on the surface temperature and (ii) the prohibition of volcanic degassing for pressures at the rock-water boundary above $\sim0.6$\,GPa. \citet{Abbot2012} found similar results and explicitly pointed out that water worlds may not be able to control their climate over geological timescales but may enter ``self-arrest'', i.e., a \textit{Moist Greenhouse} state in which such planets lose most of their water before potentially re-establishing habitability. The latter, however, was argued by \citet{Wordsworth2013} to be relatively unlikely. Another study by \citet{Foley2015} also suggests that worlds with little to no land can enter supply-limited weathering due to a deficient supply of fresh rock at their surface, which leads to very hot climates with uninhabitable conditions (with some of them potentially entering ``self-arrest''). However, more recent studies suggest that carbon cycling may still work at such planets \citep{Hoening2019,Hayworth2020}. Based on a new seafloor weathering model by \citet{KrissansenTotton2017}, \citet{Hayworth2020} even argue that a temperature-dependent \ce{CO2} feedback cycle might provide more habitable conditions than for planets with continents.

The situation certainly differs for planets that have a sufficient water content for high-pressure (HP) ice to form at the bottom of the ocean. Such worlds likely develop unstable \ce{CO2} feedback cycles \citep{Kitzmann2015}, which limit volatile and nutrient exchange with their rocky interior. In such a case, the atmospheric \ce{CO2} concentration may be solely governed by the dissolution capacity of carbon dioxide in the ocean, which can severely reduce the habitability of the planet \citep{Kitzmann2015}. However, more recent studies on subsurface-ocean-covered moons such as Ganymede and Callisto \citep{Choblet2017,Journaux2017,Journaux2020,Kalousova2018a} suggest that an exchange between the interior and the ocean is still possible via partial melting of the HP ice. Similarly, \citet{Nakayama2019} calculated that the HP ice of a tectonically active water world can theoretically be melted above the mid-ocean ridges, a pathway that may allow for the stabilization of the \ce{CO2} feedback cycle. However, \citet{Nakayama2019} further showed that such planets enter a cold snowball state rather than developing a hot or even temperate climate, because weathering is determined by the melting temperature of the HP ice rather than the surface temperature.

\citet{Kite2018} find that water worlds, where HP ice inhibits the exchange with the mantle and geochemical cycling does not occur, can nevertheless retain habitable conditions for more than 1\,Gyr, if the initial $p$CO$_2$ is within a sweet spot of $p$CO$_2 \sim 0.2-20$\,bar. As \citet{Kite2018} point out, even though the extreme pH values (with a maximum of pH\,$>$11) in such an ocean should be a solvable obstacle for life, its origin could be more restrictive\footnote{However, sites for the origin of life have been explored with highly alkaline fluids even beyond pH\,$=$\,11; see \citet{Saha2022}.}. High temperatures at such planets and high values of $p$CO$_2$ (above 0.1\,bar) might further pose significant stress to complex life.

As was recently shown by \citet{KrissansenTotton2021b}, a further crucial factor that is mostly neglected in ocean world studies, such as the ones by \citet{Cowan2014} and \citet{Hoening2019}, is the high solubility of volatiles in high-pressure melts, which severely limits magmatic outgassing for an ocean or water ice exceeding 10-100\,km (i.e., pressures of $>$0.1-1\,GPa and water inventories $w_{\rm H_2O}>$1\%). \citet{KrissansenTotton2021b} further argue that such processes not only prevent degassing of \ce{CO2} but might also suppress the outgassing of nitrogen due to N being highly undersaturated in silicate melts for $>$0.1\,GPa \citep[see also,][]{Libourel2003}. But as \citet{KrissansenTotton2021b} point out, further work is needed to quantify the partitioning and geochemical cycling of nitrogen.

An additional water world habitat, the `ice-cap zone', was proposed by \citet{Levi2017} and \citet{Ramirez2018} of which the latter showed that a temperate and stable subtropical climate can evolve for planets with dense \ce{CO2}-atmospheres (with $\gtrsim2.5$\,bar), rotation rates $\sim$3 times faster than the Earth and with orbits around M3 to G2 dwarfs. Here, the high rotation rate leads to a strong temperature gradient between poles and the subtropical region that allows for a stabilizing climate through the deposit of sea ice enriched in \ce{CO2} clathrates. However, a planet with such a high $p$\ce{CO2} does clearly not meet our definition of an EH.

All in all, most of these studies indicate that climate stability, carbon-silicate cycle, and potentially the nitrogen {and oxygen} cycles may be significantly limited on water worlds.

\vspace{\baselineskip}
\textbf{\textsl{Availability of essential nutrients:}}

A potential layer of HP ice restricts the nutrient exchange between a global ocean and the underlying rocky mantle \citep[e.g.,][]{Noack2016,Journaux2020}, even though some level of exchange could be possible due to melting and convection of the HP ice \citep{Choblet2017,Journaux2017,Journaux2020,Kalousova2018a}. For an initial nutrient budget that is sufficient for life to originate and to evolve together with a steady source of photosynthesis, however, \citet{Kite2018} emphasize that a biosphere could be sustained just by nutrient cycling alone, even if exchange with the interior is completely shut down. For this to be sustained, biomass recycling must be complete (e.g., no burial of organic matter), or resupply from the interior and/or delivery from space must happen for the biosphere not to decline \citep{Kite2018}. Nutrient recycling, however, is strongly dependent on ocean upwelling, i.e., the upward flow of deeper water to the surface, a tidal and surface wind-driven process that mostly takes place at the coastal areas of continents and the equator \citep{Olson2020}, highlighting the importance of subaerial land. The dilution of essential nutrients within a vast ocean may further constitute another relevant obstacle, not only for the maintenance of a biosphere but also for abiogenesis.

Besides dilution and restricted exchange between the water column and the silicate mantle, the absence of continents may further inhibit a proper supply of essential nutrients such as Ca, Fe, Mg, K, and particularly P \citep[e.g.,][]{Wordsworth2013,Maruyama2013,Maruyama2014,Dohm2014,Olson2020}. Soluble phosphorus, for instance, particularly in the form of phosphates, is an essential ingredient for multiple biochemical reactions without which \textit{life as we know} it would not be possible \citep[e.g.,][]{Westheimer1987,Kamerlin2013}. It also constitutes the bio-limiting nutrient for marine productivity on Earth \citep[e.g.,][]{Tyrrell1999,Filippelli2008}, and its restricted availability was probably the reason for suppressed biological activity during the Precambrian \citep[e.g.,][]{Bjerrum2002,Kipp2017}. A change in the phosphorus cycle, on the other hand, coincides with the rise of complex animal life during the Proterozoic eon \citep[e.g.,][]{Reinhard2017}. Phosphorus is further crucial for the evolution of atmospheric oxygen since photosynthesis is limited by the supply of riverine P \citep{Hao2020a,Glaser2020,Olson2020,Watanabe2021}. Since phosphorus, however, is mostly retrieved through chemical weathering of subaerial continental crust \citep[e.g.,][]{Wheat1996,Paytan2007,Glaser2020,Hao2020b}, a water world that does not have continents will have a significantly inhibited supply of phosphates compared to a planet with subaerial land. As \citet{Glaser2020} calculated, the input rate of P into the ocean on a pelagic planet might be reduced by a factor of $\sim$440 compared to Earth due to a significant decrease in physical weathering and chemical dissolution. This likely makes P not only the bio-limiting nutrient for such a world \citep[e.g.,][]{Glaser2020,Lingam2019WaterFrac}, but also a limiting factor for the rise of atmospheric oxygen \citep{Glaser2020,Watanabe2021} -- a factor that could further prohibit the emergence of complex life if $p$O$_2$ cannot rise above certain threshold levels \citep{Catling2005}.

{That a rise of oxygen may be questionable on water worlds was also suggested by \citet{Hoening2023}. They found that a lower weathering rate and the correspondingly smaller availability of critical nutrients on `ocean planets'\footnote{The same was found for `land planets', i.e., planets mostly covered by land \citep{Hoening2023}.} will reduce their bioproductivity and biomass by a third to half of the Earth's, a value that could be too low for producing a larger amount of O$_2$ than consumed by geological activity. Note, however, that `ocean planet' in \citet{Hoening2023} only refers to planets covered mostly by oceans (and not entirely), indicating that some intricate balance between subaerial land and ocean will be needed.}

There are, however, recent studies that investigated the production of bioavailable phosphorus within a purely marine ecosystem. \citet{Syverson2021} performed laboratory experiments to investigate the release of soluble phosphorus through weathering of submarine basalt under anoxic conditions and found that the release of P in such environments is comparable to the release in modern rivers. In a companion article, \citet{Filippelli2022} notes that even though the reaction rates produced in the experiment by \citet{Syverson2021} are comparable to those in modern rivers, actual weathering activity at submarine basalt in the modern Earth ocean (EO) is quite small and mostly restricted to mid-ocean ridges. This author, therefore, points out that the results by \citet{Syverson2021} can be seen as a maximum possible rate of P release, but that the entire flux rate from basalt in the whole ocean cannot reflect such narrow and extreme results. Another study by \citet{Pasek2020} reported the production of bioavailable P through the serpentinization of ultramafic rock as another pathway to phosphorus production at ocean-covered water worlds. However, \citet{Pasek2020} additionally emphasize that, even though this proposed redox reaction may be an important source of P at early Earth, ocean-covered icy moons and exoplanets, the prevalence of this process remains unclear, as only one rock outcrop was investigated by their study. {In addition, one should note that these authors chose rock types and setups that are most conducive to liberating phosphorus, implying that reality could be less favorable.}

\vspace{\baselineskip}
\textbf{\textsl{Condensation of steam atmospheres:}}

Besides \ce{H2O}, volatile-rich water worlds will also accrete high fractions of \ce{CO2} and are even proposed to accrete comet-like compositions \citep[e.g.,][]{Leger2004}, with comets containing \ce{CO2} as the second most abundant volatile at $\sim$3-30\,mol\% relative to water \citep[e.g.,][]{Mumma2011,Ootsubo2012}. While most studies \citep[e.g.,][]{Kitzmann2015,Levi2017,Ramirez2018} implicitly assume that the outgassed water vapor will soon condense to form oceans, this might not necessarily be the case. \citet{Marounina2020} modeled whether the outgassed steam atmosphere can indeed condense. They found that exoplanets with a rocky mass of 1\,M$_{\oplus}$ and 2\,M$_{\oplus}$ that host a comet-like, \ce{CO2}-rich volatile reservoir (CO$_2$ and H$_2$O) of $>$11\,wt\% and $>$6\,wt\%, respectively, cannot form a global water ocean. In such a case, the steam atmosphere will not be able to condense, thereby rendering the planet uninhabitable. A potential pathway for these planets toward habitability may be the hydrodynamic escape of their steam atmospheres to such an extent that the remaining gases can condense. However, even if one assumes a water mass fraction of 10\,wt\%, this would amount to $>$400 Earth oceans. If just 1\,wt\% needs to be lost for the rest to condense, this would still imply that $>$40 EOs have to be lost via atmospheric escape.

\vspace{\baselineskip}
\textbf{\textsl{Origin of life:}}

There are several, some of them mutually exclusive, theories on the origin of life; specifically, two of them are currently hotly debated. These are alkaline hydrothermal vents \citep[e.g.,][]{Martin2003,Martin2007,Martin2008,Lane2012,Russell2014,Russell2021} and freshwater hot spring pools \citep[e.g.,][]{Mulkidjanian2012,Damer2015,Damer2020,Deamer2019,Milshteyn2018,VanKranendonk2021b} as potential locations for life's origin. Further currently discussed and often related environments cover (non-exhaustively) different variations of Darwin's `warm little ponds' on volcanic islands \citep[e.g.,][]{Pearce2017,Bada2018,Pearce2022}, deep-reaching tectonic faults \citep[e.g.,][]{Schreiber2012,Mayer2018}, {shallow-sea alkaline hydrothermal vents \citep[e.g.,][]{Guzman2009ShallowHV,Barge2022ShallowHV},} hydrothermal lakes with a potential ice-covered surface \citep{Szostak2016}, crater lakes \citep[e.g.,][]{Osinski2021}, carbonate-rich lakes \citep{Toner2020}, nuclear geysers \citep{Ebisuzaki2017,Maruyama2019}, tidal pools and lagoons \citep{Robertson1995,Lathe2004}, intermountain dry valleys \citep{Benner2012}, floating pumice \citep{Brasier2011,Brasier2013}, hydrothermal sedimentary layers between oceanic crust and seawater \citep{Westall2018}, coastal volcanic splash pools \citep{Fox2013}, {water-air interfaces such as sea sprays and aerosols \citep{Dobson2000,Donaldson2004Aerosols,Tervahattu2004Aerosol,Griffith2012WaterAir,Griffith2013Aerosol,Deal2021WaterAir,Holden2022SeaSpray},} and panspermia \citep[e.g.,][]{Valtonen2009,Ginsburg2021Panspermia,Balbi2021Panspermia}, although the latter idea does not address the origin of life directly (as it implicitly redirects life's origin to some, mostly undefined, extraterrestrial environment; see also Section~\ref{sec:originLife}). Most of these potential solutions to the origins of life problem may need water and land, or shallow water (except of the alkaline hydrothermal vent hypothesis and, potentially, panspermia). {Even the aerosol scenario needs, or at least benefits from, the presence of subaerial land, which can provide valuable feedstock molecules for enabling the aerosols to function as miniature prebiotic chemical reactors.}

A world entirely covered by (deep) oceans may pose some problems for life to originate \citep[e.g.,][]{Mulkidjanian2012,Milshteyn2018,Damer2020}, at least within the open water column. Besides the dilution of nutrients, the lack of free energy, and the `water problem' \citep{Benner2012} related to the rapid rate of hydrolysis, several other specific problems were posited against the alkaline hydrothermal vent scenario. For example, the proposed pH-gradient across mineral membranes between the acidic ocean and the alkaline vent fluid \citep{Lane2012} was criticized for providing far too little energy \citep[e.g.,][]{Jackson2016,Ross2018,Damer2020} and no convincing mechanism for substituting the natural pH gradient to allow decoupling from the vent \citep{Jackson2017}. Also, \citet{Milshteyn2018} showed experimentally that the high concentration of ionic solutes in the seawater prevents the self-assembly of membranes and encapsulated polymers. Condensation reactions in saline seawater may provide another hurdle for polymerization \citep{Ross2018}, but hydrothermal sediments were recently proposed as a potential solution \citep{Westall2018}.

By now, no in-situ experiments at hydrothermal vents were performed. Any support for this theory is laboratory based, which gave rise to further methodological criticism \citep{Waechtershaeuser2016}. Additional experiments will therefore be needed to tackle some of the arguments listed above and to test whether life could indeed originate at hydrothermal vents. If so, ocean-covered icy moons like Enceladus and Europa, and exoplanets without any subaerial land could at least in principle provide habitats for life to emerge. Searching for biosignatures at these moons may therefore provide crucial information for better understanding the origin and prevalence of life within the solar system \citep[see also,][]{Deamer2017,Longo2020} and beyond.

However, recent studies argue for the importance of wet-dry cycles as a crucial requirement to overcome the `water problem' and for life to originate \citep[e.g.,][]{Ross2016,Becker2018,Damer2020,VanKranendonk2021b}. Continuous wet-dry cycling through hydration and dehydration of subaerial freshwater ponds or hot springs can concentrate and accumulate organic molecules as long as the rate of polymer synthesis exceeds hydrolysis \citep[e.g.,][]{Damer2015}. Chemical energy available through the concentration of organic compounds and potential reactants supports condensation reactions and synthesized polymers can be encapsulated within lipid versicles when rehydrated, forming first protocells \citep[e.g.,][]{Deamer2019}. The next dehydration cycle will further concentrate these so-called `progenotes' together with organic compounds into a `hydrogel' in which most may decompose, while others merge, exchange polymers, and potentially stabilize to form ever more complex compartments through growth and evolution \citep[e.g.,][]{Damer2015,Damer2020}. Part of this cycle, such as the formation of membranes, was already successfully tested within today's hot springs in Yellowstone, Kamchatka, and at other sides \citep[e.g.,][]{Joshi2017,Milshteyn2018,Deamer2019,Deamer2021}. Since life's origin, however, minerals within such hot springs evolved significantly through diverse chemical alterations such as oxidation during and after the GOE \citep[see, e.g.,][]{Hazen2002}. It therefore remains to be shown that conditions at early Earth could have also synthesized similar organic compounds \citep{Longo2020}.

Besides wet-dry cycles, rock-air, water-air, and water-rock interactions -- of which only the latter is shared with alkaline hydrothermal vents -- provide further potential advantages for the origin of life \citep[see, e.g.,][for a comprehensive list]{VanKranendonk2021}. Hot springs and other subaerial settings can concentrate life's building blocks, such as a wide variety of amino acids \citep[e.g.,][]{Kvenvolden1970,Cronin1983,Martins2008,Pearce2017} that are delivered by meteorites, interplanetary dust particles, or even generated by the interaction of the atmosphere with frequent SEP impacts \citep{Kobayashi2023}. Other prebiotic elements \citep[e.g.,][]{VanKranendonk2021b} and molecules such as HCN \citep[][]{Pearce2022}, phosphate \citep{Toner2020} or borate minerals \citep{Grew2011} can be concentrated and provided through chemical weathering and reactions in the early atmosphere. The evolving protocells can be exchanged between individual pools of various chemical environments that may contain different organic compounds to drive novel prebiotic chemical reactions \citep[e.g.,][]{Deamer2019}. Hot springs further produce a wide variety of mineral interfaces important for the formation of oligomers \citep[e.g.,][]{Ferris2005}, and support the formation of membranous compartments \citep[e.g.,][]{Hargreaves1977,Deamer2002}.

There are potential limitations to the hot spring hypothesis as well \citep[see, e.g.,][]{Longo2020,Russell2021} such as `the phosphate problem' \citep{Pasek2017}, i.e., the bio-availability of phosphate, which is assumed to be low within hot springs. However, it could have been abundant in carbonate-rich lakes \citep{Toner2020}, and it was recently shown that lightning can form prebiotic, reactive phosphorus that subsequently concentrates on existing landmasses \citep{Hess2021}.  While metabolism may have been possible at hydrothermal vents through some form of the Wood-Ljungdahl pathway \citep{Martin2007,Sojo2016}, no metabolic mechanism was yet shown to work in hot springs. \citet{Longo2020} further point out that most subaerial hydrothermal pools lack high amounts of trace metals whereas they are abundant at submarine vents. \citet{Miyakawa2002} further argued that HCN and formaldehyde cannot be efficiently formed in warm water, which could favor a cold origin of life on different subaerial settings. Early Earth may have also lacked continents \citep[e.g.,][]{Johnson2020}, was likely bathed in significant amounts of high energy solar radiation \citep{Cnossen2007} and potentially bombarded by frequent impacts \citep{Abramov2009} during the so-called Late Heavy Bombardment (LHB) as late as $\sim$3.8-4.0\,Ga \citep{Bottke2017}. Since the LHB likely turns out to be just an artifact \citep[e.g.,][]{Boehnke2016,Brasser2020,Zellner2017,Hartmann2019}, however, life could have even emerged as early as $\sim$4.36\,Ga \citep{Benner2020}. The high energy radiation from the young Sun might also have been less dramatic for early life than initially thought as was demonstrated by \citet{Ranjan2017}, who found that prebiotic molecules should have been shielded from high solar activity even for relatively low CO$_2$ partial pressures. Instead, UV light could have provided another energetic pathway for the production of prebiotic compounds for the origin of life \citep[e.g.,][]{Rimmer2021}.

All in all, the origin of life is an active and lively field of research, and more experiments are needed to shed light on the question of whether life can emerge without subaerial land or not. In recent years, however, focus slightly shifted from alkaline hydrothermal vents to hot springs and related `warm little pond' theories relying on wet-dry cycles and hence on water and land \citep[e.g.,][]{Damer2020,VanKranendonk2021b,Deamer2022}.

\vspace{\baselineskip}
\vspace{\baselineskip}

But did the Earth even have water, subaerial land, and an atmosphere at the origin of life and early on in the Hadean and Archean eons? \citet{Korenaga2021} provides a comprehensive review of the current state of research and argues for land having been present throughout the entire history of the Earth. Likely, Earth never has been a water world. The very early ocean probably has been shallow with the water column roughly growing from initially $<$1\,km to $\sim$6\,km towards the end of the Hadean eon due to a steady degassing of \ce{H2O} -- a process that resulted in a decline of the exposed continents \citep{Korenaga2021}. While a maximum exposure of subaerial continents of $\sim$20\% was likely present at the mid-Hadean eon \citep{Guo2020}, it seems reasonable that no continental land existed in the early Archean eon \citep{Bada2018}. However, as \citet{Korenaga2021} points out, it is not only continents that can provide areas of exposed land. Hotspot islands and subaerial oceanic plateaus should have been more abundant during the Hadean eon than at present due to an increased flux of mantle plumes driven by a hotter core \citep[e.g.,][]{ORourke2017}. Impact-induced topography likely provided another source of subaerial land since massive meteorites must have produced craters with rims of several kilometers in height \citep{Collins2005} on a frequent basis. It is therefore likely that all these processes taken together provided subaerial land over the entire history of the Earth, even though the land-water fraction might have been relatively low during the late Hadean and early Archean eons.

So, is Earth an outlier or are planets that provide both surface water together with subaerial land at the same time frequent in the Galaxy? This will be discussed in the following appended section.

\subsection{The frequency of planets with subaerial land and oceans}\label{sec:h2ofreq}

Planet formation models that include the delivery of water suggest that the H$_2$O content can vary by orders of magnitude but also that water worlds \citep[e.g.,][]{Raymond2004,Ciesla2015,Mulders2015,Tian2015,Alibert2017,Zain2018,Bitsch2019} and, for M stars in particular, dessicated planets \citep[e.g.,][]{Lissauer2007,Raymond2007,Tian2015} are common phenomena. Planets with water contents similar to that of the Earth, on the other hand, seem to be relatively rare \citep[e.g.,][]{Tian2015,Zain2018,Simpson2017,Lingam2019WaterFrac}. Such bimodal distribution is governed by an interplay of different water delivery and removal processes as briefly outlined in the following.

Water can be delivered to the planet through planetesimal \citep[e.g.,][]{Raymond2004,Ciesla2015,Ronco2018,Sanchez2018,Zain2018} and pebble accretion \citep[e.g.,][]{Sato2016,Bitsch2019,Ida2019,Schoonenberg2019,Liu2020} either in-situ within the HZ or through migration from beyond the snow line. It can also be produced directly from stellar nebula gas \citep[see, e.g.,][]{Ikoma2006,Zain2018,Kimura2020,Lammer2021,Kimura2022} through chemical reactions between the accreted hydrogen and the underlying magma ocean that form H$_2$O. Planets that grow to masses of $M_{\rm pl} \geq 0.65$\,M$_{\odot}$ in the stellar nebula may produce more than one and, potentially up to 10, additional Earth oceans (EOs) from the accreted H$_2$ \citep{Lammer2021}. Rapidly growing, and particularly massive, planets might therefore end up not only with a thick hydrogen-envelope but also as water worlds.

There are, however, counteracting processes that may reduce a planet's bulk and surface water content. Early hydrodynamic escape can lead to significant water loss \citep[e.g.,][]{Ramirez2014,Luger2015,Tian2015,Johnstone2020Water} and it was shown that planets in the HZ of M stars such as the Trappist-1 planets may be able to lose several Earth oceans into space \citep[e.g.,][]{Luger2015,Bolmont2017,Bourrier2017,Moore2024}. At the same time, however, these planets may also accrete extensive amounts of water mass fractions on the order of $x_{\rm H_2O}\sim$10\% \citep{Schoonenberg2019}, which corresponds to roughly 300 EOs for an Earth-sized planet (excluding the core). \citet{Tian2015}, who modeled planet growth together with water delivery and atmospheric escape, have shown that these intertwined processes result in a bimodal distribution with planets being either extremely dry (defined through water mass fractions $x_{\rm H_2O}<5 \times 10^{-9}$) or extremely water-rich ($x_{\rm H_2O}>0.01$), at least for M and K stars. HZ planets around G dwarfs, on the other hand, showed a relatively smooth distribution from very dry to water-rich.

Another process that may significantly affect the amount of surficial water is the partitioning of H$_2$O between mantle and surface \citep[e.g.,][]{Cowan2014,Komacek2016,Dorn2021,Guimond2022b,Miyazaki2022,Moore2024}. \citet{Cowan2014} concluded that a tectonically active planet `of any mass' can have land together with oceans for water fractions of less than $w_{\rm H_2O}\sim0.2$\% since substantial amounts of water can be stored in the interior. \citet{Komacek2016} further concluded that it will need $\gtrsim0.3$\% to have entirely inundated surfaces at a planet if volatile cycling is only dependent on temperature or seafloor pressure. If, however, degassing is mostly dependent on seafloor pressure and ingassing mainly on mantle temperature, these authors found that super-Earths may already be water worlds for water fractions similar to the Earth's. \citet{Guimond2022b} further found that for tectonically active planets the maximum mantle-water mass fraction decreases with planetary mass roughly by $M_{\rm pl}^{-0.23}$ whereas the whole-mantle water capacity scales with $\varpropto M_{\rm pl}^{0.69}$, thereby reaching an upper limit of 1-2\,EOs for $M_{\rm pl} = 0.3-3.0\,$M$_{\oplus}$. Since the maximum surface water mass fraction further decreases with $M_{\rm pl}^{-0.75}$, \citet{Guimond2022a,Guimond2022b} {speculate} that land and oceans may be less likely on larger planets than on smaller ones {since the water in excess of the planet's solid-mantle water capacity will either form deep oceans, a thick steam atmosphere, or escape into space}. {This speculation is in agreement with a very recent paper by \citet{Moore2024}, who finds that an Earth-like surface with oceans and exposed continents is less likely on super-Earth's (with $M_{\rm pl}=8$\,M$_{\oplus}$) than on Earth-mass planets, as more water becomes sequestered in the interior for larger planets with super-Earths mostly ending up either as desiccated planets or water worlds.} For stagnant-lid planets, \citet{Guimond2022a} even derived that a water mass fraction of $w_{\rm H_2O}<10^{-4}$ (that is, $\sim$0.5 EOs for an Earth-mass planet) may already be sufficient to completely inundate any land on an Earth-sized planet. Note that all these upper limits for water storage within the mantle are well below the value of 1\%  as used by \citet{Tian2015} as an upper threshold for planets with Earth-like water contents.

A problem that may emerge for planets within the HZ(CL) is the relative amount of early degassing of H$_2$O and CO$_2$. Based on planetary C/H and H$_2$O/CO$_2$ ratios, \citet{Bower2022} and \citet{Miyazaki2022} point out that degassing from the solidifying magma ocean may be strongly inhibited for C/H$>$0.1 \citep{Bower2022} and H$_2$O/CO$_2<$0.4 \citep{Miyazaki2022}, respectively. \citet{Miyazaki2022} further argue that planets with such inefficient early degassing may evolve into a stagnant-lid planet due to the missing surface water, thereby likely hindering efficient outgassing of water over its entire lifetime for an initial absorbed stellar surface flux of $S_{\rm eff} > 240$\,W\,m$^{-2}$. Planets with such ratios might therefore evolve into dry and uninhabitable bodies, even if the water content is similar to Earth's.

There will further be a minimum initial H$_2$O concentration within the solidifying mantle for an ocean to form on a planet's surface. \citet{Salvador2017} calculate for the Earth that the initially outgassed H$_2$O content must have been $\gtrsim 0.09$ EOs for a CO$_2$ concentration within the mantle of 0.06\%, a typical estimate for the CO$_2$ mass fraction within the BSE \citep[e.g.,][]{Dasgupta2010}. \citet{Miyazaki2022}, however, argue that an even higher water content might be needed by also taking into account volatiles trapped in the pore space of the melt-solid mixture. These authors \citep[see Fig.~3c of][]{Miyazaki2022} calculate that for CO$_2$ concentrations between $w_{\rm CO_2}\sim 0.013-0.065\%$, the Earth would have needed an initial H$_2$O concentration of $w_{\rm H_2O}\gtrsim 0.008-0.06\%$ (that is, $\sim 0.2-1.8$ EOs) for the ocean to immediately condense after outgassing. However, for the water to not condense during the first 5\,Gyr at Earth's orbit \citep[see Fig.~11b of][]{Miyazaki2022}, the initial water concentration could have been lower still, roughly between about $w_{\rm H_2O}\sim 0.003-0.015\%$, i.e., 0.1-0.4 EOs, which is still well above the lower threshold used by \citet{Tian2015} of $w_{\rm H_2O} = 5\times 10^{-7}\%$.

Based on these various studies, a reasonable range of water concentrations that may allow for the simultaneous existence of water and land at the surface might therefore optimistically be estimated to be $w_{\rm H_2O}=\sim$0.003-0.2\%. For comparison, \citet{Lingam2019WaterFrac} estimated such range {to be 20 times narrower}, i.e., $w_{\rm H_2O}=0.048-0.15$\%. A very recent study by \citet{Stern2024} estimates the surface water mass fraction that still allows for subaerial land and oceans to be $w_{\rm H_2O}=0.007-0.027$\%. If one adds to this the mantle water mass fraction of $w_{\rm H_2O}=0.008-0.08$\% \citep[from][]{Lecuyer1998} these authors state in their study, this amounts to a total BSE water mass fraction of $w_{\rm H_2O}=0.015-0.127$\%, which is also clearly narrower by about an order of magnitude. {By taking into account the partitioning of water into the mantle and diffusion- and energy-limited escape, another recent study by \citet{Moore2024} found that a planet with a mass of $M_{\rm pl}= 1.0$\,M$_{\oplus}$ in the HZ of a low-mass M dwarf can end up with oceans and exposed continents for an initial water content of 3-8\,EO, that is, $w_{\rm H_2O}\sim0.1 - 0.3\%$. This range is again much narrower than ours but gives a higher upper limit, which is due to the strong water loss into space such a planet experiencea around late-type M dwarfs. For higher stellar masses, the upper limit can hence be considered to be lower\footnote{Note, again, that also \citet{Moore2024} take $R_{\rm XUV}$ in the energy-limited equation to be equal with $R_{\rm pl}$ (see Sections~\ref{sec:bulge} and \ref{sec:h2oImp}), which will consequently underestimate the loss of water. Around late-type M dwarfs one may hence expect that the initial water mass fraction needed could even be higher than found by \citet{Moore2024}.}.}

To set these values in perspective, one Earth ocean, including the entire hydrosphere (i.e., all rivers, surface ices, the atmosphere and the oceans) is equivalent to $1.4 \times 10^{21}$\,kg \citep[e.g.,][]{Lecuyer1998,Bodnar2013,Peslier2017}, which amounts to a fraction of $\sim2\times 10^{-4}$ of the entire mass of the Earth; another $\sim 3 \times 10^{20}$\,kg can be found in the crust \citep{Lecuyer1998}. However, Earth also stores water within its mantle even though the exact content is not very well established. While \citet{Hirschmann2009} derived a range of 0.2-1.6 EOs through Monte Carlo simulations based on the mantle's H/C ratio, \citet{Marty2012} estimated a significantly higher value of 10$\pm$5 EOs. As pointed out by \citet{Korenaga2017}, however, the latter estimate may be problematic due to some debatable elemental ratios upon which the result of\citet{Marty2012} was established. In addition, water may have also been sequestered into the core but its content is very badly constrained \citep[e.g.,][]{Korenaga2017}. Excluding the core, the whole water mass fraction of the BSE (bulk silicate Earth) can therefore, based on the estimate by \citet{Hirschmann2009}, be expected to be in the range of $1.4-2.8$ Earth oceans, which amounts to roughly 0.05-0.1\% of the entire BSE, a value that is well within $\sim$0.003-0.2\%. Note, however, that the review on the Earth's water content by \citet{Peslier2017} lists a broader possible range of $\sim$0.04-0.25\% with an average of $\sim$0.15\%, based on different literature values.

So, how big might the fraction of exoplanets with H$_2$O mass fractions between $w_{\rm H_2O}\sim$0.003-0.2\% be? \citet{Raymond2004} simulated terrestrial planet formation and water delivery for the solar system through oligarchic growth of planetary embryos and planetesimals as a function of giant planet mass, its position and eccentricity, the position of the snow line, and the solid density within the nebula. These authors formed 43 planets that were located between 0.8 and 1.5\,AU out of which 4 were completely dry. An additional 11 showed water contents equating to a water fraction of $\leq$0.2\% for the entire planet. The rest of the bodies showed water contents that were above such threshold. However, the model by \citet{Raymond2004} did neither include planetary migration nor atmospheric escape.

\citet{Mulders2015}, who similarly did not include migration and escape, also studied planet formation and water delivery but for different stellar masses. For a solar-mass star, \citet{Mulders2015} obtained a fraction of $\sim$25\% of all HZ planets with a water concentration that is between the thresholds chosen above, i.e., between 0.003-0.2\%. This value increases towards $\sim 40\%$ for $M_{\star} = 0.6\,$M$_{\odot}$, but declines steeply for even smaller stellar masses with $\sim4$\% and $<$1\% for 0.4\,M$_{\odot}$ and 0.2\,M$_{\odot}$, respectively \citep[see Fig.~4 in ][]{Mulders2015}. However, these authors point out that, while the effect of hydrodynamic escape on the final water contents of Earth-sized planets needs to be studied further, the water concentration of their planets may nevertheless be considered to be lower limits due to planetary migration being not included in their simulations.

The above mentioned study by \citet{Tian2015} included hydrodynamic escape and planetary migration to obtain statistics on the water content of the final planets within the HZ of stars with $M_{\star}=\{1.0, 0.5, 0.3\}\,$M$_{\odot}$. For their main scenario (green bars in their Fig.~2), for which the stellar XUV surface flux scales to the bolometric luminosity with $L_{\rm XUV}/L_{\rm bol} = 10^{-4}$ and the heating efficiency was assumed to be $\epsilon = 10\%$, 91 and 45 of a total of 407 Earth-mass planets in the HZ of the solar-like star were ocean and dune planets, respectively. The remaining 271 bodies had a water content between $5\times 10^{-7}\%$ and 1\% by mass, a range considered to be `Earth-like' by \citet{Tian2015}.

While the distribution of water concentration for $M_{\star}=1.0$\,M$_{\odot}$ was yet continuous, \citet{Tian2015} found a clear bimodal distribution for 0.5 and 0.3\,M$_{\odot}$ with only 12 out of 292 and just 1 out of 55 planets being within their `Earth-like' water content range. For the other two scenarios, that is, $L_{\rm XUV}/L_{\rm bol} = 10^{-3}$ with $\epsilon = 10\%$ (red bars in their Fig.~2), and $L_{\rm XUV}/L_{\rm bol} = 10^{-5}$ with $\epsilon = 1\%$  (blue bars in their Fig.~2), the final outcomes were similar with values remaining the same for their red, and slightly increasing for their blue case.

However, the water range chosen by \citet{Tian2015} is significantly higher than the range in which a planet may have oceans and continents at the same time as assumed in our model (i.e., 0.003\%-0.2\%). By taking these tighter bounds, one can see in Fig.~4 of \citet{Tian2015} that for the red and green scenarios only 74, five, and zero planets remain for 1.0\,M$_{\odot}$, 0.5\,M$_{\odot}$, and 0.3\,M$_{\odot}$, respectively, which equates to fractions of $\sim$18\%, $\sim2$\% and 0\%. For the blue scenario, the values increase slightly towards 76, 8, and 2 planets that end up with the potentially `right' amount of water.

\citet{Kimura2022} similarly modeled planet formation by including migration, volatile delivery, and atmospheric escape from M dwarfs with $M_{\star} = \{0.1, 0.3, 0.5\}\,$M$_{\odot}$, but also implemented the formation of water from a planet's primordial atmosphere. They found higher occurrence rates for planets with Earth-like water mass fractions than \citet{Tian2015} since the additional H$_2$O prevented some of the planets from drying out completely. While \citet{Tian2015} simulated planets between $M_{\rm pl}=0.1-10.0$\,M$_{\oplus}$, \citet{Kimura2022} restricted themselves to masses between $M_{\rm pl}=0.1-3.0$\,M$_{\oplus}$ and provided additional statistics for Earth-mass bodies between $M_{\rm pl}=0.7-1.3$\,R$_{\oplus}$. Within this mass range, \citet{Kimura2022} found that 75\% of the planets have water mass fractions below 100 times that of the Earth's. Out of these, about 5\% and 10\% for stellar masses of $M_{\star}=0.3$\,M$_{\odot}$ and $M_{\star}=0.5$\,M$_{\odot}$, respectively, had H$_2$O contents above 0.001\% and were therefore assumed habitable by the authors. However, 100 times the Earth's water fraction, i.e., a water content of 2.27\% is clearly above our estimated upper limit of 0.2\% while 0.001\% is slightly below our lower limit of 0.003\%.

Besides, \citet{Kimura2022} provide further statistics on their entire set of planets between $M_{\rm pl}=0.1-3.0$\,M$_{\oplus}$. As can be seen in Fig.~3 of their study (lower panel), between $\sim$2.15\%, in case that no water forms from the primordial atmosphere (i.e, $X_{\rm H_2O} = 0$), and $\sim$4.3\%, in case that the fraction of water to hydrogen in the atmosphere reaches a value of $X_{\rm H_2O} = 0.8$, end up with a water fraction between roughly 0.003\% and 0.2\%, in case that the host star holds a mass of $M_{\star}=0.5$\,M$_{\odot}$. For $M_{\star}=0.3$\,M$_{\odot}$ (middle panel) and $M_{\star}=0.1$\,M$_{\odot}$ (upper panel), \citet{Kimura2022} find occurrence rates of $\sim$0.1-1.3\% and $\sim$0.15 - 0.3\%, respectively. These values are mostly slightly above the values of \citet{Tian2015} and again illustrate the trend that occurrence rates of planets with both oceans and subaerial land decreases toward later spectral types.

Several other planet formation studies investigated the water content of HZ planets based on planet formation models. Even though none of these included planetary migration together with atmospheric escape at the same time, several other parameters were investigated that may alter the final water content. The outcomes of these diverse studies were broadly similar by only forming a low amount of planets within the proposed range of water content.

\citet{Ciesla2015}, for instance, modeled how planetesimals resembling the high water content of comets (instead of the usually implemented carbonaceous chondrite composition) affect volatile amounts. In total, only 3 out of 48 HZ planets fell into such range; for solar-like stars, each of the 13 planets had a water mass fraction above $0.2\%$, while just 3 out of 29 planets had the right amount of H$_2$O for stars with 0.8\,M$_{\odot}$ and 0.6\,M$_{\odot}$. The remaining 6 planets were located in the HZ of stars with 0.4\,M$_{\odot}$ and 0.2\,M$_{\odot}$, and showed a water concentration below 0.003\%. HZ planets around low-mass stars may therefore still turn out to be mostly dry, even if the amount of water in the disk increases.

For solar-like stars, \citet{Sato2016} investigated how disk size and turbulence might influence the accretion of water-rich pebbles onto terrestrial embryos around solar-like stars and found that only compact disks ($<$100\,AU), in which turbulence at 1\,AU is strong and the snow line arrives at the HZ later than 2-4 Myr allow for water fractions $<0.23$\%. In a follow-up study, \citet{Ida2019} found that compact disks in the order of 30-50\,AU may lead to planets with Earth-like water fractions, depending on the mass of icy material that still resides in the outer disk at the time the snow line passes the planet. In total, \citet{Ida2019} formed 3 out of 14 HZ planets with water fractions in the range of 0.003\%-0.2\%.

\citet{Zain2018} further investigated different dynamical planet formation and migration scenarios with giant planets around solar-like stars. These authors found that most scenarios end up with water-rich bodies in the HZ. In total, no planet out of 27 that finally came to reside within the HZ showed a water mass fraction between 0.003\% and 0.2\%. These authors therefore conclude that the formation of planets with Earth-like water contents might be rare events, independent from the respective dynamical scenario. Generally, the formation of water-rich planets seems to be more common than the formation of dry ones \citep[e.g.,][]{Ronco2018}.

Besides planet formation models, one could also try to infer statistics on the frequency of dry and water worlds through Bayesian evidence. \citet{Simpson2017} defined such a model and found strong evidence that the Earth with its land and ocean coverage presents an evolutionary selection bias. Due to our planet's close proximity to the water world limit, \citet{Simpson2017} infers that most of the habitable planets shall have more than 90\% of their surface covered with water and that the clear majority of planets have to be water worlds. As can be seen in Fig.~5 of \citet{Simpson2017}, depending on the chosen parameters, between $\sim$8-40\% of all worlds are not entirely covered by ocean. Dry worlds, on the other hand, are rare within the Bayesian model developed by \citet{Simpson2017}.

Another estimate on the frequency of planets that are covered by ocean and land was put forward by \citet{Lingam2019WaterFrac} who approximated that a water fraction between 0.048-0.15\% would allow for the simultaneous presence of both oceans and continents. By assuming a uniform and a log-uniform distribution of the water concentration over the entire population of HZ planets, they roughly estimated the fraction of worlds to have both to be $\sim2 \times 10^{-3}$ and $\sim$0.1, respectively.

This short literature review illustrates that at least from the point of view of our current understanding of planet formation and atmospheric escape, terrestrial bodies with both oceans and subaerial land may be relatively rare. It, moreover, seems that G and early K-type stars may harbor the best conditions for such worlds to be formed. 
    \section{B$_{\rm env}^{\rm moon}$: The importance and occurrence rate of large moons}\label{sec:appMoon}

\subsection{The potential importance of large moons}\label{sec:moon1}

The classic argument for the importance of a large moon \citep[see, e.g.,][]{Waltham2019} is its potential role in stabilizing Earth's obliquity \citep[e.g.,][]{Laskar1993,Laskar1993b} and hence its climate \citep[e.g.,][]{Spiegel2009,Ferreira2014}. However, recent studies shed doubt on this argument and showed that the Earth may only exhibit modest variations in obliquity on relatively large timescales \citep[e.g.,][]{Lissauer2012,Li2014}. Studies on Mars \citep{Laskar2004} and exoplanets excluding a substantial moon \citep[e.g.,][]{Shan2018,Quarles2020}, however, also indicate that the obliquity of such planets can potentially vary substantially depending on different system parameters such as the initial obliquity of the body, orbital architecture, gas giants within the system, or planetary rotation rate. Studies by \citet{Quillen2018} and \citet{Podvigina2022} further found that a large moon can have either stabilizing or destabilizing effects on its host planet, again strongly depending on the initial conditions. Even for the Earth, it was pointed out that its axis may become unstable in about 1.5\,Gyr when Earth's precession rate will fall below 26\,''/year due to its gravitational interaction with the Moon, a process that would happen even earlier if the Moon would be heavier \citep{Ward1982,Waltham2019} and would have already happened for the Moon's mass being larger by just a few \citep{Waltham2004} to a maximum of 50\% \citep{Brasser2013}.

However, even if so, strong variations in obliquity might not necessarily lead to unstable climates and may even expand the outer region of the classical habitable zone \citep{Armstrong2014}. For smaller stars, the stellar host may even substitute the potential stabilizing effect of a large moon \citep[e.g.,][]{Waltham2019}. Finally, \citet{Brasser2013} point out that only about 14\% of all systems evolve towards an Earth-like system in their simulations with the others likely being less habitable.

Whether a large moon is therefore indeed needed for preventing strong obliquity variations and stabilizing the climate will critically depend on the initial conditions of the entire stellar system. {Additionally, we point out that life could in principle even evolve to be highly resilient against substantial obliquity variations, thereby weakening this argument further \citep[see also,][]{ Cockell2020Astrobiology}.  Based on such line of reasoning alone}, one can thus not conclude that a large moon serves as a necessary requirement for the evolution of EHs.

In recent years, however, several different additional arguments in favor of a large moon were put forward. Even though future studies may be needed to assess some of these aspects in greater detail, it seems at least likely that Earth would have evolved very differently, if our lunar companion would never have formed. We briefly summarize each of these additional arguments in the following.

\vspace{\baselineskip}
\textbf{\textsl{Tides aided the origin of life:}}

The moon was much closer to the Earth at the time of the origin of life, which resulted in higher and more frequent tides, as the recession of the moon slowed down the Earth's rotation toward a period of 24 hours at the present day. These conditions may have resulted in the production of lagoons and tidal pools on subaerial land with periodic tidal flooding and drying at a higher frequency than today \citep{Lathe2004}. This could have supported nutrient concentration and the production of DNA-like polymers \citep[][]{Lathe2004} with polymerization taking place during the evaporation and dissociation during the flooding phase \citep{Lathe2005}.

However, several studies placed different criticisms against the importance of tidally induced wet-dry cycles at early Earth. These are, (i) a potentially longer rotational period of the Earth at around $\sim$4\,Ga \citep{Varga2006} than the initially assumed 2-6 hours by \citet{Lathe2004}; (ii) the potential absence of subaerial land \citep[see, e.g.,][]{Lingam2018}; and (iii) that tidal cycling may not work properly without the presence of complex RNA secondary structures to reduce the ratio between DNA elongation and replication during potential tidal chain reactions \citep{Fernando2007}. However, the length of day (LOD) at the origin of life may likely have been shorter \citep{Lathe2006} than argued by \citep{Varga2006}, subaerial land presumably has been present \citep{Korenaga2021}, and more recent studies indicate that wet-dry cycles, in general, may indeed be crucial for the origin and early evolution of progenotes and first forms of primitive life \citep[e.g.,][]{Damer2015,Ross2016,Damer2020} - see also Section~\ref{sec:h2o}. Moreover, \citet{Lingam2018} point out that, even if tides were not crucial for the origin of life at Earth, it may nevertheless be of importance for its origin elsewhere in the Universe. However, tides around low-mass stars can also be initiated by the stellar host itself \citep[e.g.,][]{Lingam2018}{ and that planet-planet tides in compact multi-planet systems can themselves be quite substantial \citep[e.g.,][]{Wright2018Tides}}.

\vspace{\baselineskip}
\textbf{\textsl{Heating Earth's early climate:}}

The Sun's bolometric luminosity was only about 70\% of its present-day value during the Hadean and early Archean eons and steadily increased since then \citep[e.g.,][]{Baraffe2015}. Before $\sim$3.5\,Ga, the Earth should therefore have obtained too little energy to keep the surface from freezing \citep[e.g.,][]{SAgan1972,Kasting1988}, but liquid water existed as early as 4.4\,Ga \citep{Mojzsis2001} and no traces of surface glaciation were yet found before $\sim$2.9\,Ga \citep[][]{Young1998,VanKranendonk2012} -- a conundrum that is generally known as the `Faint Young Sun Paradox' or FYSP \citep[e.g.,][]{Sagan1997,Feulner2012}. Even though recent simulations indicate that an increased partial pressure of \ce{CO2} may have been sufficient to resolve the paradox \citep[e.g.,][]{Charnay2020}, different other factors may have contributed to avoiding a snowball Earth in the Hadean and Archean eons, which may be crucial at exoplanets as well. \citet{Heller2021} suggested that tidally induced thermal heating due to the interaction between the Earth and the newly formed Moon may have been one of these contributing factors. These authors find that the Earth-Moon system already lost 99\% of its spin-orbit energy budget, mostly through tidal heating, a process that could have increased Earth's surface temperature early-on by up to 5\,K and, as pointed out by \citet{Heller2021}, might have additionally aided hydrothermal fluid circulation in the early crust, which could explain the presence of hydrothermal chert veins at the Dresser formation at $\sim$3.49\,Ga.

\vspace{\baselineskip}
\textbf{\textsl{Aiding the rise of atmospheric O$_2$:}}

Earth's atmosphere O$_2$ showed a broad step-wise increase in partial pressure, during the GOE at $\sim$2.4\,Ga and the Neoproterozoic Oxygenation Event (NOE) at about 0.6\,Ga \citep{Och2012}. Without the increase of $p$O$_2$ toward present-day levels, complex animal life would not be possible, and sufficient oxygenation can be expected to be a necessary requirement at any exoplanet for complex life to evolve \citep[e.g.,][]{Catling2005,Canfield2007,Lyons2014,Reinhard2016Oxygen,Zhang2016Oxygen}. The main reasons and mechanisms for this sudden increase in atmospheric O$_2$ are yet debated and it may be an interrelation between biological, geochemical, and tectonic activity \citep[e.g.,][]{Och2012}. \citet{Klatt2021}, however, recently suggested an additional mechanism that is directly related to the existence of the Moon, i.e., the tidally driven decrease in our planet's rotation rate. These authors suggest that a longer length of day increases the net production of O$_2$ by Proterozoic cyanobacterial mat analogs. Even though the gross photosynthetic production is only dependent on the entire photon flux and thus independent of the LOD, the diel carbon burial - and the therewith connected net production of O$_2$ - additionally depends on the length of the daily bacterial activity cycle as longer days can build up steeper O$_2$ gradients and fluxes within the mat, which boosts carbon burial and a related increase in $p$O$_2$.

As \citet{Klatt2021} further argue, both GOE and NOE could have been directly related to the increase in LOD through lunar-induced tidal friction. During the mid-Proterozoic, the so-called `boring billion years' with no increase in pO$_2$, Earth's deceleration may have been the lowest in its entire history based on continental configuration \citep{Meyers2018} or could have even been constant due to resonant atmospheric thermal tides implying no change in LOD at all \citep{Zahnle1987,Bartlett2016,Mitchell2023DayLength,Wu2013DayLength}. \citet{Klatt2021} were able to combine their O$_2$-production model with a model of Earth's deceleration by \citet{Bartlett2016} and showed that GOE and NOE as well as the Palaeozoic Oxidation Event (POE) at $\sim$0.35\,Ga \citep[e.g.,][]{Krause2018} could have indeed been related to a change in day length. A more recent study by \citet{Farhat2022} further found that the Moon likely encountered resonances in oceanic dissipation that were accompanied with rapid changes of day length at $\sim$0.35\,Ga and $\sim$0.55\,Ga coinciding with NOE and POE, respectively. Both oxygenation events themselves are further correlated with the Cambrian explosion \citep[e.g.,][]{Wood2019} and the Great Ordovician Biodiversity Event \citep[e.g.,][]{Edwards2017}, respectively.

Based on such argument, Earth's large Moon may have been an important trigger for the oxygenation of the planet and the therewith connected origin and evolution of complex life. However, we should note a few caveats here. The results by \citet{Klatt2021} depend on step-wise jumps of the length of day and it is not entirely clear whether the increase in LOD was indeed due to jumps or more gradual, although \citet{Klatt2021} point out that the net productivity would have also increased for a more gradual increase in LOD. Other sources may also lead to a jump in $p$O$_2$ such as H$_2$ escape, continental phosphorus supply and continental growth \citep[see,][and references therein]{Klatt2021}. These authors therefore emphasize that the exact magnitude of the day length effect remains uncertain since they did not consider other potential effects on oxygen production and since their model is relying on several as of yet uncertain assumptions (e.g., the relationship between diel and long-term burial efficiency, and the pattern of LOD change). {In addition, \citet{Klatt2021} base their results on the investigation of cyanobacterial mats whose similarity to Precambrian analogs cannot be regarded as fully established.}

At exoplanets, large satellites may well play a similar role than on Earth, but planetary bodies can also experience tidal friction induced by their host stars. While such host star-induced process will be less important for K, G, and F stars, its effect may be significant for M dwarfs. In such a case, the host star could substitute the role of a large moon and potentially support the oxygenation of a planet. However, for the Earth it took $>$2\,Gyr for the GOE and $\sim$4\,Gyr for the NOE to take place while low-mass M dwarfs may synchronize HZ planets within 1\,Gyr or less \citep{Barnes2017}. For these stars, oxygenation would have to occur much faster than on Earth (with the caveat of increasingly higher $F_{\rm XUV}$ and $F_{\rm X}$ values towards the past) or rely on completely different processes.

\vspace{\baselineskip}
\textbf{\textsl{Supporting availability/transport of nutrients:}}

Oceanic tides, induced by large moons and, potentially even more so, by M dwarfs \citep{Lingam2018}, are together with winds one of the main drivers of efficient ocean mixing \citep[e.g.,][]{Munk1998,Wunsch2004}. For the present Earth, \citet{Munk1998} found that about 3.7\,TW of power is available from tidal energy, out of which 3.2\,TW stem from the Moon while an additional 1\,TW of energy may be provided by winds. For mixing the abyssal ocean, these authors further calculate that about 2\,TW are needed to keep the global abyssal density distribution. Besides 1\,TW from global winds, \citet{Egbert2000} have afterward shown that indeed $1\pm0.25$\,TW from the entire tidal energy is dissipated in the deep ocean, thereby likely assuring its large-scale circulation. Similarly, \citet{Kuhlbrodt2007} found that winds and tides are integral for ocean overturning circulation to take place. Since ocean mixing is further very important for the distribution and upwelling of nutrients into the euphotic zone \citep{Sandstrom1984,Holligan1985,MacKinnon2003,Schafstall2010,Lingam2018,Tuerena2019}, this implies a potentially crucial role of the Earth's large Moon in promoting substantial nutrient availability and supporting Earth's biodiversity.

\citet{Lingam2018} further point out that tides also regulate coastal erosion and sediment movement, which in turn may increase biological activity; the tidal cycle could for instance supply shallow regions with sufficient nutrients to support photosynthesis. A recent study by \citet{Crawford2022} shows that the stronger tidal forcing on early Earth was a main driver of biogeochemical cycling by playing a key role in weathering and the transport of nutrients to shelf areas itself and, therein, from shallower to deeper parts of the shelf. Besides, strong tidal interactions at the time of the origin of life could have supported the opening of crustal fractures, thereby promoting hydrothermal activity, and connecting different chemical and mineralogical reservoirs through subsurface fluid circulation, which could have led to a mixture and exchange of different prebiotic molecules \citep{Stueeken2013}. It can therefore be argued that the existence of the Moon may have favored the origin and evolution of life, not only through the provision of tidal pools and wet-dry cycles as described above but also through enhancing hydrothermal activity \citep[see also,][]{Heller2021} and nutrient transport.

Here, we also emphasize that M dwarfs could provide tidal energy 1 to 2 orders of magnitudes above the one provided by the Moon-Sun system \citep{Lingam2018}, which could again substitute the role of a large moon, at least as long as the respective planets are orbiting asynchronously to allow for strong tidal friction \citep[see,][ ]{Lingam2018}, and as long as high tidal friction will not overheat the planet \citep{McIntyre2022}.

\vspace{\baselineskip}
\textbf{\textsl{Initiating/sustaining a long-term magnetic dynamo:}}

Earth has a strong intrinsic magnetic field while Venus has not. \citet{Jacobson2017} suggest that the reason for this distinction may lie within the mixing of their planets' cores. In their model, stratified cores suppress large-scale convection, which in turn prohibits the onset of a substantial magnetic dynamo. The cores of accreting protoplanets, however, will grow with stable compositional stratifications as long as no high-energetic giant impact that happens at a very late stage of accretion completely homogenizes their compositions \citep{Jacobson2017}. The giant impact that formed the Moon is generally believed to be such a high-energetic event \citep[e.g.,][]{Canup2012,Cuk2012,Lock2018} and it may have happened relatively late \citep[e.g.,][]{Jacobson2014}, which would be in agreement with Earth's outer core being exceptionally well mixed \citep{Mandea2012}. This would explain, as \citet{Jacobson2017} argue, the existence of the Earth's dynamo while its absence at Venus implies no high-energetic impact happened late in its accretion history.

However, even if such a giant impact may be needed for the initiation of a strong magnetic dynamo, this scenario does neither imply the formation and need for a large moon after the impact happened nor the persistence of the generated dynamo over billions of years. But the existence and maintenance of the Earth's intrinsic magnetic field from its earliest evidence during the Hadean eon \citep{Tarduno2015,Tarduno2020} until the present day is yet relatively poorly understood. The nucleation of the inner core can presently provide most of the convective power needed to drive the dynamo \citep[e.g.,][]{Landeau2022}. Prior to the start of nucleation at about 1\,Ga, however, the core heat-flow may have been below the conductive heat-flow, which would have inhibited the presence of a strong intrinsic magnetic field based on thermal convection alone, a problem called `the new core paradox' \citep{Olson2013}. While the Earth's precession may be too weak to provide enough energy to drive the dynamo \citep{Landeau2022}, tidal distortion by the Moon could potentially induce sufficient mechanical forcing to maintain the Earth's magnetic field, particularly prior to the nucleation of the inner core \citep{Andrault2016,Landeau2022}. As \citet{Andrault2016} point out, 0.5 to 1\,TW from the entire tidal energy of 3.7\,TW could be available for the outer core to drive the dynamo - a value that may be within the energy range needed for sustaining the intrinsic magnetic field \citep{Buffett2002,Christensen2004}. Similar processes may explain the long-gone dynamos of Mars \citep{ArkaniHamed2009} and the Moon itself \citep{Dwyer2011}.

However, as of now numerical models are not advanced enough to prove the concept of tidally-driven dynamos \citep{Landeau2022}. The interconnection between the Moon and the evolution of the Earth's magnetic field therefore remains relatively poorly understood, although this discussion gives some indication of its importance. Whether an intrinsic magnetic field in itself may actually be a necessary factor for an EH to evolve will be briefly discussed in Section~\ref{sec:mag}.

\vspace{\baselineskip}
\textbf{\textsl{Protecting an early atmosphere:}}

The early Moon likely had an intrinsic magnetic field to power a Lunar paleomagnetosphere at least from $\sim$4.2\,Ga \citep[e.g.,][]{Lawrence2008,GarrickBethell2009}. Its longevity is uncertain and highly debated with cessation being suggested to be as late as 3.56\,Ga \citep{Shea2012}, 1.5\,Ga \citep{Tikoo2017}, or even 1.0\,Ga \citep{Mighani2020}. A recent re-evaluation of Apollo samples \citep{Tarduno2021}, however, indicates that no long-lived dynamo was present after $\sim$4\,Ga, but that measured magnetizations of earlier ages may be attributed to a combination of shock and impact fields. As long as the magnetic field existed though, it may helped to protect Earth's early atmosphere \citep{Green2020,Green2021}. Simulations by \citet{Green2020} show that most of the interaction with the solar wind will be focused towards the Lunar magnetosphere if located on the dayside of the Earth. In this scenario, the cusp of the Earth's magnetic field will additionally move poleward, thereby reducing atmospheric outflow through open field lines. With the Moon on the nightside, the lunar magnetosphere interrupts the Earth's plasma sheet and hinders ions to escape through the magnetospheric tail. \citet{Green2020} therefore argue that both configurations may have reduced non-thermal escape from the early Earth, illustrating another potential advantage that a large satellite could offer for its host planet. Signs of such interaction, \citet{Green2020} argue, may yet be visible through terrestrial N$_2$ that was potentially implanted onto the Lunar surface \citep[see also,][]{Ozima2005}. In a follow-up study, \citet{Green2021} further extend their work to exomoons, but give no quantitative results how such configuration may enhance an exoplanet's habitability.


\vspace{\baselineskip}
\textbf{\textsl{Initiating/maintaining plate tectonics:}}

It was already suggested in the 1970s that lunar induced tides could be a main driver of plate tectonics \citep[e.g.,][]{Solomon1975,Richardson1976}, but the induced tidal stress may be too small to drive plate motion \citep{Ray2001}. However, as a complementary model to classic mantle convection, \citet{Riguzzi2010} suggested that the tidal torque and its respective deceleration of the Earth's rotation as initiated by the misalignment between the tidal bulge and the Earth-Moon gravitational alignment may provide sufficient energy to drive plate tectonics. \citet{Hofmeister2022} further argue that force imbalances in the Earth-Moon-Sun system could be the reason for large-scale tectonic motions and the underlying mechanism for plate tectonics. They suggest that Earth's oscillating barycenter -- a consequence of the lunar drift due to the Sun's pull on the system -- induces stress onto the lithosphere, which creates mid-ocean ridges and plate segments. The same mechanism may also induce the westward drift of the entire lithosphere \citep[e.g.,][]{ Carcaterra2018,Hofmeister2022,Zaccagnino2022} and lead to westward-dipping subduction slaps reaching sufficiently deeper into the mantle than eastward-dipping slaps \citep{Riguzzi2010,Doglioni2015}.

{However, as a caveat it must be noted that the calculations of \citet{Hofmeister2022} were performed for the present-day tidal forces in the Earth-Sun-Moon system. Plate tectonics, on the other hand, could have already originated as early as during the Hadean eon\footnote{However, it was also suggested that modern plate tectonics did not start prior to the Neoproterozoic while transient subduction was present earlier than that; \citet{ONeill2018Tectonics}.} \citep[e.g.,][]{Maruyama2018,Windley2021Tectonics,Chowdhury2023Tectonics}, a time when the acting forces were very different. That said,} if plate tectonics {indeed} rely on an oscillating barycenter, this will further explain why the Earth has tectonics whereas Mercury, Venus, and Mars do not (and potentially never had). It further warrants a deeper investigation of the Pluto-Charon-Sun system to evaluate whether some tectonic regime existed (or still exists) at these dwarf planets\footnote{{Interestingly, New Horizons found indications of tectonic features at Pluto with relatively little degradation, indicating that they may be relatively young \citep{Moore2016Pluto,Hammond2016Pluto}. However, a dual-synchronous state between Pluto and Charon was likely reached very early-on \citep{Barr2015Pluto}.}} \citep{Hofmeister2022}.

{If working,} the mechanism proposed by \citet{Hofmeister2022} can only evolve in planet-moon-star systems and cannot be applied to planets without a large satellite orbiting M dwarfs. However, some studies evaluated whether tectonics can be initiated in compact M dwarf systems due to tidal heating, as this can be significantly larger compared to the Moon-Earth system itself \citep{Jackson2008,Zanazzi2019,McIntyre2022}. As \citet{Zanazzi2019} have shown, tidal stress in such systems can become comparable to the stress induced by mantle convection onto the lithosphere, which could be an alternative driver of plate subduction (while planet-planet interactions may generally be too weak for the initiation of plate tectonics). However, \citet{McIntyre2022} found that only a small portion of 767 simulated tidally locked exoplanets may indeed fall within a tidal heating regime that allows for active tectonics over geological time scales. For most of their modeled bodies, tidal interaction with the host star is either too weak or leads to such strong volcanism that the \textit{Runaway Greenhouse} limit is reached. Similarly, \citet{Jackson2008} found a wide range of different outcomes due to tidal heating ranging from the initiation of plate tectonics to Io-like uninhabitable planets. They further point out that tidally driven degassing could be an option to replenish planetary atmospheres. However, we note that tidally heated planets with high degassing rates may become devoid of any volatiles before its host star’s XUV flux decreases below the critical threshold value since M dwarfs remain highly active for billions of years. Such planets may therefore remain uninhabitable even if tidal heating may have initiated tectonics and volcanic degassing.

Finally, we note that the initiation of plate tectonics may not have happened without the giant Moon-forming impact, as \citep{Waltham2019} points out, based on the same line of reasoning as raised by \citet{Jacobson2017} for the initiation of Earth's intrinsic magnetic field. The giant impact would have disrupted the density stratification in the Earth's mantle and without such a mixing, mantle convection, and hence the initiation of plate tectonics, may have been suppressed \citep{Waltham2019}.

\vspace{\baselineskip}
\textbf{\textsl{Delivery of water:}}

The Moon was formed by a giant impact late during Earth's accretion \citep[e.g.,][]{Canup2001,Canup2012,Lock2020}. Based on the molybdenum isotopic composition of the Earth, \citet{Budde2019} suggest that the impacting protoplanet Theia either had a purely carbonaceous chondritic (CC) or a mixed non-carbonaceous chondritic (NC)-CC composition. In either case, it had a high water concentration. Since Theia delivered up to 0.1\,M$_{\oplus}$ of CC-material, the Moon-forming impact could therefore account for up to the entire water budget of the Earth \citep{Budde2019}. Without this impact, our planet would have consequently remained relatively dry and potentially uninhabitable. However, even if this is the case for the Earth, one has to note that stochastic accretion within other systems may lead to entirely different evolutionary paths with planets accreting a proper amount of water even without a late giant impact.

\vspace{\baselineskip}
\textbf{\textsl{Triggering evolution of complex life/tetrapods:}}

The following argument does not relate to complex life in general but more specifically to the emergence of aquatic tetrapodomorphs and land-dwelling tetrapods, i.e., four-limbed vertebrate land animals from which all mammals on this planet evolved. This may hence be specifically important for the emergence of intelligent life.

It was \citet{Romer1933}, who first emphasized that the development of chiridian limbs of aquatic tetrapodomorphs may well be related to the periodic emergence and evaporation of shallow tidal pools. As the tide abated, fish got stranded within drying pools -- a recurring process that not only favors the evolution of air-breathing organs but also selects for extremities strong enough to allow movement on land. \citet{Balbus2014} points to the potential importance of tidal modulation due to the specific architecture of the Sun-Moon-Earth-system, which creates an entire network of isolated tidal pools with different evaporation times. This author therefore speculates that the comparable angular size of the Sun and the Moon may even constitute an anthropic bias as this may not be based on pure coincidence. Moreover, \citet{Balbus2014} showed through paleogeographic reconstructions that the distribution of land during the Devonian period, the time of the first emergence of tetrapods, was specifically conducive for producing a large range of tidal amplitudes. In a more recent study, \citet{Byrne2020} used tidal modelling to confirm large tidal ranges of over 4 meters that were present from the late Silurian, the time at which the earliest members of the osteichtyan group (bony fish) emerged, until the late Devonian, the date of the oldest tetrapod fossils. The entire transition from fish to tetrapods was hence accompanied by strong tidal dynamics, highlighting its importance as a driver for evolutionary processes, as \citet{Byrne2020} conclude.

That tides induced by the Moon could have been important for the evolution of tetrapods, and thus intelligent life, is also supported by tetrapod fossils and their conserved tracks, which were found in habitats related to tidal pools \citep[e.g.,][]{Niedwiedzki2010,Clack2012,Balbus2014}. \citet{Sallan2018} further found that early vertebrate clades originated in shallow intertidal-subtidal environments. However, an origin of tetrapods in other environments such as coastal lagoons or freshwater environments are also discussed in the scientific literature \citep[e.g.,][]{Long2004,Retallack2011,Narkiewicz2015}. \citet{Mai2019}, who studied the present-day land invasion by some species of mudskipper in fresh and saline waters of the Mekong river, conclude that -- analogous to this mudskipper's present land invasion -- the emergence of tetrapods may first have evolved at coastal regions (which includes tidal pools), where they evolved aerial respiration and the possibility to enter land prior to adopting to freshwater environments such as deltaic regions. However, \citet{Mai2019} point out that this hypothesis needs to be tested further in the future.

{Finally, we note that a similar task of tidal modulation in favor of tetrapods is unlikely to be performed by low-mass M dwarfs. \citet{Lingam2022} calculated that stars around $M_{\star}\sim0.1\,$M$_{\odot}$ are unlikely to induce strong tidal modulations in planets with a similar mass and insulation as the Earth.}

\vspace{\baselineskip}
\textbf{\textsl{Altering a planet's evolution:}}

Finally, we emphasize that the Earth could have ended up as an entirely different planet if the Moon-forming impact did not take place. \citet{Jellinek2015}, for instance, argue that the Earth developed plate tectonics because collisional erosion \citep[see, e.g., also][]{ONeill2008} removed part of the incompatible radioactive isotopes from the Earth's protocrust. Venus, on the other hand, did experience less collisional erosion and ended up with higher abundances of $^{235}$U, $^{238}$U, $^{232}$Th and $^{40}$K, making the planet hotter, thereby preventing plate tectonics. The Moon-forming impact could have also altered the geochemical composition and/or oxidation state of the Earth \citep[e.g.,][]{Wade2016}, which could affect subsequent degassing. However, giant impacts may not only alter a planet's {state} towards more habitable conditions. The reverse could entirely be possible as well, and any negative or positive outcome will depend on the specific characteristics of each system.

\vspace{\baselineskip}
\vspace{\baselineskip}

All these arguments do not indisputably prove a large moon to be a necessary requirement for the evolution of EHs, as most of these need additional testing and research. It does, however, suggest that their importance cannot be neglected just by refuting the obliquity argument alone.

\subsection{The frequency of suitable large Moons}\label{sec:moonFreq}

Based on specific impact requirements derived from giant-impact simulations of \citet{Canup2004}, \citet{OBrien2006} found that potential moon-forming impacts were relatively frequent. \citet{Raymond2009}, on the other hand, produced giant impacts in their simulations that meet such criteria on a relatively rare basis, i.e., about 4\% of all impactors may have been able to form planet-satellite systems similar to the Earth's. \citet{Elser2011} specifically studied the frequency of moon-forming impacts to reproduce the Solar System based on simulation runs performed by \citet{Morishima2010}. They found that giant impacts with the required energy and orbital parameters for forming an Earth-like moon occurred in about 1 out of 12 cases, i.e., with a frequency of roughly 8\%. They further derived the uncertainty of these simulations to range from 1 out in 45 planets to 1 out in 4, i.e., from $\sim 2 - 25$\%.

All these simulations were performed within numerical simulations for solar-mass stars. They also cannot tell anything about the subsequent evolution of the system, e.g., whether the satellite will migrate away or toward the host planet, and about the timescale of its orbital stability. The evolution of the emerged planet-moon system, however, was studied by \citet{Brasser2013}, who showed that only in 14\% of all simulated cases the emerging system was comparable to the Earth's with rotation periods between 12 and 48 hours and a planetary obliquity below 40$^{\circ}$ or above 140$^{\circ}$. As \citet{Brasser2013} point out, systems with diverging parameters either end up in a double synchronous state or with the moon colliding onto the planet -- both scenarios may prevent habitability. Combining the outcome of \citet{Brasser2013} with those of \citet{Elser2011} gives an occurrence rate of Earth-like planet-moon systems of below 2\% with an entire range of $\sim 0.3 -3.5$\% for solar-like stars. \citet{Waltham2019} further adopted the model of \citet{Brasser2013}, and found that the stability of the host planet's obliquity decreases with increasing moon mass, so that long-term stability may be the exception for satellite masses $\gtrsim2.5$\,M$_{\rm Moon}$.

The stability of planet-moon systems was investigated by a broad range of different studies, which generally found that these systems either end up with the satellite escaping, finding a stable orbit, or colliding/merging with the planet \citep[e.g.,][]{Counselman1973,Ward1973,Ward1982,Touma1994,Barnes2002,Sasaki2012}. The longevity of large moons in the HZ of different stars, however, was first studied in great detail by \citet{Sasaki2014}. They simulated minimum stellar masses for a large moon to survive for at least 5\,Gyr by mostly varying stellar mass, satellite and planet mass, orbital/rotational period, planet composition and the tidal dissipation value $Q_{\rm pl}$. The latter describes the ratio of the total kinetic energy of the tide to the energy dissipated during one tidal period \citep[see, e.g.,][]{Goldreich1963}. For the Earth, this is estimated to be $Q_{\rm pl} \approx 12$ \citep{Murray2000}, which hinges towards strong tidal dissipation that mostly takes place in the shallow waters of the oceans, as pointed out by \citet{Sasaki2014} -- a potentially crucial factor for a biosphere's nutrient supply as discussed in the previous subsection. A high value of $Q_{\rm p}$, on the other hand, may apply to water worlds (with deep oceans) or planets without any oceans \citep[e.g., Mars has $Q_{\rm pl} \approx 86$;][]{Murray2000}. Low values for $Q_{\rm pl}$, which can hence be considered to be more reasonable for EHs than high ones, lead to planets losing their moons faster. \citet{Sasaki2014} calculated that the minimum stellar mass for which a large Moon can survive for at least 5\,Gyr in the HZ with $Q_{\rm pl} = 10$ ranges from $M_{\star}=0.6$\,M$_{\odot}$ to $M_{\star}=0.8$\,$M_{\odot}$, depending on the initial condition. For an Earth-like composition, these authors found a minimum mass of $M_{\star}=0.64$\,M$_{\odot}$, a value that decreases to $M_{\star}=0.49$\,M$_{\odot}$ for $Q_{\rm pl} = 100$. According to this study, one could therefore conclude that no large moons surviving as long as the Earth's can be found in the HZ of stars with $M_{\star}\leq0.64$\,M$_{\odot}$.

However, \citet{Piro2018}, who investigated the tidal interaction of star-planet-moon systems around low-mass stars with a focus on Earth-Moon-like systems, points out the importance of the initial rotation rate, $\Omega_{\rm pl,0}$, of the host planet. For stellar masses of $M_{\star}=0.2$\,M$_{\odot}$ and $M_{\star}=0.5$\,M$_{\odot}$, this author found survival timescales of 10$^6$ - 10$^{10}$ years depending on $\Omega_{\rm pl,0}$, orbital separation and stellar mass. While for $M_{\star}=0.5$\,M$_{\odot}$ the survival time could reach 10$^{10}$ years, the survival time scale in the HZ of stars with $M_{\star}=0.2$\,M$_{\odot}$ only ranges between roughly 10$^6$ and 10$^7$ years. The potential survival of a large moon may hence be insufficiently long for EHs to evolve around late M dwarfs. At least for tightly packed systems, this is further supported by \citet{Kane2017}, who showed that planets within compact arrangements such as Trappist-1 cannot host moons on stable orbits.

\citet{MartinezRodriguez2019} investigated the potential stability of moons in the HZ of M dwarfs by specifically looking at already discovered planetary systems. They found that out of 33 exoplanet candidates located in the HZ, only 4 can in principle host Moon- to Titan-mass exomoons for longer than the Hubble time. A follow-up study by \citet{Tokadjian2020} investigated the longevity of large moons around the entirety of Kepler planets with two different tidal dissipation models, i.e., the `constant time lag' (CTL) and `constant phase lag' (CPL) models \citep[for details see,][]{Tokadjian2020}. Out of 482 rocky exoplanets assumed with $R_{\rm pl}<1.75$\,R$_{\oplus}$ and $Q_{\rm pl}=12$, they found just 2 planets able of hosting a satellite with a lifetime above 1\,Gyr (and an additional one to be close to 1\,Gyr by applying the CPL model). One of them -- Kepler-452b --  is stable for 15\,Gyr whereas the other is tidally disrupted by the planet after $<10$\,Gyr, respectively. Of the entire 482 planets, however, only 6 are located in the optimistic HZ with Kepler-452b being one of them. This planet has a radius of $R_{\rm pl}\sim$1.6\,R$_{\oplus}$ \citep{Jenkins2015} and orbits a G-type star with a mass slightly higher than the Sun. By using their CPL model, \citet{Tokadjian2020} conclude that Kepler-452b is the only so far discovered `rocky' HZ planet that may be able to host a large moon over an extended period.

Another study by \citet{Dobos2021} also found that the so-called `survival rate' (defined as the probability to either survive as long as the age of the Sun or the age of the respective star, if lower) of a large moon is strongly dependent on the host planet's orbital period. Based on their model, they conclude that out of the entire 36 HZ planets (including planets with $R_{\rm pl}>1.75$\,R$_{\oplus}$) that \citet{Tokadjian2020} found to be stable for $>1$\,Gyr, 29 have a survival rate higher than 50\%. Kepler-452b, however, is none of them and was estimated by \citet{Dobos2021} to have a survival rate of only 41\%. This study, similarly, found contrasting survival rates for the 4 planets identified by \citet{MartinezRodriguez2019} of 0\%, 0\%, 0\%, and 6\%, potentially indicating that the frequency of stable moons around low-mass stars may even be lower than derived from \citet{MartinezRodriguez2019}.

As written above, a large moon can either i) migrate away from the planet, thereby leaving the planet's Hill sphere to become unbound, ii) find a stable orbit around its host, or iii) migrate inwards to either get tidally disrupted or collide with the planet. \citet{Hansen2022} further followed the fate of unbound moons and found that all of them will eventually return and collide with their former host, mostly within just 10\,Myr and, on average, faster in multi-planet systems than in single ones. While moons initially migrating toward the host planet may mostly be tidally disrupted to form rings \citep{Piro2018} or smaller moons \citep{Hesselbrock2017}, unbound satellites that will return to the host planet will preferentially collide due to their higher relative speed \citep{Hansen2022}. A collision between the moon and its former host could significantly hinder, reduce or even completely eradicate the habitability of its former host planet. If already inhabited, it could sterilize the planet entirely \citep{Hansen2022} and potentially render a later origin of life impossible since the post-impactor conditions is certainly different than the initial conditions. The oxidation of the surface and atmosphere, for instance, could have significantly changed, and the availability of prebiotic molecules might be strongly reduced).

    \section{Other Factors Affecting the Maximum Number of Earth-like Habitats}\label{sec:other}

Here, we discuss and assesses a (non-exhaustive) list of additional criteria that may or may not be necessary for the emergence and evolution of EHs, and that may or may not be correlated with each other (see Section~\ref{sec:caveats} for a discussion). Some of the parameters listed in this section should become quantifiable in the next years due to new developments and scientific results, particularly due to advanced exoplanet statistics gathered from novel and upcoming ground- and space-based instrumentation. Here, one should also note that the following discussion of the different potential requirements will most certainly be incomplete in terms of topics and publications that we cite since it is almost impossible to keep track of any new research published within such wide and interdisciplinary fields. {This is specifically true for any biological factors feeding into $B_{\rm life}$ since this is not our major field of expertise.}

\subsection{Additional factors affecting $\eta_{\star}$}\label{sec:otherEtaStar}

\subsubsection{Potential requirements feeding into $A_{\rm GHZ}$}\label{sec:otherAGHZ}

\paragraph{Sterilization events other than supernovae:}
Besides supernovae, planets may particularly be sterilized by GRBs \citep[see,][for a review]{Spinelli2023}. Besides for the galactic bulge (see Section~\ref{sec:MWcomp}), GRBs may also present a threat in the galactic disc, specifically if they correlate with low-metallicity regions \citep[e.g.,][]{Jimenez2013} as predominantly found in the outer parts of the disc.

\citet{Annis1999} were the first who suggested that GRBs might be the reason for the `Great Silence' by periodically sterilizing the Milky Way. \citet{Karam2002} rejected such Galaxy-wide sterilization, but calculated them to be able to damage DNA and to alter biological evolution. By applying galactic metallicity data and distribution rates of GRBs, however, \citet{Piran2014} found that the likelihood of GRBs being the source of mass extinctions at Earth is actually high. Indeed, different studies indicate that the late Ordovician extinction event at about 440\,Ma was at least partially triggered by a GRB \citep{Melott2004,RodriguezLopez2021}. \citet{Piran2014} further found that the probability of a lethal GRB in the Milky Way is highest within $r=4$\,kpc of the galactic center with a chance of 95\% to affect planets during the last 0.5\,Gyr. This compares to a 50\% and $<$50\% chance at the Sun's location and beyond 10\,kpc from the galactic center, respectively. The highest risk of GRBs in general, according to \citet{Piran2014}, may be found in low metallicity galaxies, however, particularly for redshifts of $z > 0.5$, for which \textit{life as we know it} may not be able to exist. Another study by \citet{Gowanlock2016} found that the highest surface density of potential biospheres that could presently survive a GRB would be located toward the inner disk irrespective of whether GRBs mostly follow the SFH or the metallicity gradient of the galaxy. \citet{Gowanlock2016} further found that at least 65\% of today's stars were not strongly affected by GRBs during the last 1\,Gyr, thereby concluding that they may not pose a significant danger for life in the Galaxy and that lifeforms could have even survived GRBs for redshifts of $z > 0.5$. At least compared to SNs, the findings by \citet{Gowanlock2016} agree with a study by \citet{Dayal2016} who showed that SN-types SNII and SNIa dominate the total radiation budget on a galactic scale and at any cosmic epoch, whereas GRBs together with AGNs only contribute $\lesssim1\%$. {Here, we point out that future X-ray observatories such asNewAthena can play an important role in understanding the effects of GRBs on galactic habitability, since they can observe GRBs back towards very high redshifts.\footnote{See, e.g., {https://www.cosmos.esa.int/web/athena/home} for NewAthena.}}

\citet{Spinelli2023} modeled lethal effects from GRBs and SNe and found that the galactic disk within a galactocentric distance of $r=10$\,kpc was unsuitable for life during the first 6\,Gyr because of the frequent occurrence rate of high-energetic events, mostly short GRBs toward the center and long GRBs farther outwards\footnote{For a description of short and long GRB characteristics see, e.g., \citet{Ghirlanda2009}; for short ones see, e.g., also \citet{Berger2014}.}. Since then, they conclude that long GRBs became the most lethal source, particularly toward the outer parts of the Galaxy whereas the galactic disk between $r=2-8$\,kpc may have been the best region for complex life to evolve. \citet{Spinelli2023} finally estimate that one long GRB occurred during the last $\sim$1\,Gyr within a few tens of kpc around the Sun, an event that could indeed be associated with the late Ordovician extinction.

Another potential high-energy source, the supermassive black hole in the center of our galaxy \citep[e.g.,]{Balbi2017,Lingam2019}, may not pose a realistic threat to the present-day galactic disk, since Sagittarius A$^{\star}$ shows only modest activity with outbursts occurring every $10^7 - 10^8$ years \citep[e.g.,][]{Hopkins2006}. However, Sgr\,A$^{\star}$ can still be occasionally lethal for planets in the galactic bulge due to tidal disruption events or a sudden increase in activity \citep[]{AmaroSeoane2019,Pacetti2020,Ambrifi2022}. Active galactic nuclei in general, can pose significant threats to potentially habitable planets. \citet{Wislocka2019}, for instance, calculated that the radiation during the active phase of Sgr\,A$^{\star}$ could have eroded up to several Earth-like atmospheres with planets being save only at a distance of $>$7\,kpc. \citet{Forbes2018} further conclude that about 0.2\% of all planets in the Universe may lose an Earth ocean worth of hydrogen due to AGNs, and even 10\% of all planets can lose an equivalent of Earth's present atmosphere. Typically, however, such studies do not consider XUV-induced extended upper atmospheres, which will generally lead to an underestimate of atmospheric losses.

A final source for high-energetic radiation events are giant flares from so-called soft gamma repeaters (SGR), i.e., a class of neutron stars with very strong magnetic fields. \citet{Horvath2012} find that an SGR within 10\,pc can induce a population reduction of 90\% of even extremophile lifeforms such as \textit{d. Radiodurans,}, which indicates that SGR can indeed pose a risk for complex life. However, \citet{Horvath2012} also find that at the galactic orbital radius of the Sun such events will only occur at a frequency of about one event per 5\,Gyr within a radial distance of 20\,pc. Compared to SN events, it may therefore be relatively negligible. In general, high-energetic events other than SNe may reduce the potential amount of Hs relatively insignificant, at least by only considering the present-day galactic disk. If one expands our estimate to the galactic bulge, the past, and the Universe in general, GRBs, AGNs, and potentially also SGR, cannot be neglected. {Finally, we note again that these high-energetic events must not sterilize the entire planet to be detrimental; it is sufficient to destroy and hinder the evolution of a global biosphere and complex life.}

\paragraph{Binary stars:}\label{sec:binaries}

About 50\% of all stars are part of binary or multi,-star systems \citep[e.g.,][]{Duquennoy1991,Duchene2013,Raghavan2010,ElBadry2021} which likely present entirely different habitability conditions than single stars \citep[see, e.g.,][for an overview]{PilatLohinger2019}. For binaries, planets can either orbit {their} host stars on circumbinary (P-type) or circumstellar (S-type) orbits. For both, the HZ boundaries change significantly compared to a single star \citep[e.g.,][]{Haghighipour2013,Kaltenegger2013,Eggl2020} with P-type binaries showing a highly dynamic HZ \citep{Haghighipour2013} and S-type systems being more stable \citep{Kaltenegger2013}. In addition, and contrary to what one may expect intuitively, HZs in P-type systems can either be smaller than for the individual stars taken separately or be even non-existing, particularly if the two stellar masses are non-similar \citep[e.g.,][]{Wang2019,Cuntz2020,Kong2022}. For S-type systems, \citet{Eggl2013} analyzed 19 close-by stellar systems and found that 17 out of these would allow for dynamically stable HZs.

\citet{Simonetti2020} simulated a high random number of binary systems and found that circumbinary HZs are generally rare since $\sim95\%-97\%$ of binaries show a lack of overlap between radiative HZ and dynamically stable regions. Even though circumbinary HZs therefore only exist in about $\sim4\%$ of all cases, \citet{Simonetti2020} further show that the rate of P-type HZs will be significantly higher if the stellar separations are $a_{\rm b}\lesssim0.2\,$Au. Conversely, circumstellar HZs are possible in $\sim78\%-94\%$ of all binary systems, although these become rare for $a_{\rm b}\gtrsim1.0$\,AU. \citet{Simonetti2020} further found, in agreement with other studies \citep[e.g.,][]{Eggl2012,Cuntz2015,Wang2017Binaries}, that the habitability of binary systems is strongly dependent on the eccentricity, $e_{\rm b}$, of the stellar orbits, and that F and G dwarfs have a slightly higher fraction of HZs than later-type stars. Both results are further supported by \citet{Graham2021}, who showed that the stellar eccentricity of short-period binaries influences the existence of a stable circumbinary HZ, with the corresponding stellar mass limits significantly increasing for increasing $e_{\rm b}$. While for $e_{\rm b} = 0$ no HZ can exist for stars whose component masses are $M_{\star}\leq$0.12\,M$_{\odot}$ (with part of the HZ being unstable up to even $M_{\star}=0.28$\,M$_{\odot}$), these values increase towards $M_{\star}=\{0.135,0.345\}$\,M$_{\odot}$ for $e_{\rm b} = 0.1$, $M_{\star}=\{0.16,0.42\}$\,M$_{\odot}$ for $e_{\rm b} = 0.3$, and $M_{\star}=0.18,0.47\}$\,M$_{\odot}$ for $e_{\rm b} = 0.5$, respectively \citep{Graham2021}.

Tidal interactions between closely orbiting binaries may further reduce the frequency of planets due to an increase in dynamical instabilities \citep{Fleming2018}, a potential reason for the lower frequency of planets around very close binary pairs \citep[e.g.,][]{Bonavita2020}. \citet{Kraus2016} further investigated binaries of solar-like stars and found that below a semimajor axis cutoff of $a_{\rm b}=47$(+59/-23)\,AU, the planet occurrence rate decreases to $\sim$0.34 times the occurrence rate of wider binaries or single stars. As \citet{Kraus2016} conclude, this would indicate that one-fifth of all solar-like stars in the Galaxy cannot host planetary systems due to their binary companion.


Due to the HZs around binaries being to some extent dynamic or non-circular, planets may periodically leave the respective HZs or at least experience relatively strong variations in the received stellar surface flux. However, \citet{Wolf2020Binaries} found that Earth-like planets at such orbits could experience circumbinary-driven global mean surface temperature variations up to $\sim$5\,K with maximum daytime temperature changes up to even $\sim$12\,K on a seasonal basis. Based on these results, \citet{Wolf2020Binaries} conclude that even though such variation may pose some challenges for extant life, true climate catastrophes can be avoided and habitability remains intact. \citet{HaqqMisra2019Binaries} further found that planets with large ocean fractions should experience only modest variation whereas those with large land fractions could experience large temperature variations, which might inhibit continuous habitability\footnote{Another restriction for the appropriate water mass fraction a planet can possess?}. Climate extremes in binary systems can also be enhanced due to large obliquity variations as likely experienced within thse systems \citep{Quarles2020}.

\citet{SanzForcada2014} and \citet{Johnstone2019Binaries} have shown that planets on circumbinary orbits in tight binary systems can receive an increased total amount of XUV surface flux, even as high as a factor of 50 \citep{Johnstone2019Binaries} compared to single star systems, due to tidal interactions between the binaries. One can therefore expect that rocky planets with Earth-like atmospheres around binaries will be even more susceptible to complete atmospheric erosion than in single star systems. However, \citet{Johnstone2019Binaries} also point out that fast rotators in binary systems may spin down faster, which could decrease total atmospheric erosion. \citet{Mason2013} further found that tidal rotational breaking in close binaries can reduce magnetic activity and hence decrease XUV radiation and stellar mass loss. Similar results were found by \citet{Zuluaga2016} for Kepler-16, Kepler-47, and Kepler-453. However, \citet{Johnstone2015Binaries} emphasize that the wind-wind interaction of close binaries leads to strong shock waves through which a HZ planet needs to pass several times on each orbit. This will reduce its magnetosphere \citep{Johnstone2015Binaries} and increase non-thermal escape. Based on the referenced studies, one may expect that a decrease or increase in XUV flux and stellar winds will strongly depend on the initial parameters of the system.

Moreover, we note some potential positive effects of binary systems. As found by \citet{Simonetti2020}, at least a small subset of their binary systems may evolve an HZ wider than for a single star with the same stellar mass as the combined mass of both binaries. \citet{Shevchenko2017} even concludes that binary systems are more conducive to the evolution of life with the Earth being an outlier. This author argues that several aspects of binary systems may favor the emergence of a biosphere. Some specific binary constellations -- particularly among M dwarfs -- will produce mean motion resonances that are closely inside or within the respective HZ, which might place a planet naturally inside the HZ in contrast to single stars. \citet{Shevchenko2017} estimates the percentage of binaries meeting the needed properties to be around $\sim$10\%.

Other potentially positive habitability effects pointed out by \citet{Shevchenko2017} include generically arising seasonal climate variations due to insolation changes, an effect that may foster climate stability, at least according to \citet{Olsen2016}; an additional climate stabilization by reducing obliquity similar as the Moon does for the Earth; tides, for (i) aiding the origin of life as first argued by \citet{Lathe2004}, and (ii) as a heat source for plate tectonics; aiding water delivery due to a planet's passage through the 6/1, 7/1 or any higher-order mean motion resonance with the central binary star and the therewith-connected disturbance of icy bodies in the outer disk. Based on a study by \citet{Forgan2015Photosynthesis}, \citet{Shevchenko2017} also points out that luminosity variations and the therewith-connected stellar eclipses may lead to a variety of forcing timescales that may foster photosynthesis.

The manifold arguments by \citet{Shevchenko2017} potentially show that binary systems do not necessarily have to decrease habitability. However, due to the reduced stability of P-type HZs,  one can expect that taking binary systems into account will likely decrease the maximum number of EHs. If one even excluded all binary stars, such strict criterion would reduce the number of EHs by roughly 50\% as about every second star finds itself in binary or multi-stellar systems. However, the close binary fraction (with $a_{\rm b}<10$\,AU) of solar-like stars seems to be strongly anti-correlated with metallicity with binary fractions of $f_{\rm b,close}=\{53\pm12,40\pm6,24\pm4,10\pm3\}\%$ for mean field metallicities of [Fe/H]\,=\,$\{-3.0,-0.1,-0.2,+0.5\}$ \citep{Moe2019}.

\paragraph{Galactic distribution of $^{26}$Al (and $^{60}$Fe):}\label{sec:26Al}

The short-lived radionuclides $^{26}$Al and $^{60}$Fe are important radioactive heat sources at the very first stages of planet formation and crucial for early differentiation of planetesimals \citep[e.g,][]{Grimm1993,Fu2014,Lichtenberg2016a,Eatson2024}. The initial abundance of $^{26}$Al is particularly important for the early evolution of planetary embryos since it alters the chemical bulk abundance of volatiles substantially \citep[e.g.,][]{Adams2021,Lichtenberg2021Fractionation} and helps dehydrate planetesimals \citep{Lichtenberg2021Water}. However, the abundance of $^{26}$Al was elevated in the protoplanetary disc of the solar system by 3-20 times from the galactic mean \citep[e.g.,][]{Lugaro2018}, indicating that the ISM cannot be its source \citep[e.g.,][]{Cote2019}. The isotope $^{26}$Al is produced in massive stars \citep[e.g.,][]{Limongi2006} and is believed to be distributed either through the winds of these stars, through supernovae \citep[e.g.,][]{Lugaro2018} {and to a substantial amount through novae \citep{Vasini2022}}. The Solar System's enrichment could therefore be explained by an early enrichment of the molecular cloud the Sun formed in, either through a close encounter with {a nova, }supernova or the capture of $^{26}$Al-enriched stellar winds of massive stars \citep[e.g.,][]{Parker2020}.

Such conditions, however, could be rare. \citet{Gounelle2015} argue that pollution through SNe is unlikely (by resulting in wrong $^{26}$Al/$^{60}$Fe ratios) and that the Solar System has formed within an $^{26}$Al wind-enriched shell of a star with a mass of $M_{\star} > 32$\,M$_{\odot}$, which would result in a probability of just 1\% for any stellar system to be born within such a scenario. Similarly, \citet{Dwarkadas2017} show that the $^{26}$Al enrichment could have been produced if the Solar System formation was triggered at the edge of a bubble formed through the stellar winds of a Wolf-Rayet star. These massive stars with intense stellar winds are relatively rare and \citet{Dwarkadas2017} estimate that only 1\%-16\% of all solar-like stars could have formed in such regions and hence show similar $^{26}$Al abundances. \citet{PortegiesZwart2019}, who studied a similar formation process, conclude that a system equivalent to the Sun's forms only every 30\,Myr giving about $\sim$40\,000 such systems currently existing in the entire Milky Way. However, \citet{Reiter2020}, who investigated the galactic distribution of $^{26}$Al and its inclusion into protoplanetary discs, conclude that about $\sim$25\% of all systems may be enriched in $^{26}$Al if $\gtrsim$50\% of all stars are born in high-mass regions. {How the recent finding of novae being a fundamental contribution to $^{26}$Al injection into the disk, as recently found by \citet{Vasini2022} and previously suggested by \citet{Bennett2013Novae}, may increase the chances of a stellar system receiving sufficient amounts of $^{26}$Al, has not yet been investigated. We note, however, that the galactic nova rate of around 30\,yr$^{-1}$ \citep[e.g.,][]{Kawash2022} is significantly higher than the galactic SN rate of roughly 1\,century$^{-1}$ \citep[e.g.,][]{Tammann1994}.}

Another process for $^{26}$Al enrichment was suggested by \citet{Gaches2020}, who show that $^{26}$Al can be enriched through spallation processes at the surface of a protostellar disc via irradiation with high-energetic cosmic rays, at least if specific conditions can be met. A study by \citet{Curry2022} further investigated the prevalence of short-lived radionuclides by studying polluted white dwarfs. These authors conclude that a substantial fraction of exoplanetesimals have formed an iron core which is interpreted to be best explained via $^{26}$Al enrichment and that $^{26}$Al could therefore be common in star-forming regions, as, e.g., suggested by \citet{Reiter2020}. \citet{Curry2022} conclude that their white dwarf observations indicate differentiated exoplanetesimals within a minimum of 4\% of all exoplanetary systems, but, as they point out, more likely within tens of percent of systems.

Although $^{26}$Al enrichment might be more common than initially expected, one can still expect that implementing this factor will further reduce the probability of EHs to form. {If so, however, it is important to also include galactic stellar migration into any GHZ model that wants to take account of $^{26}$Al. As previously mentioned, the Sun with its elevated primordial abundance of short-lived radioactive nuclei was potentially born at a galactocentric distance of only $r\sim$5\,kpc \citep[e.g.,][]{Baba2023} and migrated outwards afterwards\footnote{Interestingly, \citet{Vasini2022} found $^{26}$Al injection rates to be highest for galactocentric distances of $r=5-7$\,kpc.}, profiting from higher $^{26}$Al injection rates early-on, but also from lower SN rates during later ages. Four different possibilities exist in such a case, i.e., (i) a star was always at its present position at, e.g., 8\,kpc, (ii) it migrated from its birth location at, e.g., 5\,kpc to 8\,kpc, (iii), it remained at 5\,kpc, or (iv) it migrated inwards from 8\,kpc to 5\,kpc. In view of SN events during the star's evolution and $^{26}$Al injection right after its birth, only scenario (i) would have positive effects on habitability (high $^{26}$Al injection rates at birth, low SN rates later-on), whereas scenarios (ii-iv) would tend to have negative effects.}

\paragraph{Orbital perturbations:}\label{sec:perturb}

Orbital perturbations from a close encounter with nearby stars or stellar remnants are another important factor, which can have various detrimental effects on a system, i.e.,
\begin{enumerate}[label=(\roman*)]
        \item removing a planet from the system \citep[e.g.,][]{Cai2019,Li2019,VanElteren2019,Rickman2023},
        \item altering a planet's orbit \citep[e.g.,][]{McTier2020,Arbab2021},
        \item perturbing a stellar system's Oort cloud \citep[e.g.,][]{Hut1987,Fouchard2009,Kaib2009},
        \item and stunting planet formation through photoevaporation or tidal disruption of the protoplanetary disk \citep[e.g.,][]{Vincke2018,Winter2018a,Winter2018b}.
    \end{enumerate}
Such perturbations are specifically important for clustered environments \citep[e.g.,][]{Ovelar2012,JimenezTorres2013,Kane2018Cluster,Cai2019,FlamminiDotti2019,Fujii2019,VanElteren2019,Arbab2021,Rickman2023} and for the bulge \citep[e.g.,][]{JimenezTorres2013,Balbi2020,McTier2020,Arbab2021} whereas stellar encounters in less dense parts of the disk -- such as the Solar Neighborhood -- are relatively unlikely. \citet{Arbab2021}, for instance, calculate the chance of a detrimental stellar encounter with the Solar System to be $\sim4.3\times 10^{-4}$ in 4\,Gyr, whereas it is 5.5 times higher for the galactic bulge. For typical star-forming regions, a close encounter within 1000\,AU was recently calculated to happen to 50\% of all protoplanetary disk hosting stars \citep{Cuello2023}. For the bulge, about 80\% of stars have at least one of such encounters within 1\,Gyr \citep{McTier2020}, as in Section~\ref{sec:MWcomp}.

About 30\% of all disk stars form in compact stellar clusters \citep[e.g.,][]{Kruijssen2012,Chandar2017} from which only a few percent remain within relatively long-lived bound clusters such as the Pleiades \citep[e.g.,][]{Lada2003,Chandar2017}. Within these dense stellar environments, stable planetary orbits may be significantly reduced. From more than 4000 exoplanets discovered by July 2019 only 30 were indeed located in stellar clusters and only one planet was found to reside within a dense globular cluster \citep{Cai2019}. A study by \citet{Ovelar2012} further indicates that the lifetime of a planet within the HZ decreases with increasing stellar density and stellar mass. Our Solar System could therefore not have formed in a dense environment with $>10^3$ stars per pc$^3$, unless such a cluster dispersed in a relatively short time of $\leq100$\,Myr. However, recent studies found that only a small fraction of planets at orbits comparable to the HZ will ultimately be ejected from their stellar system within clustered environments \citep[e.g.,][]{Fujii2019,Cai2019,FlamminiDotti2019}. If one simply assumes that about 30\% of stars are born in dense clusters with less than $\sim$5\% remaining in bound clusters, the ratio of planets discovered in clusters versus planets discovered around field stars will be quite similar to the ratio of stars within clusters versus actual field stars \citep[by taking the numbers from][]{Cai2019}. This implies that the planet occurrence rates of both may not diverge significantly, which was indeed found to be the case \citep{Meibom2013}.

Accounting for orbital perturbations could hence remove only a small fraction of stars from our sample. We further note that a potentially reduced planet frequency around stars in bound stellar clusters may already be accounted for indirectly, at least to a certain extent, in various occurrence rate estimates since the Kepler FOV also contains stellar clusters. A positive aspect may additionally come into play, as orbital perturbations can lead to an increased ejection rate of rocky material from stellar systems. This could in principal aid undirected panspermia in dense stellar regions (see Section~\ref{sec:originLife} for a discussion).

Perturbations of a star's Oort cloud may further trigger increased comet showers in the inner system that could potentially lead to mass extinctions. At least for Earth, however, induced comet showers are thought to have had little impact in the past, without triggering extinctions since the Cambrian explosion \citep{Kaib2009,Feng2014,Kaib2018}. Whether more densely populated galactic regions pose a greater risk of catastrophic comet showers remains unclear. It was, however, found that Oort clouds in more densely populated regions evolve to be more compact, implying that closer stellar encounters will be needed than for the Solar neighborhood \citep{Kaib2011,Kaib2018}.

The capture of a giant planet from a passing stellar system can provide an additional pathway for an increase in asteroid and comet showers in the HZ. \citet{Kokaia2022} studied the effect of these captures on Kuiper-belt-like regions and found that in 20\% of their simulations, the impactor flux increased by 5 to 10 times compared to the Earth's for periods up to several Gyr. In 40\% of the cases, these captures induced cataclysmic events with impactor fluxes increasing several thousand times on a 10\,Myr timescale, while they additionally destabilized inner planets for another 40\% of the cases \citep{Kokaia2022}.

Wide binary companions (with a periastron of $a_{\rm b,p}\lesssim100$\,AU) perturbed by nearby stars or galactic tides pose another risk \citep[see,][for a detailed discussion]{Kaib2018}. Although perturbed wide binaries only affect outer giant planets initially, the induced perturbations can easily cascade toward the inner system \citep{Zakamska2004}. \citet{Kaib2013} showed that a solar binary with $M_{\star}=0.1$\,M$_{\odot}$ and an initial periastron of $a_{\rm b,p}>1000$\,AU can be driven toward $a_{\rm b,p}\sim100$\,AU within only $\sim$1\,Gyr through chance perturbations from the galactic disk environment -- an orbital distance sufficiently close to destabilize the Sun's giant planets \citep{Malmberg2011}. After 10\,Gyr, the probability of triggering a planet-planet scattering is about 20-30\%, whereas it is even 50-70\% for a 1\,M$_{\odot}$ companion \citep{Kaib2013}. Such a scattering most likely leads to a planet ejection, an increase in eccentricity for the surviving ones \citep[e.g.,][]{Rasio1996,Kaib2018} and may further increase comet and asteroid impacts onto HZ planets \citep{Kaib2018}. Recently, \citet{Brown2022} found that a perturbation of Neptune's orbit of only $\sim$0.1\% leads to an order of magnitude probability increase of destabilizing the Solar System within 5\,Gyr.

In denser stellar environments, the detrimental effects of wide binary stars on habitability will further increase compared to the Solar Neighborhood \citep{Kaib2013,Kaib2018}. Wide companions can therefore significantly threaten the evolution and stability of EHs, another factor that adds to the discussion on whether binaries may reduce the maximum amount of EHs in the Milky Way (see Section~\ref{sec:binaries}). About $\sim15-20$\% of solar-like stars have a wide binary companion with a semi-major axis of $a_{\rm b}\gtrsim200$\,AU, irrespective of metallicity \citep{Moe2019}.

\subsubsection{Potential requirements feeding into $A_{\rm at}$}\label{sec:otherAat}

\paragraph{The stellar plasma environment:}

As already discussed in Section~\ref{sec:XUV}, stellar winds and CMEs will lead to a diverse range of non-thermal escape processes; see, e.g., \citet[][]{Chamberlain1963} and, more recently, \citet[]{Gronoff2020} for a review on non-thermal escape processes. Importantly, stellar winds are expected to be significantly stronger during the early evolution of stars of all masses \citep[e.g.,][]{Wood2005,Johnstone2015WindI,Johnstone2015WindII,Airapetian2016,Vidotto2021,Wood2021} as might be the frequency of CMEs \citep[e.g.,][]{Odert2017,Kay2019} that often accompany stellar flares. In addition, HZ planets around M dwarfs are particularly interacting with stellar wind ram pressures that are by orders of magnitude higher than the solar wind ram pressure at the Earth's orbit \citep[e.g.,][]{Garaffo2016}, thereby leading to extreme atmospheric erosion by the wind \citep[e.g.,][]{Airapetian2017,RodriguezMozos2019,Dong2020TOI} and by CMEs \citep[e.g.,][]{Khodachenko2007} -- even for very old M dwarf systems such as the $\sim$10\,Gyr old Barnard's star \citep{France2020}.

However, \citet{Wood2021} recently found through observations of astrospheres that the stellar mass loss from M stars might be lower than initially expected. These authors report that the flare/CME relation seems not to hold for M dwarfs, implying that CMEs may be less frequent. The mass loss per surface area, however, is comparable to the mass loss of the Sun, as \citet{Wood2021} report, which indicates that the stellar wind ram pressure in the HZ of M dwarfs will still be significantly higher than in the HZ of G stars. We further note that, even if CMEs at M dwarfs are to some extent suppressed, such a full or partial suppression may then lead to an accompanied increase in the star's integrated X-ray flux by up to an order of magnitude for a duration of up to several hours \citep{AlvaradoGomez2019,AlvaradoGomez2022}, an effect that would further enhance thermal escape.

As we already discussed stellar wind induced non-thermal escape relatively detailed in Section~\ref{sec:XUV}, we only point out some important additional takeaways:
\begin{enumerate}
  \item Besides planets around M (and K) dwarfs, even those around solar-like stars will likely suffer significant atmospheric erosion early on during the system's evolution, either through non-thermal ion escape \citep[e.g.,][]{Lichtenegger2010,Lichtenegger2022} or polar outflow \citep[e.g.,][]{Kislyakova2020,Grasser2023}. This can even lead to the loss of an Earth-like 1\,bar atmosphere in $\lesssim$10\,Myr, likely even for magnetized planets \citep{Lichtenegger2010}. Also here, higher CO$_2$ mixing ratios will reduce atmospheric escape due to a smaller expansion of the thermosphere.
  \item Non-thermal escape {will remain an important loss process over larger timescales} than thermal escape. Particularly at an age, when the XUV surface flux decreases to such an extent that thermal escape becomes negligible, non-thermal losses will still remain elevated for some extended time \citep[e.g.,][]{Scherf2021,Lichtenegger2022}. There will therefore be a transition period in which a N$_2$-dominated atmosphere can no longer be thermally lost but still non-thermally.
  \item Even if Earth-like atmospheres around M dwarfs did not thermally escape, strong non-thermal losses would still be capable of rapidly eroding them \citep[e.g.,][]{Airapetian2017,RodriguezMozos2019,France2020,Dong2020TOI}. Atmospheric escape, particularly at low-mass stars and young stellar systems, depends on both the plasma and radiation environment of a star. High XUV surface fluxes ionize large parts of planetary thermospheres leading to a decrease of the so-called Jeans escape parameter, $\lambda_{\rm exo}$\footnote{The Jeans escape parameter is the rate of the thermal energy of a particle to the gravitational energy of a planet at the planet's exobase, $r_{\rm exo}$ i.e., $\lambda_{\rm exo}= GM_{\rm pl}m_{i}/(k_{\rm B} r_{\rm exo} T_{\rm exo})$, where $m_{i}$ is the particle mass, $k_{\rm B}$ is the Boltzmann constant, and $T_{\rm exo}$ is the exobase temperature.  For values of $\lambda_{\rm exo}\lesssim2-3$, atmospheres were shown to escape hydrodynamically \citep[e.g..][]{Volkov2011,Erkaev2015}.} \citep[e.g.,][]{Bauer2004}, and thus to increased non-thermal and thermal escape rates.
  \item Non-thermal escape is not only governed by a star's quiet stellar wind. It may further be enhanced significantly by frequent CMEs that provide a much higher ram pressure than quiet stellar winds \citep[e.g.,][]{Khodachenko2007,Lammer2007,AlvaradoGomez2022AUMic}. Around M dwarfs, these CMEs may be able to erode tens to hundreds of bars of N$_2$, or even CO$_2$-dominated atmospheres \citep{Lammer2007}\footnote{One should note that the results of \citet{Lammer2007} are based on CME frequency estimates that could be too high if the flare/CME relation does not hold for M dwarfs, as discussed above.} {But also around solar-like stars, strong and frequent CMEs will lead to substantial losses. \citet{Dong2017Dehydration}, for example, found that `Carrington-type' space weather events can enhance H$_2$O$^+$ escape rates by about two orders of magnitude above the Earth's present-day level.}
  \item Solar/stellar energetic particles (SEP) from flare-induced stellar proton events may provide another detrimental factor, as they might crucially increase the radiation dose at the surface of a planet. For M dwarfs, SEP events can increase ionizing radiation by several orders of magnitude in dependence of atmospheric density, magnetic field shielding, and stellar activity \citep[e.g.,][]{Atri2020,Hu2022}. SEPs are also driving ozone depletion, which can induce further subsequent ramifications such as an increase in UV surface flux and biotic damage \citep[e.g.,][]{Lingam2017Superflare,Tilley2019}. On the other hand, however, solar energetic particles can be important for the production of prebiotic molecules and hence aid the origin of life \citep[e.g.,][]{Lingam2018SEP,Kobayashi2023}. SEPs further initiate N$_2$ fixation either through direct dissociation of N$_2$ or the initiation of lightning. While this is a positive aspect in terms of prebiotic chemistry \citep{Lingam2018SEP}, {SEP induced N$_2$ fixation may be able to severely deplete atmospheric N$_2$ around low-mass stars. As \citet{Grenfell2020} highlight, SEP events on Earth can lead to a formation rate of $\sim2\times10^{33}$ NO molecules/day \citep[see also][]{Jackman2005}. If one assumes that such SEP events are continuous around active M dwarfs, this would result in an atmospheric nitrogen removal rate of $\sim$17\,Tg\,N/year\footnote{Note that \citet{Grenfell2020} calculates 0.04\,Tg NO/year. However, if one calculates the production of NO/year based on $\sim2\times10^{33}$ NO molecules/day, this gives $\sim$40 Tg NO/year, i.e., a nitrogen removal rate of 17 TG N/year. The subsequent calculation in \citet{Grenfell2020} may hence be erroneous.}. However, as \citet{Grenfell2020} further point out, the cosmic ray flux of superflaring stars could be up to $10^5$ times higher than SEPs on Earth \citep[see also, e.g.,][]{Davenport2016}, which could initiate substantial N$_2$ fixation. If one assumed continuous superflaring, this fixation rate would lead to a nitrogen removal rate of $17\times10^5$\,Tg\,N/year, implying a complete fixation of an Earth-like atmosphere with 800 mbar N (i.e., $\sim4\times10^9$\,Tg N)  within only a few 1000 years. Although such an extreme value may not be realistic as superflares are no continuous phenomenon, it highlights the strong potential of SEPs to deplete Earth-like atmospheres around M dwarfs.}
\end{enumerate}


\paragraph{Flares:}\label{sec:flares}

Flares and superflares are another factor feeding into (i) atmospheric stability through enhancing thermal escape \citep[e.g.,][]{DoAmaral2022,France2020}, and (ii) the photochemistry of the upper atmosphere by either depleting protective ozone \citep[e.g.,][]{Vida2017,Yamashiki2019,Howard2018}  - a negative effect - or by potentially enhancing prebiotic chemistry - a positive effect. The potential deleterious effects of (super)flares (atmospheric erosion and ozone depletion) were already discussed in Section~\ref{sec:XUV}, so here we focus on their potential positive effects.

Several studies note that frequent flares can increase the UV flux \citep[e.g.,][]{Bogner2022,Maas2022Flares} and the visible photons \citep[e.g.,][]{Mullan2018} a planet receives from its host star. This can be crucial as the quiescent UV energy received from low-mass stars might be too low to kick-start or even maintain sufficient prebiotic chemistry \citep[e.g.,][]{Rugheimer2015,Rimmer2018,Ranjan2017MStars,Spinelli2023UVHZ} and photosynthesis \citep{Mullan2018}. Around M dwarfs, flares could therefore be essential for abiogenesis and for sustaining a biosphere \citep[e.g.,][]{Ranjan2017MStars}. \citet{Mullan2018}, for instance, found that flares can increase oxygenic photosynthesis effectiveness (PSE) at mid-M dwarf by a factor of 5-20 above quiescent conditions, which may increase effectiveness by up to 50-60\% of the Earth's photosynthesis. For late-type M dwarfs, however, PSE remains less effective by almost one order of magnitude. {In addition to PSE, however, flaring duration and frequency are of significant importance, as it affects the time-averaged photosynthetic flux. This implies that flares should occur often and over extended periods of time to increase their effect on photosynthesis. Although \citet{Mullan2018} suggest a selection effect in favor of life for stars adhering to these conditions, one could also argue that it is exactly these stars that pose the highest risk for ozone depletion and atmospheric erosion.}

\citet{Bogner2022} analyzed TESS light curves of 112 M dwarfs, and although 2\,532 flares were detected on 35 stars, none exhibited high enough energy to trigger either ozone depletion or chemical reactions necessary for the production of RNA. \citet{Guenther2020}, who provide a relation to estimating the flare frequency needed to trigger and maintain prebiotic chemistry, further calculated that out of 401 early-type and 271 mid- to late-type M dwarfs in their sample only 0.7\% and 3.0\%, respectively, show flare frequencies to be sufficiently high. For Trappist-1 specifically, \citet{Ducrot2020}, \citet{Guenther2020}, \citet{Glazier2020}, and \citet{Maas2022Flares} found that the UV flux produced via flares is too low to allow for prebiotic chemistry. \citet{Jackman2023}, however, reanalyzed empirical flare models and suggested that the aforementioned studies may underestimate the respective UV flux produced by such flares and therefore slightly corrected the results by \citet{Guenther2020} toward 4.0\% and 13\%, respectively. These authors therefore point out that Trappist-1 could in principle produce flares, albeit rarely, with energies high enough for prebiotic chemistry to happen, but further caution that these would be accompanied by damaging far-ultraviolet (FUV) irradiation. Another recent study by \citet{Brasseur2023} also suggests that UV flux calculations based on white light flare measurements might underestimate the accompanied UV flux. {Based on flare observations in the NUV and FUV from the GALEX Space Telescope, \citet{Berger2014} even found that 98\% of their investigated flares are usually underestimated in the FUV flux by 3 to 12 times in many studies that assess the effect of flares on planetary habitability since these mostly assume a constant 9000\,K black body irradiation.} Further research is therefore needed to better understand the ffect of flares on prebiotic chemistry, sterilization, and enhanced atmospheric escape.

Other, potentially positive effects of flares recently mentioned in the scientific literature are the associated interior heating of a planet to promote long-lived volcanic degassing \citep{Grayver2022}, and the production of abiotic ozone for shielding the surface from damaging UV irradiation \citep{Ridgway2023}. \citet{Grayver2022} calculated that ohmic dissipation within a planet's interior initiated through the magnetic flux carried by flare-associated CMEs can produce up to 20\,TW of heat within the mantle, a value comparable to the Earth's present-day heat production rate; if the planet also hosts an intrinsic magnetic field, additional Joule heating may further increase this value. These authors point out that the associated elevated volcanism and outgassing may replenish an escaping atmosphere. However, such an active and continuous degassing might lead to enhanced volatile loss, as already pointed out earlier. Any N$_2$, H$_2$O, and CO$_2$ might hence already be lost, when a star's XUV flux finally decreases below the threshold value for atmospheric stability. In addition, Joule heating will also affect the upper atmosphere's stability by increasing its temperature, expansion, and escape \citep[e.g.,][]{Bates1973,Bates1974,Cohen2024JouleHeating}. The enhanced ozone shielding, pointed out by \citet{Ridgway2023}, may additionally impede prebiotic chemistry by reducing the scarcely available UV flux even further. {Flares may even lead to elevated surface ozone concentrations that are lethal for \textit{life as we know it} \citep{Cooke2024}.} Finally, one may mention that flares at active low-mass stars can predominantly occur at high stellar latitudes \citep{Ilin2021} thereby decreasing the irradiation received by planets in the equatorial plane. While this could diminish the effect of flares on atmospheric erosion, it also decreases the available energy needed for prebiotic chemistry.


\paragraph{UV habitability and photosynthesis effectiveness:}\label{sec:UVHab}

As briefly discussed in the last section, UV irradiation onto a planet can be crucial for the emergence and evolution of life. The related concept of a UV habitable zone (UVHZ) was first pointed out and calculated by \citet{Buccino2006} and \citet{Buccino2007}, and then further discussed by, e.g., \citet{Guo2010}, \citet{Oishi2016b}, and \citet{Spinelli2023UVHZ}. \citet{Buccino2006} were the first to calculate such UVHZ for a sample of 17 stars in the G to F-type spectral classes and found that the UVHZ is mostly located on closer orbital locations than the traditional HZ. A total of 41\% of the investigated stars show no overlap between both HZ concepts, thereby indicating that life may not be possible on planets around these stars.

While the UVHZ is mostly located closer to the star than the traditional HZ for low-mass stars, \citet{Buccino2006} also found that for both investigated F stars a region within the HZ exists, where complex life will be destroyed by the strong incident UV irradiation. In a follow-up study, \citet{Buccino2007} further investigated the UVHZ of 5 specific M dwarfs (i.e., HIP\,74995, HIP\,109388, HIP\,113020, AD\,Leo, EV\,Lac) and found that, during quiescent phases, none of them produces a sufficiently high UV surface flux to allow for biogenic processes.

\citet{Guo2010} were the first to simulate continuous UVHZs for stellar masses of$M_{\star}=0.08-4.00$\,M$_{\odot}$. They found that for stellar effective temperatures below $T_{\rm eff}=4\,600$\,K, the UVHZ lies closer to the star than the traditional HZ, whereas it is farther away for host stars with $T_{\rm eff}>7\,137$\,K. An overlap between both HZ concepts is therefore only possible for stars with temperatures in-between these boundary values. Although \citet{Guo2010} repeat the argument of stellar flares potentially aiding prebiotic chemistry for stars with $T_{\rm eff} < 4\,600$\,K, they also point out that the probability of life `decreases dramatically' for stars with $T_{\rm eff} > 7\,137$\,K due to DNA destruction and damage to proteins and lipids via the enhanced UV surface flux.

Recently, however, \citet{Ahlers2022} point out that prior studies on the UVHZ did not take into account the rapid rotation of stars with $M_{\star} \gtrsim 1.3$\,M$_{\odot}$. For these stellar masses, they found that the HZ moves further inwards toward the star due to rapid stellar rotation, and because of gravity darkening substantially reducing stellar UV emission. \citet{Ahlers2022} therefore suggest that the UVHZ could extend to higher stellar masses than originally thought.

\citet{Oishi2016b} further studied the evolution of the UVHZ as a function of metallicity for $Z = \{1.0,0.1,0.01\}$\,Z$_{\odot}$. Interestingly, they find that the UVHZ will move inward if the metallicity decreases, because the emitted UV photons from metals within the stellar atmosphere are going to decrease in unison. For $Z = 1.0$\,Z$_{\odot}$, \citet{Oishi2016b} find HZ overlaps only for $M_{\star}\sim0.8-1.5$\,M$_{\odot}$ for at least some part of the main-sequence lifetime. This shifts toward $M_{\star}\sim0.5-1.2$\,M$_{\odot}$ and $M_{\star}\sim0.5-1.0$\,M$_{\odot}$ for $Z = 0.1$\,Z$_{\odot}$ and $Z = 0.01$\,Z$_{\odot}$, respectively. Note, however, that \citet{Oishi2016b} give much narrower ranges as they only consider stellar masses for which the overlap exists over the entire main-sequence lifetime. For such a restricted requirement, no overlap can be found for $Z = 0.001$\,Z$_{\odot}$ whereas only narrow ones exist around $M_{\star} \sim 1.2\,$M$_{\odot}$ and $M_{\star}\sim0.8-1.0$\,M$_{\odot}$ for $Z = 0.1$\,Z$_{\odot}$ and $Z = 1.0$\,Z$_{\odot}$, respectively.

A recent study by \citet{Spinelli2023UVHZ} derived an empirical relation between the near UV (NUV) luminosity and $T_{\rm eff}$ for calculating the UVHZ of MKGG-type stars. From 17 stars between stellar effective temperatures $T_{\rm eff}=2550-5146$\,K that host 23 planets within their respective HZs, \citet{Spinelli2023UVHZ} find that 18 indeed orbit outside their host star's UVHZ. These authors further suggest that only stars with $T_{\rm eff}\gtrsim 3900$\,K and $T_{\rm eff}\gtrsim 4600$\,K might indeed provide an overlap of both HZ concepts for atmospheres being 100\% and 10\% UV transparent, respectively. These authors further emphasize that they considered a quite liberal UV flux threshold for triggering abiogenesis with a 50\% yield of photochemical products \citep[from,][]{Rimmer2018}, a value that might be too small for resulting in a sufficient concentration of prebiotic molecules. Higher yields will lead to a shift toward higher stellar effective temperatures and vice versa \citep{Spinelli2023UVHZ}.

M dwarfs may also provide too little visible light for photosynthesis. As already briefly discussed above, PSE could reach about half the efficiency of the present-day Earth through further energy input via flares around mid-type M dwarfs, whereas it would be at minimum one order of magnitude lower for late-type M dwarfs  \citep{Mullan2018}. Another study by \citet{Lehmer2018} found that photosynthesis at planets around M dwarfs with $T_{\rm eff}=2300-4200$\,K would be light-limited instead of nutrient-limited as on Earth, even if photosynthesis extended toward infrared wavelengths -- an extant biosphere comparable to the Earth's might therefore not be possible. \citet{Lingam2019Photosynthesis} similarly conclude that photosynthesis at stars with $M_{\star}\lesssim 0.2$\,M$_{\rm \odot}$ might receive too little photons to maintain an Earth-like biosphere. However, a recent laboratory study by \citet{Claudi2022} shows that different cyanobacteria are able to photosynthesize under simulated M dwarf irradiation in a similar way than under solar light, which highlights the importance of simulating photosynthesis under exotic light conditions. {For K dwarfs, on the other hand, a recent experimental study finds that garden cress and cyanobacteria exhibit comparable and even significantly higher photosynthetic efficiency under synthetic K dwarf irradiation compared to solar conditions \citep{Vilovic2024}.}

{The studies in the last paragraph mostly deal with Earth-like photosynthetic systems that are specifically evolved for operating under a solar-like star. Some recent work began to move beyond this paradigm such as \citet{Duffy2023} and \citet{Chitnavis2023}, who modeled variations in photosynthetic antennae, or \citet{Lingam2021Photo} and \citet{Illner2023}, who investigated the hypothesis that the absorption maxima of photopigments are close to the peak spectral photon flux of the host star \citep[as initially proposed by][]{Kiang2007PhotoA}. In the latter case, photopigment structures could in principle be very different allowing for their absorption maximum to coincide with the host star's peak photon flux. For this to happen, however, \citet{Lingam2021Photo} found that the number of excited electrons must significantly increase per putative photopigment (i.e., from about 20 electrons per pigment for the Sun to almost 50 in the case of Trappist-1). These authors, however, remain agnostic on whether such pigments can in principle exist, and if so, how they could be structured. In a related study, \citet{Illner2023} further theorize that the red (and blue) shift of the peak wavelength could be achieved by the evolution of photopigments comprising metals with lower (and higher) electronegativity. \citet{MarcosArenal2022} further developed a metric\footnote{The metric developed by \citet{MarcosArenal2022} is called ExoPhot and can be found at \url{https://github.com/ExoPhotProject}.} for investigating the fitness of various (existing and theoretical) photopigments under different exoplanetary conditions based on the so-called total absorption rate, a quantification of the spectral overlap between stellar \& atmospheric spectroscopic model with the absorption spectra of said pigments. These are important steps for moving beyond aforementioned paradigm.}

Another study by \citet{Covone2021} also simulated PSE for stellar effective temperatures of $T_{\rm eff}=2600-7200$\,K, finding similar results as \citet{Lehmer2018} and \citet{Lingam2019Photosynthesis} that indicate biospheres around cool stars
to be light-limited. As pointed out by \citet{Lingam2019Photosynthesis}, such worlds might not be able to sufficiently build up O$_2$ toward detectable levels. Indirectly, this fits relatively well with \citet{Duffy2023}, who applied a model of photosynthetic antennae to a range of model stellar spectra. Their findings indicate that low-mass stars below $T_{\rm eff}\sim$3300\,K might select for anoxygenic photosynthetic microbes that do not produce O$_2$, although they will not necessarily be light-limiting. Because of this, the biological build-up of substantial amounts of O$_2$ seems indeed unfavorable. If one additionally considers O$_2$ limits for complex life \citep{Catling2005}, the emergence of a complex biosphere is highly unlikely around low-mass M dwarfs, even if an N$_2$-dominated atmosphere could be retained. However, another recent study by \citet{Chitnavis2023} that follow-up the work of \citet{Duffy2023}, shows that oxygenic photosynthesis might in principle also be possible around low-mass M dwarfs by introducing an enthalpic gradient into their photosynthetic antenna model. Interestingly, recent findings showed that cyanobacteria \citep{Battistuzzi2023} {and even microalgae \citep{Battistuzzi2023Microalgae}} could adapt to the light of a typical M dwarf, results that \citet{Chitnavis2023} suggest to be in support of their outcomes. {In addition, cyanobacteria growing in caves in extreme low-light environments, where the near-infrared portion of the spectrum is enriched, produce pigments that can harvest near-infrared light, highlight their potential to grow under M dwarf conditions \citep{Jung2023}.} Finally, \citet{Hall2023} recently defined a photosynthetic habitable zone (PHZ) and calculated that it does not necessarily have to overlap with the traditional HZ. Depending on their assumptions, such an overlap optimally covers the entire M to F spectral classes or could even be restricted to heavier G and F stars. {The PHZ of \citet{Hall2023}, however, is purely based on conventional Earth-like photosynthesis.}

This summary illustrates that UV radiation and visible light photons, are important to consider for investigating habitable environments around other types of stars. \citet{Scharf2019} hence points out that photoautotrophic organisms around M dwarfs might be even doubly disadvantaged -- not only by the low flux of photons employable for photosynthesis, but also because of the low exergy of the system, i.e., the maximum work that can be performed from the balance of energy gain from the host star and energy loss from the planet. The availability of energy for life was consequently also investigated by \citet{HaqqMisra2019} by assuming the existence of a relation between stellar mass and the timescale for the development of complex life, an assumption first put forward by \citet{Livio1990} and further refined by \citet{HaqqMisra2018PET}. Comparing the stellar energy available at Earth with the one available around other spectral classes, \citet{HaqqMisra2019} find that the biologically available power at a planet around a star with $M_{\star}=0.3$\,M$_{\odot}$ is about an order of magnitude less than for stars with, e.g., $M_{\star}$=0.8\,M${_\odot}$ and $M_{\star}$=1.0\,M${_\odot}$. By calculating the total energy $E_{\rm bio}^{\oplus}$ available at the top of the atmosphere over the entire time, $t_{\rm evo}$, needed from Earth's emergence until the evolution of complex life, and by assuming a similar amount of energy is required at other planets, \citet{HaqqMisra2019} derives an equation for calculating $t_{\rm evo}$ for planets orbiting other stars, i.e.,
\begin{equation}\label{eq:tevo}
  t_{\rm evo} = t_{\rm cl,\odot}\exp{\left[4\left(1-\frac{M_{\star}}{\mathrm{M}_{\odot}}\right)\right]},
\end{equation}
where $t_{\rm cl,\odot}$ is the typical time needed for a Sun-like star to evolve complex life (assumed to be equivalent to the Earth's value of $t_{\rm evo} \sim 4$\,Gyr). As one can see, $t_{\rm evo}$ exceeds the Hubble time, $t_0 \sim 13.8$\,Gyr, for $M_{\star}<0.7$\,M$_{\odot}$. {This also impacts the GHZ -- if $t_{\rm evo}$ indeed scales with $M_{\star}$, planets around low-mass stars will need longer periods without sterilizing events than higher-mass stars for complex life to evolve. Generally assuming 4\,Gyr for planets around any stellar mass could therefore be too optimistic (although this can be counteracted by faster evolution or higher resilience against radiation).}

So, if complex life on other planets requires a comparable amount of total energy to evolve as on Earth, stars with $M_{\star}<0.7$\,M$_{\odot}$ will still be devoid of complex life or its evolution will at least be restricted/unlikely, which fits well to the arguments of \citet{Scharf2019}. {This also fits well with a recent study showing that the free energy available in a system to sustain a complex biosphere increases with stellar mass \citep{Petraccone2024Entropy}.} The oxygenation time \citep{Catling2005} itself might likewise depend on the free energy available on a planet, since this in turn depends on, e.g., photosynthesis and photochemistry, and thus on the availability of incident photons \citep{HaqqMisra2019}. Relatively young planets, not only around M dwarfs, might thus still need time for atmospheric O$_2$ levels to rise and for complex life to evolve. The earliest stellar age at which EHs can evolve will therefore, besides $\alpha_{\rm at}^{\rm ll}$, also depend on the available stellar energy and, hence, again on a star's mass.


\paragraph{Intrinsic climate cooling:}

Another interesting aspect, `intrinsic climate cooling', was put forth by \citet{Waltham2019Climate}, who argues that the evolution of M dwarfs might be too slow for Earth-like planets to avoid glaciation. If the geophysical forcing of a planet in the HZ of a red dwarf star is similar to the Earth's, its climate may cool at a faster rate, since the much slower increase in the effective temperature of the host star cannot compensate for the weathering of greenhouse gases from the planet's atmosphere (if existing). \citet{Waltham2019Climate} calculate that such a planet becomes globally glaciated within a few Gyr. If so, many planets around M dwarfs would likely be glaciated before its host star's XUV flux decreases below its lower limit, $\alpha_{\rm at}^{\rm ll}$. However, whether a planet around another star will indeed follow the same `intrinsic climate cooling' than the Earth might be contentious since the carbon-silicate cycle shows a stabilizing negative feedback mechanism against surface temperature \citep[e.g.,][]{Walker1981,Kasting2019}. This may indicate that climate cooling can proceed slower around lower-mass stars for balancing weaker increases in stellar luminosity.


\subsection{Additional factors affecting $\eta_{\rm EH}$}\label{sec:otherEtaEH}

\subsubsection{Potential requirements feeding into $B_{\rm pc}$}\label{sec:otherBCHNOPS}

\paragraph{Planetary accretion speed:}\label{sec:acc}

As discussed in Paper~I, protoplanets that accrete a certain planetary mass until the end of the stellar nebula will accrete H$_2$ from the protoplanetary gas disk. If the protoplanet exceeds a mass of $M_{\rm pl}\sim0.5$\,M$_{\oplus}$ \citep[e.g.,][]{Johnstone2015Atmospheres} at the time of nebula dissipation, a primordial atmosphere remains at the planet, but is lost subsequently through hydrodynamic escape. If the planetary mass will be too high, however, the primordial atmosphere cannot escape entirely and the planet will evolve into an EH. For a slowly rotating solar-like star, this critical mass will be around $M_{\rm pl}\sim$0.8\,M$_{\oplus}$ \citep[e.g.,][]{Lammer2020a,Erkaev2022}. An Earth-mass planet accreting entirely in the gas disk will therefore remain uninhabitable.

This planetary mass limit increases for lower-mass stars and vice versa due to the differences in the XUV surface flux in the stars’ respective HZCLs. This effect could therefore be particularly significant for higher-mass stars and for systems with long disk lifetimes\footnote{For comparison, the Solar Nebula dissipated after about 3-4\,Myr \citep{Bollard2017,Wang2017Nebula,Weiss2021Nebula} whereas the characteristic nebula lifetime is $\sim2.5$\,Myr \citep[e.g.,][]{Mamajek2009}.}.

Besides the additional formation of H$_2$O from the primordial atmosphere \citep[e.g.,][]{Ikoma2006,Kimura2020,Lammer2021}, we finally note that planetary accretion can strongly affect the subsequent evolution of the planet including its chemical and elemental composition, either through devolatilization \citep{Benedikt2020,Wang2022PlanetComp}, giant impacts or other chemical or physical processes in the disk and during subsequent accretion such as the influence of the accretion disk's magnetic field on the final composition of a planet \citep{McDounough2021}.

\paragraph{C/O and Mg/Si ratios as tracers of mineralogy:}\label{sec:CO}

So-called carbon-rich planets are hypothesized to exist around stars with carbon-to-oxygen ratios of $\rm C/O\geq0.8$ \citep[e.g.,][]{Bond2010,Moriarty2014}, or even as low as $\rm C/O>0.65$\footnote{The Sun has a C/O ratio of $\rm C/O\sim 0.55$ \citep[e.g.,][]{Asplund2009}.} \citep[e.g.,][]{Moriarty2014}. For C/O ratios above these values, C acts as a refractory element and silicon carbide (SiC) forms within protoplanetary disks instead of rock-forming minerals such as SiO$_2$ \citep[e.g.,][]{Bond2010}. Planets accreting within such an environment accumulate large amounts of SiC, which subsequently transforms into quartz and particularly graphite \citep{Hakim2018}, thereby leading to a graphite crust that acts like a rapidly cooling stagnant lid \citep{Hakim2019b}. This will significantly impact geophysical processes such as volatile degassing, and carbon and water cycles, directly affecting habitability\footnote{{However, frequent impacts could in principle disrupt the graphite-layer and allow for some interior-surface exchange, but frequent impacts would pose a separate risk for habitability.}} \citep{Unterborn2014,Hakim2019b}. Planets with a graphite layer that completely isolates a silicate layer below cannot likely degas a secondary atmosphere \citep{Hakim2019b}. If the layer is thin enough to potentially allow outgassing, the atmospheres of carbon-rich planets will nevertheless lack oxygen-rich gases, but could instead be rich in reducing species such CH$_4$ and H$_2$ \citep{AllenSutter2020}. For graphite layers of several 100\,km in depth, any mantle recycling to the surface may further be inhibited \citep{Hakim2019b}. The evolution of `carbide planets' toward EHs might therefore strongly be restricted.

But how often might such planets occur? \citet{DelgadoMena2010} and \citet{Petigura2011CO} investigated the C/O ratio of 370 FGK stars from the HARPS sample and 457 FG stars from the California Planet Survey, respectively, and found large fractions with C/O ratios above 0.8, i.e., 24\% and 29\%, respectively. These values were later corrected downwards by \citet{Fortney2012} to be $10-15$\%, and further criticized by \citet{Nissen2013} who could not confirm any C/O ratios above 0.8 for 33 solar-type stars from the HARPS sample. \citet{Brewer2016} further concluded that there are no stars with C/O$>0.7$ out of a sample of 849 stars within the solar neighborhood. Similarly, \citet{SuarezAndres2018} did not find any solar-type stars with C/O$>$0.8 out of 99 planet-hosting stars from the HARPS sample. However, \citet{Santos2017} noted that 10 stars in their sample of 371 Solar Neighborhood stars have C/O ratios above 0.8, i.e., less than 4\%. Another recent study of 249 FGK stars in the Northern hemisphere by \citet{Stonkute2020} found that out of 78 stars, for which the C and O abundance were determined, 12\% show a C/O ratio higher than 0.65, but none are above 0.8.

It hence seems that planet-hosting stars with C/O ratios above 0.8 are very rare. However, up to $\sim$10\% of stars could indeed have C/O ratios between 0.65 and 0.8, and may therefore be able to form carbon-rich planets \citep{Moriarty2014}. However, the fraction of carbide-planets might be rarer still since \citet{Moriarty2014} found that carbon-rich planetesimals around solar-like stars will mostly be restricted to orbital distances $\lesssim$0.5-0.6\,AU if the C/O ratio is between 0.65 and 0.8; for $\rm C/O>0.8$, on the other hand, they could form in the entire disk.

Another important tracer is the stellar Mg/Si ratio \citep[e.g.,][]{CarterBond2012MgSi,Spaargaren2020,Mojzsis2021}\footnote{The Earth has Mg/Si$\sim$1.16 \citep[e.g.,][]{Dauphas2015}.}. Besides influencing the general mantle mineralogy, high ratios with $\rm Mg/Si\gtrsim1.0$ lead to faster planetary cooling and mechanically weaker ferropericlase-rich mantles, whereas low ratios lead to slower cooling and mechanically stronger mantles \citep[e.g.,][]{Spaargaren2020,Spaargaren2022,Mojzsis2021}. As \citet{Spaargaren2020} have calculated, Mg/Si controls the competition of de- and ingassing for plate tectonics planets, and degassing for their stagnant lid counterparts. While degassing at stagnant lid planets can continue for at least 4\,Gyr in case of high Mg/Si ratios, it can take up to $\sim$8\,Gyr for the opposite case. For plate tectonic planets, $\sim$95\% of their water budget is degassed in the first Gyr in any of the cases studied by \citet{Spaargaren2020}, but the Mg/Si ratio will determine whether ingassing (high Mg/Si) or outgassing (low Mg/Si) dominates afterward. Such a ratio can hence be crucial for whether a planet might still retain some volatiles after its host star falls below $\alpha_{\rm at}^{\rm ll}$ (see Section~\ref{sec:C-N-Availability}). Due to the differences in mineralogy, viscosity, and cooling rates, the likelihood of plate tectonics will also vary for different Mg/Si ratios but its exact effect is contentious \citep[e.g.,][]{Mojzsis2021}.

Based on thermodynamic phase-equilibria simulations to calculate mantle and melt composition, \citet{Unterborn2017} tentatively estimated the fraction of planets that might be able to evolve plate tectonics\footnote{The importance of plate tectonics for biological evolution is, e.g., elaborated in \citet{Stern2024}} for two samples of stars, i.e., for 609 FGK stars as studied by \citet{Adibekyan2012}, and for 89 stars known to host exoplanets with a radius below $R_{\rm pl} \leq 1.6$\,R$_{\oplus}$ that were analyzed through APOGEE \citep[][]{Wilson2010}. Based mostly on their Mg, Si, and Fe composition, they found that about one-third of these stars could host planets with a composition able to support plate tectonics. Due to the limited knowledge of the parameters controlling plate tectonics, these results may be contentious, but they still provide a tentative hint that elemental composition and the Mg/Si ratio indeed matter for the evolution of plate tectonics. In addition, the results of \citet{Unterborn2017} do not imply that all planets with the `right' composition will indeed evolve plate tectonics, but that one-third might only pose an upper limit.

Regarding the stellar fraction in the Galaxy with Mg/Si ratios below and above 1.0, \citet{SuarezAndres2018} found, based on the HARPS sample, that all investigated stars with low-mass planets (below $M_{\rm pl}=3$0\,M$_{\oplus}$) have Mg/Si ratios between 1.0 and 2.0, whereas this is 85\% for stars with high-mass planetary companions. For the Solar Neighborhood, \citet{Brewer2016} found that about 60\% of all stars show $1\leq\rm Mg/Si<2$. \citet{Hinkel2018} further investigated the potential planet mineralogy around the 10 closest stars to the Sun by assuming that the refractory element composition of these stars is similar to their planets'. Although they found that their mineralogy may potentially be similar to the Earth's, \citet{Hinkel2018} point out that the precision in measuring the respective molar fractions within these stars is too small to determine different mineralogical pathways. The assumption that Mg/Si (as well as Fe/Si) is indeed essentially identical between planets and their host stars, however, was already shown by \citet{Thiabaud2015}. These authors further found that, contrary to Mg/Si, the C/O ratio of a planet depends only weakly on the star's C/O ratio. This, however, does not affect the argument that high C/O ratios in stars will lead to carbon-rich planets since the C/O ratio dictates the chemistry in the protoplanetary disk, which in turn favors the formation of carbide planets.

Although the investigated stars by \citet{Hinkel2018} might host mineralogically similar planets compared to our Solar System, there are hints that other mineralogies indeed exist in the Galaxy. By investigating 23 polluted white dwarfs, \citet{Putirka2021}, for instance, found that just one of them contained traces of planetary debris that has a similar composition to the Earth's based on Mg, Si, Fe, and Ca. The rest, however, showed exotic compositions and mineralogies. {Because of galactic evolution, Mg/Si additionally varies over time, which may point towards a certain window in time, where the emergence of plate tectonics (and habitability) could be specifically favored \citep{Mojzsis2021}.}

Finally, it has to be mentioned that there are additional compositional characteristics or chemical processes that will shape the mineralogy of a planet such as its oxygen fugacity \citep[e.g.,][]{Mojzsis2021} or Si/Fe ratio \citep[e.g.,][]{Frank2014,ONeill2020Tectonics} but a further discussion on these issues is beyond the scope of this study. However, there is one specific aspect related to composition that we briefly address in the next section.

\paragraph{Radioactive heat budget:}

The radioactive heat budget a planet inherits from the protoplanetary disk is an important parameter for its subsequent evolution since it determines its cooling history and affects volcanic degassing and its tectonic regime \citep[see, e.g.,][]{ONeill2020Review,ONeill2020Tectonics,Nimmo2020}. It is also strongly related with other requirements such as the long-term stability of carbon-silicate and nitrogen cycles. We will not discuss the importance of the radioactive heat budget in detail here since we presented its relevance already to some extent in Paper~I. However, we point out that the abundance of U, Th, and K varies over galactic age \citep{Frank2014,ONeill2020Review,ONeill2020Tectonics} and space \citep[e.g.,][]{Unterborn2015,Botelho2019}. Based on these variations, \citet{Nimmo2020} calculated that the radiogenic heat production of a planet can vary between $\sim$30\% to $\sim$300\% of the Earth's value. As \citet{Nimmo2020} emphasize, such a variation will have strong effects on degassing, but also on the evolution of planetary magnetic fields. The evolution and distribution of radioactive isotopes in the Galaxy can therefore be an important tracer for habitability, even though the final abundance of radioactive isotopes within a planet will dependent on various parameters such as stellar composition, devolatilisation, and even atmospheric escape \citep{Erkaev2022}. Crucially, not all planets will end up providing a long-term habitable environment, also because of the different heat budgets they might inherit.

\paragraph{Availability of CO$_2$ and N$_2$:}\label{sec:C-N-Availability}

Within our current model, we simply assume that every planet can still evolve into an Earth-like Habitat after the stellar activity of its host star decreases below the XUV threshold value. This, however, is certainly a significant simplification and overestimate. In reality, different factors will determine whether a planet can indeed evolve habitable conditions afterwards.

Besides geological activity, a planet certainly needs an atmosphere. Before the star falls below the XUV threshold value (Section~\ref{sec:XUVEffect}), which can take billions of years, an N$_2$-O$_2$-dominated atmosphere is either not stable or dominated by high CO$_2$ mixing ratios to counteract thermal/non-thermal escape. Until the atmosphere becomes finally stable, however, nitrogen can either be lost into space or retain in the mantle. Similarly, CO$_2$ escapes if the stellar XUV flux is too strong \citep[see, e.g.,][]{Tian2009,Johnstone2021}. This indicates that, after the stellar XUV flux decreases below the threshold value, a planet must still (i) be geologically active with functioning carbon-silicate and nitrogen cycles, and (ii) have sufficient N$_2$ (and CO$_2$) available that can be degassed from its interior. If CO$_2$ dominates the atmosphere and was not lost into space during the star's active phase, any excess CO$_2$ must additionally be weathered into the surface to allow the evolution of an N$_2$-O$_2$-dominated atmosphere with minor CO$_2$ mixing ratios that are below the toxicity limits for complex life.

This sounds trivial. However, since volatile budgets (including H$_2$O and O$_2$) are restricted on any planet, it might easily happen that the host star's radiation and plasma environment finally allows for an N$_2$-O$_2$-dominated atmosphere to be stable, but that there are simply no volatiles left to be degassed from its interior. Such planets will end up either completely barren and/or with a very thin atmosphere, potentially even dominated by heavy noble gases\footnote{Heavy noble gases (e.g., Ar, Kr, Xe) are so heavy that they are lost much harder than any other volatiles. Noble gases may therefore remain and form a, potentially thin, but entirely exotic kind of atmosphere.}. Even if the planet still keeps some volatiles in its interior, if geological activity already ceased, they will simply remain trapped.

Degassing through flare-CME-induced interior heating \citep[][]{Grayver2022}, induction heating from the host star's magnetic field \citep{Kislyakova2017,Kislyakova2018,Kislyakova2020Induction}, or tidal interaction with the host star and/or a large moon will not help either. If a planet, instead of geological activity due to radioactive heating, experiences any prolonged externally induced processes that lead to strong and frequent volcanic degassing, it can lose all volatiles prior to its host star becoming inactive. This can be detrimental for planets around highly active stars.

For how long a stagnant lid planet can indeed degas was simulated by \citet{Unterborn2022} based on a planet's radiogenic heat budget and galactic evolution. They found that the lifetime of mantle degassing increases with planetary mass but decreases with galactic age \citep[due to the gagactic availability of radioactive isotopes decreasing over time; see, e.g.,][]{Frank2014,ONeill2020Review}. \citet{Unterborn2022} derived two relations for the maximum age a stagnant lid planet can degas \textit{today} in the Galaxy, i.e.,
\begin{equation}\label{eq:U2022a}
  Age_{\rm max}^{\rm avg} = -7.1 + 8.9 \left(\frac{M_{\rm pl}}{\rm M_{\oplus}}\right)^{0.09}~\mbox{Gyr},
\end{equation}
and
\begin{equation}\label{eq:U2022a}
  Age_{\rm max}^{\rm UL95\%CI} = -1.7 + 5.4 \left(\frac{M_{\rm pl}}{\rm M_{\oplus}}\right)^{0.21}~\mbox{Gyr},
\end{equation}
where $Age_{\rm max}^{\rm avg}$ is the maximum age based on the average of their calculated degassing lifetimes and $Age_{\rm max}^{\rm UL95\%CI}$ is the upper bound based on the 95\% confidence interval of their degassing simulations. With these equations, \citet{Unterborn2022} calculate the maximum degassing age for a stagnant lid planet with $M_{\rm pl}=1.0$\,M$_{\oplus}$ to be $Age_{\rm max}^{\rm avg}\sim 1.8$\,Gyr and $Age_{\rm max}^{\rm UL95\%CI}=3.7$\,Gyr, respectively. However, \citet{Unterborn2022} emphasize that the real maximum age could potentially be higher due to several chemical and geophysical arguments, but also due to tidal heating. Planets with plate tectonics were further not considered by \citet{Unterborn2022}, whose degassing lifetimes might be substantially different. Note that the Earth with an age of $\sim$4.5\,Gyr is still volcanically active and will remain so for quite some time.

However, the simulations by \citet{Unterborn2022} illustrate that volcanic degassing timescales have a limit, thereby profoundly affecting the prevalence of EHs. The longer the activity of the host star remains high and the lower a planet's N$_2$, CO$_2$ and H$_2$O budgets are, the likelier it is that a planet will have no atmosphere and remain barren. Finally, O$_2$ also needs to be accounted for. There must be some availability of O$_2$, for instance through surface water that can be decomposed into O$_2$ via photosynthesis. If all water escapes from the planet or oxygenic photosynthesis never evolves, it will not matter whether some CO$_2$ or N$_2$ remains on the planet, as it can never evolve into an EH due to the lack of O$_2$.

\paragraph{Availability of bioessential elements:}

Various other elements besides O, H, C, and N are crucial for the origin and evolution of complex life. Phosphorus is the obvious example since it is essential for \textit{life as we know it} and is often, besides nitrogen \citep{Tyrrell1999}, considered as the limiting nutrient \citep[e.g.,][]{Pasek2017,Lingam2018Phosphorus} -- potentially even for humanity \citep{Elser2012Phosphorus}. During the Precambrian, P was likely limiting the biological productivity, which might have contributed to the delay in the rise of atmospheric O$_2$ \citep[e.g.,][]{Kipp2017,Reinhard2017}. A change in the phosphorus cycle, however, occurred between 800 and 635\,Ma accompanied by an increase in atmospheric O$_2$ levels and the rise of animals \citep{Reinhard2017,Cox2018}. Also, the GOE could be linked with a rise in sedimentary phosphorus recycling \citep{Alcott2022} and the change in phosphorus availability could in-turn be linked to long-term mantle cooling and, thus, plate tectonics \citep{Cox2018}. At Earth, the availability of P was therefore likely a limiting factor, not only for complex life, but also for the emergence of our N$_2$-O$_2$-dominated atmosphere.

Besides geochemical availability, P also must be present at a planet in the first place. \citet{Hinkel2020} emphasize the role of stellar phosphorus abundance for exoplanets and our understanding of astrobiology. As they point out, only $\sim$1\% of all stars and exoplanet hosts have their P abundances presently known through observations, with the Sun having a relatively high abundance of P. Since phosphorus is strongly partitioned into the core, planets around stars with significantly less P could therefore lack surficial P and, consequently, life \citep{Hinkel2020}.

Several other elements play a crucial role in biology. Besides alkali metals such as K and Na \citep[e.g.,][]{Sigel2016}, transition metals like Fe, Cu, Mn, and Mg are essential for electron-transfer processes in aerobic metabolism \citep[e.g.,][]{Moore2017}. \citet{Covone2022} therefore argue for the availability of these transition metals to be an essential feature that must be factored into the search for life, as their distribution in the Universe will not be uniform, but a function of SN frequency and galactic dynamics.

\subsubsection{Potential requirements feeding into $B_{\rm env}$}\label{sec:otherBenv}

\paragraph{Carbon-silicate \& nitrogen cycles:}\label{sec:C-N}

The importance of both carbon-silicate and nitrogen cycles is discussed in Paper~I. However, here we emphasize a few additions to our discussion. First, this certainly necessary requirement for building and keeping N$_2$-O$_2$-dominated atmospheres with minor amounts of CO$_2$ stable over geological timescales is obviously dependent on other criteria that were already discussed within this study, such as the amount of water available at a planet, the radioactive heat budget, a planet's mineralogy, and probably even the existence of a large moon. Moreover, the availability of N$_2$ and CO$_2$ itself is obviously necessary for both cycles to work. It is therefore important to note that these factors cannot be implemented into any model independently without crucially underestimating the prevalence of EHs.

Second, through investigating specific (isolated) parameters such as volcanic degassing \citep[e.g.,][]{Noack2014,Noack2017,Tosi2017}, carbon-silicate cycling \citep[e.g.,][]{Valencia2018,Foley2018,Foley2019,Hoening2019} or water degassing \citep{Godolt2019} various studies report that, under specific circumstances, stagnant lid planets may provide habitable conditions. However, no study investigated several criteria in unison or whether nitrogen cycling could in principal work on stagnant lid planets. It might, hence, be possible that stagnant lid planets indeed provide habitable conditions under specific circumstances, but it remains unknown whether they can also form EHs. However, neither can the occurrence rate of plate tectonics be reasonably estimated.

This opens two possibilities. Most worlds with subaerial water and land that host N$_2$-O$_2$-dominated atmospheres, could be either plate tectonic, or even stagnant lid worlds if the frequency of plate tectonics is exceedingly rare. This is why we define such a requirement via the existence of working carbon-silicate and nitrogen cycles and not via the existence of plate tectonics\footnote{{Note, however, that plate tectonics may further be crucial for habitability in several other aspects such as the long-term stability of a planet's hydrosphere and lithosphere \citep[e.g.,][]{Wang2023Tectonics}, and for biological evolution \citep{Stern2024}.}}. In any case, however, it seems reasonable to assume that the occurrence rate of planets where both cycles function could be relatively rare. If we consider the estimate by \citet{Unterborn2017} to be reasonable and plate tectonics to be the dominating state for meeting this requirement, its occurrence rate will at best be $\beta_{\rm env}^{\rm cycle}\lesssim$0.3, but probably much lower. However, \citet{Weller2018} argue that plate tectonics is less a characteristic of specific rheologies but of dynamic, temporal variation, meaning that planets migrate through different tectonic states at similar thermal states, but due to feedback mechanisms and random variations, at different evolutionary times \citep[see also, e.g.,][]{ONeill2016,Wong2016}. If that is so, and there is evidence that the Earth has indeed migrated through various tectonic states \citep[e.g.,][]{Debaille2013,ONeill2014}, it will be less a question of whether a planet develops plate tectonics, but more of a question of when it develops it and whether it develops it at the \textit{right time} compared to other important processes and timescales.

\paragraph{Orbital eccentricity \& planetary obliquity:}

Planets may orbit their stars on highly eccentric orbits so, thereby not entirely moving inside the HZ(CL). For example, \citet{Kane2012} found that of 489 planets for which complete orbital solutions were known, 187 had eccentricities greater than $e_{\rm pl}=0.2$. However, most of the planets with eccentric orbits are located in single-planet systems (as far as known) and are gas giants. While the latter may arise from observational bias \citep{Kane2012}, the higher mean eccentricity of single-planet systems was confirmed by subsequent studies \citep[e.g.,][]{Limbach2014,Xie2016Eccentricity,VanEylen2019}. \citet{Xie2016Eccentricity}, for instance, found a dichotomy between single and multi-planet systems -- whereas singles show an average eccentricity of $e_{\rm pl}\sim$0.3, multiples have almost circular orbits with $e_{\rm pl}\sim$0.04.

Different models further found that the climate of HZ planets with highly eccentric orbits can be still habitable on a broad parameter range \citep[e.g.,][]{Kane2021Eccentricity,Dressing2010,Bolmont2016,Way2017,Bolmont2016,Way2017}. Around M dwarfs, eccentricity variations are less severe than around G stars where highly eccentric planets can be partially frozen on their orbits \citep{Linsenmeier2015,Bolmont2016}. \citet{Linsenmeier2015} found that a planet with an Earth-like atmosphere and an eccentricity of $e_{\rm pl}=0.5$ can experience habitable conditions even for a higher range of incident stellar surface fluxes, albeit not on the entire orbit and with conditions being uninhabitable on the whole surface for part of the orbit. Interestingly, they also found that the habitable time of planets on eccentric orbits decreases with increasing obliquity. This was also found by \citet{Deitrick2018} who showed that an Earth-like planet orbiting a G star can experience Milankovich cycles for large orbital and obliquity variations, which reduces habitability by inducing runaway glaciation. These authors therefore conclude that planets receiving the same insulation can be either habitable or completely ice-covered based on obliquity and orbital variations. A combination of higher eccentricity and obliquity might therefore be particularly susceptible to triggering at least  partially uninhabitable surface conditions. Even if putative organisms evolved into being relatively resilient against obliquity variations, this would still pose a serious risk for survival.

Some other risks relate to eccentricity and/or obliquity. \citet{Palubski2020}, for example, find that eccentricities above $e_{\rm pl}=0.55$ and $e_{\rm pl}=0.38$ can initiate a \textit{Runaway Greenhouse} state on planets orbiting G and M dwarfs, respectively, which will also lead to extreme water loss. An entire Earth ocean can thus be lost within $\sim$3.6\,Gyr for around a G star, whereas it only takes $\sim$690\,Myr around an M dwarf. Eccentricity can further influence the stability of a moon around a planet \citep[e.g.,][]{Hamilton1992}, thereby potentially decreasing their occurrence rate. {Eccentric orbits may also induce strong tidal heating, leading to detrimental effects for a biosphere \citep[e.g.,][]{McIntyre2022}.}

\paragraph{The existence of an intrinsic magnetic field:}\label{sec:mag}

In the prevailing paradigm, intrinsic planetary magnetic fields are considered to be essential for a habitable planet \citep[e.g,][]{Lammer2009,Tarduno2014,Cockell2016} mostly because of two persuasive reasons. First, a magnetosphere protects a planet from non-thermal atmospheric erosion from the solar wind \citep[e.g.,][]{Dehant2007,Lundin2007} and frequent CMEs \citep[e.g.,][]{Khodachenko2007}. Second, it protects the surface from SEPs and galactic cosmic rays (GCR), which both can have deleterious effects on surficial organisms \citep[e.g.,][]{Griessmeier2005,Griessmeier2009,Brack2010,Belisheva2012}. During the last years, however, this paradigm became debated \citep[e.g.,][]{Blackman2018,Gunell2018,Egan2019,Ramstad2021,Way2022}, since ion loss through the Earth's polar caps constitutes an important escape mechanism that approximately equals the escape rates from Venus and Mars, both of which do not have intrinsic magnetic fields today \citep[e.g.,][]{Gunell2018}. Polar outflow could, even have been significantly higher early in Earth's history \citep{Kislyakova2020,Grasser2023} and it was shown that it can provide a significant escape channel at magnetized planets around M dwarfs \citep[e.g.,][]{Dong2019}. Even for the present-day Earth, escape rates initiated by CMEs impacting the magnetosphere could be higher than initially thought \citep{Zhang2022}.

Whether intrinsic magnetic fields provide some protection, particularly for the extreme conditions during the early solar system and for planets around M and K dwarfs, however, remains poorly constrained, as no studies compare planets with the same atmospheric composition and incident XUV flux, but with and without intrinsic magnetic field \citep[see discussion in][]{Way2022}. A comparison between Mars, Venus, and Earth \citep[see, e.g.,][]{Gunell2018,Ramstad2021} necessarily includes escape from atmospheres with vastly different compositions and therefore lack comparability. Giving Mars or Venus an N$_2$-dominated atmosphere will change their thermal and non-thermal loss rates significantly, as will also the case on Earth with a CO$_2$-dominated atmosphere. This was also shown by \citet{Dong2020TOI} theoretically, who found highly different loss rates for N$_2$- and CO$_2$-dominated atmospheres at the Venus-analog TOI-700~d. However, some studies investigated the protection of planetary magnetic fields against atmospheric escape around M dwarfs, and found that it cannot prevent the atmosphere from complete erosion \citep[e.g.,][]{GarciaSage2017,Airapetian2017,RodriguezMozos2019}, although escape will significantly  be diminished if an intrinsic magnetic field indeed is present \citep{Dong2020TOI}. Whether this could also be the case around G and K dwarfs remains unclear.

Intrinsic magnetic fields were also proposed to be essential for protecting a planet from GCRs and SEPs \citep[e.g.,][]{Brack2010,Griessmeier2005,Griessmeier2009}. Associated secondary particle showers that reach the surface can severely affect the DNA and mutation rates of unprotected lifeforms \citep[e.g.,][]{Belisheva2012}. A thick atmosphere ($\geq$1\,bar) together with an induced magnetic field, however, might be sufficient to prevent most of the energetic particles from reaching the surface \citep[e.g.,][]{Griessmeier2016}. A magnetosphere can also prevent the destruction of a potential ozone layer by energetic particles, a crucial component for avoiding UV radiation damage \citep[e.g.,][]{Glassmeier2010}. The almost complete vanishing of the Earth's magnetic field due to a polar reversal at the end of the Ediacaran period at $\sim$550 Ma was already proposed to be the reason for the End-Ediacaran extinction event \citep{Meert2014,Wei2014}. Recent studies even highlight the possibility that the Laschamp magnetic excursion in the late Quaternary could have contributed to the extinction of large mammals and potentially even Homo Neanderthalensis \citep[e.g,][]{Channell2019,Cooper2021,Erdmann2021}. \citet{Lingam2019Magnetosphere} investigated this issue further for the Earth's last $\sim$1\,Gyr and found that the magnetosphere might have only slightly protected from energetic particles and atmospheric escape by reducing both by a factor of $\sim$2. For M stars \citep{Lingam2019Magnetosphere}, and eventually also for the early Earth, the protection effect could be larger though.

The occurrence rate of planets with intrinsic magnetic fields, at least, was recently estimated by \citet{McIntyre2019} based on the dipole moment scaling from \citet{Olson2006}. They found that $44\pm13\%$ of all known rocky HZ planets with $R_{\rm pl}\leq 1.23$\,R$_{\oplus}$ might possess non-negligible intrinsic magnetic fields. {Another recent study by \citet{Atkinson2024} investigated whether 1601 known exoplanets lie outside their star's Alfv\'{e}n surface -- which would induce strong interactions with the host star -- but inside their HZ. They found 84 planets meeting both criteria out of which 11 planets have a mass of $M_{\rm pl}=0.5-3.0\,$M$_{\oplus}$. By comparing their results with the ones from \citet{McIntyre2019}, \citet{Atkinson2024} identified only four planets -- TOI-700d, e, K2-3d and Kepler-186f -- that meet their habitability criteria and may possess a significant magnetic field. Note that only Kepler-186f, however, has a radius below $R_{\rm pl} = 1.23$\,R$_{\oplus}$, while it is $R_{\rm pl}\sim 1.46$\,R$_{\oplus}$ for K2-3d \citep{DiamondLowe2022}.} If magnetospheres turn out to be essential, this could therefore have a profound effect on the prevalence of EHs.

\paragraph{Tidally locked planets:}\label{sec:locked}

\citet{Barnes2017} found that rocky HZ planets could be tidally locked within just $\sim$1\,Gyr and $\sim$10\,Gyr for stars below $M_{\star}\sim0.7$\,M$_{\odot}$ and up to $M_{\star}\sim1.0$\,M$_{\odot}$, respectively. However, asynchronous rotation can be prolonged or even maintained for planets with atmospheres via thermal tides, indicating that Earth-like planets with a 1\,bar atmosphere can avoid tidal locking for stellar masses above $M_{\star}\sim0.5-0.7$\,M$_{\odot}$ \citep{Leconte2015}. Tidal interactions with moons and/or closely orbiting planets can further perpetuate or prolong asynchronous rotation. Still, most planets around M dwarfs will be tidally locked.

Synchronously rotating planets could even have advantageous characteristics. The HZ for these planets might be extended \citep[e.g.,][]{Kopparapu2016Tides,Yang2013}, tectonic and volcanic activity can be induced \citep[e.g.,][]{McIntyre2022}, and other positive effects similar to those from large moons, e.g., nutrient upwelling \citep{Lingam2018}, could prevail, at least as long as tides are prevalent at these bodies. If tidal forces are strong enough, on the other hand, tidal heating can render planets uninhabitable, as found for $\sim$18\% of all presently known tidally locked planets \citep{McIntyre2022}.

Synchronous rotation was long assumed to render climates uninhabitable due to the atmosphere freezing out on the dark side \citep[e.g.,][]{Joshi1997}. Since then, however, many studies suggest this scenario to be unsubstantial and that stable and habitable climate states will exist \citep[e.g.,][]{Joshi1997,Joshi2003,Yang2013,Boutle2017,Checlair2017,Turbet2016,Turbet2018,Hu2020}, if either \textit{Runaway Greenhouse} \citep[e.g.,][]{Kite2011,Kane2018} or atmospheric collapse \citep[e.g.,][]{Wordsworth2015,AuclairDesrotour2020,Turbet2018,AuclairDesrotour2022} can be avoided. Since atmospheres at tidally locked planets need to balance the extreme difference in insulation between the dark hemisphere and the sub-stellar point, a high abundance of water vapor or CO$_2$ might be needed to compensate this difference via radiative transport. However, \citet{Turbet2016} showed that the trapping of water at the poles or the nightside can easily be avoided even with low to negligible amounts of CO$_2$ since water vapor can regulate the temperature. For planets with small to negligible water reservoirs relatively high amounts of CO$_2$ are needed, e.g., $p$CO$_2\sim0.5$\,bar for the case of Proxima Centauri B, a value toxic for complex life. However, whether a planet that always shares the same hemisphere with its host star can indeed evolve into an EH is questionable.

\paragraph{Planetary catastrophes:}

As discussed in Section~\ref{sec:moon1}, large moons tend to be drawn toward their host planets via tidal interaction, leading either to a tidal breakup of the moon or to a catastrophic collision with the planet -- events that could indeed be quite common \citep[e.g.,][]{Brasser2013,Tokadjian2020}. Any planet that suffers from such a collision will likely be sterilized. This is what we understand as planetary catastrophes.

There are various other catastrophic events that can happen to an evolving EH. Impacts of giant asteroids and orbital instabilities that lead to a dramatic orbital change or a collision with another planet could be some of them. Such an instability likely happened early in the Solar System's history and was the potential reason for the ejection of a third ice giant \citep[e.g.,][]{Nesvorny2018}. Asteroid impacts already initiated at least one of the Earth's mass extinctions \citep[e.g.,][]{Alvarez1980} and more massive and/or frequent impacts are certainly not impossible somewhere else in the galaxy, especially since some stellar systems have significantly larger debris disks than the Sun \citep[e.g.,][]{Greaves2004}.

Large-scale volcanism in the form of large igneous provinces (LIP) can be another planetary-scale catastrophe that terminates a planet's habitability. The chance of a pair of events similar in size to the largest known LIPs simultaneously happening in Earth's recorded history is $\sim$30\% and even $\sim$50\% for smaller LIPs \citep{Way2022Volcano}. Depending on the planet's climate state, simultaneous LIPs can be sufficient to initiate the \textit{Runaway Greenhouse} effect, a catastrophe that could have beset Venus \citep{Way2022Volcano}. In general, climate shifts from habitable toward either a snowball or \textit{Runaway Greenhouse} state might be crucially important, as recent studies have shown that the climatic long-term habitability of the Earth could just have been pure random chance \citep{Tyrrell2020,Arnscheidt2022}.

Another realistic option for a planetary catastrophe is the induced destruction of a planet due to a biological or artificial species either intentionally or unintentionally, even though such catastrophe must be statistically rare if EHs themselves are already uncommon (and ETIs consequently even more so). Its rareness, however, will not prevent it from happening here on Earth, at least not through mankind itself.

\paragraph{The role of Jupiter- and Saturn-like planets:}\label{sec:jup}

The danger of catastrophic asteroid impacts opens the popular debate on whether Jupiter and/or Saturn-like planets beyond the HZ are required to enable habitable conditions on a rocky exoplanet by diminishing the frequency of impact events \citep[e.g.,][]{Wetherill1994,Ward2000}. However, several studies \citep[e.g.,][]{Horner2008,Horner2009,Grazier2016} shed doubt on this assumption and some indicate that the impact frequency may even increase for certain system parameters \citep[e.g.,][]{Horner2009}.

Besides, the existence of both Jupiter- and Saturn-like giant planets beyond the ice line could nevertheless be a crucial factor for the evolution of EHs, since their very existence might hinder the growth of the inner planetary embryos toward super-Earths \citep[e.g.,][]{Batygin2015,Izidoro2018,Raymond2020,Raymond2022} and prevent the delivery of too much or too little water \citep[e.g.,][]{Grazier2016,Raymond2020}. For the Solar System, the co-existence of both giants together could have even avoided \citep[e.g.,][]{Masset2001,Morbidelli2007,Zhang2010a,Zhang2010b,DAngelo2012,Raymond2020} that Jupiter underwent its Type II migration \citep[e.g.,][]{Lin1986,Lin1993} deeper into the inner system, which could have disturbed the growth and orbits of the terrestrial planets.

There are no hard numbers on how frequent such a combination of gas giants may be, but \citet{Raymond2020} estimate that the frequency of Jupiter-like planets with correct parameters for orbital separation and eccentricity could be as low as 1.0\% for Sun-like stars and 0.1\% for the entire stellar spectrum, respectively, an estimate that does not consider the combined existence of Jupiter- and Saturn-like giant planets. Such a rareness is somehow confirmed by \citet{Chametla2020}, who found that the parameters allowing for a reversed migration via resonance capturing between Jupiter and Saturn might be a low-probability scenario and cannot be expected to happen in other planetary systems.  \citet{Chametla2020} therefore conclude that the therewith connected Grand Tack \citep[e.g.,][]{Tsiganis2005,Walsh2011} generally seems to be an unlikely scenario. These potential effects, however, might not directly feed into $B_{\rm env}$ as these can mostly be subsumed within the planet occurrence rate, $\beta_{\rm HZCL}$, and $\beta_{\rm pc}^{\rm H_2O}$, and $\beta_{\rm pc}^{\rm accr}$, respectively.

A close-by giant planet can also affect the long-term orbital stability of a HZ planet \citep[e.g.,][]{PilatLohinger2008,PilatLohinger2008b,Georgakarakos2018,Horner2020}. A recent study by \citet{Georgakarakos2018}, for example, found that massive giant planet neighbours can reduce the habitability of a planet due to an induced orbital change, even though in some rare cases the habitability may be enhanced. \citet{Bailey2022} similarly concluded that giant companions mostly reduce habitability but with some specific parameters increasing it. Another study by \citet{Horner2020} investigated how changes in Jupiter's orbit might alter Earth's climate due to changes in its orbital evolution. Most tweaks on the parameters ended with negligible influences, indicating that Earth's specific conditions do not favor the ``Rare Earth'' argument, as originally formulated by \citet{Ward2000}, at least not related to the need of Jupiter. However, the architecture of a planetary system may indeed be an important factor to consider.

\paragraph{Dark Matter:}

This factor might as well be subsumed within any other category as dark matter still presents itself as an enigma. However, some studies nevertheless discuss the potential effect of different forms of dark matter on planetary habitability. \citet{Hooper2012} point out that the annihilation of dark matter within a planet larger than the Earth and in the inner galaxy, i.e., in regions with higher dark matter density, could potentially maintain liquid water even in the absence of starlight. While this might be a beneficial effect, \citet{Randall2014} propose that the existence of a dark matter disk at the mid-plane of the Galaxy and Earth's crossings through this disk can be an explanation for a potential periodicity in crater recording and connected mass extinctions due to perturbations of Oort cloud comets. This 26-30\,Myr cycle and its potential connection with a dark matter disk was also emphasized by \citet{Rampino2015} and \citet{Rampino2020}, who further point out that interactions with weakly interacting massive particles (WIMPs) in the galactic plane could lead to heating in the Earth's interior and a related increase in geologic activity. {The Cretaceous-Paleogene (K-Pg) extinction event at 66\,Ma coincides {relatively well} with the emergence of the Deccan Traps LIP, which {was even suggested} to have on-itself been sufficient to trigger the related mass extinction \citep{Keller2020,Cox2023}}. {However, there is a broad scientific consensus that the K-Pg extinction was almost exclusively triggered by the Chicxulub impactor \citep{Chiarenza2020Chicxulub,Hull2020Chicxulub}.} Besides, the impactor has also been suggested to enhance the Deccan Traps LIP \citep{Richards2015} and to trigger further volcanism \citep{Byrnes2018}.

\citet{Kramer2022} revisits the dark matter theory and also finds some correlation between impacts, mass extinctions, the galactic oscillation of the Solar System, and the crossing of the spiral arms. Their best-fit values even predict the Chicxulub impact at 66\,Ma correctly. Since dark matter is still very poorly understood, however, such correlations and its effect on planetary habitability certainly can neither ruled out nor factored in.

\subsubsection{Potential requirements feeding into $B_{\rm life}$}\label{sec:otherBlife}

This section only gives a brief and non-exhaustive overview of potential requirements feeding into $B_{\rm life}$. There are certainly additional factors such as the emergence of denitrification and/or similar nitrogen cycling processes \citep[e.g.,][]{Zerkle2017Cycle}, but it is beyond the scope of our study to discuss potential biological requirements in great detail. The following can therefore be seen as pure examples. We also note that some of the already discussed parameters such as the availability of bioessential elements\footnote{Since the availability of bioessential elements - at least its principle existence at the planet - is a question of planetary composition and its initial set-up, it makes sense to include it in $B_{\rm pc}$. Whether these elements are then indeed biochemically available for, e.g., metabolism is another topic that may indeed fit better into $B_{\rm life}$. That is why we logically include the P cycle into $B_{\rm live}$.}, may also fit, at least to some extent, within this requirement cluster.

\paragraph{Origin of life and panspermia:}\label{sec:originLife}

Any planet that evolves an N$_2$-O$_2$-dominated atmosphere likely experienced the origin of life, which can therefore be considered as a necessary requirement for EHs to evolve. However, the Earth presently provides the only sample we know that life originated and evolved. Any estimate on how frequent abiogenesis may be in the Galaxy can therefore mostly be based on speculation. There are, however, some parameters that one can theoretically consider. \citet{Lingam2020Oxygen} emphasized the toxicity of high O$_2$ abundances for life, which is congruent with laboratory experiments indicating that the emergence of life is favored at reduced environments \citep[e.g.,][]{Benner2020,Sasselov2020}. On worlds where O$_2$ builds up abiotically due to water loss, the origin of life might correspondingly be less likely. \citet{Lingam2020Oxygen} even proposed that abiotic O$_2$ build-up can shift the HZ outwards, complementary to CO$_2$ toxicity shifting the outer boundary of the HZ inwards, an effect that will particularly affect M dwarfs. This may even lead to an intersection of O$_2$ and CO$_2$ toxicity boundaries and, if so, to a lack of HZCLs around the affected stars.

\citet{Lichtenberg2022} further suggest that reducing environments for the origin of life are indeed favored around FGK dwarfs. They performed simulations of the late bombardment in M dwarf systems and found that young planets around stars with $M_{\star}\leq$0.4\,M$_{\odot}$ do not experience enough impacts for producing and delivering prebiotically relevant concentrations of reduced molecules needed for the origin of life. {In addition, \citet{Anslow2023} found that planets around low-mass stars suffer from a significantly higher number of high-velocity impacts, indicating that the delivery and survival of prebiotic molecules may become impossible around M dwarfs, at least for loosely packed systems.}

Besides the origin of life on a planet, we also need to mention panspermia, more specifically lithopanspermia, which describes the transport of living organisms from one stellar system to another by way of meteoroids or other minor bodies \citep[e.g.,][]{Hinteregger1981,Melosh2003,Adams2005Panspermia,Wesson2010,Lingam2022Panspermia}. This could in principle increase the number of inhabited planets without the need of life's origin taking place on each of them separately. It could even seed life on planets where environmental conditions strongly decrease or even hinder the chance of abiogenesis to happen (e.g., water worlds?). Although panspermia was neglectfully treated for many decades \citep[e.g.,][]{Balbi2021Panspermia}, its scientific relevance is increasing during recent years due to a manifold of different arguments. Not only has it now been very well proven that material interchanges between Solar System bodies such as Earth and Mars \citep[e.g.,][]{Nyquist2001,Worth2013}, recent years also witnessed the detection of the first interstellar objects, namely I2/Borisov \citep{Jewitt2019} and I1/`Oumuamua \citep{Meech2017}, {and the potential first detection of an interstellar meteor \citep{Siraj2022IM}\footnote{Note that this claim is disputed; see \citet{Brown2023IMCritique}.}, illustrating the feasibility of material transport between stellar systems.}

{In addition, experiments show that organisms such as extremophile microbes \citep[e.g.,][]{Horneck1994,Horneck2001} and tardigrades \citep[e.g.,][]{Erdmann2017} can sustain in shielded space environments for extended periods of time \citep[see, e.g.,][for a discussion]{Lingam2021Life} and survive the high pressures involved in hypervelocity impacts \citep[e.g.,][]{Burchell2004,Horneck2008,Price2013} and ejections \citep[e.g.,][]{FajardoCavazos2009}. However, survival rates for impacts with $v_{\rm imp}\gtrsim5$\,km/s are relatively low for microbes \citep[i.e., in the order of 10$^{-5}$, see, e.g.,][]{Burchell2004,FajardoCavazos2009} and tardigrades did not survive impact velocities $v_{\rm imp}>0.9$\,km/s \citep{Traspas2021}. We emphasize that the escape velocity of Earth-mass planets is well beyond $v_{\rm imp}\sim$5\,km/s and that impact velocities exceeding twice the respective escape velocity are typically needed for ejecting material from a planet \citep[e.g.,][]{Lingam2021Life}. So, one could expect that survival and ejection rates are very low or negligible for heavier, Earth-mass bodies, implying that panspermia is more feasible for lower-mass planets comparable to Mars. However, early atmospheric erosion is an especially significant problem for low-mass planets and could affect the origin of life itself.}

{Setting aside potential problems with ejection and impact velocities, survival rates and timescales, how likely might a coincident transport of life within or between stellar systems actually be? Whereas the transfer of ejecta between bodies inside a stellar system \citep{Krijt2017,Lingam2017Panspermia} is relatively likely on timescales of roughly $t_{\rm et}\sim10^4 - 10^6$ years within the Solar System \citep[e.g.,][]{Worth2013} or even $t_{\rm et}\sim10^2 - 10^3$ years in the Trappist-1 system \citep[e.g.,][]{Krijt2017}, the probability of lithopanspermia between different stellar systems is mostly found to be almost negligible among galactic field stars \citep[e.g.,][]{Melosh2003,Adams2005Panspermia,Adams2022Panspermia}. This was recently estimated to be in the order of $\sim10^{-10}-10^{-13}$ objects transferred per year and life-bearing world \citep{Lingam2022Panspermia}\footnote{Note that this estimate does not take into account that only a fraction, $f_{\rm bio}$, of the objects transferred will also contain a population of organisms that can indeed seed the recipient world, see \citet{Lingam2022Panspermia}.}. If one considers dense stellar environments such as the galactic bulge \citep[e.g.,][]{Balbi2020} or stellar birth clusters \citep[e.g.,][]{Adams2005Panspermia,Belbruno2012,Adams2022Panspermia} such probability can rise significantly toward order of unity per Myr for stellar clusters \citep{Lingam2022Panspermia}.}

{Such estimate seems promising at first sight but holds a significant caveat that needs to be overcome. Although a relatively high number of stars are indeed born in stellar clusters \citep[e.g.,][]{Kruijssen2012,Chandar2017}, most disperse within a few tens of Myr \citep[e.g.,][]{PortegiesZwart2010,Pfalzner2013}. A transfer of life must therefore occur during a relatively short timescale leaving two possibilities: (i) either life must have originated and transferred in the stellar cluster within $t_{\rm ol}<$100\,Myr after cluster formation, or (ii) life must have reached the stellar cluster from a non-related galactic field star. The latter was shown to be more plausible than the former \citep{Adams2005Panspermia}, although a planet providing habitable conditions must already have formed within the cluster that can indeed be seeded with life.}

For (i) to happen, it is important to note that the planet must not only form within less than 100\,Myr, it must also evolve toward providing conditions for the origin of life to actually happen and for this life to be transferred to another habitable planet -- all of which must occur within the same $<$100\,Myr. Whether this can indeed happen during such a short timescale is essentially unknown although the earliest possible emergence of RNA formation on Earth was estimated to be around $\sim4.36\pm0.1$\,Ga, i.e., $\sim$100-300\,Myr after the Sun's formation \citep{Bennett2022}. In addition, we highlight that the XUV flux of such young stars may prevent the existence of a stable atmosphere, a factor that could further reduce the chance of successful panspermia in the cluster\footnote{As a side note, the background FUV radiation of a stellar cluster itself may present another interesting factor related to the origin and evolution of life as this is typically $\sim$3000 times larger than the average interstellar value of $1.6\times10^{-3}$\,erg\,cm$^{-2}$\,s$^{-1}$ \citep{Fatuzzo2008,Adams2022Panspermia}.} {If the target planet has no atmosphere or does generally not provide accurate habitable conditions for the impacting species, the survival chances will be negligible. Panspermia can hence only be successful if the microbes survive the impact and indeed crash onto an appropriate planet to assure long-term survival, circumstances that may be rare.}

{Besides lithopanspermia, we can also mention the possibility of directed panspermia \citep[e.g.,][]{Shklovskii1966,Crick1973}, i.e., the intentional delivery of life from one planet to another through ETIs. For this to happen, however, ETIs must either exist in the Milky Way or humankind must seed other planets \citep[e.g.,][]{Milligan2016}. Although ETIs will necessarily be less common than microbial lifeforms (see Section~\ref{sec:discussion}), directed panspermia can in principle be much more effective than lithopanspermia, since the seeds of life could be directly targeted at another appropriate planet or stellar system. Interestingly, several studies finally suggest that panspermia can in principle be discernable from localized origins of life as it can produce specific spatial patterns in the galactic environment \citep[e.g.,][]{Lin2015,Grimaldi2021}.}

\paragraph{Phosphorus cycle:}

The availability of phosphorus is a necessary requirement for the phosphorus cycle \citep[e.g.,][]{Filippelli2008,JustinoMaldonado2022} to properly work on a planet. As is the case for the carbon-silicate and nitrogen cycles, however, availability alone does not assure that the cycle indeed works and will not be a limiting factor for the origin and evolution of (complex) life. {In addition, one can expect that different planets will accrete various amounts of P depending on their host star's composition, location within the galactic disk, and their own planetary accretional history}.

{That said, P was already suggested to be a potential limiting factor for the origin of life \citep{Walton2021Phosphorus}}, and as already discussed, {its restricted availability could have suppressed biological activity during the Precambrian \citep[e.g.,][]{Bjerrum2002,Kipp2017}}. On the other hand, P may have aided the rise of atmospheric O$_2$ \citep{Hao2020a,Glaser2020,Olson2020,Watanabe2021} and the emergence of complex life, as this likely coincided with a change in the phosphorus cycle \citep[e.g.,][]{Reinhard2017}. {Related to this, we emphasize a study by \citet{JustinoMaldonado2022}, which investigates Earth's P cycle without biology, on the one hand, and biology's distinct influence, on the other. It finds the most significant impact of organisms on the phosphorus cycle to trigger an increase in continental weathering, a process that likely started with the first terrestrial ecosystems, potentially as early as $\sim$2.6\,Ga \citep{Watanabe2000} but certainly with the rise of the first embryophyte land plants around $\sim0.7-0.46$\,Ga \citep[e.g.,][]{Heckman2001}. Since this process substantially increases the bioavailability of P, \citet{JustinoMaldonado2022} suggest that life's complexification can happen on any planet where biology evolves to efficiently extract P from its surroundings, regardless of the actual abundance of phosphorus.}

Finally, we again highlight chemical weathering of subaerial continental crust to be an important phosphorus supplier on Earth \citep[e.g.,][]{Wheat1996,Paytan2007,Glaser2020,Hao2020b} and most likely also on exoplanets \citep[e.g.,][]{Glaser2020,Lingam2019WaterFrac}, thereby implying that the phosphorus cycle on a water world might be severely limited. This certainly implies some correlation between the phosphorus cycle and the requirement, $B_{\rm pc}^{\rm H_2O}$, of possessing the right amount of water (Sections~\ref{sec:h2o} and \ref{sec:h2oEffect}). Besides this brief discussion, it is beyond the scope of our study to discuss the phosphorus cycle, as well as the life-essential oxygen and sulfur cycles, in any more detail.

\paragraph{Origin of oxygenic photosynthesis \& oxygenation time:}\label{sec:oxy}

The origin of oxygenic photosynthesis is another crucial need for the emergence of an EH. Photosynthesis, in turn, will {besides {biogenic \citep{Catling2001} and abiotic \citep[e.g.,][]{Segura2007,Wordsworth2014,Harman2015}} build-up of O$_2$ via UV photolysis,} also be related to the aforementioned oxygenation time \citep[e.g.,][]{Knoll1986,McKay1996,Catling2005}, which was $t_{\rm O_2}\sim4$\,Gyr on Earth. If other worlds need comparable or even longer timescales for the planet's oxygenation to such an extent that it allows for the emergence of complex life, $t_{\rm O_2}$ itself will be a fundamental restriction on the prevalence of EHs in the Galaxy. At least the emergence of oxygenic photosynthesis may not be possible around low-mass stars \citep[e.g.,][]{Duffy2023}.

{However, oxygenation times might be substantially different for other planets. \citet{McKay1996}, for instance, suggested that $t_{\rm O_2}$ for Mars would have been significantly faster than for Earth due to its small mass, a lack of plate tectonics, and its smaller initial water abundance. Key evolutionary steps such as oxygenic photosynthesis, endosymbiosis and multicellularity could therefore have been reached much faster, if Mars did not desiccate early in its history. Besides planetary parameters that may affect $t_{\rm O_2}$, stellar parameters are certainly influencing the oxygenation time as well. In analogy to the Earth's evolution, this effect was recently derived by \citet{Lingam2021Life} as a function of stellar mass by assuming that the characteristic timescale {for oxygenation to reach post-GOE levels} will be smaller on planets receiving a higher UV flux in the appropriate wavelength range for the photolysis of H$_2$O (by taking the Lyman-$\alpha$ flux as a proxy). By decomposing the oxygenation of a planet into three distinct regimes, i.e., (i) the origin of life, (ii) the origin of oxygenic photosynthesis, and (iii) the subsequent timescale for the oxygenation of the atmosphere via UV photolysis, \citet{Lingam2021Life} arrive at the following simple relation, i.e.,}
\begin{eqnarray}\label{eq:oxygenation}
t_{\rm O_2} \sim
    \begin{cases}
      1.8\,\mathrm{Gyr}\left(\frac{M_{\star}}{M_{\rm \odot}}\right)^{-3} +0.3\,\mathrm{Gyr}\left(\frac{M_{\star}}{M_{\rm \odot}}\right)^{2.3} & M_{\star}\lesssim M_{\odot}\\
      1.8\,\mathrm{Gyr}\left(\frac{M_{\star}}{M_{\rm \odot}}\right)^{-1} +0.3\,\mathrm{Gyr}\left(\frac{M_{\star}}{M_{\rm \odot}}\right)^{-3.3} & M_{\star}\gtrsim M_{\odot}.
    \end{cases}
\end{eqnarray}
{With this, one finds that planets around lower-mass stars need longer to be sufficiently oxidized, which is due to the lower UV flux the planets receive from their host. Interestingly, $t_{\rm O_2}$ takes longer than the Hubble time for stars with masses of $M_{\star}\lesssim 0.5$\,M$_{\odot}$, another criterion that hinders the emergence of complex life on M dwarf planets in a similar manner as $t_{\rm evo}$, for which Equation~\ref{eq:tevo} finds a limit of $M_{\star}< 0.7$\,M$_{\odot}$.}

{However, we highlight that the actual oxygenation of worlds around active M dwarfs will likely proceed much faster because of the abiotic XUV induced photolysis of H$_2$O, a process through which substantial amounts of water and oxygen can be lost from the planet with part of the O$_2$ oxidizing the mantle and accumulating in the atmosphere within relatively short geological timescales \citep[e.g.,][]{Johnstone2020Water,Lingam2020Oxygen}. Although this process may lead to a much faster oxygenation of the planet than estimated with Equation~\ref{eq:oxygenation}, this does not signify good news for the evolution of complex life. On the contrary, the origin and early evolution of life will be severely hindered since reducing conditions are believed to be necessary for life to originate \citep[e.g.,][]{McCollom2005,Luisi2016,Lingam2020Oxygen}.}

Finally, it is interesting to note that this criterion could also be interrelated with the necessary requirement $B_{\rm pc}^{\rm H_2O}$, a requirement that could equate with the need of having subaerial land and oceans at the same time (as described in previous sections). Recently, \citet{Stevenson2023} proposed that the timing of the evolution of oxygenic photosynthesis and the related diversification of cyanobacteria was strongly connected to the geological evolution of the Earth in the Meso- and Neoarchean, more specifically, to the growth of subaerial continental crust,  {thereby potentially supporting the need for the simultaneous existence of subaerial land and oceans. However, one must be cautious with this reasoning since error bars on the emergence ages of photosynthesis and subaerial continental crust on Earth, and hence any causal relation between them, are still relatively high.}


    \newpage
\clearpage
\section{List of Abbreviations/Acronyms}\label{app:acronyms}

\begin{itemize}[labelwidth=2cm,align=left,itemindent=2cm]
    \item[\textbf{AGN}]  active galactic nucleus
    \item[\textbf{APOGEE}] Apache Point Observatory Galactic Evolution Experiment
    \item[\textbf{BD}] brown dwarf
    \item[\textbf{BH}] black hole
    \item[\textbf{BSE}] bulk silicate Earth
    \item[\textbf{C-Si}] carbon-silicate
    \item[\textbf{CC}] carbonaceous chondritic
    \item[\textbf{CETI}] communicating extraterrestrial intelligence
    \item[\textbf{CHNOPS}] carbon, hydrogen, nitrogen, oxygen, phosphorus, sulfur
    \item[\textbf{CME}] coronal mass ejection
    \item[\textbf{CPL}] constant phase lag
    \item[\textbf{CTL}] constant time lag
    \item[\textbf{CLHZ}] complex life habitable zone
    \item[\textbf{DNA}] deoxyribonucleic acid
    \item[\textbf{EH}] Earth-like Habitat
    \item[\textbf{ELT}] Extremely Large Telescope
    \item[\textbf{EO}] Earth ocean
    \item[\textbf{epos}] Exoplanet Population Observation Simulator
    \item[\textbf{ETI}] extraterrestrial intelligence
    \item[\textbf{EUV}] extreme ultraviolet
    \item[\textbf{FOV}] field of view
    \item[\textbf{FUV}] far ultraviolet
    \item[\textbf{FYSP}] Faint Young Sun Paradox
    \item[\textbf{GCR}] galactic cosmic rays
    \item[\textbf{Ga}] billion years ago
    \item[\textbf{Gaia DR2}] Gaia Data Release 2
    \item[\textbf{GHZ}] galactic habitable zone
    \item[\textbf{GHZ}] great oxygenation event
    \item[\textbf{GRB}] gamma-ray burst
    \item[\textbf{Gyr}] billion years
    \item[\textbf{HabEx}] Habitable Exoplanet Observatory
    \item[\textbf{HARPS}] High Accuracy Radial Velocity Planet Searcher
    \item[\textbf{HBA}] hydro-based approximation
    \item[\textbf{HWO}] Habitable Worlds Observatory
    \item[\textbf{HP}] high pressure
    \item[\textbf{HZ}] habitable zone
    \item[\textbf{HZCL}] habitable zone for complex life
    \item[\textbf{IMF}] initial mass function
    \item[\textbf{ISM}] interstellar medium
    \item[\textbf{Kepler DR25}] Kepler Data Release 25
    \item[\textbf{K-Pg}] Cretaceous-Paleogene (extenction event)
    \item[\textbf{KOI}] Kepler object of interest
    \item[\textbf{JWST}] James Webb Space Telescope
    \item[\textbf{LIP}] large igneous province
    \item[\textbf{LIFE}] Large Interferometer for Exoplanets
    \item[\textbf{LUVOIR}] Large UV Optical Infrared Surveyor
    \item[\textbf{Ma}] million years ago
    \item[\textbf{MDF}] metallicity distribution function
    \item[\textbf{METI}] messaging extraterrestrial intelligence
    \item[\textbf{Myr}] million years
    \item[\textbf{NC-CC}] non-carbonaceous chondritic
    \item[\textbf{NOE}] Neoproterozoic Oxygenation Event
    \item[\textbf{NUV}] near ultraviolet
    \item[\textbf{PAL}] present atmospheric level
    \item[\textbf{POE}] Paleozoic Oxygenation Event
    \item[\textbf{ppm}] parts per millions
    \item[\textbf{PSE}] photosynthesis effectiveness
    \item[\textbf{RNA}] ribonucleic acid
    \item[\textbf{SCUSS}] South Galactic Cap u-band Sky Survey
    \item[\textbf{SDSS}] Sloan Digital Sky Survey
    \item[\textbf{SEP}] solar energetic particles
    \item[\textbf{SETI}] search for extraterrestrial intelligence
    \item[\textbf{SFH}] star formation history
    \item[\textbf{SFR}] star formation rate
    \item[\textbf{SGR}] soft gamma repeater
    \item[\textbf{Sgr\,A$^{\star}$}] Sagittarius A$^{\star}$
    \item[\textbf{SMBH}] supermassive black hole
    \item[\textbf{SN}] supernova
    \item[\textbf{SNIa}] supernova type Ia
    \item[\textbf{SNII}] supernova type II
    \item[\textbf{TDE}] tidal disruption event
    \item[\textbf{TMT}] Thirty Meter Telescope
    \item[\textbf{TOI}] TESS object of interest
    \item[\textbf{UV}] ultraviolet
    \item[\textbf{UVHZ}] ultraviolet habitable zone
    \item[\textbf{WIMP}] weakly interacting massive particle
    \item[\textbf{WD}] with dilution
    \item[\textbf{XUV}] X-ray and extreme ultraviolet
\end{itemize}

    \newpage
\clearpage
\section{List of Variables \& Symbols}\label{app:acronyms}

\begin{itemize}[labelwidth=1cm,align=left,itemindent=1cm]
    \item[$a$] semi-major axis
    \item[$A$] normalization factor for $n_{\rm p}$
    \item[$A_{\rm at}(\prod_{i=1}^n\alpha_{\rm at}^{i})$] requirement cluster on atmospheric stability
    \item[$A_{\rm GHZ}(\prod_{i=1}^n \alpha_{\rm GHZ}^{i})$]  requirement cluster on GHZ
    \item[$a_{\mathrm{H_2O},i}$] fitting parameter for deriving $\beta_{\rm pc}^{H_2O}$ for each stellar mass
    \item[$a_{\mathrm{HZCL}}$] fitting parameter for deriving $\beta_{\rm HZCL}$ for each stellar mass
    \item[$a_{\rm moon}$] fitting parameter for deriving $\beta_{\rm env}^{\rm moon}$ for each stellar mass
    \item[$a_{\rm b}$] orbital separation between two binary stars
    \item[$a_{\rm b,p}$] periastron of a binary star
    \item[$a_p$] fitting parameter for the planet orbital period function
    \item[$a_r$] fitting parameter for the planet radius function
    \item[$Age_{\rm max}^{\rm avg}$] maximum degassing age of stagnant lid planets
    \item[$Age_{\rm max}^{\rm UL95\%CI}$] upper bound on maximum degassing age (95\% confidence)
    \item[$B_{\rm env}(\prod_{i=1}^n \beta_{\rm env}^i )$] requirement cluster on long-term environmental stability
    \item[$B_{\rm life}(\prod_{i=1}^n \beta_{\rm life}^i )$] requirement cluster on the origin and co-evolution of life
    \item[$b_{\mathrm{H_2O},i}$] fitting parameter for deriving $\beta_{\rm pc}^{H_2O}$ for each stellar mass
    \item[$b_{\rm HZCL}$] fitting parameter for deriving $\beta_{\rm HZCL}$ for each stellar mass
    \item[$b_{\rm moon}$] fitting parameter for deriving $\beta_{\rm env}^{\rm moon}$ for each stellar mass
    \item[$b_p$] fitting parameter for the planet orbital period function
    \item[$b_r$] fitting parameter for the planet radius function
    \item[$B_{\rm pc}(\prod_{i=1}^n \beta_{\rm pc}^i )$] requirement cluster on planetary compositional set-up
    \item[C/O] carbon to oxygen ratio
    \item[$d_{\rm EH,d}$] average distance between EHs in the galactic disk
    \item[$d_{\rm EH,d,1\%}$] $d_{\rm EH,d}$ for $x_{CO_2,max}=1\%$
    \item[$d_{\rm EH,d,10\%}$] $d_{\rm EH,d}$ for $x_{CO_2,max}=10\%$
    \item[$d_{\rm EH,SN}$] average minimum distance between EHs in the Solar Neighborhood
    \item[$d_{\rm EH,SN,1\%}$] $d_{\rm EH,SN}$ for $x_{CO_2,max}=1\%$
    \item[$d_{\rm EH,SN,10\%}$] $d_{\rm EH,SN}$ for $x_{CO_2,max}=10\%$
    \item[$d_{\rm HZ(CL)}$] orbital distance of the HZ(CL) boundaries
    \item[$d_{\rm pl}$] orbital distance of a planet
    \item[$d_{\rm \langle HZCL \rangle}$] mean HZCL orbital distance
    \item[$e_{\rm b}$] orbital eccentricity in stellar binary systems
    \item[$E_{\rm bio}^{\oplus}$]  total energy available ontop of Earth's atm. from its emergence to the evolution of complex life
    \item[$e_{\rm pl}$] orbital eccentricity of a planet
    \item[$f(P_{\rm pl})$] planet orbital period function
    \item[$f(R_{\rm pl})$] planet radius function
    \item[$f_{\rm b,close}$] close binary fraction
    \item[$f_{\rm bio}$] transferred object fraction containing seedable organisms
    \item[$f_{\rm civ}$]  fraction of planets with communicating ETIs
    \item[$f_{\rm EH}$] ratio between $\eta_{\rm EH}$ and $\beta_{\rm HZCL}$
    \item[$f_{\rm EH,\star}$] ratio between $\eta_{\rm EH}\times\eta_{\star}$ and $\beta_{\rm HZCL}$
    \item[$f_{\rm ETI}$] ratio between the (maximum) number of ETIs in this study vs the one in \citet{Westby2020}
    \item[$f_{\rm HZ}$] fraction of rocky exoplanets in the HZ \citep[from][]{Westby2020}
    \item[$f_{\rm in}$]  fraction of planets with intelligent life
    \item[$f_{\rm L}$] fraction of stars older than 5 Gyr \citep[from][]{Westby2020}
    \item[$f_{\rm life}$]  fraction of planets with life
    \item[$f_{\rm M}$] fraction of stars with sufficient metallicity \citep[from][]{Westby2020}
    \item[$f_{\rm pl}$]  fraction of planet hosting stars
    \item[$F_{\rm X}$] stellar X-ray surface flux at a planet's orbit
    \item[F$_{\rm X,\oplus}$] $F_{\rm X}$ at present-day Earth
    \item[$F_{\rm X,max}$] stellar X-ray surface flux threshold
    \item[$F_{\rm EUV}$] stellar EUV surface flux at a planet's orbit
    \item[$F_{\rm XUV}$] stellar XUV surface flux at a planet's orbit
    \item[$F_{\rm XUV,max}$] stellar XUV surface flux threshold
    \item[$F_{\oplus}$] absorbed solar surface flux in the Earth's atmosphere
    \item[$K$] factor accounting for Roche-lobe effects
    \item[$k_{\rm B}$] Boltzmann constant
    \item[$L$]  average lifetime of ETIs
    \item[L$_{\rm bol,\odot}$] solar bolometric luminosity
    \item[$L_{\rm bol}$] stellar bolometric luminosity
    \item[L$_{\rm X}$] stellar X-ray luminosity
    \item[$L_{\rm X,\odot}$] solar X-ray luminosity
    \item[$L_{\rm XUV}$] stellar XUV luminosity
    \item[$L_{\rm XUV,\odot}$] solar XUV luminosity
    \item[$L_{\rm EUV}$] stellar EUV luminosity
    \item[$L_{\rm EUV,\odot}$] solar EUV luminosity
    \item[$M_{\rm acc}$] accretion disk mass
    \item[$M_{\rm d}$] total stellar mass of the galactic disk
    \item[$M_{\rm d, thick}$] total stellar mass of the galactic thick disk
    \item[$M_{\rm d, thin}$] total stellar mass of the galactic thin disk
    \item[$m_{i}$] particle mass
    \item[M$_{\rm Jup}$] one jupiter mass
    \item[$M_{\rm pl}$] planetary mass
    \item[$M_{\rm pl,max}$] maximum planetary mass
    \item[$M_{\rm pl,min}$] minimum planetary mass
    \item[$M_{\rm sat}$] the entire satellite mass of a gas giant
    \item[M$_{\odot}$] Sun's mass
    \item[M$_{\oplus}$] Earth's mass
    \item[$M_{\star}$]  stellar mass
    \item[$\bar{M}_{\star}$] average stellar mass
    \item[$M_{_{\star}\rm low}$] lower stellar mass 
    \item[$M_{_{\star}\rm up}$] upper stellar mass 
    \item[M$_{\odot}$]  one solar mass
    \item[$\dot{M_{\rm en}}$] atmospheric mass-loss rate of a certain species
    \item[$N_{\rm civ}$]  number of communicating ETIs in the Galaxy
    \item[$n_{\rm e}$]  fraction of planets suitable for life
    \item[$N_{\rm EH}$]  maximum number of EHs in the galactic disk
    \item[$N_{\rm EH,1\%}$]  $N_{\rm EH}$ with atmospheric mixing ratio of $x_{CO_2,max}=1\%$ in the galactic disk
    \item[$N_{\rm EH,10\%}$]  $N_{\rm EH}$ EHs with atmospheric mixing ratio of $x_{CO_2,max}=10\%$ in the galactic disk
    \item[$N_{\rm EH,SN}$] maximum number of EHs in the Solar Neighborhood
    \item[$N_{\rm ETI}$] maximum number of extraterrestrial intelligence
    \item[$N_i$] maximum number of habitats of category $i$
    \item[$N_{\rm Hab}$] maximum number of habitable planets (of any kind of habitats)
    \item[$N_{\rm pl}$] number of planets per star over a specific range in period
    \item[$N_{\rm pl,tot}$] number of planets per star
    \item[$N_{\rm TH}$] maximum number of Titan-like habitats
    \item[$N_{\rm TH}$] maximum number of subsurface ocean worlds
    \item[$N_{\star}$]  number of stars in the galactic disk
    \item[$n_{\star}$]  stellar number density
    \item[$\tilde{N}_{\rm EH}$] hypothetical number of EHs
    \item[$\tilde{N}_{\rm EH,1\%}$] $\tilde{N}_{\rm EH}$ with $x_{CO_2,max}=1\%$
    \item[$\tilde{N}_{\rm EH,10\%}$] $\tilde{N}_{\rm EH}$ with $x_{CO_2,max}=10\%$
    \item[$P_{\rm pl}$]  orbital period of a planet
    \item[$P_{\rm pl,break}$]  break in orbital period distribution
    \item[$P_{\rm pl,max}$]  maximum orbital period of a planet
    \item[$P_{\rm pl,max}$]  minimum orbital period of a planet
    \item[$p$CO$_2$]  partial atmospheric pressure of CO$_2$
    \item[$p$N$_2$]  partial atmospheric pressure of N$_2$
    \item[$p$O$_2$]  partial atmospheric pressure of O$_2$
    \item[$\mathcal{P}(N_{\rm EH,1\%})$] probability of finding an EH with $x_{CO_2,max}=1\%$
    \item[$\mathcal{P}(N_{\rm EH,10\%})$] probability of finding an EH with $x_{CO_2,max}=10\%$
    \item[$\mathcal{P}(\rm GGP)$] probability of hosting a gas giant
    \item[$\mathcal{P}(Z_{\rm min,SN})$] probability to be above $Z_{\rm min}$ in the Solar Neighborhood
    \item[$Q_{\rm pl}$] tidal dissipation value
    \item[$r$]  radial distance from the galactic center
    \item[$R$]  Galactic star formation rate (Drake Equation)
    \item[$r_0$]  inner radius boundary for the galactic disk model
    \item[$r_1$]  outer radius boundary for the galactic disk model
    \item[$r_{\rm lb}$] radius of the lookback time sphere
    \item[$r_{\rm d}$] scale length of the galactic disk
    \item[$r_{\rm EH,SN}$] radius of the average galactic spherical volume occupied by an EH
    \item[$r_{\rm exo}$] exobase
    \item[$r_{\rm H}$] Hill radius
    \item[$R_{\rm ph}$] photospheric radius
    \item[$R_{\rm pl}$] planetary radius
    \item[$R_{\rm pl,break}$] break in planetary radius distribution
    \item[$R_{\rm pl,max}$]  maximum radius of a planet
    \item[$R_{\rm pl,min}$]  minimum radius of a planet
    \item[$R_{\rm XUV}$] atmospheric XUV absorption radius
    \item[$r_{\odot}$]  the Sun's radial distance from the galactic center
    \item[R$_{\oplus}$]  Earth's radius
    \item[$S_{\rm eff}$] stellar effective surface flux
    \item[$S_{\rm eff,max}$] stellar effective surface flux threshold
    \item[$S_{\rm eff,\odot}$] solar effective surface flux at various HZ boundaries
    \item[S$_{\rm eff,\oplus}$] solar effective surface flux at the Earth's orbit
    \item[$S_{\rm top}$] incident stellar surface flux on top of a planet's atmosphere
    \item[S$_{\rm top,\oplus}$] incident solar surface flux on top of the Earth's atmosphere
    \item[$t$]  time
    \item[$t_0$]  Hubble time, i.e., $\sim$13.8\,Gyr, and/or `today'
    \item[$t_{\rm cl,\oplus}$] typical time needed for solar-like star to evolve complex life
    \item[$t_{\rm lb}$] lookback time
    \item[$T_{\rm eff}$] stellar effective temperature
    \item[$T_{\rm exo}$] exobase temperature
    \item[$t_{\rm evo}$] time needed from Earth's emergence to the evolution of complex life
    \item[$t_{\rm p}$]  time of phase transition
    \item[$T_{\rm pl}$] planetary atmospheric mean temperature
    \item[$t_{\rm et}$] time for ejecta transfer
    \item[$t_{\rm O_2}$] oxygenation time
    \item[$t_{\rm ol}$] time of origin of life
    \item[$T_{\star}$] stellar effective temperature minus solar effective temperature
    \item[$t_{\oplus}$] solar age
    \item[$t_{\star}$] stellar age
    \item[$V_{\rm d}$]  spatial volume of the galactic disk
    \item[$w_{\rm H_2O}$] water mass fraction at the planet
    \item[$w_{\rm CO_2}$] CO$_2$ fraction at the planet
    \item[$x_{\rm CO_2}$] atmospheric CO$_2$ mixing ratio
    \item[$x_{\rm CO_2,max}$] maximum atmospheric CO$_2$ mixing ratio
    \item[$X_{\rm H_2O}$] atmospheric water mass fraction
    \item[$x_{\rm O_2}$] atmospheric O$_2$ mixing ratio
    \item[$v_{\rm imp}$] impact velocity
    \item[$V_{\rm SN}$] volume of the Solar neighborhood
    \item[$z$]  galactocentric height
    \item[$z_1$]  upper galactocentric height boundary for the galactic disk model
    \item[$z_{\rm d}$] scale height of the galactic disk
    \item[$Z_{\rm min}$] stellar metallicity cutoff
    \item[$z_{\odot}$]  the Sun's galactocentric height
    \item[$Z_{\odot}$]  solar metallicity
    \item[$Z_{\star}$]  stellar metallicity
    \item[$Z$] stellar to solar metallicity
    \item[$\alpha$]  IMF power law index
    \item[$\alpha_{\rm at}^{\rm ll}$] lower atmosphere stability limit
    \item[$\alpha_{\rm at}^{\rm ul}$] upper atmosphere stability limit
    \item[$\alpha_{\rm at}^{\rm flare}$] flare survival requirement
    \item[$\alpha_{\rm at}^{S_{\rm eff}}$] stellar surface flux limit
    \item[$\alpha_{\rm at}^{\rm sw}$] stellar wind requirement
    \item[$\alpha_{\rm at}^{\rm XUV}$] XUV threshold requirement
    \item[$\alpha_{\rm GHZ}^{\rm met}$]  metallicity requirement
    \item[$\alpha_{\rm GHZ}^{\rm SN}$]  supernova requirement
    \item[[$\alpha$/Fe]  $\alpha$-process elements to iron ratio
    \item[$\beta_{\rm env}^{\rm cycle}$] working C-Si and N-cycle requirement
    \item[$\beta_{\rm env}^{\rm moon}$] potential large moon requirement
    \item[$\beta_{\rm env}^{\rm tl}$] potential requirement of not being tidally locked
    \item[$\beta_{\rm life}^{\rm origin}$] origin of life requirement
    \item[$\beta_{\rm pc}^{\rm accr}$] accretion requirement
    \item[$\beta_{\rm pc}^{\rm C/O}$] C/O ratio requirement
    \item[$\beta_{\rm pc}^{\rm H_2O}$] water requirement
    \item[$\beta_{\rm HZCL}$] rocky planet occurrence rate in the HZCL
    \item[$\beta_{\rm HZCL}$] planet occurrence rate in HZ
    \item[$\Gamma_{\oplus}$] differential planet occurrence rate of Earth-analogs
    \item[$\delta_{\rm HZCL}$] scaling factor between HZ and HZCL
    \item[$\epsilon$] heating efficiency
    \item[$\langle\theta_{\rm i.r,1\%}$] mean rate of implemented requirements for $x_{CO_2,max}=1\%$
    \item[$\langle\theta_{\rm i.r,10\%}$] mean rate of implemented requirements for $x_{CO_2,max}=10\%$
    \item[$\langle\theta_{\rm n.r,1\%}$] mean rate of neglected requirements to meet $N_{\rm,EH,1\%}$
    \item[$\langle\theta_{\rm n.r,10\%}$] mean rate of neglected requirements to meet $N_{\rm,EH,10\%}$
    \item[$\eta_{\rm EH}$]  frequency of EHs in the HZCL
    \item[$\eta_{\rm EH,1\%}$]  $\eta_{\rm EH}$ with atmospheric mixing ratio of $x_{CO_2,max}=1\%$ in the HZCL
    \item[$\eta_{\rm EH,10\%}$]  $\eta_{\rm EH}$ with atmospheric mixing ratio of $x_{CO_2,max}=10\%$ in the HZCL
    \item[$\eta_{\rm EM}$] frequency of habitable Earth-like moons
    \item[$\eta_{\rm ETI}$] frequency of ETIs
    \item[$\eta_{\rm ETI,i}$] frequency of ETIs originating on any habitat $i$
    \item[$\eta_{i}$] frequency of habitats of category $i$
    \item[$\eta_{\rm gp}$] occurrence rate of giant planets
    \item[$\eta_{\rm gp,HZCL}$] occurrence rate of giant planets in the HZCL
    \item[$\eta_{\rm TH}$] frequency of Titan-lie habitats
    \item[$\eta_{\rm SSOW}$] frequency of subsurface ocean worlds
    \item[$\eta_{\star}$]  frequency of stars able to host EHs
    \item[$\eta_{\star,1\%}$]  $\eta_{\star}$ with atmospheric mixing ratio of $x_{CO_2,max}=1\%$
    \item[$\eta_{\star,10\%}$]  $\eta_{\star}$ with atmospheric mixing ratio of $x_{CO_2,max}=10\%$
    \item[$\eta_{\star,i}$] frequency of stars able to host habitats of category $i$
    \item[$\eta_{\star,\rm TH}$] frequency of stars able to host Titan-lie habitats
    \item[$\eta_{\star,\rm SSOW}$] frequency of stars able to host subsurface ocean worlds
    \item[$\eta_{\oplus}$]  \textit{Eta-Earth} (frequency of rocky exoplanets in the HZ of solar-like stars)
    \item[$\delta M_{\star}$] stellar mass bin
    \item[$\delta N_{\star}$] number of stars, N$_{\star}$ within stellar mass bin, $\delta M_{\star}$
    \item[$\lambda_{\rm exo}$] Jeans escape parameter
    \item[$\rho_{\rm d}$]  density distribution of the galactic disk
    \item[$\rho_{\star}$]  stellar mass density
    \item[$\Pi_{\rm bio}\left(\prod_{i=1}^n \pi_{\rm bio}^i\right)$] biological requirement cluster for ETIs to evolve
    \item[$\Pi_{\rm civ}\left(\prod_{i=1}^n \pi_{\rm civ}^i\right)$] social/technological requirement cluster for ETIs to evolve
    \item[$\Sigma_0$] central surface density of the galactic disk
    \item[$\tau_{\rm EH}$] average lifetime of an EH
    \item[$\tau_{\rm ETI}$] average lifetime of ETIs
    \item[$\tau_{\rm MS}$] stellar main-sequence lifetime
    \item[$\tau'$] average age of the stars in the Galaxy minus 5\,Gyr \citep[from][]{Westby2020}
    \item[$\xi_0$] IMF normalization constant
    \item[$\zeta_{\oplus}$] planet frequency with lower and higher boundary conditions being within 20\% of the Earth's values
    \item[$\Omega_{\rm pl,0}$] initial rotation rate of a planet
    \item[[Fe/H]] a star's Fe to H ratio relative to the Sun
    \item[[Fe/H]$_{\rm crit}$] critical metallicity threshold for forming rocky planets
    \item[$\beta_{\mathrm{HZCL},Z_{\rm min}}$] mean occurrence rate, $\beta_{\rm HZCL}$, over the entire stellar mass range
    \item[$(\eta_{\star}\eta_{\rm EH})_{1\%}$] $\eta_{\star}\times\eta_{\rm EH})$ for $x_{CO_2,max}=1\%$
    \item[$(\eta_{\star}\eta_{\rm EH})_{10\%}$] $\eta_{\star}\times\eta_{\rm EH})$ for $x_{CO_2,max}=10\%$
    \item[$(\eta_{\star}\eta_{\rm EH})_{\rm SN,1\%}$] $\eta_{\star}\times\eta_{\rm EH})$ in the Solar Neighborhood for $x_{CO_2,max}=1\%$
    \item[$(\eta_{\star}\eta_{\rm EH})_{\rm SN,10\%}$] $\eta_{\star}\times\eta_{\rm EH})$ in the Solar Neighborhood for $x_{CO_2,max}=10\%$
    \item[$\beta_{\mathrm{HZCL},Z_{\rm min}}$] metallicity weighted rocky planet occurrence rate in the HZCL
\end{itemize}

\end{appendix}

\section*{Acknowledgments}

MS acknowledges the FWF project FWF-ESPRIT $\mathrm{D-1522P33620}$ and also thanks Daria Kubyshkina for fruitful discussions. {We also thank four anonymous referees for their thoughtful and valuable comments and suggestions that helped to improve the article significantly.}


\bibliographystyle{mnras}
\bibliography{refs_ee}
\label{lastpage}

\end{document}